\documentclass[12pt]{article}
\usepackage{jheppub}

\pdfoutput=1

\usepackage{amsmath,bbm,array,amsfonts,graphicx,wrapfig,lscape,float,mathtools,multirow,longtable}
\usepackage[dvipsnames]{xcolor}
\usepackage{array}

\newcommand{\be}{\begin{equation}}
\newcommand{\ee}{\end{equation}}
\newcommand{\beq}{\begin{equation}}
\newcommand{\beql}[1]{\begin{equation}\label{#1}}
\newcommand{\eeq}{\end{equation}}
\newcommand{\ba}{\begin{array}}
\newcommand{\ea}{\end{array}}
\newcommand{\bea}{\begin{eqnarray}}
\newcommand{\beal}[1]{\begin{eqnarray}\label{#1}}
\newcommand{\eea}{\end{eqnarray}}
\newcommand{\ben}{\begin{enumerate}}
\newcommand{\een}{\end{enumerate}}
\newcommand{\bean}{\begin{eqnarray*}}
\newcommand{\eean}{\end{eqnarray*}}
\newcommand{\eref}[1]{(\ref{#1})}
\newcommand{\sref}[1]{\S\ref{#1}}
\newcommand{\tref}[1]{Table~\ref{#1}}
\newcommand{\nn}{\nonumber}

\newcommand{\fref}[1]{Figure \ref{#1}}
\newcommand{\btab}[1]{\begin{tabular}{#1}}
\newcommand{\etab}{\end{tabular}}

\newcommand{\comment}[1]{}

\newcommand{\ud}{\mathrm{d}}

\newcommand{\qed}{\nobreak \ifvmode \relax \else
      \ifdim\lastskip<1.5em \hskip-\lastskip
      \hskip1.5em plus0em minus0.5em \fi \nobreak
      \vrule height0.75em width0.5em depth0.25em\fi}

\newcommand{\Tr}{\text{Tr}}

\definecolor{darkspringgreen}{rgb}{0.09, 0.45, 0.27}
\definecolor{forestgreen}{rgb}{0.13, 0.55, 0.13}

\usepackage{array}
\usepackage{physics}
\usepackage{subcaption}
\usepackage{tikz,tikz-3dplot}
\usepackage[colorlinks=true]{hyperref}
\newcolumntype{C}[1]{>{\centering\let\newline\\\arraybackslash\hspace{0pt}}m{#1}}

\definecolor{yellow2}{rgb}{0.98, 0.80, 0.20}

\title{Brane Tilings and \\ Polygons with Two Interior Points} 

\author{Rak-Kyeong Seong}

\affiliation{
Department of Mathematical Sciences, and 
Department of Physics,\\ 
Ulsan National Institute of Science and Technology,\\
50 UNIST-gil, Ulsan 44919, South Korea
}
\emailAdd{seong@unist.ac.kr}

\preprint{
\begin{flushright}
UNIST-MTH-26-RS-07\\
\end{flushright}
}

\abstract{
We fully classify the brane tilings
and the $4d$ $\mathcal{N}=1$ supersymmetric quiver gauge theories
corresponding to toric Calabi-Yau $3$-folds
whose toric diagrams are convex lattice polygons with two interior lattice points.
There are $45$ such polygons up to $GL(2,\mathbb{Z})$ transformations
and lattice translations: $5$ triangles, $19$ quadrilaterals,
$16$ pentagons and $5$ hexagons.
We identify all $323$ toric phases corresponding to these toric Calabi-Yau $3$-folds.
For every toric phase, we present the brane tiling with its quiver and superpotential,
the perfect matching matrix and the $F$- and $D$-term charge matrices.
For each brane tiling, we study the mesonic moduli space of the corresponding
abelian $4d$ $\mathcal{N}=1$ theory through its Hilbert series and plethystic logarithm,
and identify its generators and their leading relations.
We also identify the mesonic global symmetry for each brane tiling
and determine the superconformal $U(1)_R$ charges.
}

\begin{document}

\maketitle

\section{Introduction}\label{s_intro}

The worldvolume theory of a stack of D3-branes probing a singular non-compact Calabi-Yau 3-fold is 
a $4d$ $\mathcal{N}=1$ supersymmetric gauge theory. 
Its matter content is encoded in a graph known as the \textit{quiver} and its interactions in a superpotential.
For an interesting subset of these $4d$ $\mathcal{N}=1$ theories, 
we can restrict the non-compact Calabi-Yau 3-fold to be toric, 
where the geometry is encoded entirely by a convex lattice polygon on $\mathbb{Z}^2$ known as the \textit{toric diagram} \cite{Leung:1997tw}. 
The large family of $4d$ $\mathcal{N}=1$ supersymmetric gauge theories
associated with toric Calabi-Yau 3-folds
is represented by periodic bipartite graphs on a 2-torus $T^2$ 
known as \textit{brane tilings} or \textit{dimer models} \cite{Hanany:2005ve,Franco:2005rj,Franco:2005sm,Hanany:2005ss,Hanany:2006nm,Kennaway:2007tq}.
Every consistent brane tiling encodes the Lagrangian of the corresponding $4d$ $\mathcal{N}=1$ theory
 \cite{Hanany:2005ss,Hanany:2006nm,Kennaway:2007tq}.

Convex lattice polygons on $\mathbb{Z}^2$
with a single interior point are known as \textit{reflexive polygons} and
form a unique family of toric Calabi-Yau 3-folds.
Given that two convex lattice polygons related by a $GL(2,\mathbb{Z})$ transformation are considered to be equivalent by corresponding to the same toric Calabi-Yau 3-fold,
there are precisely $16$ reflexive polygons \cite{Kreuzer:1995cd,Kreuzer:1998vb,Kreuzer:2000qv,Kreuzer:2000xy}.
These reflexive polygons appear in string theory through \textit{mirror symmetry} \cite{Candelas:1989hd,morrison-1993-6,Batyrev:1993oya,Batyrev:1994ju,Batyrev:1996hiq,cox1999mirror,Hori:2003ic} 
and correspond to toric Calabi-Yau 3-folds that are the complex cones over the Gorenstein toric Fano surfaces, 
among them the del Pezzo surfaces, whose associated $4d$ $\mathcal{N}=1$ theories have been studied extensively \cite{Feng:2000mi,Feng:2001xr,Feng:2002fv,Feng:2002zw,Davey:2011mz}. 
In \cite{Hanany:2012hi}, all $30$ brane tilings  associated to the $16$ reflexive polygons
were identified with their corresponding $4d$ $\mathcal{N}=1$ theories.
Due to \textit{Seiberg duality} \cite{Seiberg:1994pq,Feng:2001bn,Beasley:2001zp},
certain toric Calabi-Yau 3-folds have more than one associated brane tiling, also referred to as \textit{toric phases} \cite{Feng:2000mi,Feng:2001xr,Feng:2001bn, Feng:2002zw,Franco:2003ea}.

Reflexive polygons come in pairs under polar duality \cite{Batyrev:1993oya, Batyrev:1996hiq, Kreuzer:1995cd}.
Even though polar dual polygons exist not just for reflexive polygons, 
only for reflexive polygons the polar dual is again a convex lattice polygon on $\mathbb{Z}^2$.
For an associated brane tiling, the \textit{mesonic moduli space} \cite{Butti:2007jv,Feng:2007ur,Benvenuti:2006qr} of the abelian $4d$ $\mathcal{N}=1$ theory with $U(1)$ gauge groups is the toric Calabi-Yau 3-fold itself.
When the toric diagram is reflexive, the convex hull of the mesonic flavor-charge vectors of its generators, expressed in a suitable lattice basis, is the polar dual polygon \cite{Hanany:2012hi}.
Brane tilings associated with reflexive toric diagrams can also be related by \textit{specular duality} \cite{Hanany:2012vc}
and have the same \textit{master spaces} \cite{Forcella:2008bb,Forcella:2008eh}.

The following work takes the next step and considers convex lattice polygons in $\mathbb{Z}^2$ that have exactly two interior lattice points.
Up to $GL(2,\mathbb{Z})$ transformations and lattice translations, there are exactly $45$ such polygons \cite{wei2012lattice}.
They correspond to a family of $45$ toric Calabi-Yau 3-folds whose toric diagrams have at most six extremal vertices.
These $45$ toric diagrams consist of $5$ triangles, $19$ quadrilaterals, $16$ pentagons and $5$ hexagons in $\mathbb{Z}^2$ as summarized in \tref{t_polycount}.

In order to identify the brane tilings corresponding to the $45$ convex lattice polygons with two interior lattice points, we recall that partial resolution of a toric Calabi-Yau 3-fold corresponds to the
\textit{Higgs mechanism} in the associated $4d$ $\mathcal{N}=1$ theory \cite{Feng:2002fv}.
In principle, by starting with brane tilings associated with abelian orbifolds of $\mathbb{C}^3$ \cite{Hanany:2010cx,Davey:2010px,Hanany:2010ne,Davey:2011dd} 
whose toric diagrams contain the $45$ polygons, one can remove extremal vertices of the toric diagrams via Higgsing to obtain brane tilings associated with toric diagrams of any convex polygonal shape. 
This technique was used to obtain the $30$ brane tilings corresponding to the $16$ reflexive polygons in \cite{Hanany:2012hi}.
Since toric Seiberg duality preserves the toric diagram, we generate the toric phases for each of the $45$ polygons by repeatedly applying Seiberg duality to quadrilateral faces of the brane tilings obtained via Higgsing,
integrating out massive fields and identifying equivalent brane tilings, until no new toric phases appear.

The result that we present in this work is the complete classification
of all $323$ toric phases associated with the $45$ convex lattice polygons
with two interior lattice points.\footnote{
Our classification was completed in 2017, but never published. 
The results have only recently been organized into the present manuscript with the assistance of AI tools.
Our work completes the partial results in \cite{Bao:2020kji}. 
}
When we count the distinct toric phases in our classification, we have $5$ toric phases for the triangles, 
$57$ for the quadrilaterals, 
$137$ for the pentagons and 
$124$ for the hexagons. 
Many of the toric Calabi-Yau 3-folds that have toric diagrams with two internal points are familiar Calabi-Yau geometries,
among them the family of $Y^{3,q}$ \cite{Benvenuti:2004dy,Benvenuti:2004wx}, members of $X^{p,q}$ \cite{Hanany:2005hq}, 
of $Z^{p,q}$ \cite{Oota:2006eg} and of $L^{a,b,c}$ \cite{Franco:2005sm}, 
and the abelian orbifolds of $\mathbb{C}^3$ \cite{Hanany:2010cx,Davey:2010px,Hanany:2010ne,Davey:2011dd}.
Given that we also discover the brane tilings associated with toric Calabi-Yau 3-folds that are part of no previously known family of geometries,
we introduce new families and naming conventions of toric Calabi-Yau 3-folds whose toric diagrams are convex lattice polygons with 2 internal points.

Each toric phase in our classification is presented with its brane tiling, 
including the quiver and superpotential of the associated $4d$ $\mathcal{N}=1$ theory,
the perfect matching matrix $P$, and the $F$- and $D$-term charge matrices. 
By taking the abelian $4d$ $\mathcal{N}=1$ theory with $U(1)$ gauge groups corresponding to each of the $323$ brane tilings in our classification,
we study the corresponding mesonic moduli spaces and toric Calabi-Yau 3-folds
by computing their \textit{Hilbert series} \cite{Benvenuti:2006qr,Hanany:2006uc,Feng:2007ur,Butti:2007jv,Forcella:2009vw}.
The plethystic logarithm \cite{Benvenuti:2006qr,Feng:2007ur,Hanany:2007zz} of the Hilbert series provides information about the generators of the mesonic moduli space and the leading relations among them.
For each toric phase and corresponding brane tiling,  we express the generators of the mesonic moduli space as gauge invariant products of bifundamental chiral fields of the $4d$ $\mathcal{N}=1$ theories.
We also present the generators with their global symmetry charges, including their mesonic flavor charges and 
superconformal $U(1)_R$ charges, which we obtain via $a$-maximization \cite{Intriligator:2003jj,Butti:2005vn,Butti:2005ps}.
We note that some of the $45$ toric Calabi-Yau 3-folds permit one of the $U(1)$ mesonic flavor factors to enhance to an $SU(2)$ mesonic flavor symmetry.
However, we see no further enhancement involving a mesonic flavor symmetry with two $SU(2)$ factors or an $SU(3)$ factor.
\\

The brane tilings presented in this work can also be explored through the accompanying online database at:
\begin{center}
\url{https://rakkyeongseong.github.io/branetilings/}.
\end{center}
An archived version of the accompanying data is deposited on Zenodo \cite{Zenodo:24398417}.
\\

\begin{figure}[H]
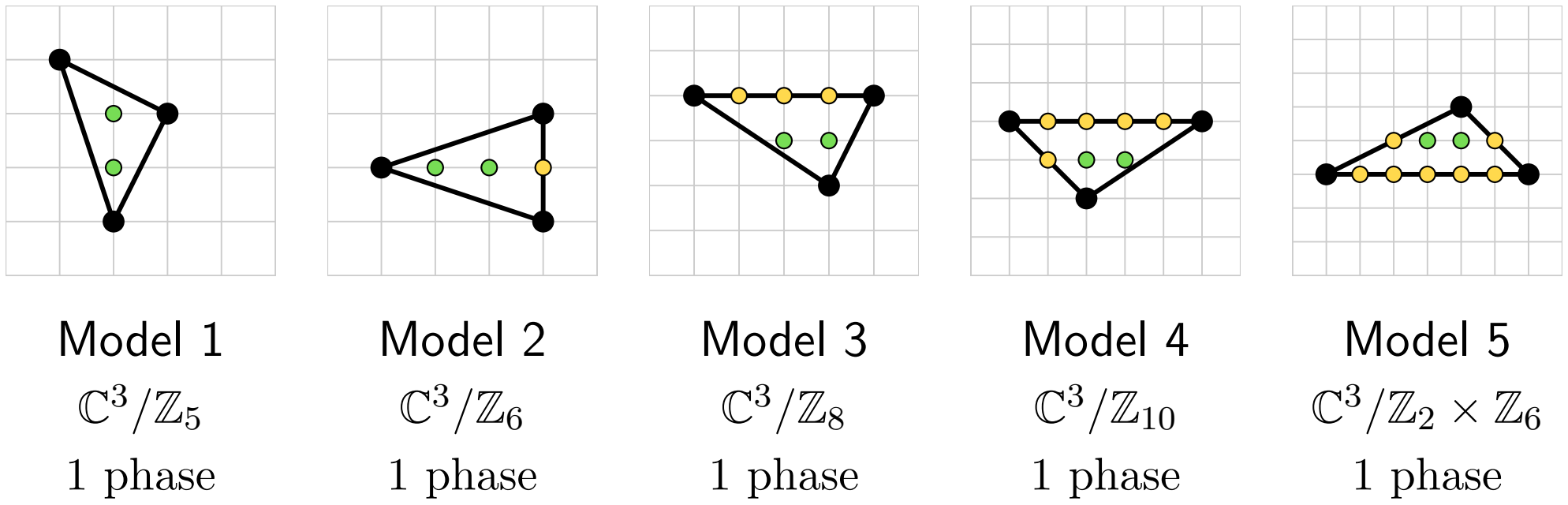

\centering
\hypersetup{linkcolor=black}
\setlength{\tabcolsep}{3pt}
\begin{tabular}{ccccc}
\includegraphics[width=0.16\textwidth]{figures/sum_01.pdf} & \includegraphics[width=0.16\textwidth]{figures/sum_02.pdf} & \includegraphics[width=0.16\textwidth]{figures/sum_03.pdf} & \includegraphics[width=0.16\textwidth]{figures/sum_04.pdf} & \includegraphics[width=0.16\textwidth]{figures/sum_05.pdf} \\
\hyperref[s_m1]{\textsf{Model 1}} & \hyperref[s_m2]{\textsf{Model 2}} & \hyperref[s_m3]{\textsf{Model 3}} & \hyperref[s_m4]{\textsf{Model 4}} & \hyperref[s_m5]{\textsf{Model 5}} \\
{\small \hyperref[s_m1]{$\mathbb{C}^3/\mathbb{Z}_{5}$}} & {\small \hyperref[s_m2]{$\mathbb{C}^3/\mathbb{Z}_{6}$}} & {\small \hyperref[s_m3]{$\mathbb{C}^3/\mathbb{Z}_{8}$}} & {\small \hyperref[s_m4]{$\mathbb{C}^3/\mathbb{Z}_{10}$}} & {\small \hyperref[s_m5]{$\mathbb{C}^3/\mathbb{Z}_{2} \times \mathbb{Z}_{6}$}} \\
{\small 1 phase} & {\small 1 phase} & {\small 1 phase} & {\small 1 phase} & {\small 1 phase} \\
\end{tabular}
\caption{The 5 triangles with two interior points. The model number and the number of toric phases are shown under each toric diagram.}
\label{f_sum_c3}
\end{figure}

\begin{figure}[H]
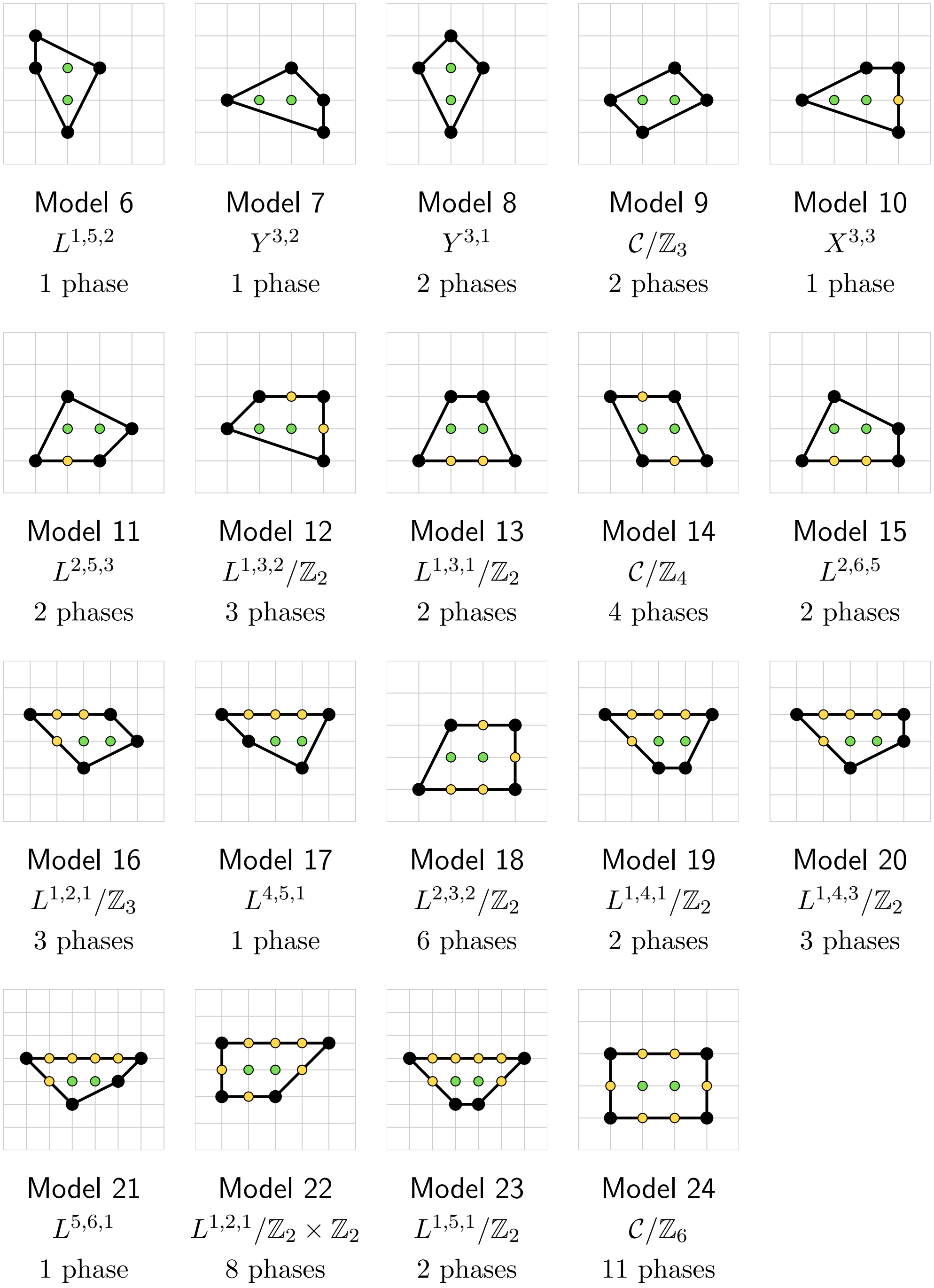

\centering
\hypersetup{linkcolor=black}
\setlength{\tabcolsep}{3pt}
\begin{tabular}{ccccc}
  \includegraphics[width=0.16\textwidth]{figures/sum_06.pdf} & \includegraphics[width=0.16\textwidth]{figures/sum_07.pdf} & \includegraphics[width=0.16\textwidth]{figures/sum_08.pdf} & \includegraphics[width=0.16\textwidth]{figures/sum_09.pdf} & \includegraphics[width=0.16\textwidth]{figures/sum_10.pdf} \\
  \hyperref[s_m6]{\textsf{Model 6}} & \hyperref[s_m7]{\textsf{Model 7}} & \hyperref[s_m8]{\textsf{Model 8}} & \hyperref[s_m9]{\textsf{Model 9}} & \hyperref[s_m10]{\textsf{Model 10}} \\
  {\small \hyperref[s_m6]{$L^{1,5,2}$}} & {\small \hyperref[s_m7]{$Y^{3,2}$}} & {\small \hyperref[s_m8]{$Y^{3,1}$}} & {\small \hyperref[s_m9]{$\mathcal{C}/\mathbb{Z}_{3}$}} & {\small \hyperref[s_m10]{$X^{3,3}$}} \\
  {\small 1 phase} & {\small 1 phase} & {\small 2 phases} & {\small 2 phases} & {\small 1 phase} \\
  \\[-1.5ex]
  \includegraphics[width=0.16\textwidth]{figures/sum_11.pdf} & \includegraphics[width=0.16\textwidth]{figures/sum_12.pdf} & \includegraphics[width=0.16\textwidth]{figures/sum_13.pdf} & \includegraphics[width=0.16\textwidth]{figures/sum_14.pdf} & \includegraphics[width=0.16\textwidth]{figures/sum_15.pdf} \\
  \hyperref[s_m11]{\textsf{Model 11}} & \hyperref[s_m12]{\textsf{Model 12}} & \hyperref[s_m13]{\textsf{Model 13}} & \hyperref[s_m14]{\textsf{Model 14}} & \hyperref[s_m15]{\textsf{Model 15}} \\
  {\small \hyperref[s_m11]{$L^{2,5,3}$}} & {\small \hyperref[s_m12]{$L^{1,3,2}/\mathbb{Z}_{2}$}} & {\small \hyperref[s_m13]{$L^{1,3,1}/\mathbb{Z}_{2}$}} & {\small \hyperref[s_m14]{$\mathcal{C}/\mathbb{Z}_{4}$}} & {\small \hyperref[s_m15]{$L^{2,6,5}$}} \\
  {\small 2 phases} & {\small 3 phases} & {\small 2 phases} & {\small 4 phases} & {\small 2 phases} \\
  \\[-1.5ex]
  \includegraphics[width=0.16\textwidth]{figures/sum_16.pdf} & \includegraphics[width=0.16\textwidth]{figures/sum_17.pdf} & \includegraphics[width=0.16\textwidth]{figures/sum_18.pdf} & \includegraphics[width=0.16\textwidth]{figures/sum_19.pdf} & \includegraphics[width=0.16\textwidth]{figures/sum_20.pdf} \\
  \hyperref[s_m16]{\textsf{Model 16}} & \hyperref[s_m17]{\textsf{Model 17}} & \hyperref[s_m18]{\textsf{Model 18}} & \hyperref[s_m19]{\textsf{Model 19}} & \hyperref[s_m20]{\textsf{Model 20}} \\
  {\small \hyperref[s_m16]{$L^{1,2,1}/\mathbb{Z}_{3}$}} & {\small \hyperref[s_m17]{$L^{4,5,1}$}} & {\small \hyperref[s_m18]{$L^{2,3,2}/\mathbb{Z}_{2}$}} & {\small \hyperref[s_m19]{$L^{1,4,1}/\mathbb{Z}_{2}$}} & {\small \hyperref[s_m20]{$L^{1,4,3}/\mathbb{Z}_{2}$}} \\
  {\small 3 phases} & {\small 1 phase} & {\small 6 phases} & {\small 2 phases} & {\small 3 phases} \\
  \\[-1.5ex]
  \includegraphics[width=0.16\textwidth]{figures/sum_21.pdf} & \includegraphics[width=0.16\textwidth]{figures/sum_22.pdf} & \includegraphics[width=0.16\textwidth]{figures/sum_23.pdf} & \includegraphics[width=0.16\textwidth]{figures/sum_24.pdf} &  \\
  \hyperref[s_m21]{\textsf{Model 21}} & \hyperref[s_m22]{\textsf{Model 22}} & \hyperref[s_m23]{\textsf{Model 23}} & \hyperref[s_m24]{\textsf{Model 24}} &  \\
  {\small \hyperref[s_m21]{$L^{5,6,1}$}} & {\small \hyperref[s_m22]{$L^{1,2,1}/\mathbb{Z}_{2} \times \mathbb{Z}_{2}$}} & {\small \hyperref[s_m23]{$L^{1,5,1}/\mathbb{Z}_{2}$}} & {\small \hyperref[s_m24]{$\mathcal{C}/\mathbb{Z}_{6}$}} &  \\
  {\small 1 phase} & {\small 8 phases} & {\small 2 phases} & {\small 11 phases} &  \\
\end{tabular}
\caption{The 19 quadrilaterals with two interior points. The model number and the number of toric phases are shown under each toric diagram.}
\label{f_sum_c4}
\end{figure}

\begin{figure}[H]
\centering
\hypersetup{linkcolor=black}
\setlength{\tabcolsep}{3pt}
\begin{tabular}{ccccc}
  \includegraphics[width=0.16\textwidth]{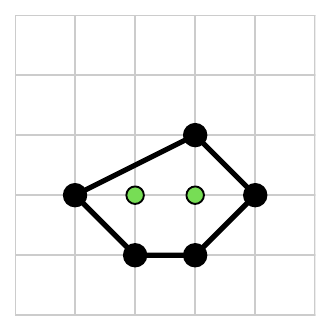} & \includegraphics[width=0.16\textwidth]{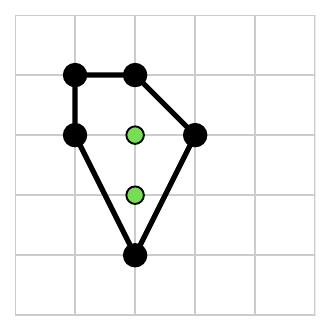} & \includegraphics[width=0.16\textwidth]{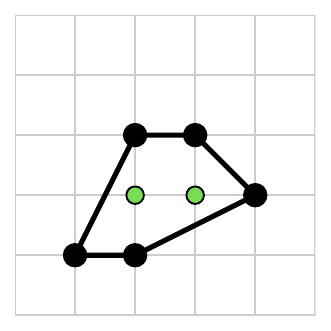} & \includegraphics[width=0.16\textwidth]{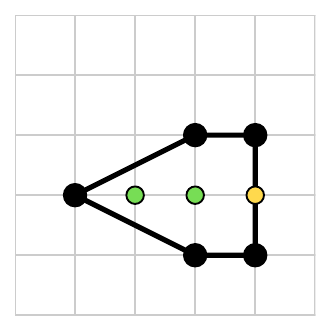} & \includegraphics[width=0.16\textwidth]{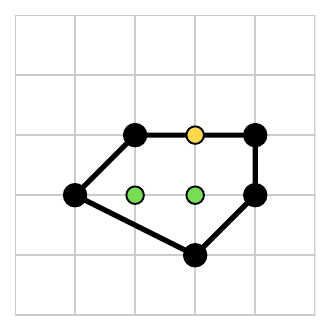} \\
  \hyperref[s_m25]{\textsf{Model 25}} & \hyperref[s_m26]{\textsf{Model 26}} & \hyperref[s_m27]{\textsf{Model 27}} & \hyperref[s_m28]{\textsf{Model 28}} & \hyperref[s_m29]{\textsf{Model 29}} \\
  {\small \hyperref[s_m25]{$X^{3,1}$}} & {\small \hyperref[s_m26]{$X^{3,2}$}} & {\small \hyperref[s_m27]{$H^{1,3,1,1}$}} & {\small \hyperref[s_m28]{$H^{1,4,1,1}$}} & {\small \hyperref[s_m29]{$H^{2,2,1,1}$}} \\
  {\small 5 phases} & {\small 3 phases} & {\small 3 phases} & {\small 4 phases} & {\small 8 phases} \\
  \\[-1.5ex]
  \includegraphics[width=0.16\textwidth]{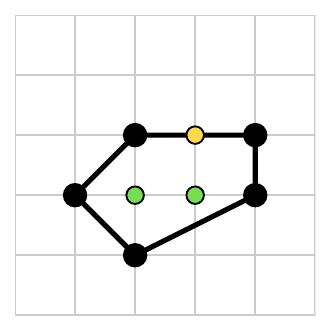} & \includegraphics[width=0.16\textwidth]{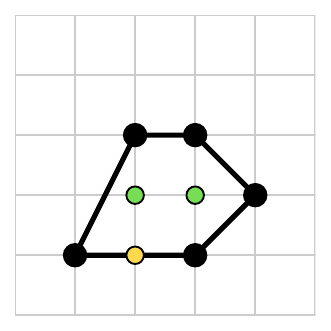} & \includegraphics[width=0.16\textwidth]{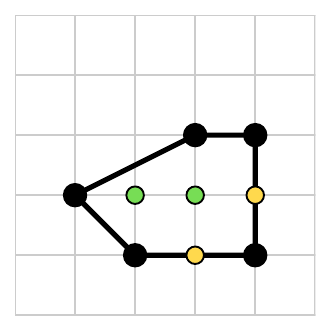} & \includegraphics[width=0.16\textwidth]{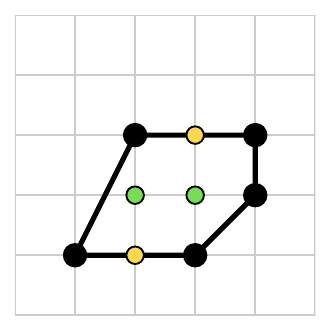} & \includegraphics[width=0.16\textwidth]{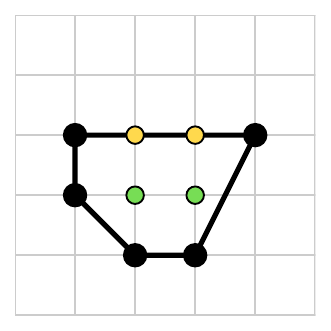} \\
  \hyperref[s_m30]{\textsf{Model 30}} & \hyperref[s_m31]{\textsf{Model 31}} & \hyperref[s_m32]{\textsf{Model 32}} & \hyperref[s_m33]{\textsf{Model 33}} & \hyperref[s_m34]{\textsf{Model 34}} \\
  {\small \hyperref[s_m30]{$J^{1,4,3,2}$}} & {\small \hyperref[s_m31]{$G^{4,2}$}} & {\small \hyperref[s_m32]{$H^{1,3,1,2}$}} & {\small \hyperref[s_m33]{$J^{4,3,1,1}$}} & {\small \hyperref[s_m34]{$G^{4,3}$}} \\
  {\small 8 phases} & {\small 7 phases} & {\small 10 phases} & {\small 9 phases} & {\small 5 phases} \\
  \\[-1.5ex]
  \includegraphics[width=0.16\textwidth]{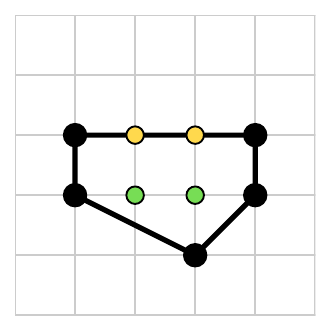} & \includegraphics[width=0.16\textwidth]{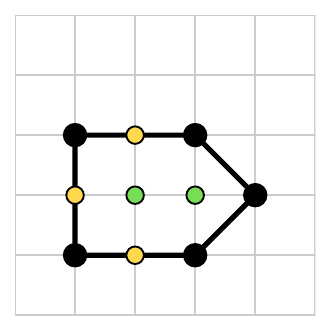} & \includegraphics[width=0.16\textwidth]{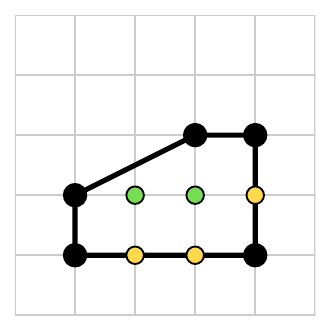} & \includegraphics[width=0.16\textwidth]{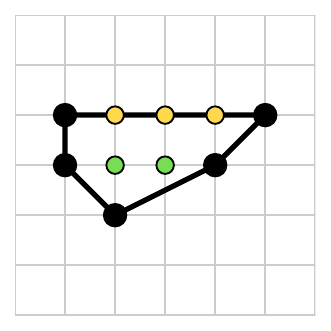} & \includegraphics[width=0.16\textwidth]{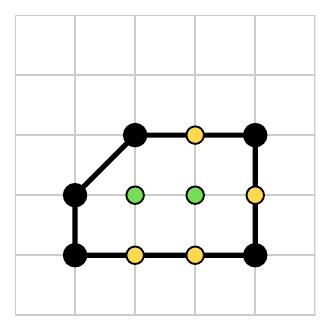} \\
  \hyperref[s_m35]{\textsf{Model 35}} & \hyperref[s_m36]{\textsf{Model 36}} & \hyperref[s_m37]{\textsf{Model 37}} & \hyperref[s_m38]{\textsf{Model 38}} & \hyperref[s_m39]{\textsf{Model 39}} \\
  {\small \hyperref[s_m35]{$H^{2,3,1,1}$}} & {\small \hyperref[s_m36]{$X^{2,1}/\mathbb{Z}_{2}$}} & {\small \hyperref[s_m37]{$G^{5,3}$}} & {\small \hyperref[s_m38]{$J^{1,5,3,4}$}} & {\small \hyperref[s_m39]{$H^{2,2,1,2}$}} \\
  {\small 8 phases} & {\small 19 phases} & {\small 12 phases} & {\small 4 phases} & {\small 25 phases} \\
  \\[-1.5ex]
  \includegraphics[width=0.16\textwidth]{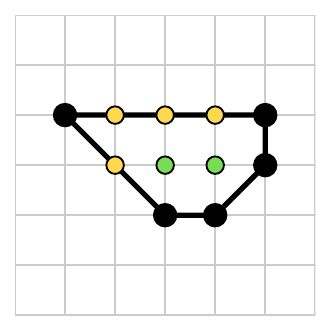} &  &  &  &  \\
  \hyperref[s_m40]{\textsf{Model 40}} &  &  &  &  \\
  {\small \hyperref[s_m40]{$G^{5,4}$}} &  &  &  &  \\
  {\small 7 phases} &  &  &  &  \\
\end{tabular}
\caption{The 16 pentagons with two interior points. The model number and the number of toric phases are shown under each toric diagram.}
\label{f_sum_c5}
\end{figure}

\begin{figure}[H]
\centering
\hypersetup{linkcolor=black}
\setlength{\tabcolsep}{3pt}
\begin{tabular}{ccccc}
  \includegraphics[width=0.16\textwidth]{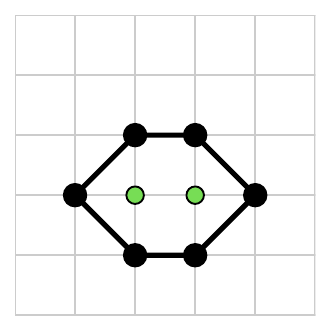} & \includegraphics[width=0.16\textwidth]{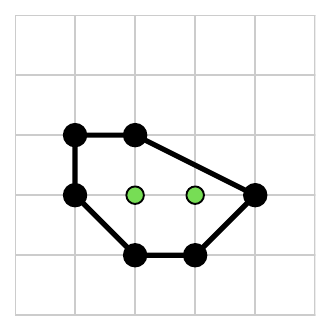} & \includegraphics[width=0.16\textwidth]{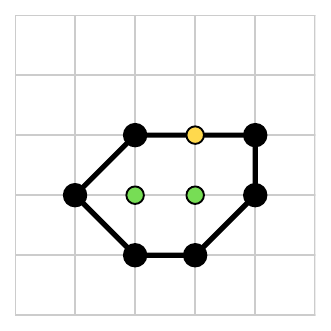} & \includegraphics[width=0.16\textwidth]{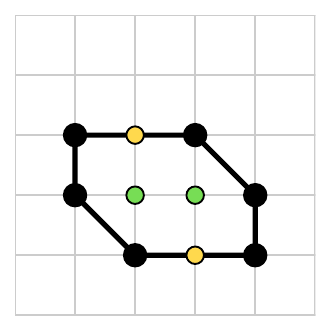} & \includegraphics[width=0.16\textwidth]{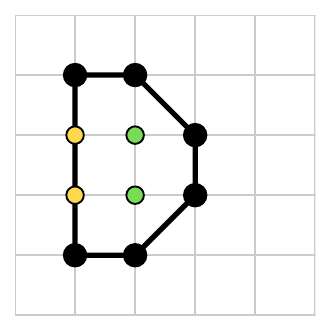} \\
  \hyperref[s_m41]{\textsf{Model 41}} & \hyperref[s_m42]{\textsf{Model 42}} & \hyperref[s_m43]{\textsf{Model 43}} & \hyperref[s_m44]{\textsf{Model 44}} & \hyperref[s_m45]{\textsf{Model 45}} \\
  {\small \hyperref[s_m41]{$Z^{3,1}$}} & {\small \hyperref[s_m42]{$Z^{3,2}$}} & {\small \hyperref[s_m43]{$H^{1,1,1,2}$}} & {\small \hyperref[s_m44]{$N^{1,3,3,2}$}} & {\small \hyperref[s_m45]{$N^{1,2,3,3}$}} \\
  {\small 13 phases} & {\small 13 phases} & {\small 42 phases} & {\small 33 phases} & {\small 23 phases} \\
\end{tabular}
\caption{The 5 hexagons with two interior points. The model number and the number of toric phases are shown under each toric diagram.}
\label{f_sum_c6}
\end{figure}

The work is structured as follows. 
Section \sref{s_back} reviews the background and the motivation
for this work, including the set of $45$ convex lattice polygons with two internal points in section \sref{s_poly}, 
the brane tiling and the forward algorithm in section \sref{s_tiling}, 
the mesonic moduli space and its Hilbert series in section \sref{s_hilbert}, 
the global symmetry and the superconformal $U(1)_R$ charges in section \sref{s_symmetry}, 
the families of toric Calabi-Yau $3$-folds after which the models are named in section \sref{s_families},
and the conditions for equivalence between brane tilings in section \sref{s_conj}. 
Sections \sref{s_c3} to \sref{s_c6} then contain the complete classification, 
where each section opens with a summary figure of the toric diagrams with 2 internal points
\fref{f_sum_c3} to \fref{f_sum_c6}, 
with the model number of each toric Calabi-Yau 3-fold and its number of toric phases. 
We conclude with a discussion section and a summary of future research directions.
\section{Background and Motivation \label{s_back}}

The $4d$ $\mathcal{N}=1$ supersymmetric gauge theories 
classified in this work are the worldvolume theories of 
D3-branes probing toric Calabi-Yau 3-folds 
whose toric diagram is a convex lattice polygon with exactly two interior points. 
There are $45$ such polygons up to $GL(2,\mathbb{Z})$ equivalence 
and the classification in the following sections presents the $323$ brane tilings associated with these polygons,
including distinct toric phases corresponding to the same toric Calabi-Yau 3-fold by Seiberg duality \cite{Feng:2000mi,Feng:2001xr,Feng:2002zw,Seiberg:1994pq,Feng:2001bn,Beasley:2001zp,Franco:2003ea}.
The following sections review the background material which the classification relies on. 
\\

\subsection{Lattice Polygons with Two Interior Points \label{s_poly}}

\paragraph{Toric Calabi-Yau 3-folds and Lattice Polygons.} 
The worldvolume theories on D3-branes probing toric Calabi-Yau 3-folds
form a large family of $4d$ $\mathcal{N}=1$ theories
realized by brane tilings.
When we have a single probe D3-brane 
and a $4d$ $\mathcal{N}=1$ theory given by a brane tiling with only $U(1)$ gauge groups, 
then the mesonic moduli space $\mathcal{M}^{mes}$ is exactly the probed toric Calabi-Yau 3-fold.
For a stack of $N$ D3-branes,
the mesonic moduli space is the $N$-th symmetric product of $\mathcal{M}^{mes}$.

The combinatorial object which encodes the geometry 
of a toric Calabi-Yau 3-fold
is a convex lattice polygon known as the toric diagram \cite{Leung:1997tw}. 
A \textit{convex lattice polygon} is the convex hull of a finite collection of points on the lattice $\mathbb{Z}^2$,
\beal{es01a01}
\Delta = \text{Conv}\{w_1,\dots,w_k\} \subset \mathbb{R}^2 ~,~ w_m \in \mathbb{Z}^2 ~,~
\eea
and throughout this work the toric diagram $\Delta$ is taken to be $2$-dimensional. 
Let us summarize a collection of properties of $\Delta$, 
which we are going to make use of in this work:
\begin{itemize}
\item The \textit{lattice points} of $\Delta$ are the elements of $\Delta \cap \mathbb{Z}^2$. 
They split into the points in the interior of the polygon and the points on its perimeter, with
\beal{es01a02}
I = | \text{int}(\Delta) \cap \mathbb{Z}^2 | ~,~
B = | \partial \Delta \cap \mathbb{Z}^2 | ~,~
n = | \Delta \cap \mathbb{Z}^2 | = I + B ~.~
\eea
The $I$ points of $\text{int}(\Delta) \cap \mathbb{Z}^2$ are called \textit{interior} or internal points, 
and the $B$ points of $\partial \Delta \cap \mathbb{Z}^2$ are called \textit{external} points.

\item The \textit{extremal} points of $\Delta$ are the corners of the convex lattice polygon.
They are denoted by,
\beal{es01a03}
v_a \in \mathbb{Z}^2 ~,~ a=1,\dots,n_e ~,~
\eea
where $n_e$ is the number of extremal points, which is also called the order of $\Delta$. 
Every lattice point of $\Delta$ is of exactly one of three types: extremal, non-extremal external, or interior. 
Note that all extremal points are external, and that $n_e \leq B$.

\item Two toric diagrams are equivalent when they are related by a $GL(2,\mathbb{Z})$ transformation combined with a possible lattice translation as follows,
\beal{es01a04}
\Delta \sim \Delta' ~~\Leftrightarrow~~ \Delta' = M \cdot \Delta + u ~,~ M \in GL(2,\mathbb{Z}) ~,~ u \in \mathbb{Z}^2 ~.~
\eea
Following the convention in the brane tiling literature, 
the resulting equivalence is referred to simply as $GL(2,\mathbb{Z})$ equivalence. 
The integers $I$, $B$, $n$ and $n_e$, as well as the area $A$ of the toric diagram, 
are invariants of \eref{es01a04}.
Two $GL(2,\mathbb{Z})$ equivalent toric diagrams are known to correspond to the same toric Calabi-Yau 3-fold.

\end{itemize}

\paragraph{Normalized Area and Pick's Theorem.} 
The area of a toric diagram 
is measured in units in which the smallest lattice triangle, 
the triangle with three external and no interior points, 
has area $1$. 
This \textit{normalized area} 
is twice the Euclidean area and is always a positive integer. 
By Pick's theorem, 
the Euclidean area $A$ of a convex lattice polygon is fixed by its lattice point count as follows,
\beal{es01a05}
A = I + \frac{B}{2} - 1 ~,~
\eea
so that by \eref{es01a05} the normalized area is given by,
\beal{es01a10}
G \equiv 2 A = 2 I + B - 2 ~.~
\eea
Here, we note that for every consistent brane tiling whose toric diagram is $\Delta$, 
the normalized area $G$ of $\Delta$ is also the number of $U(N)$ gauge groups of the corresponding $4d$ $\mathcal{N}=1$ theory.
It also corresponds to the number of faces of the brane tiling on $T^2$. 
\\

\paragraph{One Interior Point: Reflexive Polygons.} 
For $I=1$, we have convex lattice polygons that are known as \textit{reflexive} polygons \cite{Kreuzer:1998vb,Kreuzer:2000qv,Kreuzer:2000xy}. 
Placing the unique interior point at the origin, we can define the polar dual of a reflexive polygon $\Delta$ as follows,
\beal{es01a10b}
\Delta^{\circ} = \{ u \in \mathbb{R}^2 ~|~ \langle u , w \rangle \geq -1 ~~\forall~ w \in \Delta \}
~.~
\eea
One of the defining features of reflexive polygons is that 
their polar dual is another reflexive polygon,
while for convex lattice polygons with $I>1$, the polar dual is not
a lattice polygon in $\mathbb{Z}^2$.
There are exactly $16$ reflexive polygons up to $GL(2,\mathbb{Z})$ equivalence  \cite{Kreuzer:1998vb,Kreuzer:2000qv,Kreuzer:2000xy}. 
The $30$ brane tilings whose toric diagrams are the $16$ reflexive polygons were first fully identified in \cite{Hanany:2012hi}. 
In \cite{Hanany:2012hi}, it was explicitly shown that, in an appropriate lattice basis, 
the convex hull of the mesonic flavor charge vectors of the generators of each of the $30$ $4d$ $\mathcal{N}=1$ theories is the polar dual
polygon of the corresponding reflexive toric diagram.
Brane tilings associated with reflexive toric diagrams can also be related by \textit{specular duality} \cite{Hanany:2012vc}, under which the \textit{master spaces} are the same \cite{Forcella:2008bb,Forcella:2008eh}.
\\

\begin{figure}[http!!]
\centering
\hypersetup{linkcolor=black}
\setlength{\tabcolsep}{3pt}
\begin{tabular}{cccccc}
  \includegraphics[width=0.1521\textwidth]{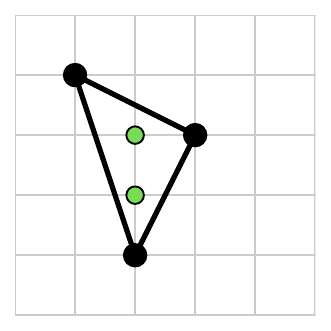} & \includegraphics[width=0.1521\textwidth]{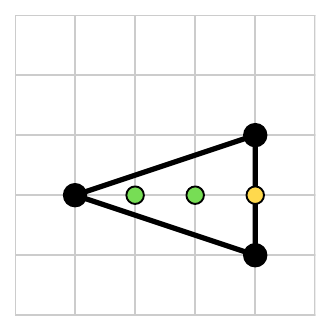} & \includegraphics[width=0.1521\textwidth]{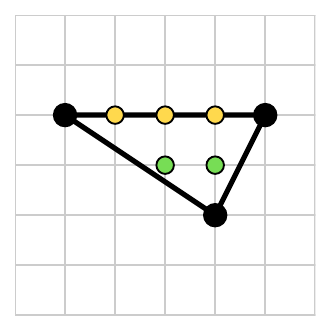} & \includegraphics[width=0.1521\textwidth]{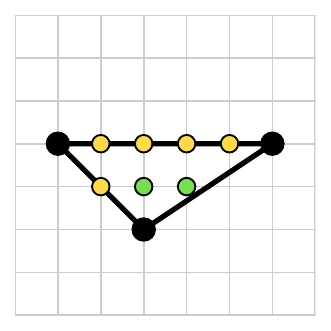} & \includegraphics[width=0.1521\textwidth]{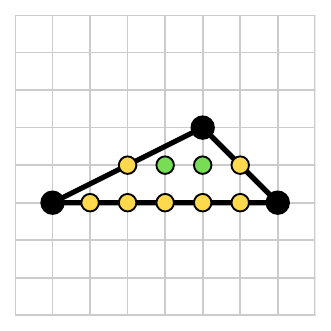} &  \\
  \hyperref[s_m1]{\textsf{Model 1}} & \hyperref[s_m2]{\textsf{Model 2}} & \hyperref[s_m3]{\textsf{Model 3}} & \hyperref[s_m4]{\textsf{Model 4}} & \hyperref[s_m5]{\textsf{Model 5}} &  \\
  {\small $\mathbb{C}^3/\mathbb{Z}_{5}$} & {\small $\mathbb{C}^3/\mathbb{Z}_{6}$} & {\small $\mathbb{C}^3/\mathbb{Z}_{8}$} & {\small $\mathbb{C}^3/\mathbb{Z}_{10}$} & {\small $\mathbb{C}^3/\mathbb{Z}_{2} \times \mathbb{Z}_{6}$} &  \\
\end{tabular}
\caption{The 5 polygons with two interior points whose convex hull is a triangle.}
\label{f_bg_c3}
\end{figure}

\begin{figure}[http!!]
\centering
\hypersetup{linkcolor=black}
\setlength{\tabcolsep}{3pt}
\begin{tabular}{cccccc}
  \includegraphics[width=0.1521\textwidth]{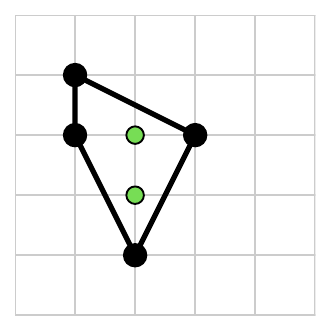} & \includegraphics[width=0.1521\textwidth]{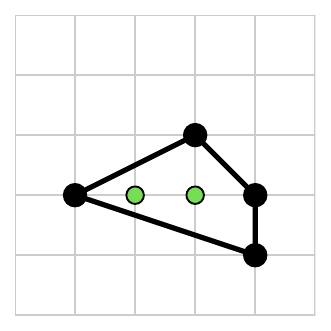} & \includegraphics[width=0.1521\textwidth]{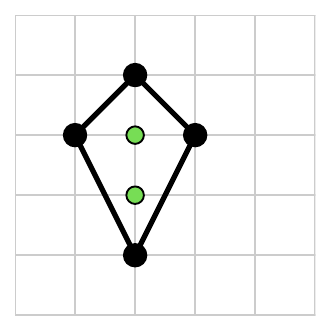} & \includegraphics[width=0.1521\textwidth]{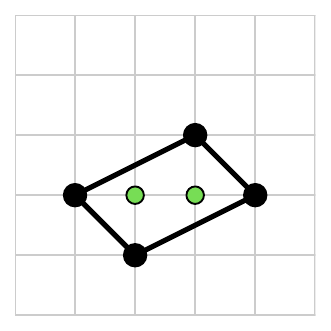} & \includegraphics[width=0.1521\textwidth]{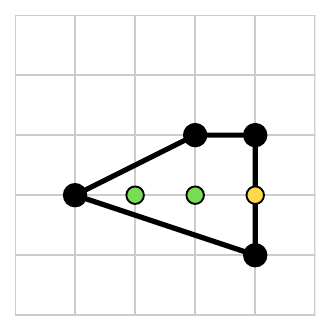} & \includegraphics[width=0.1521\textwidth]{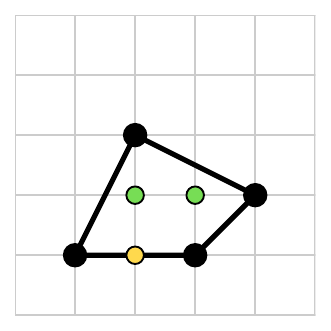} \\
  \hyperref[s_m6]{\textsf{Model 6}} & \hyperref[s_m7]{\textsf{Model 7}} & \hyperref[s_m8]{\textsf{Model 8}} & \hyperref[s_m9]{\textsf{Model 9}} & \hyperref[s_m10]{\textsf{Model 10}} & \hyperref[s_m11]{\textsf{Model 11}} \\
  {\small $L^{1,5,2}$} & {\small $Y^{3,2}$} & {\small $Y^{3,1}$} & {\small $\mathcal{C}/\mathbb{Z}_{3}$} & {\small $X^{3,3}$} & {\small $L^{2,5,3}$} \\
  \\[-1.5ex]
  \includegraphics[width=0.1521\textwidth]{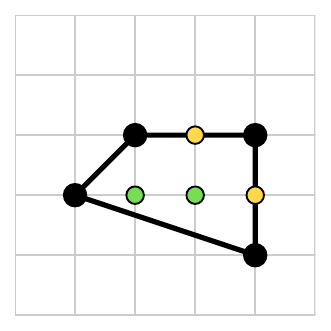} & \includegraphics[width=0.1521\textwidth]{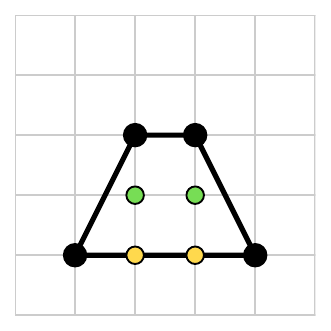} & \includegraphics[width=0.1521\textwidth]{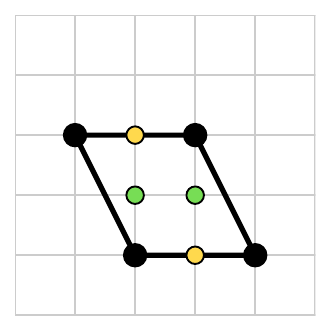} & \includegraphics[width=0.1521\textwidth]{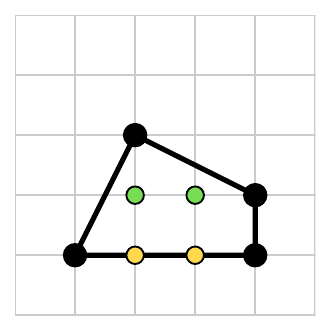} & \includegraphics[width=0.1521\textwidth]{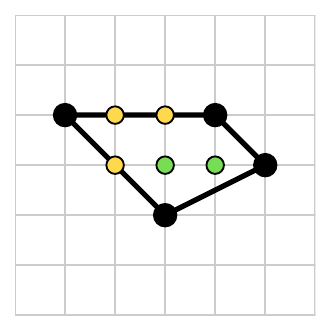} & \includegraphics[width=0.1521\textwidth]{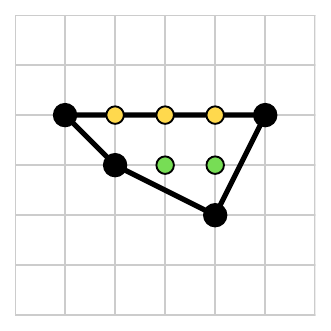} \\
  \hyperref[s_m12]{\textsf{Model 12}} & \hyperref[s_m13]{\textsf{Model 13}} & \hyperref[s_m14]{\textsf{Model 14}} & \hyperref[s_m15]{\textsf{Model 15}} & \hyperref[s_m16]{\textsf{Model 16}} & \hyperref[s_m17]{\textsf{Model 17}} \\
  {\small $L^{1,3,2}/\mathbb{Z}_{2}$} & {\small $L^{1,3,1}/\mathbb{Z}_{2}$} & {\small $\mathcal{C}/\mathbb{Z}_{4}$} & {\small $L^{2,6,5}$} & {\small $L^{1,2,1}/\mathbb{Z}_{3}$} & {\small $L^{4,5,1}$} \\
  \\[-1.5ex]
  \includegraphics[width=0.1521\textwidth]{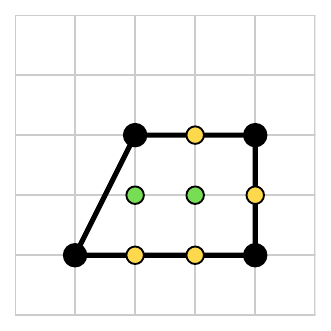} & \includegraphics[width=0.1521\textwidth]{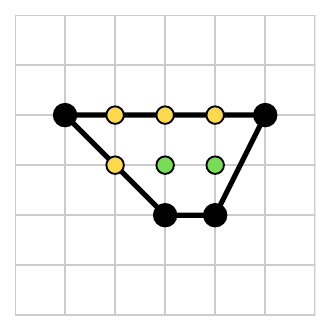} & \includegraphics[width=0.1521\textwidth]{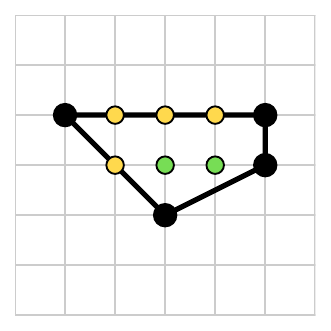} & \includegraphics[width=0.1521\textwidth]{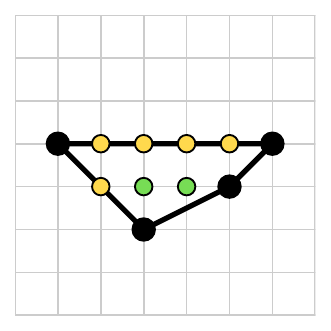} & \includegraphics[width=0.1521\textwidth]{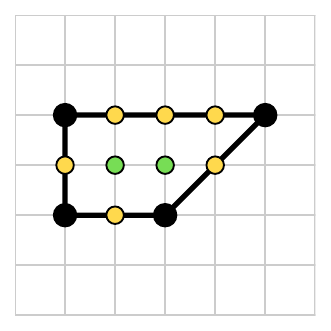} & \includegraphics[width=0.1521\textwidth]{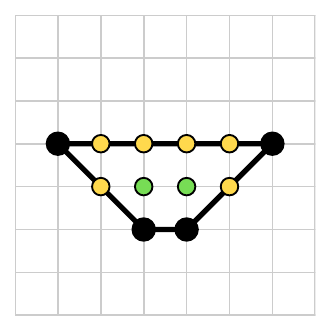} \\
  \hyperref[s_m18]{\textsf{Model 18}} & \hyperref[s_m19]{\textsf{Model 19}} & \hyperref[s_m20]{\textsf{Model 20}} & \hyperref[s_m21]{\textsf{Model 21}} & \hyperref[s_m22]{\textsf{Model 22}} & \hyperref[s_m23]{\textsf{Model 23}} \\
  {\small $L^{2,3,2}/\mathbb{Z}_{2}$} & {\small $L^{1,4,1}/\mathbb{Z}_{2}$} & {\small $L^{1,4,3}/\mathbb{Z}_{2}$} & {\small $L^{5,6,1}$} & {\small $L^{1,2,1}/\mathbb{Z}_{2} \times \mathbb{Z}_{2}$} & {\small $L^{1,5,1}/\mathbb{Z}_{2}$} \\
  \\[-1.5ex]
  \includegraphics[width=0.1521\textwidth]{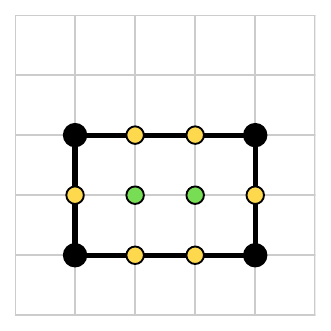} &  &  &  &  &  \\
  \hyperref[s_m24]{\textsf{Model 24}} &  &  &  &  &  \\
  {\small $\mathcal{C}/\mathbb{Z}_{6}$} &  &  &  &  &  \\
\end{tabular}
\caption{The 19 polygons with two interior points whose convex hull is a quadrilateral.}
\label{f_bg_c4}
\end{figure}

\begin{figure}[http!!]
\centering
\hypersetup{linkcolor=black}
\setlength{\tabcolsep}{3pt}
\begin{tabular}{cccccc}
  \includegraphics[width=0.1521\textwidth]{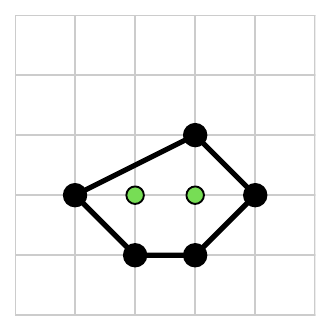} & \includegraphics[width=0.1521\textwidth]{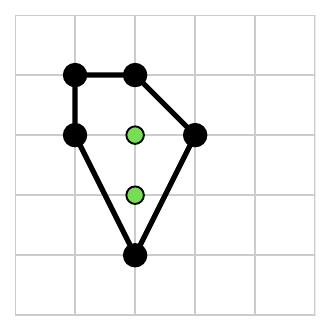} & \includegraphics[width=0.1521\textwidth]{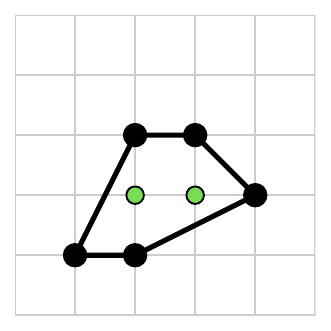} & \includegraphics[width=0.1521\textwidth]{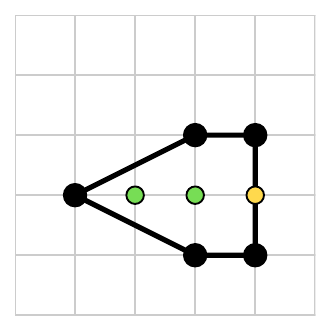} & \includegraphics[width=0.1521\textwidth]{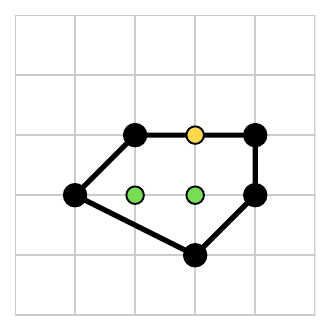} & \includegraphics[width=0.1521\textwidth]{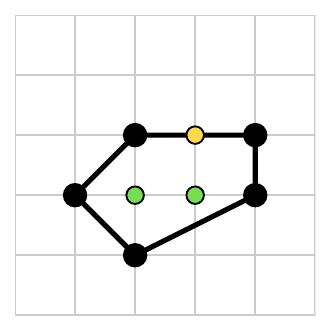} \\
  \hyperref[s_m25]{\textsf{Model 25}} & \hyperref[s_m26]{\textsf{Model 26}} & \hyperref[s_m27]{\textsf{Model 27}} & \hyperref[s_m28]{\textsf{Model 28}} & \hyperref[s_m29]{\textsf{Model 29}} & \hyperref[s_m30]{\textsf{Model 30}} \\
  {\small $X^{3,1}$} & {\small $X^{3,2}$} & {\small $H^{1,3,1,1}$} & {\small $H^{1,4,1,1}$} & {\small $H^{2,2,1,1}$} & {\small $J^{1,4,3,2}$} \\
  \\[-1.5ex]
  \includegraphics[width=0.1521\textwidth]{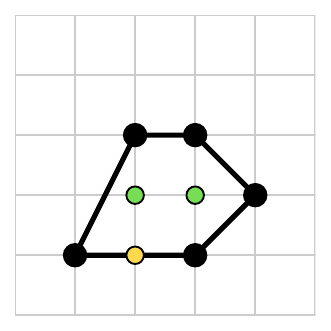} & \includegraphics[width=0.1521\textwidth]{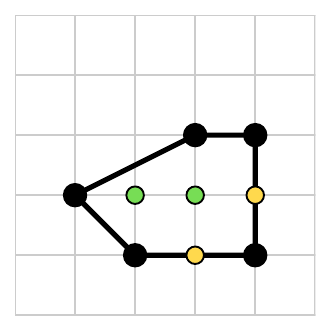} & \includegraphics[width=0.1521\textwidth]{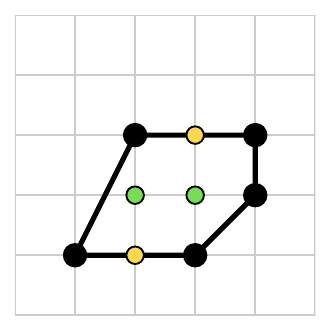} & \includegraphics[width=0.1521\textwidth]{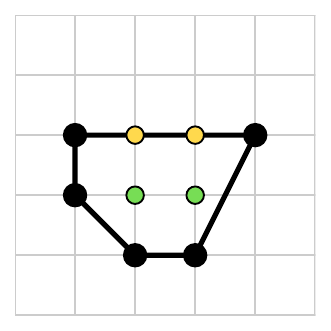} & \includegraphics[width=0.1521\textwidth]{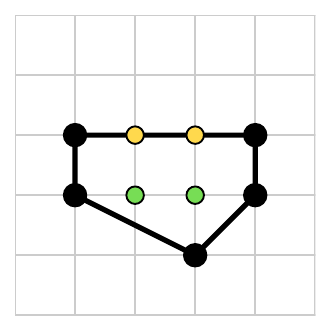} & \includegraphics[width=0.1521\textwidth]{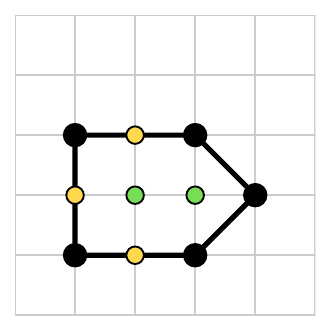} \\
  \hyperref[s_m31]{\textsf{Model 31}} & \hyperref[s_m32]{\textsf{Model 32}} & \hyperref[s_m33]{\textsf{Model 33}} & \hyperref[s_m34]{\textsf{Model 34}} & \hyperref[s_m35]{\textsf{Model 35}} & \hyperref[s_m36]{\textsf{Model 36}} \\
  {\small $G^{4,2}$} & {\small $H^{1,3,1,2}$} & {\small $J^{4,3,1,1}$} & {\small $G^{4,3}$} & {\small $H^{2,3,1,1}$} & {\small $X^{2,1}/\mathbb{Z}_{2}$} \\
  \\[-1.5ex]
  \includegraphics[width=0.1521\textwidth]{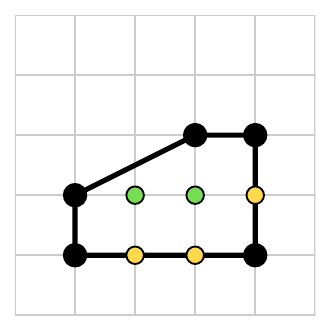} & \includegraphics[width=0.1521\textwidth]{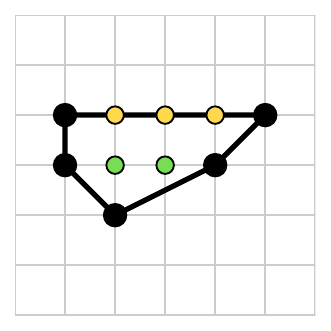} & \includegraphics[width=0.1521\textwidth]{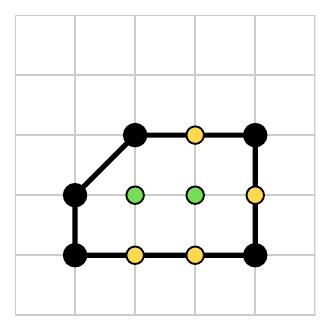} & \includegraphics[width=0.1521\textwidth]{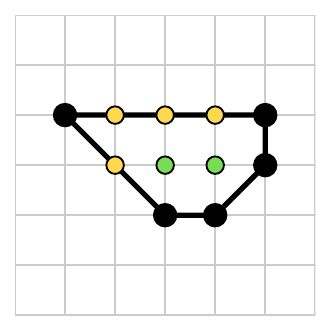} &  &  \\
  \hyperref[s_m37]{\textsf{Model 37}} & \hyperref[s_m38]{\textsf{Model 38}} & \hyperref[s_m39]{\textsf{Model 39}} & \hyperref[s_m40]{\textsf{Model 40}} &  &  \\
  {\small $G^{5,3}$} & {\small $J^{1,5,3,4}$} & {\small $H^{2,2,1,2}$} & {\small $G^{5,4}$} &  &  \\
\end{tabular}
\caption{The 16 polygons with two interior points whose convex hull is a pentagon.}
\label{f_bg_c5}
\end{figure}

\begin{figure}[httpp!!]
\centering
\hypersetup{linkcolor=black}
\setlength{\tabcolsep}{3pt}
\begin{tabular}{cccccc}
  \includegraphics[width=0.1521\textwidth]{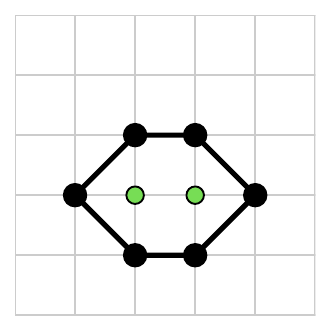} & \includegraphics[width=0.1521\textwidth]{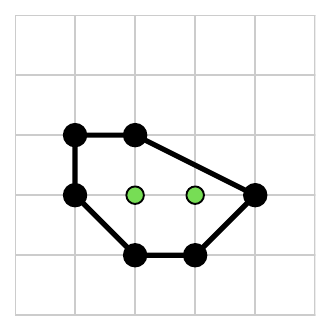} & \includegraphics[width=0.1521\textwidth]{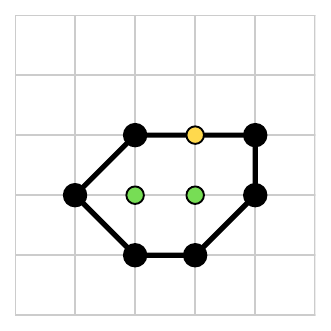} & \includegraphics[width=0.1521\textwidth]{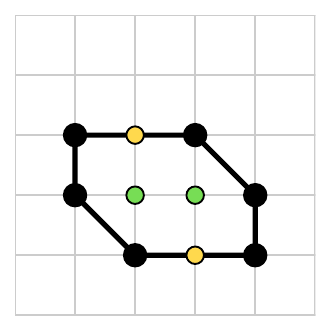} & \includegraphics[width=0.1521\textwidth]{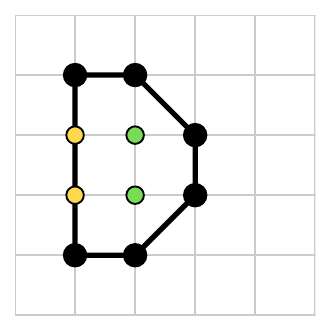} &  \\
  \hyperref[s_m41]{\textsf{Model 41}} & \hyperref[s_m42]{\textsf{Model 42}} & \hyperref[s_m43]{\textsf{Model 43}} & \hyperref[s_m44]{\textsf{Model 44}} & \hyperref[s_m45]{\textsf{Model 45}} &  \\
  {\small $Z^{3,1}$} & {\small $Z^{3,2}$} & {\small $H^{1,1,1,2}$} & {\small $N^{1,3,3,2}$} & {\small $N^{1,2,3,3}$} &  \\
\end{tabular}
\caption{The 5 polygons with two interior points whose convex hull is a hexagon.}
\label{f_bg_c6}
\end{figure}

\paragraph{Two Interior Points.} 
For $I=2$, we have convex lattice polygons
that according to \eref{es01a10} satisfy,
\beal{es01a11}
G = 2 A = B + 2 = I + B  ~,~
\eea
which is simply the 
total number of lattice points of the toric diagram with $I=2$.
The minimum $G=5$ occurs for the unique polygon with $B=3$, 
which is the toric diagram of the abelian orbifold of the form $\mathbb{C}^3/\mathbb{Z}_{5}$ with orbifold action $(1,1,3)$, as discussed in section \sref{s_m1}.
All of the distinct convex lattice polygons in $\mathbb{Z}^2$
with $I=2$ internal points have been classified by
Wei and Ding in \cite{wei2012lattice}, 
and it is now known that:
\begin{center}
\textit{There are exactly $45$ convex lattice polygons with $I=2$ interior points, up to $GL(2,\mathbb{Z})$ equivalence.}
\end{center}
These convex lattice polygons themselves are presented for triangles in \fref{f_bg_c3}, 
for quadrilaterals in \fref{f_bg_c4}, 
for pentagons in \fref{f_bg_c5} and 
for hexagons in \fref{f_bg_c6}, 
organized by the number of vertices and labelled by model number. 
Here, the five triangles are the toric diagrams of abelian orbifolds of $\mathbb{C}^3$ of order $G$ \cite{Hanany:2010cx,Davey:2010px,Hanany:2010ne,Davey:2011dd,Hanany:2011iw}.
The counting of distinct convex lattice polygons with $I=2$
for triangles, quadrilaterals, pentagons and hexagons is summarized in \tref{t_polycount}.

\begin{table}[httpp!!]
\centering
\begin{tabular}{|c|c|c|c|c|c|}
\hline
convex hull & $n_e$ & polygons & $G$ & toric phases & section \\
\hline\hline
triangle & 3 & 5 & $5,6,8,10,12$ & 5 & \ref{s_c3} \\
\hline
quadrilateral & 4 & 19 & $6,\dots,12$ & 57 & \ref{s_c4} \\
\hline
pentagon & 5 & 16 & $7,\dots,11$ & 137 & \ref{s_c5} \\
\hline
hexagon & 6 & 5 & $8,9,10$ & 124 & \ref{s_c6} \\
\hline\hline
total & $3,\dots,6$ & 45 & $5,\dots,12$ & 323 & \\
\hline
\end{tabular}
\caption{The $45$ convex lattice polygons with two interior points, organized by the number of extremal points $n_e$. For each class, the table gives the number of inequivalent polygons, the values taken by the number of $U(N)$ gauge groups $G = 2A$, the total number of toric phases presented in this work, and the section in which the corresponding brane tilings are classified.} \label{t_polycount}
\end{table}

\subsection{Brane Tilings and the Forward Algorithm \label{s_tiling}}

The worldvolume theory of a stack of $N$ D3-branes probing 
a singular
non-compact Calabi-Yau $3$-fold is a 
$4d$ $\mathcal{N}=1$
supersymmetric gauge theory. 
The gauge theory data 
relevant here are the $U(N)$ gauge groups, 
chiral matter content and superpotential $W$.
The probed Calabi-Yau $3$-fold is toric and is recovered as the mesonic moduli
space of the corresponding abelian $4d$ $\mathcal{N}=1$ theory with $U(1)$ gauge groups. 
The process of obtaining the geometry 
of the toric Calabi-Yau $3$-fold
from the information of the abelian $4d$ $\mathcal{N}=1$ theory
is known as the \textit{forward
algorithm} \cite{Feng:2000mi}. 
\\

\paragraph{Quiver $\mathcal{Q}$.} 
The $U(N)$ gauge groups and the matter content
of the $4d$ $\mathcal{N}=1$ theory 
are encoded in a directed graph known as the \textit{quiver}. 
It consists of the following components:
\begin{itemize}
\item \textit{Vertices} in $\mathcal{Q}$ correspond to the $U(N)_i$ gauge
groups with $i=1,\dots,G$. 
Throughout this work all gauge groups are $U(N)$
and when we calculate the mesonic moduli space $\mathcal{M}^{mes}$
we refer to the abelian case with $N=1$ for all gauge groups.
In the case of $N=1$, the mesonic moduli space $\mathcal{M}^{mes}$ is precisely the corresponding toric Calabi-Yau 3-fold.

\item \textit{Edges} in $\mathcal{Q}$ correspond to the bifundamental chiral fields
$X_{ij}$. They are bifundamental, transforming in the fundamental of $U(N)_i$
and the anti-fundamental of $U(N)_j$, 
and hence carry an orientation
$i\rightarrow j$ on the quiver. 
The total number of chiral fields is denoted by $E$. 
Non-abelian anomaly cancellation requires the number of incoming and outgoing edges at
every quiver vertex to be equal in the equal-rank $4d$ $\mathcal{N}=1$ theory. 
Moreover, every chiral field appears precisely
twice in the superpotential $W$, 
once in a positive and once in a negative term,
with the number of positive and negative terms in $W$ being the same. 
This is known as the \textit{toric condition} \cite{Feng:2002zw}
for $4d$ $\mathcal{N}=1$ theories corresponding to toric Calabi-Yau 3-folds.

\item The \textit{incidence matrix} $d_{G\times E}$ encodes the quiver $\mathcal{Q}$. 
Its entries are given by,
\beal{es01a20}
d_{im} =
\begin{cases}
+1 & \text{for } X_m = X_{ij} ~, \\
-1 & \text{for } X_m = X_{ji} ~, \\
0 & \text{otherwise} ~,~
\end{cases}
\eea
where $i$ runs over the $G$ gauge groups and $m$ over the $E$ chiral fields.
Since the rows of $d_{G\times E}$ sum to zero, only $G-1$ of them are
independent. 
The independent rows are collected in the reduced quiver incidence matrix
$\bar{d}_{(G-1)\times E}$.
\end{itemize}

\begin{figure}[http!!]
\centering
\includegraphics[width=0.9\textwidth]{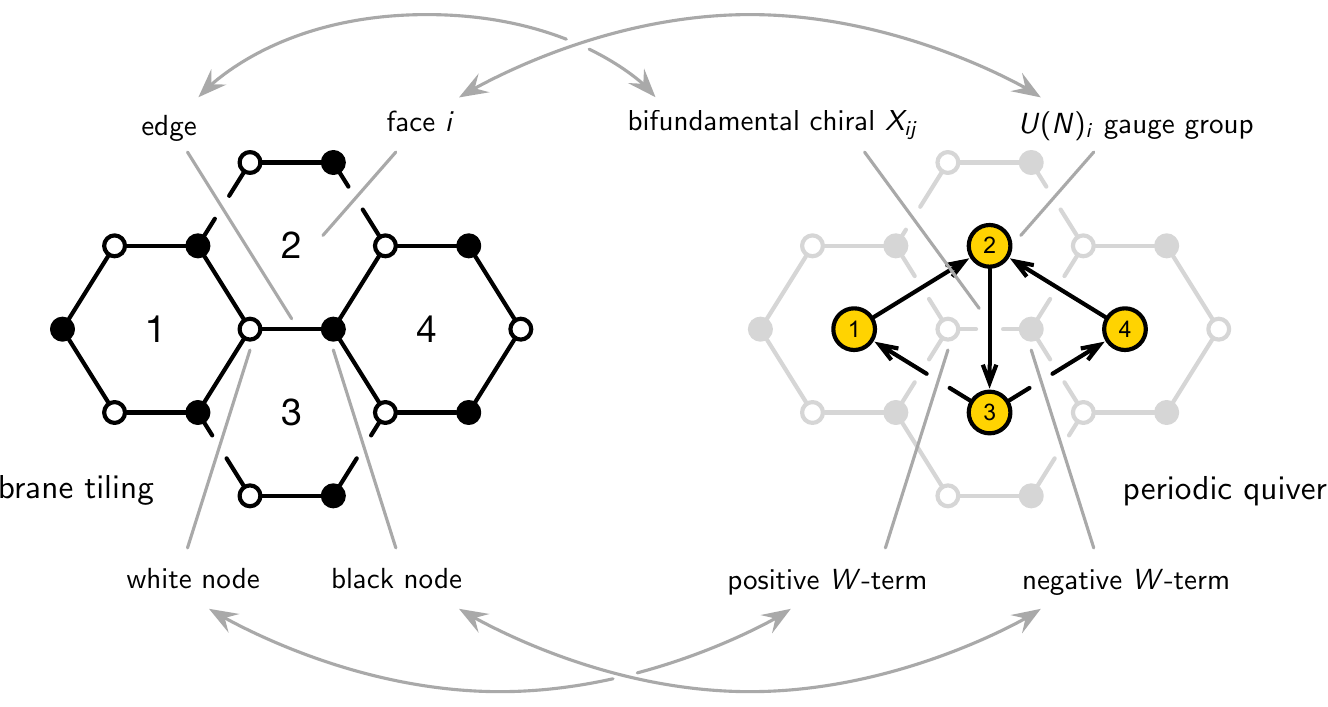}
\caption{The dictionary between a brane tiling and its corresponding $4d$ $\mathcal{N}=1$ supersymmetric gauge theory.}
\label{fig_dict}
\end{figure}

\paragraph{Brane Tilings (Dimer Models).} The quiver and the superpotential can be
combined into a single object which represents the corresponding $4d$ $\mathcal{N}=1$ supersymmetric gauge theory
and toric Calabi-Yau 3-fold.
This object is known as a brane tiling
or a dimer model \cite{Hanany:2005ve,Franco:2005rj,Franco:2005sm}. 
It is a periodic bipartite graph on a $2$-torus $T^2$ 
with the following
components:
\begin{itemize}
\item \textit{White (resp. black) nodes} 
correspond to positive (negative) terms in the superpotential. 
They carry a clockwise (anti-clockwise) orientation. 
The valence of a node is 
the number of bifundamental chiral fields in the
corresponding superpotential term.

\item \textit{Edges} connect a white node with a black node
and correspond to the bifundamental chiral fields. 
Going around a node 
along its orientation
reproduces the chiral fields 
of the associated superpotential term in the
correct cyclic order. 
Because the brane tiling is bipartite, every edge ends on a white and a black node, 
which implies that every
corresponding bifundamental chiral field occurs once in a positive and once in a negative term of $W$.
This is also known as the \textit{toric condition} \cite{Feng:2002zw}
for brane tilings and corresponding $4d$ $\mathcal{N}=1$ theories associated to toric Calabi-Yau 3-folds.

\item \textit{Faces} correspond to the $U(N)_i$ gauge groups. 
Every edge of a brane tiling 
separates two faces, 
corresponding to gauge groups $U(N)_i$ and $U(N)_j$,
and represents a bifundamental field $X_{ij}$
that transforms in the fundamental of $U(N)_i$
and the anti-fundamental of $U(N)_j$. 
The quiver orientation of $X_{ij}$
is determined by the orientations around the black and white nodes at the two ends
of the edge.
\end{itemize}
\fref{fig_dict} summarizes the above components of a brane tiling.
\\

\paragraph{Consistency.} 
Not every bipartite graph on $T^2$
corresponds to a consistent $4d$ $\mathcal{N}=1$ supersymmetric gauge theory 
whose mesonic moduli space with $U(1)$ gauge groups
is a toric Calabi-Yau $3$-fold. 
A brane tiling can be identified as \textit{consistent}
\cite{Hanany:2005ss,Hanany:2006nm,Kennaway:2007tq,Gulotta:2008ef}
if its correspondence to a $4d$ $\mathcal{N}=1$ supersymmetric gauge theory 
and a toric Calabi-Yau 3-fold
satisfies the following conditions:
\begin{itemize}
\item The brane tiling is a \textit{bipartite} graph 
where every edge connects to a white and a black node
and hence satisfies the \textit{toric condition}.
This has the added consequence that 
by going around a face of a brane tiling, 
the nodes alternate in color. 
This in turn implies that every face is bounded by an even number of edges, whose corresponding arrows in the quiver alternate between incoming and outgoing.
Consequently, every quiver vertex has an equal number of incoming and outgoing arrows,
and the non-abelian gauge \textit{anomalies} cancel in the equal-rank $4d$ $\mathcal{N}=1$ theory.

\item Because the brane tiling lives on a 2-torus $T^2$, 
the numbers of faces, edges and nodes satisfy the \textit{Euler relation},
\beal{es01a21}
G - E + V = 0 ~,~
\eea
where $G$ is the number of $U(N)$ gauge groups, 
$E$ is the number of bifundamental chiral fields and 
$V$ is the number of superpotential terms. 

\item The \textit{zig-zag paths}
are paths
along edges of a brane tiling, 
which turn maximally left at a white node
and maximally right at a black node. 
A consistent brane tiling has only closed
zig-zag paths on the 2-torus $T^2$. 
Moreover, every zig-zag path is homologically non-trivial. 
On the universal cover, no zig-zag path self-intersects and no pair intersects more than once in the same direction.

\item For a consistent brane tiling the number of $U(N)$ gauge groups $G$ 
is equal to the normalized area $2A$ of the toric diagram $\Delta$.
\end{itemize}
In this work, we classify all brane tilings that satisfy the above consistency conditions
and correspond to toric Calabi-Yau $3$-folds whose toric diagrams are convex lattice polygons
in $\mathbb{Z}^2$ with exactly two interior lattice points.
Recently, machine learning techniques, in particular explainable AI, have been used in \cite{Lee:2026avt} in order to 
study the consistency of brane tilings and corresponding $4d$ $\mathcal{N}=1$ theories.
\\

\begin{figure}[http!!]
\centering
\includegraphics[width=0.5\textwidth]{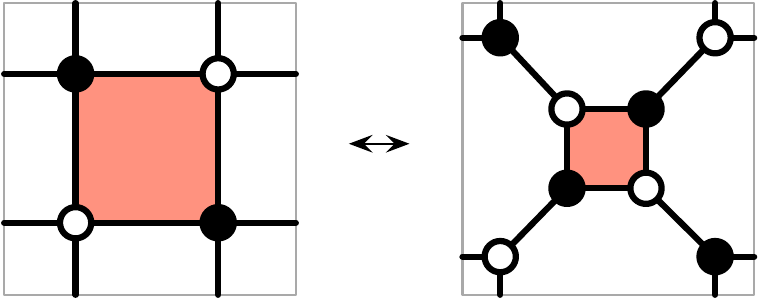}
\caption{Local mutation on a quadrilateral face (red) of a brane tiling, known as a spider move or urban renewal \cite{kenyon2000treesmatchings}, that realizes Seiberg duality.}
\label{fig_seiberg}
\end{figure}

\paragraph{Seiberg Duality.} 
Seiberg duality \cite{Seiberg:1994pq}
relates distinct $4d$ $\mathcal{N}=1$ theories 
that flow to the same IR fixed point under renormalization group flow.
For $4d$ $\mathcal{N}=1$ theories 
realized by brane tilings,
this correspondence is also referred to as \textit{toric duality} \cite{Feng:2001bn}.
Brane tilings that are related by Seiberg duality
have the same mesonic moduli space 
and hence are associated with the same toric Calabi-Yau 3-fold.
Because of this, 
multiple distinct brane tilings can correspond to the same 
toric Calabi-Yau 3-fold.
We refer to such brane tilings and $4d$ $\mathcal{N}=1$ theories 
corresponding to the same toric Calabi-Yau 3-fold
as
\textit{toric phases} \cite{Feng:2000mi,Feng:2001xr}.

A pair of brane tilings related by a single Seiberg duality transformation
is connected by a local mutation of the bipartite graph on the 2-torus
known as a \textit{spider move} or \textit{urban renewal} \cite{kenyon2000treesmatchings}.
This mutation, as illustrated in \fref{fig_seiberg},
acts on a quadrilateral face of the brane tiling corresponding to a $U(N)$ gauge group.
When all gauge groups have equal rank,
as is the case for our work here, 
the 4 bifundamental chiral fields incident on this quadrilateral face
give $N_f = 2 N$ flavors
associated with the $U(N)$ gauge group.
As a result, under Seiberg duality, 
the dual gauge group then has the following rank, 
\beal{es01a22}
N_f - N = N ~,~
\eea
and therefore under such a duality, 
the resulting brane tilings have again gauge groups of equal rank.

By labelling the gauge group on which Seiberg duality is performed by $k$, 
the bifundamental chiral fields $X_{ik}$ and $X_{kj}$
corresponding to electric quarks of the original $4d$ $\mathcal{N}=1$ theory
are reversed to give the following dual quarks, 
\beal{es01a23}
X_{ik} \longrightarrow \widetilde{X}_{ki}
~,~
X_{kj} \longrightarrow \widetilde{X}_{jk}
~.~
\eea
For each product of the form, 
\beal{es01a24}
X_{ik} X_{kj} ~,~
\eea
we obtain a mesonic chiral field in the dual theory as follows, 
\beal{es01a24m1}
M_{ij} \sim 
X_{ik} X_{kj} ~.~
\eea

Under Seiberg duality, 
the original superpotential $W$
is modified by replacing every occurrence of the product of the form $X_{ik} X_{kj}$
with the corresponding meson $M_{ij}$.
Furthermore, 
the superpotential obtains additional coupling terms of the form, 
\beal{es01a24m2}
W_{\text{mag}}
= \sum_{i,j} M_{ij} \widetilde{X}_{jk} \widetilde{X}_{ki}
~,~
\eea
which result in the superpotential of the dual theory given by, 
\beal{es01a24m3}
W^\prime
= 
W \Big|_{X_{ik}X_{kj} \rightarrow M_{ij}} + W_{\text{mag}}
~.~
\eea
We note here that the new superpotential may contain massive quadratic terms of the form $M_{ij}X_{ji}$, 
which can be integrated out by solving the corresponding $F$-term relations and substituting the solutions back into $W^\prime$.

\fref{fig_seiberg} illustrates Seiberg duality on brane tilings in terms of the spider move
on a quadrilateral face of the bipartite graph on $T^2$.
The aim of the following work is to identify all toric phases 
associated with toric Calabi-Yau 3-folds
whose toric diagrams have 2 internal points.
For the $45$ toric Calabi-Yau 3-folds,
we identify in total $323$ toric phases
as summarized in \tref{t_polycount}.
\\

\paragraph{Perfect Matchings, GLSM Fields, and $F$- and $D$-terms.} 
A new set of fields can be defined from the set of bifundamental chiral fields of a brane tiling. 
The purpose of this new basis of fields is to describe the $F$-term and the $D$-term constraints of the $4d$ $\mathcal{N}=1$ theory 
in a common setting, in which the toric geometry of the mesonic moduli space $\mathcal{M}^{mes}$ of the abelian $4d$ $\mathcal{N}=1$ theory 
becomes manifest. 
These are gauged linear sigma model (GLSM) \cite{Witten:1993yc} fields and are represented by perfect
matchings of the corresponding brane tiling. 
They have the following properties:
\begin{itemize}
\item A \textit{perfect matching} is a subset of the edges of a brane tiling such that every node of the brane tiling is the endpoint of precisely 
one edge in the subset. 
Equivalently, it is a subset of the bifundamental chiral fields which covers every superpotential term exactly once. 
The number of perfect matchings is given by $c$ and the perfect matchings are labelled by $a=1,\dots,c$. 

\item The perfect matchings are collected in the \textit{perfect matching matrix} $P$ with entries given by, 
\beal{es01a25}
P_{ma} =
\begin{cases}
1 & \text{if the chiral field } X_m \text{ is an edge of the perfect matching } p_a ~,~ \\
0 & \text{otherwise} ~.
\end{cases}
\nn\\
\eea
Reading the matrix along a row expresses a chiral field $X_m$ as a product of GLSM
fields $p_a$ as follows,
\beal{es01a26}
X_m = \prod_{a=1}^{c} p_{a}^{P_{ma}} ~,
\eea
where $p_a$ denotes both a perfect matching and its corresponding GLSM field in a brane tiling.
The map in \eref{es01a26} solves the $F$-terms of the superpotential $W$, 
which is the reason why the GLSM fields form a natural basis to describe the mesonic moduli space $\mathcal{M}^{mes}$ of a brane tiling
and its corresponding abelian $4d$ $\mathcal{N}=1$ theory.

\item \textit{$F$-terms.} 
In a brane tiling, 
the perfect matching parametrization solves the $F$-term equations
on the coherent component of the master space.
The charges of the GLSM fields $p_a$ under the $F$-term constraints are given by the kernel
of the perfect matching matrix $P$ as follows,
\beal{es01a26b}
Q_{F\,(c-G-2)\times c} = \ker\left(P_{E\times c}\right) ~.
\eea

\item \textit{$D$-terms.} 
The $D$-terms of the $4d$ $\mathcal{N}=1$ 
theory take the following form \cite{Witten:1993yc},
\beal{es01a27}
D_i = -e^2\left(\sum_{m} d_{im}\,|X_m|^2 - \zeta_i\right) ~,
\eea
where $m$ runs over the chiral fields $X_m$ labelling the columns of the incidence matrix $d_{G\times E}$ in \eref{es01a20}, 
$i$ runs over the gauge groups of the quiver, 
$e$ is the gauge coupling and $\zeta_i$ is the Fayet-Iliopoulos parameter. 
In terms of the GLSM fields $p_a$,
the $D$-term constraints are encoded in the reduced quiver matrix $\bar{d}_{(G-1)\times E} $ as follows, 
\beal{es01a28}
\bar{d}_{(G-1)\times E} = Q_{D\,(G-1)\times c} \cdot P^{t}_{c\times E} ~,
\eea
which defines the \textit{$D$-term charge matrix} $Q_{D\,(G-1)\times c}$.
The Fayet-Iliopoulos parameters $\zeta_i$ of \eref{es01a27} are set to zero throughout, 
so that the mesonic moduli space $\mathcal{M}^{mes}$ for the abelian $4d$ $\mathcal{N}=1$ theory 
is the affine toric Calabi-Yau $3$-fold associated with the corresponding brane tiling.
\end{itemize}

\paragraph{The Toric Diagram.} 
The $Q_F$ and $Q_D$ charge matrices corresponding to
the $F$- and $D$-terms, respectively, 
can be concatenated 
into the \textit{total charge matrix}
of the following form, 
\beal{es01a29}
Q_t = \left(
\begin{array}{c}
Q_F \\
Q_D
\end{array}
\right) ~,~
\eea
which is of size $(c-3)\times c$. 
The kernel of the total charge matrix takes the following form,
\beal{es01a30}
G_t = \ker\left(Q_t\right) ~,~
\eea
which is of size $3\times c$.
The columns of $G_t$ are the coordinates of the points of the \textit{toric diagram} of the toric Calabi-Yau 3-fold.
Every GLSM field, and hence every corresponding perfect matching, 
is thereby assigned a point of the toric diagram. 
By a suitable $GL(3,\mathbb{Z})$ transformation the last row of $G_t$ can always be brought to the form $(1,1,\dots,1)$, 
which is the statement that all points of the toric diagram lie on a plane at unit distance from the origin and
hence that the mesonic moduli space $\mathcal{M}^{mes}$ is Calabi-Yau.
The toric diagram is then read off from the first two rows of $G_t$ as a convex lattice polygon in $\mathbb{Z}^2$. 
In the following sections $G_t$ is always presented in this form.

We note that several perfect matchings are in general mapped to the same point in the toric diagram. 
The number of perfect matchings over a given lattice point is the \textit{multiplicity} of that point. 
Since all perfect matchings associated with a given lattice point have identical coordinates, 
the corresponding columns of $G_t$ coincide and are collapsed into a single column per lattice point 
for a compact presentation of the brane tiling classification in this work.
\\

\subsection{Mesonic Moduli Space, Hilbert Series and Generators \label{s_hilbert}}

For abelian $4d$ $\mathcal{N}=1$ theories
with $U(1)$ gauge groups that are realized by brane tilings, 
their mesonic moduli spaces $\mathcal{M}^{mes}$
are the toric Calabi-Yau $3$-folds probed by the D3-branes. 
We can study and characterize the mesonic moduli space
by calculating the generating function which counts the mesonic gauge invariant operators
on $\mathcal{M}^{mes}$.
This generating function is known as the \textit{Hilbert series} \cite{Benvenuti:2006qr,Hanany:2006uc,Feng:2007ur,Butti:2007jv,Forcella:2009vw}. 
Information about the degrees and multiplicities of the \textit{generators} of the mesonic moduli space and their leading relations can be extracted
from the Hilbert series using \textit{plethystics} \cite{Hanany:2007zz}.
\\

\paragraph{The Mesonic Moduli Space.} 
We can express the mesonic moduli space $\mathcal{M}^{mes}$
of a brane tiling corresponding to an abelian $4d$ $\mathcal{N}=1$ theory with $U(1)$ gauge groups
in terms of the following \textit{symplectic quotient}, 
\beal{es01a30b}
\mathcal{M}^{mes} = (\mathbb{C}^c // Q_F) // Q_D ~.~
\eea
Here, we note that
the symplectic quotient ${}^{\text{Irr}}\mathcal{F}^{\flat} = \mathbb{C}^c // Q_F$ is the coherent component of the \textit{master space} \cite{Forcella:2008bb,Forcella:2008eh,Forcella:2008ng,Zaffaroni:2008zz,Forcella:2008tqe,Hanany:2010zz}.
We note here that throughout this work, $\mathcal{M}^{mes}$ denotes the branch obtained from the coherent component of the master space.
The quotient in \eref{es01a30b} removes the $c-G-2$ charges in $Q_F$ and the $G-1$ charges in $Q_D$ from the $c$ GLSM fields, 
leaving the following, 
\beal{es01a31}
\dim_{\mathbb{C}} \mathcal{M}^{mes} = c - (c-G-2) - (G-1) = 3 ~,~
\eea
which is the dimension 
of the mesonic moduli space $\mathcal{M}^{mes}$.
For the $4d$ $\mathcal{N}=1$ theory
with $U(N)$ gauge groups, 
the mesonic moduli space becomes the $N$-th symmetric product $\mathcal{M}^{mes}_N = \text{Sym}^{N} \mathcal{M}^{mes}$.
\\

\paragraph{The Hilbert Series.} 
The Hilbert series for the mesonic moduli space $\mathcal{M}^{mes}$ of a brane tiling
is the generating function for the mesonic gauge invariant operators.
Given the charge matrices $Q_F$ and $Q_D$, 
we can make use of the \textit{Molien integral formula} given by, 
\beal{es01a32}
g(y_a ; \mathcal{M}^{mes}) =
\prod_{i=1}^{c-3} \oint_{|z_i|=1} \frac{\mathrm{d} z_i}{2 \pi i z_i}
\prod_{a=1}^{c}
\frac{1}{1 - y_a \prod_{j=1}^{c-3} z_j^{(Q_t)_{ja}}} ~,~
\eea
to obtain the Hilbert series for $\mathcal{M}^{mes}$.
Here,
$y_a$ is the fugacity for the GLSM field $p_a$,
and $Q_t$ is the total charge matrix in \eref{es01a29} of the brane tiling. 

We can also unrefine the Hilbert series by setting $y_a=t$ for each GLSM field $p_a$
associated with an extremal point of the toric diagram and $y_a=1$ for all remaining fields.
The resulting single variable generating function $g(t;\mathcal{M}^{mes})$ grades mesonic gauge invariant operators by their total degree in the extremal perfect matchings.
\\

\paragraph{Plethystic Logarithm.} 
Information about the degrees and multiplicities of the generators of the mesonic moduli space $\mathcal{M}^{mes}$ and their leading relations
is encoded in the \textit{plethystic logarithm} \cite{Hanany:2007zz} of the Hilbert series.
The plethystic logarithm for 
the Hilbert series $g(y_a;\mathcal{M}^{mes})$ takes the following form,
\beal{es01a35}
PL\left[g(y_a;\mathcal{M}^{mes})\right] = \sum_{k=1}^{\infty} \frac{\mu(k)}{k} \log \left[ g(y_a^k ; \mathcal{M}^{mes}) \right] ~,~
\eea
where $\mu(k)$ is the M\"obius function. 
If the mesonic moduli space $\mathcal{M}^{mes}$ is a
\textit{complete intersection}, the expansion of \eref{es01a35} is finite.
An infinite expansion therefore implies that $\mathcal{M}^{mes}$ is not a complete intersection.
The higher-order terms encode information about algebraic dependencies, including relations among relations, known as \textit{syzygies}
\cite{Feng:2007ur,Hanany:2007zz}.

In the following work,
our classification of brane tilings 
gives the plethystic logarithm of the Hilbert series for $\mathcal{M}^{mes}$ 
refined by a fugacity $t_a$ for every extremal perfect matching $p_a$.
Our work lists the generators of every mesonic moduli space $\mathcal{M}^{mes}$ 
explicitly, both in terms of the extremal GLSM fields $p_a$ and the associated bifundamental chiral fields for every brane tiling.
We present explicit first order relations among the generators with degrees and charges consistent with the refined plethystic logarithm.
\\

\subsection{Global Symmetry and $U(1)_R$ charges \label{s_symmetry}}

The generators of the mesonic moduli space $\mathcal{M}^{mes}$
carry charges under the global symmetry of the $4d$ $\mathcal{N}=1$ theory and brane tiling.
This includes the mesonic flavor symmetry and the $U(1)_R$ symmetry.
\\

\paragraph{Global Symmetries.} 
The mesonic moduli space $\mathcal{M}^{mes}$ of an abelian $4d$ $\mathcal{N}=1$ theory and brane tiling
with $U(1)$ gauge groups
is a toric Calabi-Yau 3-fold.
Its isometry is associated with the mesonic flavor symmetry and the $U(1)_R$ symmetry of the $4d$ $\mathcal{N}=1$ theory.
The possible choices of the mesonic flavor symmetry with the $U(1)_R$ symmetry are as follows,
\begin{itemize}
\item $U(1)_{f_1} \times U(1)_{f_2} \times U(1)_R$
\item $SU(2)_x \times U(1)_f \times U(1)_R$
\item $SU(2)_x \times SU(2)_y \times U(1)_R$
\item $SU(3)_{x_1,x_2} \times U(1)_R$~,~
\end{itemize}
where the subscripts on the mesonic flavor symmetry factors label the fugacities
used to refine the Hilbert series with respect to the corresponding flavor charges.
The $U(1)_R$ symmetry has a fugacity given by $\bar{t}$, whereas, as discussed above, the fugacity $t$ is used to count the total degree in the extremal GLSM fields $p_a$.
Here, we note that the enhancement of $U(1)$ mesonic flavor symmetry factors
to an $SU(2)$ or $SU(3)$ factor is indicated by repeated columns in the total
charge matrix $Q_t$ in \eref{es01a29}.
\\

\paragraph{$U(1)_R$ charges from $a$-maximization.} 
The superconformal $U(1)_R$ charges are determined by $a$-maximization
\cite{Intriligator:2003jj,Butti:2005vn,Butti:2005ps}.
This procedure identifies the superconformal $U(1)_R$ symmetry
by maximizing the trial central charge
\beal{es01a40}
a = \frac{3}{32} \left(3 \Tr(R^3) - \Tr(R)\right) ~.~
\eea

We assign to each bifundamental chiral field $X_{ij}$
a trial $U(1)_R$ charge $R_{ij}$.
For every superpotential term $W_a$, these charges satisfy,
\beal{es01a41}
\sum_{X_{ij} \in W_a} R_{ij} = 2 ~.~
\eea
Furthermore, 
the vanishing NSVZ beta function for each $SU(N)_k$
gauge factor at the superconformal fixed point \cite{Intriligator:2003jj}
leads to the following constraint at each face $F_k$ of the brane tiling, 
\beal{es01a42}
\sum_{X_{ij} \in F_k} (1-R_{ij}) = 2 ~.~
\eea
Accordingly, for a brane tiling, the trial $a$-function in \eref{es01a40}
takes the form, 
\beal{es01a42b1}
a(R_{ij}) =
\frac{3N^2}{32}
\left[
2G 
+ 
3\sum_{X_{ij}}
(R_{ij}-1)^3
-
\sum_{X_{ij}}
(R_{ij}-1)
\right]~,~
\eea
which simplifies to,
\beal{es01a42b2}
a(R_{ij}) =
\frac{9N^2}{32}
\left[
G 
+ 
\sum_{X_{ij}}
(R_{ij}-1)^3
\right]~,~
\eea
whose maximization determines the exact superconformal $U(1)_R$ charges for chiral fields $X_{ij}$.
\\

\paragraph{$U(1)_R$ charges from volume minimization.} 
We recall that the mesonic moduli space $\mathcal{M}^{mes}$ of the abelian $4d$ $\mathcal{N}=1$ theory and brane tiling
is a toric Calabi-Yau 3-fold.
It is a non-compact cone over a Sasaki-Einstein 5-manifold $X_5$ 
whose volume is related under minimization to the superconformal $U(1)_R$ charges
of the \textit{divisors} of the toric geometry \cite{Martelli:2005tp,Martelli:2006yb}.
The toric divisors are related to the extremal points of the toric diagram and the extremal GLSM fields of the brane tiling.

The volume of the Sasaki-Einstein manifold $X_5$ is obtained by minimizing the volume function extracted from the Hilbert series of $\mathcal{M}^{mes}$,
\beal{es01a50}
\text{vol}(b_a; X_5) =
\frac{8\pi^3}{27} \lim_{\mu\rightarrow 0} \mu^3
~g(t_a = e^{-\mu b_a}; \mathcal{M}^{mes}) ~,~
\eea
where $g(t_a; \mathcal{M}^{mes})$ is the Hilbert series given in \eref{es01a32}, and $t_a$ are the fugacities associated with the extremal GLSM fields $p_a$.
The parameters $b_a$ are the trial $U(1)_R$ charges assigned to these fields, while all other GLSM fields are assigned zero $U(1)_R$ charge in our parametrization.
These trial charges determine a \textit{Reeb vector}, which encodes the geometric action of the trial $U(1)_R$ symmetry on the Calabi-Yau cone.
The volume function is minimized subject to $\sum_a b_a=2$ \cite{Martelli:2005tp,Martelli:2006yb}.

In parallel, we can also write down the Hilbert series associated with the divisor $D_a$ and the corresponding extremal GLSM field $p_a$.
By modifying the Molien integral formula in \eref{es01a32}, we have the following Hilbert series expression for $D_a$, 
\beal{es01a51}
g^{D_a}(t_m; \mathcal{M}^{mes})
=
\prod_{i=1}^{c-3} \oint_{|z_i|=1} 
\frac{\ud z_i}{2\pi i z_i} 
\left[
t_a \prod_{k=1}^{c-3} z_k^{(Q_t)_{ka}}
\right]^{-1}
\prod_{s=1}^{c}
\frac{
1
}
{
1-t_s
\prod_{j=1}^{c-3} z_j^{(Q_t)_{js}}
}
~.~
\eea
Using the Hilbert series expression for the divisor $D_a$, 
we can write down the trial $U(1)_R$ charge expression for the extremal GLSM field $p_a$ as follows, 
\beal{es01a52}
R_a = \lim_{\mu \rightarrow 0} \frac{1}{\mu}
\left[
\frac{
g^{D_a}(t_m=e^{-\mu b_m}; \mathcal{M}^{mes})
}{
g(t_m=e^{- \mu b_m}; \mathcal{M}^{mes})
}
- 1
\right]
~,~
\eea
which becomes the superconformal $U(1)_R$ charge for $p_a$
when the parameters $b_m$ take the critical values $b_m^*$ when $\text{vol}(b_a; X_5)$ reaches a global minimum value.
\\

\subsection{Families of Toric Calabi-Yau 3-folds \label{s_families}}

The toric diagrams given by the $45$ convex lattice polygons in $\mathbb{Z}^2$ are named after families of toric Calabi-Yau $3$-folds. 
In our work, a family of toric Calabi-Yau $3$-folds is specified by a toric diagram whose corners are given in terms of a set of integer parameters. 
Established examples include the families $Y^{p,q}$ \cite{Benvenuti:2004dy,Benvenuti:2004wx}, $X^{p,q}$ \cite{Hanany:2005hq}, $Z^{p,q}$ \cite{Oota:2006eg} and $L^{a,b,c}$ \cite{Franco:2005sm}. 
Besides these families,  we also have the abelian orbifolds of $\mathbb{C}^3$ \cite{Hanany:2010cx,Davey:2010px,Hanany:2010ne,Davey:2011dd}
whose toric diagrams are lattice triangles,
abelian orbifolds of the conifold $\mathcal{C}$ \cite{Hanany:2010cx}, 
as well as abelian orbifolds of $L^{a,b,c}$ and $X^{p,q}$.
In this work, we also use the previously introduced $H^{a,b,c,d}$ family \cite{Lee:2026tju}
and introduce new families $G^{p,q}$, $J^{a,b,c,d}$ and $N^{a,b,c,d}$,
which allow us to completely name the toric Calabi-Yau 3-folds
associated with the $45$ convex lattice polygons in $\mathbb{Z}^2$ with 2 internal points.
Here, the $G^{p,q}$ family is a subfamily of
the $K^{a,b,c,d}$ family, which was first introduced in \cite{Bao:2020kji}.
\tref{t_families} summarizes every family of toric Calabi-Yau 3-folds, 
which we make use of to describe the $45$ toric Calabi-Yau 3-folds whose toric diagrams are convex lattice polygons with $2$ internal points.
\\

\begin{table}[httpp!!]
\centering
\begin{tabular}{|c|c|c|}
\hline
Family & Figure & Models \\
\hline\hline
$Y^{p,q}$ & \ref{f_fam_Y} & \hyperref[s_m2]{2}, \hyperref[s_m7]{7}, \hyperref[s_m8]{8}, \hyperref[s_m9]{9} \\
\hline
$X^{p,q}$ & \ref{f_fam_X} & \hyperref[s_m10]{10}, \hyperref[s_m25]{25}, \hyperref[s_m26]{26} \\
\hline
$Z^{p,q}$ & \ref{f_fam_Z} & \hyperref[s_m41]{41}, \hyperref[s_m42]{42} \\
\hline
$L^{a,b,c}$ & \ref{f_fam_L} & \hyperref[s_m1]{1}, \hyperref[s_m3]{3}, \hyperref[s_m6]{6}, \hyperref[s_m11]{11}, \hyperref[s_m15]{15}, \hyperref[s_m17]{17}, \hyperref[s_m21]{21} \\
\hline
$G^{p,q}$ & \ref{f_fam_G} & \hyperref[s_m27]{27}, \hyperref[s_m28]{28}, \hyperref[s_m31]{31}, \hyperref[s_m32]{32}, \hyperref[s_m34]{34}, \hyperref[s_m37]{37}, \hyperref[s_m40]{40} \\
\hline
$H^{a,b,c,d}$ & \ref{f_fam_H} & \hyperref[s_m27]{27}, \hyperref[s_m28]{28}, \hyperref[s_m29]{29}, \hyperref[s_m32]{32}, \hyperref[s_m35]{35}, \hyperref[s_m39]{39}, \hyperref[s_m41]{41}, \hyperref[s_m43]{43} \\
\hline
$J^{a,b,c,d}$ & \ref{f_fam_J} & \hyperref[s_m30]{30}, \hyperref[s_m33]{33}, \hyperref[s_m36]{36}, \hyperref[s_m38]{38} \\
\hline
$N^{a,b,c,d}$ & \ref{f_fam_N} & \hyperref[s_m41]{41}, \hyperref[s_m42]{42}, \hyperref[s_m43]{43}, \hyperref[s_m44]{44}, \hyperref[s_m45]{45} \\
\hline\hline
$\mathbb{C}^3/\Gamma$ & -- & \hyperref[s_m1]{1}, \hyperref[s_m2]{2}, \hyperref[s_m3]{3}, \hyperref[s_m4]{4}, \hyperref[s_m5]{5} \\
\hline
$\mathcal{C}/\Gamma$ & -- & \hyperref[s_m9]{9}, \hyperref[s_m14]{14}, \hyperref[s_m24]{24} \\
\hline
$L^{a,b,c}/\Gamma$ & \ref{f_fam_L} & \hyperref[s_m12]{12}, \hyperref[s_m13]{13}, \hyperref[s_m16]{16}, \hyperref[s_m18]{18}, \hyperref[s_m19]{19}, \hyperref[s_m20]{20}, \hyperref[s_m22]{22}, \hyperref[s_m23]{23} \\
\hline
$X^{p,q}/\Gamma$ & \ref{f_fam_X} & \hyperref[s_m36]{36} \\
\hline
\end{tabular}
\caption{
The families of toric Calabi-Yau $3$-folds after which the models of this work are named.
}
\label{t_families}
\end{table}

\begin{figure}[http!!]
\centering
\begin{subfigure}[b]{0.46\textwidth}
  \centering
  \includegraphics[scale=0.4]{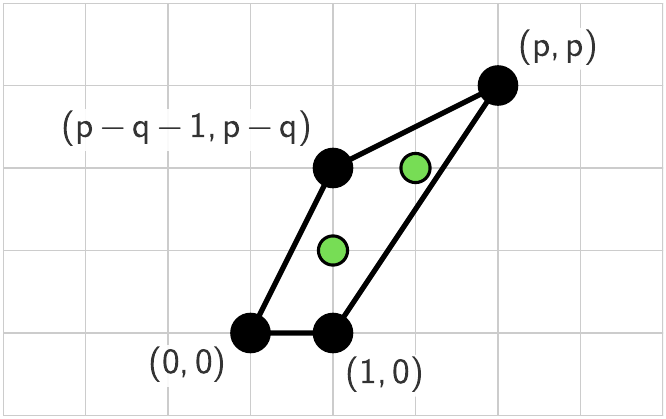}
  \caption{$Y^{p,q}$}
  \label{f_fam_Y}
\end{subfigure}
\hfill
\begin{subfigure}[b]{0.46\textwidth}
  \centering
  \includegraphics[scale=0.4]{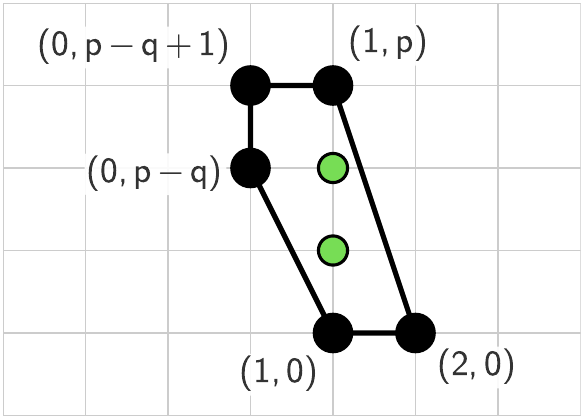}
  \caption{$X^{p,q}$}
  \label{f_fam_X}
\end{subfigure}
\par\medskip
\begin{subfigure}[b]{0.46\textwidth}
  \centering
  \includegraphics[scale=0.4]{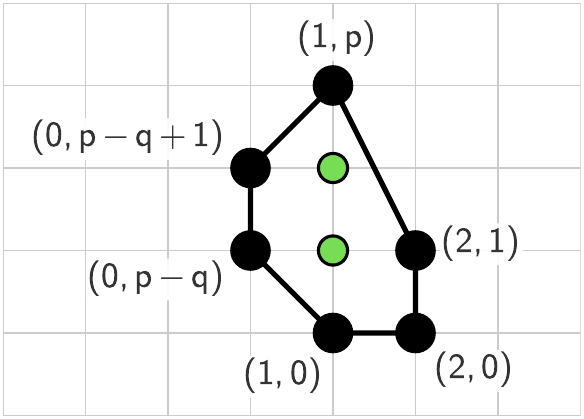}
  \caption{$Z^{p,q}$}
  \label{f_fam_Z}
\end{subfigure}
\hfill
\begin{subfigure}[b]{0.46\textwidth}
  \centering
  \includegraphics[scale=0.4]{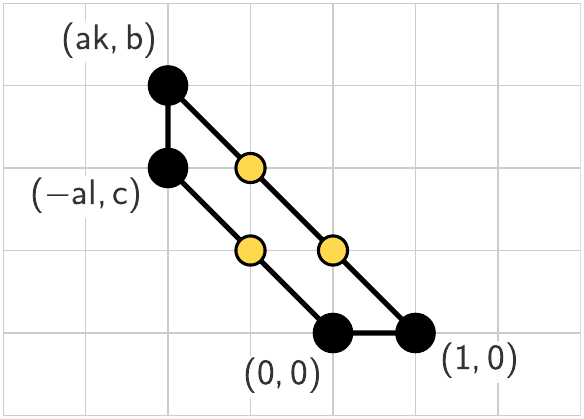}
  \caption{$L^{a,b,c}$}
  \label{f_fam_L}
\end{subfigure}
\caption{The general toric diagrams of (a) $Y^{p,q}$, (b) $X^{p,q}$, (c) $Z^{p,q}$ and (d) $L^{a,b,c}$, with the extremal points labelled by their coordinates.}
\label{f_fam_pub}
\end{figure}

\paragraph{Abelian Orbifolds.}
The abelian orbifolds in our classification are of the form $\mathcal{X}/\Gamma$,
where $\mathcal{X}$ is $\mathbb{C}^3$, the conifold $\mathcal{C}$, a member of the $L^{a,b,c}$ family or $X^{2,1}$.
$\Gamma$ takes the form $\mathbb{Z}_n$ or $\mathbb{Z}_{n_1}\times\mathbb{Z}_{n_2}$.
Let $z_1,\dots,z_d$ denote the generators of $\mathcal{X}$, expressed in terms of the extremal GLSM fields $p_a$.
The orbifold action under $\Gamma$
is specified by
$a_i=(a_{i1},a_{i2})$, where $a_{ij}\in\mathbb{Z}_{n_j}$.
An element $s=(s_1,s_2)\in\Gamma$ acts as follows,
\beal{es01a59b}
z_i \mapsto \exp\left[2\pi i\sum_{j=1}^{2}\frac{a_{ij}s_j}{n_j}\right]z_i ~.~
\eea
For $\Gamma=\mathbb{Z}_n$, we set $n_1=n$, $n_2=1$ and write $a_{i1}=a_i$.
The orbifold action satisfies the following conditions:
\begin{itemize}
\item \textit{Preservation of the defining relations.}
For every binomial relation of the form $R_l=\prod_{u\in I_l^+}z_u-\prod_{v\in I_l^-}z_v=0$,
with repeated factors counted with multiplicity, 
we require
\beal{es01a59bb1}
\sum_{u\in I_l^+}a_{uj}\equiv\sum_{v\in I_l^-}a_{vj}\pmod{n_j}
~,~ j=1,2 ~.~
\eea

\item \textit{Calabi-Yau condition.}
The holomorphic top form of $\mathcal{X}$ must be invariant under $\Gamma$.
Writing the product of extremal GLSM fields as $\prod_{a=1}^{n_e}p_a=\prod_{i=1}^{d}z_i^{c_i}$,
with $c_i\in\mathbb{Z}_{\geq0}$, we require
\beal{es01a59bb2}
\sum_{i=1}^{d}c_i a_{ij}\equiv0\pmod{n_j} ~,~ j=1,2 ~.~
\eea

\item \textit{Faithfulness.}
The only element acting trivially on all generators is the identity,
\beal{es01a59bb3}
\sum_{j=1}^{2}\frac{a_{ij}s_j}{n_j}\in\mathbb{Z}
~~\text{for all }i
\quad\Longrightarrow\quad s=0\text{ in }\Gamma ~.~
\eea
For $\Gamma=\mathbb{Z}_n$, this condition is equivalent to
$\gcd(a_1,\dots,a_d,n)=1$.
\end{itemize}
For product groups, the rows of the displayed orbifold action matrices correspond to the cyclic factors in their stated order, 
and the columns correspond to the generators.

For $\mathbb{C}^3$, the toric diagram $\Delta$ has extremal points $(0,0)$, $(1,0)$ and $(0,1)$,
with generators $z_1=p_1$, $z_2=p_2$ and $z_3=p_3$ and no defining relations.
In addition to faithfulness, the action $(a_1,a_2,a_3)$ must satisfy
$a_1+a_2+a_3\equiv0\pmod n$.
The abelian orbifolds of $\mathbb{C}^3$ \cite{Davey:2010px,Hanany:2010ne,Davey:2011dd} appearing in our classification are of the following form,
\beal{es01a59bb}
&&
\mathbb{C}^3/\mathbb{Z}_5~(1,1,3) ~,~ \quad
\mathbb{C}^3/\mathbb{Z}_6~(1,1,4) ~,~ \quad
\mathbb{C}^3/\mathbb{Z}_8~(1,3,4) ~,~
\nn\\
&&
\mathbb{C}^3/\mathbb{Z}_{10}~(1,4,5) ~,~ \quad
\mathbb{C}^3/\mathbb{Z}_2\times\mathbb{Z}_6~
\left(\begin{array}{ccc}
0 & 1 & 1 \\
1 & 0 & 5
\end{array}\right) ~,~
\eea
whose toric diagrams are triangles with $2$ internal points in $\mathbb{Z}^2$.

For the conifold $\mathcal{C}$, the extremal points of the toric diagram are
$(0,0)$, $(1,0)$, $(1,1)$ and $(0,1)$, with GLSM fields $p_1,\dots,p_4$
assigned in this order.
The generators of the mesonic moduli space are
expressed in terms of the extremal GLSM fields $p_a$ as follows,
\beal{es01a59c}
z_1=p_1p_2 ~,~ z_2=p_2p_3 ~,~ z_3=p_3p_4 ~,~ z_4=p_4p_1 ~,~
\eea
and satisfy $z_1z_3=z_2z_4$. Preservation of this relation requires $a_1+a_3\equiv a_2+a_4\pmod n$.
The Calabi-Yau condition requires the product of extremal GLSM fields
$p_1p_2p_3p_4=z_1z_3=z_2z_4$ to be invariant, giving
\beal{es01a59d}
a_1+a_3\equiv a_2+a_4\equiv0\pmod n ~.~
\eea
Accordingly, the abelian orbifolds of the conifold in our classification are of the following form,
\beal{es01a59e}
\mathcal{C}/\mathbb{Z}_3~(1,1,2,2) ~,~ \quad
\mathcal{C}/\mathbb{Z}_4~(1,2,3,2) ~,~ \quad
\mathcal{C}/\mathbb{Z}_6~(2,3,4,3) ~,~
\eea
whose toric diagrams are quadrilaterals with $2$ internal points in $\mathbb{Z}^2$.
\\

\paragraph{$Y^{p,q}$ Family.}
For integers $p\geq1$ and $0\leq q\leq p$, the toric diagram of $Y^{p,q}$
\cite{Benvenuti:2004dy,Benvenuti:2004wx} is the convex hull of the following points, 
\beal{es01a60}
(0,0) ~,~ (1,0) ~,~ (p,p) ~,~ (p-q-1,p-q) ~,~
\eea
as illustrated in \fref{f_fam_Y}. The area of the toric diagram is given by
\beal{es01a61}
A=p ~.~
\eea
At $q=p$, the point $(0,0)$ lies on the edge joining $(-1,0)$ and $(1,0)$.
The diagram therefore becomes a triangle corresponding to the abelian orbifold of the form $\mathbb{C}^3/\mathbb{Z}_{2p}$ with orbifold action $(1,1,-2)$
\cite{Davey:2010px,Hanany:2010ne,Davey:2011dd}.
At $q=0$, the toric diagram corresponds to an abelian orbifold of the form $\mathcal{C}/\mathbb{Z}_p$ with orbifold action $(1,1,-1,-1)$.
The toric diagram has $I=p-1$ interior points, so the members with $I=2$ have $p=3$.
They are $Y^{3,3}$ equivalent to $\mathbb{C}^3/\mathbb{Z}_6~(1,1,4)$,
$Y^{3,2}$, $Y^{3,1}$ and
$Y^{3,0}$ equivalent to $\mathcal{C}/\mathbb{Z}_3~(1,1,2,2)$.
\\

\paragraph{$X^{p,q}$ Family.}
For integers $1\leq q\leq p$, the toric diagram of $X^{p,q}$
\cite{Hanany:2005hq} is the convex hull of the following set of points,
\beal{es01a62}
(1,0) ~,~ (2,0) ~,~ (1,p) ~,~ (0,p-q+1) ~,~ (0,p-q) ~,~
\eea
as illustrated in \fref{f_fam_X}. The area of the toric diagram is given by
\beal{es01a63}
A=p+\frac{1}{2} ~.~
\eea
For $q=p$, the point $(1,0)$ lies on the edge joining $(0,0)$ and $(2,0)$ and the toric diagram is a quadrilateral.
Extending the range to $q=0$ introduces no new toric diagrams, since the polygons for $X^{p,0}$ and $X^{p,1}$ are $GL(2,\mathbb{Z})$ equivalent.
The toric diagram has $I=p-1$ interior points.
The members appearing in our classification are therefore the quadrilateral $X^{3,3}$ and the pentagons $X^{3,1}$ and $X^{3,2}$.

Our classification also contains an abelian orbifold of $X^{2,1}$.
Assigning extremal GLSM fields $p_1,\dots,p_5$ to the points in \eref{es01a62} at $p=2$, $q=1$ in the displayed order, the generators of
$X^{2,1}$ can be written as follows, 
\beal{es01a63b}
&&
z_1=p_1p_2^2p_3 ~,~
z_2=p_2p_3^2p_4 ~,~
z_3=p_3^2p_4^2p_5 ~,~
z_4=p_1p_3p_4^2p_5^2 ~,~
\nn\\
&&
z_5=p_1^2p_4^2p_5^3 ~,~
z_6=p_1^2p_2p_4p_5^2 ~,~
z_7=p_1^2p_2^2p_5 ~,~
z_8=p_1p_2p_3p_4p_5 ~.~
\eea
Their defining ideal is generated by $14$ quadratic binomials equating
products with the same GLSM monomial, including
$z_1z_4=z_2z_6=z_3z_7=z_8^2$.
Compatibility with these relations and invariance of the product of extremal GLSM fields
$z_8=p_1p_2p_3p_4p_5$ give the following constraints on the orbifold action, 
\beal{es01a63c}
&&
a_3\equiv a_2-a_1 ~,~
a_4\equiv-a_1 ~,~
a_5\equiv-a_1-a_2 ~,~
\nn\\
&&
a_6\equiv-a_2 ~,~
a_7\equiv a_1-a_2 ~,~
a_8\equiv0
\pmod n ~.~
\eea
All congruences are understood modulo $n$.
The orbifold appearing in our classification is of the following form,
\beal{es01a63d}
X^{2,1}/\mathbb{Z}_2~(1,1,0,1,0,1,0,0) ~.~
\eea
\\

\paragraph{$Z^{p,q}$ Family.}
For integers $0<q<p$, the toric diagram of $Z^{p,q}$ \cite{Oota:2006eg}
is the hexagon with extremal points given by, 
\beal{es01a65}
(1,0) ~,~ (2,0) ~,~ (2,1) ~,~ (1,p) ~,~ (0,p-q+1) ~,~ (0,p-q) ~,~
\eea
as illustrated in \fref{f_fam_Z}.
It is obtained by adding $(2,1)$ to the toric diagram of $X^{p,q}$, corresponding to unhiggsing the associated brane tiling.
The area of the toric diagram is given by
\beal{es01a66}
A=p+1 ~.~
\eea
The toric diagram has $I=p-1$ interior points.
Accordingly, we have
$2$ members appearing in our classification,
$Z^{3,1}$ and $Z^{3,2}$.
\\

\paragraph{$L^{a,b,c}$ Family.}
The toric diagram of $L^{a,b,c}$ \cite{Franco:2005sm} is the convex hull of
the following set of points,
\beal{es01a67}
(0,0) ~,~ (1,0) ~,~ (ak,b) ~,~ (-al,c) ~,~
\eea
where $a,b,c$ are positive integers with $\gcd(b,c)=1$, and $k,l\in\mathbb{Z}$
satisfy $ck+bl=1$.
The diagram is illustrated in \fref{f_fam_L}.
Writing $d=a+b-c$, the four points are extremal points of a quadrilateral when $d>0$.
The area of the toric diagram is given by, 
\beal{es01a68}
A=\frac{1}{2}\max(a+b,c) ~.~
\eea
In our classification, the triangular cases are
$L^{1,1,5}$ equivalent to $\mathbb{C}^3/\mathbb{Z}_5~(1,1,3)$ and
$L^{1,3,8}$ equivalent to $\mathbb{C}^3/\mathbb{Z}_8~(1,3,4)$.
The members $L^{1,5,2}$, $L^{2,5,3}$, $L^{2,6,5}$, $L^{4,5,1}$ and $L^{5,6,1}$ have $d>0$ and quadrilateral toric diagrams.

We also encounter orbifolds of $L^{1,2,1}$, $L^{1,3,1}$, $L^{1,4,1}$, $L^{1,5,1}$, $L^{2,3,2}$, $L^{1,3,2}$ and $L^{1,4,3}$.
For $c=a$, we assign extremal GLSM fields $p_1,\dots,p_4$ to the $4$ points in \eref{es01a67} in the given order.
The generators of $L^{a,b,a}$ are given by, 
\beal{es01a68c}
z_1=p_1^ap_2^b ~,~ z_2=p_2p_3 ~,~
z_3=p_3^bp_4^a ~,~ z_4=p_4p_1 ~,~
\eea
with defining relation $z_1z_3=z_2^bz_4^a$.
At $a=b=1$, these reduce to the generators of the conifold in \eref{es01a59c}.
Preservation of the binomial relation and invariance of the product of extremal GLSM fields
$z_2z_4=p_1p_2p_3p_4$ give us the following constraints on the orbifold action,
\beal{es01a68d}
a_1+a_3\equiv b\,a_2+a\,a_4\pmod n ~,~\qquad
a_2+a_4\equiv0\pmod n ~.~
\eea

The toric diagrams of $L^{1,3,2}$ and $L^{1,4,3}$ are reflexive
and the generators correspond to lattice points of the respective polar dual polygons \cite{Hanany:2012hi}.
Using the same ordering of extremal GLSM fields as above, 
the generators for $L^{1,3,2}$ are given by,
\beal{es01a68e}
&&
z_1=p_1p_2^2 ~,~
z_2=p_2^2p_3 ~,~
z_3=p_2p_3^2p_4 ~,~
z_4=p_3^3p_4^2 ~,~
z_5=p_1p_3^2p_4^2 ~,~
\nn\\
&&
z_6=p_1^2p_3p_4^2 ~,~
z_7=p_1^3p_4^2 ~,~
z_8=p_1^2p_2p_4 ~,~
z_9=p_1p_2p_3p_4 ~.~
\eea
Their defining ideal is generated by $20$ binomial relations, including $z_1z_5=z_2z_6=z_3z_8=z_9^2$.
Compatibility with these relations and invariance of $z_9=p_1p_2p_3p_4$ give the following constraints on the orbifold action
modulo $n$,
\beal{es01a68f}
&&
a_3\equiv a_2-a_1 ~,~
a_4\equiv a_2-2a_1 ~,~
a_5\equiv-a_1 ~,~
a_6\equiv-a_2 ~,~
\nn\\
&&
a_7\equiv a_1-2a_2 ~,~
a_8\equiv a_1-a_2 ~,~
a_9\equiv0
\pmod n ~.~
\eea
Similarly, the generators for $L^{1,4,3}$ are as follows,
\beal{es01a68g}
&&
z_1=p_1p_2^2 ~,~
z_2=p_2^3p_3 ~,~
z_3=p_2^2p_3^2p_4 ~,~
z_4=p_2p_3^3p_4^2 ~,~
\nn\\
&&
z_5=p_3^4p_4^3 ~,~
z_6=p_1p_3^2p_4^2 ~,~
z_7=p_1^2p_4 ~,~
z_8=p_1p_2p_3p_4 ~.~
\eea
Their defining ideal is generated by $14$ quadratic binomials, including
$z_1z_6=z_3z_7=z_8^2$.
Compatibility with these relations and invariance of
$z_8=p_1p_2p_3p_4$ give the following constraints on the orbifold action, 
\beal{es01a68h}
&&
a_3\equiv a_2-a_1 ~,~
a_4\equiv a_2-2a_1 ~,~
a_5\equiv a_2-3a_1 ~,~
\nn\\
&&
a_6\equiv-a_1 ~,~
a_7\equiv a_1-a_2 ~,~
a_8\equiv0
\pmod n ~.~
\eea

Accordingly,
the abelian orbifolds that appear in our classification are as follows,
\beal{es01a68b}
&&
L^{1,3,2}/\mathbb{Z}_2~(0,1,1,1,0,1,0,1,0) ~,~ \quad
L^{1,3,1}/\mathbb{Z}_2~(1,1,1,1) ~,~
\nn\\
&&
L^{1,2,1}/\mathbb{Z}_3~(0,1,1,2) ~,~ \quad
L^{2,3,2}/\mathbb{Z}_2~(0,1,1,1) ~,~ \quad
L^{1,4,1}/\mathbb{Z}_2~(0,1,1,1) ~,~
\nn\\
&&
L^{1,4,3}/\mathbb{Z}_2~(1,1,0,1,0,1,0,0) ~,~ \quad
L^{1,2,1}/(\mathbb{Z}_2\times\mathbb{Z}_2)~
\left(\begin{array}{cccc}
0 & 1 & 1 & 1 \\
1 & 0 & 1 & 0
\end{array}\right) ~,~
\nn\\
&&
L^{1,5,1}/\mathbb{Z}_2~(0,1,0,1) ~.~
\eea
All have quadrilateral toric diagrams.
\\

\begin{figure}[http!!]
\centering
\begin{subfigure}[b]{0.46\textwidth}
  \centering
  \includegraphics[scale=0.4]{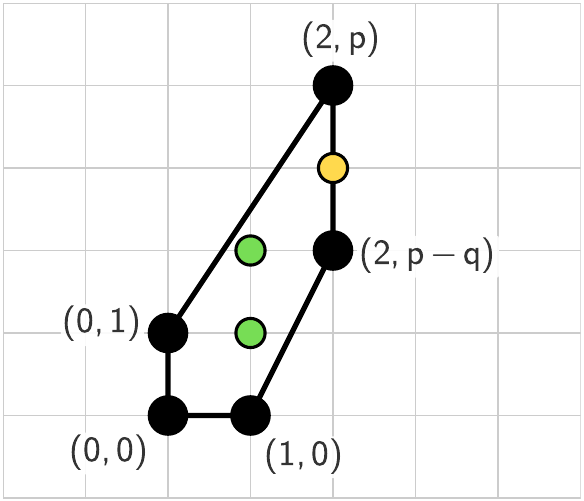}
  \caption{$G^{p,q}$}
  \label{f_fam_G}
\end{subfigure}
\hfill
\begin{subfigure}[b]{0.46\textwidth}
  \centering
  \includegraphics[scale=0.4]{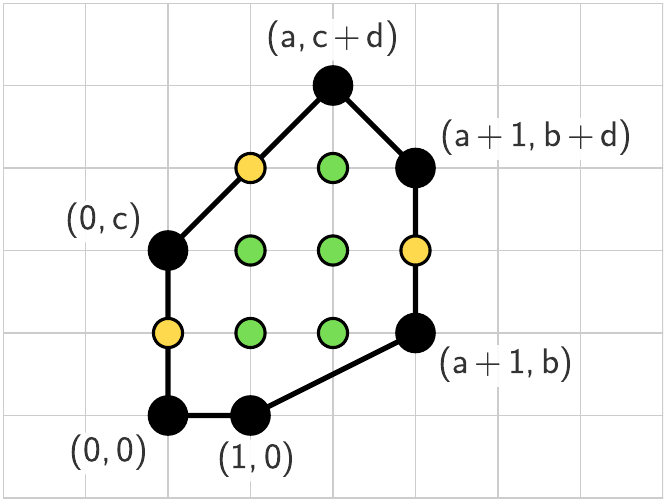}
  \caption{$H^{a,b,c,d}$}
  \label{f_fam_H}
\end{subfigure}
\caption{Toric diagrams for the (a) $G^{p,q}$ and (b) $H^{a,b,c,d}$ families.
The extremal points are labelled by their coordinates.}
\label{f_fam_GH}
\end{figure}

\paragraph{$G^{p,q}$ Family.}
We introduce the notation $G^{p,q}$ for a family of toric Calabi-Yau $3$-folds that accounts for several of the $45$ toric diagrams in our classification.
For integers $0<q<p$, the corresponding toric diagram is the pentagon with extremal points given by,
\beal{es01a69}
(0,0) ~,~ (1,0) ~,~ (2,p-q) ~,~ (2,p) ~,~ (0,1) ~,~
\eea
as shown in \fref{f_fam_G}, with area
\beal{es01a71}
A=\frac{1}{2}(p+q+2) ~.~
\eea
At $q=p$, the point $(1,0)$ lies on the edge joining $(0,0)$ and $(2,0)$ and the toric diagram is a quadrilateral corresponding to
$L^{1,p,1}/\mathbb{Z}_2$ with orbifold action $(0,1,p+1,1)$.
At $q=0$, the toric diagram is also a quadrilateral with $G^{p,0}$ equivalent to $L^{2,p,1}$.
The toric diagram has $I=\lfloor p/2\rfloor$ interior points, so $I=2$ requires $p=4$ or $p=5$.
The pentagons that we obtained are given by
$G^{4,2}$, $G^{4,3}$, $G^{5,3}$ and $G^{5,4}$.
The remaining members are
$G^{4,1}=H^{1,3,1,1}$, $G^{5,1}=H^{1,4,1,1}$ and
$G^{5,2}=H^{1,3,1,2}$, and are labelled using the $H^{a,b,c,d}$ family below.

The $G^{p,q}$ family is a subfamily of the $K^{a,b,c,d}$ family introduced in \cite{Bao:2020kji}.
For positive integers $a,b,c$ with $\gcd(b,c)=1$ and $0\leq d\leq b$, the latter is described by the convex hull of the following points,
\beal{es0a70}
(0,0) ~,~ (1,0) ~,~ (ak,b-d) ~,~ (ak,b) ~,~ (-am,c) ~,~
\eea
where $m,k\in\mathbb{Z}$ satisfy $bm+ck=1$.
For $d>0$, the polygon depends on the chosen solution $(m,k)$.
Here, we fix this ambiguity by requiring $0\leq m<c$.
For $c=1$, this gives $(m,k)=(0,1)$ and hence we have, 
\beal{es01a72}
G^{p,q}=K^{2,p,1,q} ~.~
\eea
The $7$ pentagons in our classification belonging to this family are
$K^{2,4,1,1}$, $K^{2,4,1,2}$, $K^{2,4,1,3}$,
$K^{2,5,1,1}$, $K^{2,5,1,2}$, $K^{2,5,1,3}$ and $K^{2,5,1,4}$
in \cite{Bao:2020kji}.
\\

\paragraph{$H^{a,b,c,d}$ Family.}
The $H^{a,b,c,d}$ family was introduced in \cite{Lee:2026tju} in the study
of brane tilings related by \textit{quiver-invariant dualities} and
\textit{tilting mutations}.
For positive integers $a,b,c,d$, its toric diagram is the convex hull of
the following points,
\beal{es01a73b}
(0,0) ~,~ (1,0) ~,~ (a+1,b) ~,~ (a+1,b+d) ~,~ (a,c+d) ~,~ (0,c) ~,~
\eea
as illustrated in \fref{f_fam_H}.
The point $(a,c+d)$ is non-extremal precisely when,
\beal{es01a73bb}
d\leq a(b-c) ~,~\qquad a(c+d)\geq(a-1)b ~.~
\eea
In this case, the toric diagram is a pentagon.
The point lies strictly inside the pentagon when both inequalities are strict.
At $d=a(b-c)$, it lies on the edge joining $(a+1,b+d)$ and $(0,c)$, and 
at $a(c+d)=(a-1)b$, it lies on the edge joining $(1,0)$ and $(a+1,b)$.
Otherwise, the point is an additional extremal point and the toric diagram is a hexagon.
The two hexagonal regimes are $d>a(b-c)$ and $a(c+d)<(a-1)b$.

The area of the toric diagram can be written as follows, 
\beal{es01a74}
&&
A=\frac{1}{2}\Big[b+c+(a+1)d+ac+\max\big(0,d-a(b-c)\big)
\nn\\
&&\hspace{2cm}
+\max\big(0,(a-1)b-a(c+d)\big)\Big] ~.~
\eea
For the pentagons this reduces to $A=\frac{1}{2}[b+c+(a+1)d+ac]$.
The two hexagonal regimes have areas
$\frac{1}{2}[b+c+(a+2)d+a(2c-b)]$ and
$\frac{1}{2}(ab+c+d)$, respectively.
The members of this family appearing in our classification are the hexagons
$H^{2,1,1,1}=Z^{3,1}$ and $H^{1,1,1,2}$, together with the pentagons
$H^{1,3,1,1}$, $H^{1,4,1,1}$, $H^{2,2,1,1}$,
$H^{1,3,1,2}$, $H^{2,3,1,1}$ and $H^{2,2,1,2}$.
\\

\begin{figure}[httpp!!]
\centering
\includegraphics[scale=0.4]{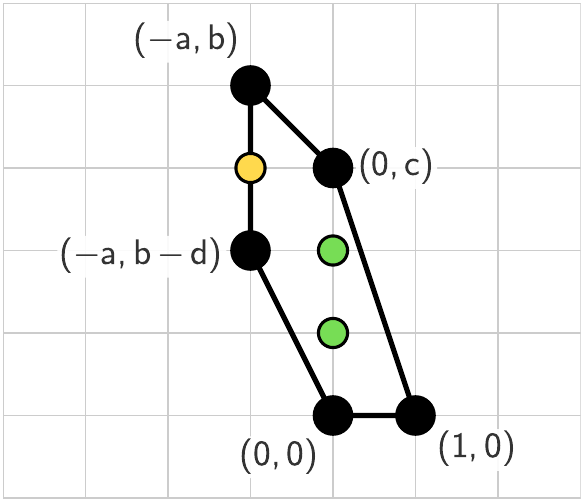}
\caption{Toric diagram for the $J^{a,b,c,d}$ family, with the extremal points labelled by their coordinates.}
\label{f_fam_J}
\end{figure}

\paragraph{$J^{a,b,c,d}$ Family.}
We introduce a further family $J^{a,b,c,d}$ to describe $4$ pentagons in our classification.
For positive integers $a,b,c,d$ with $d<b$, the corresponding toric diagram is the convex hull of
the following points, 
\beal{es01a77}
(0,0) ~,~ (1,0) ~,~ (0,c) ~,~ (-a,b) ~,~ (-a,b-d) ~,~
\eea
as shown in \fref{f_fam_J}.
The polygon is a pentagon when $(a+1)c>b$ and a quadrilateral otherwise.
At $(a+1)c=b$, the point $(0,c)$ lies on the edge joining $(1,0)$ and $(-a,b)$, and
for $(a+1)c<b$, it lies inside the quadrilateral formed by the other $4$ points.
The area of the toric diagram is given by,
\beal{es01a78}
A=\frac{1}{2}\left[b+ad+\max\big(0,(a+1)c-b\big)\right] ~.~
\eea
The $4$ members appearing in our classification are $J^{1,4,3,2}$, $J^{4,3,1,1}$, $J^{2,4,2,2}$ and $J^{1,5,3,4}$.
All have pentagonal toric diagrams.
In particular,
$J^{2,4,2,2}$ is equivalent to the abelian orbifold of the form $X^{2,1}/\mathbb{Z}_2~(1,1,0,1,0,1,0,0)$.
\\

\begin{figure}[http!!]
\centering
\includegraphics[scale=0.4]{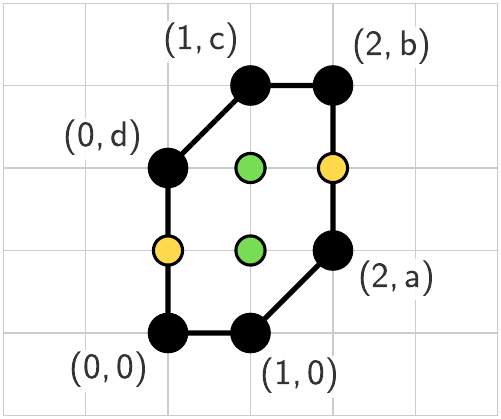}
\caption{Toric diagram for the $N^{a,b,c,d}$ family, with the extremal points labelled by their coordinates.}
\label{f_fam_N}
\end{figure}

\paragraph{$N^{a,b,c,d}$ Family.}
Finally, we introduce the $N^{a,b,c,d}$ family, whose toric diagrams are
the convex hulls of
the following set of points,
\beal{es01a79}
(0,0) ~,~ (1,0) ~,~ (2,a) ~,~ (2,b) ~,~ (1,c) ~,~ (0,d) ~,~
\eea
where the parameters are integers satisfying
$1\leq a\leq b$, $d\geq1$ and $2c>b+d$.
The toric diagram is illustrated in \fref{f_fam_N}.
For $a<b$, the $6$ points are the extremal points of a hexagon.
For $a=b$, the points $(2,a)$ and $(2,b)$ coincide and the toric diagram is a pentagon.
In both cases, the area is given by,
\beal{es01a80}
A=c+\frac{1}{2}(b-a+d) ~.~
\eea
All $5$ hexagonal toric diagrams in our classification belong to this family.
They are $N^{2,3,3,1}$, $N^{1,2,3,1}$, $N^{1,3,3,1}$, $N^{1,3,3,2}$ and $N^{1,2,3,3}$.
\\

\subsection{Charge Conjugation, Seiberg Duality and Equivalence of Brane Tilings}
\label{s_conj}

We identify $323$ inequivalent toric phases associated with the $45$ convex lattice polygons containing exactly two interior lattice points.
Each toric phase is specified by a brane tiling. 
Below we define the equivalence relation used to distinguish these toric phases and their corresponding brane tilings.
\\

\paragraph{Charge Conjugation and Sign Flip.}
Charge conjugation $\mathsf{C}$ reverses all bifundamental chiral fields $X_{ij}$
and the cyclic order of the chiral fields in every superpotential term, while preserving its overall sign.
We define the operation as follows,
\beal{es01a81}
\mathsf{C} ~:~ 
\quad X_{ij}
& \longmapsto & \bar{X}_{ji} ~,~ 
\nn\\
X_{i_1i_2} \cdots X_{i_ni_1}
& \longmapsto &
\bar{X}_{i_1i_n}\cdots \bar{X}_{i_2i_1}
~.~
\eea
Here $\bar{X}_{ji}$ is a chiral field in the conjugate bifundamental representation of $X_{ij}$. 

We also introduce an operation that reverses the overall sign of the superpotential
without changing the set of bifundamental chiral fields in the superpotential.
The sign flip operation is defined as follows,
\beal{es01a82}
\mathsf{S} ~:~ 
\quad (\mathcal{Q},W)
\longmapsto (\mathcal{Q},-W) ~,~
\eea
where $\mathcal{Q}$ is the quiver of the brane tiling. 
The sign flip is equivalent to
changing the sign of every chiral field belonging to a chosen perfect matching of the brane tiling.

The charge conjugation $\mathsf{C}$
and sign flip $\mathsf{S}$ 
operations obey the following relations,
\beal{es01a83} 
\mathsf{C}^2=\mathsf{S}^2=1
~,~
\mathsf{C}\mathsf{S}=\mathsf{S}\mathsf{C}
~.~
\eea
With the conventions for brane tilings outlined in section \sref{s_tiling}, 
$\mathsf{C}$ reflects the brane tiling by reversing the orientations around the white and black nodes
while keeping their colors, 
and $\mathsf{S}$ reflects it and exchanges the
colors, 
and $\mathsf{C}\mathsf{S}$ exchanges the colors without reflection.
We note that all these operations preserve the mesonic moduli space
and the associated toric Calabi-Yau 3-fold.
\\

\paragraph{Equivalence between Brane tilings.}
Given that a brane tiling encodes the quiver $\mathcal{Q}$
and superpotential $W$ of the associated $4d$ $\mathcal{N}=1$ theory,
let us denote a brane tiling by 
$T=(\mathcal{Q},W)$.
Accordingly, we count two brane tilings as the same toric
phase precisely when
we have the following, 
\beal{es01a84}
T\sim T'
\quad
\Longleftrightarrow\quad
T' \cong \mathsf{C}^{a}\mathsf{S}^{b}(T)
\quad
\text{for some } a,b\in\{0,1\} ~.
\eea
Here $\cong$ denotes simultaneous relabellings of the gauge groups and
chiral fields in the brane tiling
while preserving the quiver $\mathcal{Q}$ and the superpotential $W$.
This includes permutations of chiral fields connecting the same ordered pair 
of $U(N)$ gauge groups while allowing for cyclic permutations within the superpotential terms.
The comparison between two brane tilings therefore uses the full pair $(\mathcal{Q},W)$
and hence
isomorphic quivers can represent distinct toric phases when their
superpotentials are not related by \eref{es01a84}.
\\

\paragraph{Seiberg Duality and Classification.}
Seiberg duality for brane tilings as described in section \sref{s_tiling}
preserves the mesonic moduli space $\mathcal{M}^{mes}$
but can change the quiver and the superpotential of the $4d$ $\mathcal{N}=1$ theory.
Seiberg duality as a mutation on a quadrilateral face of the brane tiling commutes with $\mathsf{C}$ and
$\mathsf{S}$ up to relabelling of gauge groups and chiral fields.
As a result, Seiberg duality acts consistently on the equivalence classes of brane tilings defined by \eref{es01a84}.
In the following classification of brane tilings associated to convex lattice polygons with 2 internal points,
we retain brane tilings related by Seiberg duality as distinct toric phases whenever
\eref{es01a84} does not identify them as equivalent.

For each toric Calabi-Yau 3-fold whose toric diagram is one of the $45$ convex lattice polygons with 2 internal points, 
we start from one brane tiling, dualize every quadrilateral face, integrate out all massive fields and compare each resulting reduced brane tiling
with all previously obtained toric phases using \eref{es01a84}.
An inequivalent brane tiling is added to the classification and its quadrilateral faces are examined in turn. 
The procedure terminates when no new equivalence class of brane tilings is found for the given toric Calabi-Yau 3-fold.
We check that every brane tiling satisfies the consistency conditions in section \sref{s_tiling} and that the forward algorithm reproduces the required toric diagram up to the equivalence in \eref{es01a04}.
The classification results presented in sections \sref{s_c3} to \sref{s_c6} contain one representative brane tiling of each resulting equivalence class associated to one of the $45$ toric Calabi-Yau 3-folds that we consider in this work.
\\

\section{Triangles}\label{s_c3}

\subsection{Model 1: $\mathbb{C}^3/\mathbb{Z}_{5}~(1,1,3)$}\label{s_m1}

\begin{figure}[H]
\centering
\begin{subfigure}[b]{0.31\textwidth}
  \centering
  \includegraphics[width=\linewidth]{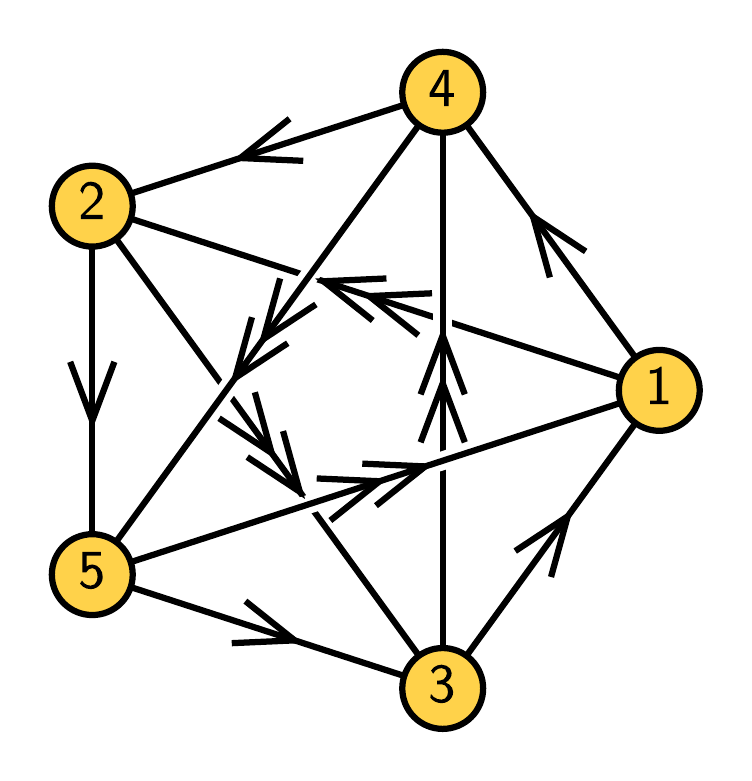}
  \caption{}
  \label{f_model_01_1_quiver}
\end{subfigure}
\hfill
\begin{subfigure}[b]{0.31\textwidth}
  \centering
  \resizebox{\linewidth}{!}{%
  \begin{tikzpicture}[x=1bp,y=1bp]
    \node[anchor=south west,inner sep=0pt,outer sep=0pt] at (0,0) {\includegraphics{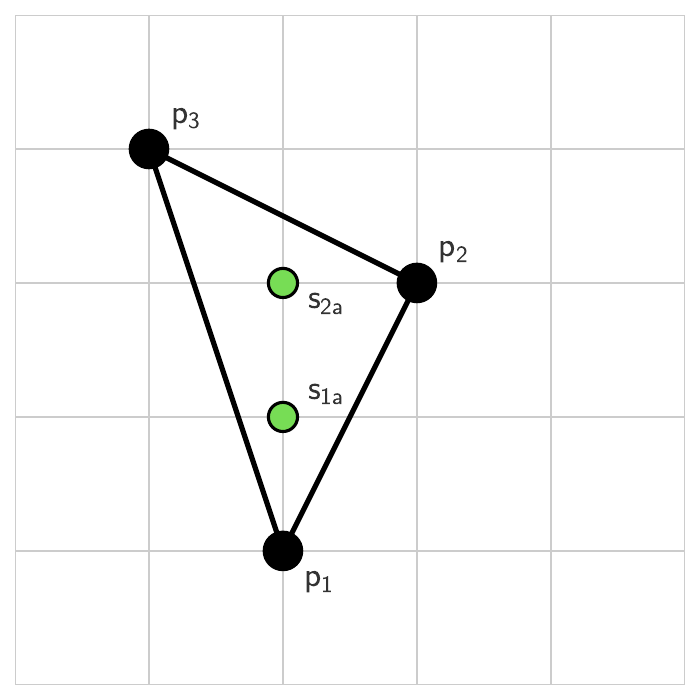}};
    \node[anchor=base west,inner sep=0pt,outer sep=0pt,text=black,overlay] at (147.94,121.33) {\font\pmcountfont=cmss10 at 16bp\pmcountfont 5};
    \node[anchor=base west,inner sep=0pt,outer sep=0pt,text=black,overlay] at (147.94,205.83) {\font\pmcountfont=cmss10 at 16bp\pmcountfont 5};
  \end{tikzpicture}}
  \caption{}
  \label{f_model_01_1_toric}
\end{subfigure}
\hfill
\begin{subfigure}[b]{0.31\textwidth}
  \centering
  \includegraphics[width=\linewidth]{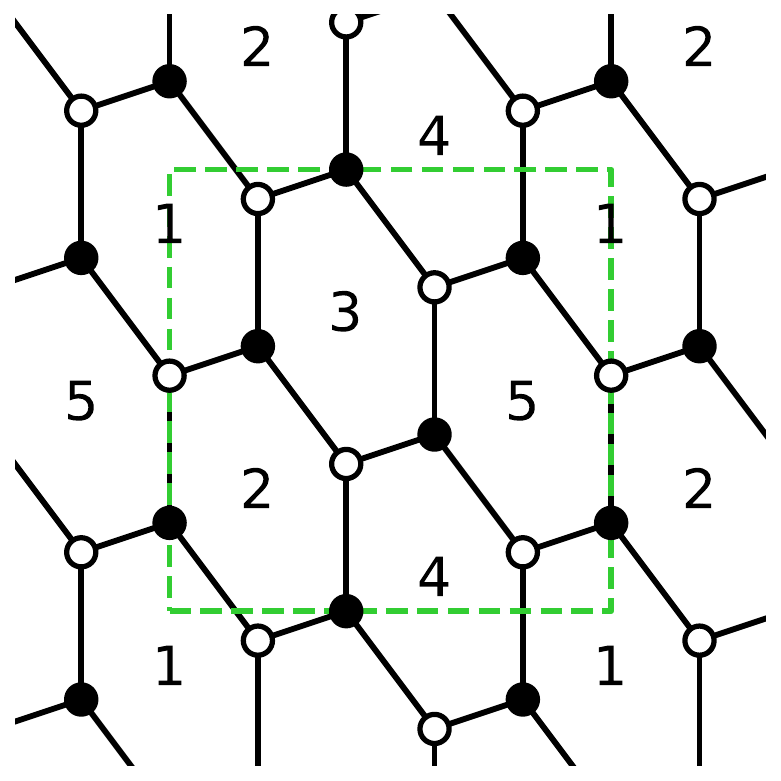}
  \caption{}
  \label{f_model_01_1_tiling}
\end{subfigure}
\caption{(a) Quiver, (b) toric diagram and (c) brane tiling for Model~1 ($\mathbb{C}^3/\mathbb{Z}_{5}\,(1,1,3)$). The numbers in (b) indicate the multiplicity of each internal point.}
\label{f_model_01_1}
\end{figure}

The superpotential of Model~1 ($\mathbb{C}^3/\mathbb{Z}_{5}\,(1,1,3)$) is given by,
\beal{es01a001}
\begin{aligned}
  W \;=\; & X^{1}_{12}\,X^{2}_{23}\,X_{31} + X^{1}_{23}\,X^{2}_{34}\,X_{42} + X^{1}_{34}\,X^{2}_{45}\,X_{53} \\
  & + X^{1}_{45}\,X^{2}_{51}\,X_{14} + X^{1}_{51}\,X^{2}_{12}\,X_{25} - X^{1}_{12}\,X_{25}\,X^{2}_{51} \\
  & - X^{1}_{23}\,X_{31}\,X^{2}_{12} - X^{1}_{34}\,X_{42}\,X^{2}_{23} - X^{1}_{45}\,X_{53}\,X^{2}_{34} \\
  & - X^{1}_{51}\,X_{14}\,X^{2}_{45} ~.~
\end{aligned}
\eea
The brane tiling has no quadrilateral faces, so that it has no toric dual and the model has a single toric phase.
Using the forward algorithm, we obtain the $P$-matrix as follows,
\beal{es01a002}
\resizebox{\ifdim\width>\linewidth\linewidth\else\width\fi}{!}{$
P \;=\; {\tiny
\left(
\begin{array}{c|ccc|ccccc|ccccc}
   & p_{1} & p_{2} & p_{3} & s_{11} & s_{12} & s_{13} & s_{14} & s_{15} & s_{21} & s_{22} & s_{23} & s_{24} & s_{25} \\
  \hline
  X^{1}_{12} & 0 & 0 & 1 & 1 & 0 & 0 & 0 & 0 & 1 & 1 & 0 & 0 & 0 \\
  X^{1}_{23} & 0 & 0 & 1 & 0 & 1 & 0 & 0 & 0 & 0 & 0 & 1 & 1 & 0 \\
  X^{1}_{34} & 0 & 0 & 1 & 0 & 0 & 1 & 0 & 0 & 1 & 0 & 0 & 0 & 1 \\
  X^{1}_{45} & 0 & 0 & 1 & 0 & 0 & 0 & 1 & 0 & 0 & 1 & 1 & 0 & 0 \\
  X^{1}_{51} & 0 & 0 & 1 & 0 & 0 & 0 & 0 & 1 & 0 & 0 & 0 & 1 & 1 \\
  X^{2}_{12} & 0 & 1 & 0 & 1 & 0 & 0 & 0 & 0 & 1 & 1 & 0 & 0 & 0 \\
  X^{2}_{23} & 0 & 1 & 0 & 0 & 1 & 0 & 0 & 0 & 0 & 0 & 1 & 1 & 0 \\
  X^{2}_{34} & 0 & 1 & 0 & 0 & 0 & 1 & 0 & 0 & 1 & 0 & 0 & 0 & 1 \\
  X^{2}_{45} & 0 & 1 & 0 & 0 & 0 & 0 & 1 & 0 & 0 & 1 & 1 & 0 & 0 \\
  X^{2}_{51} & 0 & 1 & 0 & 0 & 0 & 0 & 0 & 1 & 0 & 0 & 0 & 1 & 1 \\
  X_{14} & 1 & 0 & 0 & 1 & 1 & 1 & 0 & 0 & 1 & 0 & 0 & 0 & 0 \\
  X_{25} & 1 & 0 & 0 & 0 & 1 & 1 & 1 & 0 & 0 & 0 & 1 & 0 & 0 \\
  X_{31} & 1 & 0 & 0 & 0 & 0 & 1 & 1 & 1 & 0 & 0 & 0 & 0 & 1 \\
  X_{42} & 1 & 0 & 0 & 1 & 0 & 0 & 1 & 1 & 0 & 1 & 0 & 0 & 0 \\
  X_{53} & 1 & 0 & 0 & 1 & 1 & 0 & 0 & 1 & 0 & 0 & 0 & 1 & 0 \\
\end{array}
\right)}~,~
$}
\eea
where $p_a$ are GLSM fields corresponding to the extremal vertices in the toric diagram.
From the $P$-matrix we obtain the $F$-term charge matrix as follows,
\beal{es01a003}
\resizebox{\ifdim\width>\linewidth\linewidth\else\width\fi}{!}{$
Q_{F} \;=\; {\tiny
\left(
\begin{array}{ccc|ccccc|ccccc}
  p_{1} & p_{2} & p_{3} & s_{11} & s_{12} & s_{13} & s_{14} & s_{15} & s_{21} & s_{22} & s_{23} & s_{24} & s_{25} \\
  \hline
  0 & 0 & 0 & -1 & 1 & 0 & 0 & 0 & 0 & 1 & -1 & 0 & 0 \\
  0 & 0 & 0 & -1 & 0 & 1 & 0 & 0 & 0 & 1 & -1 & 1 & -1 \\
  0 & 0 & 0 & -1 & 0 & 0 & 1 & 0 & 1 & 0 & -1 & 1 & -1 \\
  0 & 0 & 0 & -1 & 0 & 0 & 0 & 1 & 1 & 0 & 0 & 0 & -1 \\
  0 & 1 & 1 & 1 & 0 & 0 & 0 & 0 & -1 & -1 & 0 & -1 & 0 \\
  1 & 0 & 0 & -2 & 0 & 0 & 0 & 0 & 1 & 1 & -1 & 1 & -1 \\
\end{array}
\right)}~.~
$}
\eea
We also have the $D$-term charge matrix, which is given by,
\beal{es01a004}
\resizebox{\ifdim\width>\linewidth\linewidth\else\width\fi}{!}{$
Q_{D} \;=\; {\tiny
\left(
\begin{array}{ccc|ccccc|ccccc}
  p_{1} & p_{2} & p_{3} & s_{11} & s_{12} & s_{13} & s_{14} & s_{15} & s_{21} & s_{22} & s_{23} & s_{24} & s_{25} \\
  \hline
  0 & 0 & 0 & 0 & 0 & 0 & 0 & 0 & -1 & 0 & 0 & 0 & 1 \\
  0 & 0 & 0 & 0 & 0 & 0 & 0 & 0 & 0 & 1 & -1 & 0 & 0 \\
  0 & 0 & 0 & 0 & 0 & 0 & 0 & 0 & 0 & 0 & 0 & 1 & -1 \\
  0 & 0 & 0 & 0 & 0 & 0 & 0 & 0 & 1 & -1 & 0 & 0 & 0 \\
\end{array}
\right)}~.~
$}
\eea
The toric diagram of the $\mathbb{C}^3/\mathbb{Z}_{5}$ model is given by,
\beal{es01a005}
\resizebox{\ifdim\width>\linewidth\linewidth\else\width\fi}{!}{$
G_t \;=\; {\small
\left(
\begin{array}{ccc|c|c}
  p_{1} & p_{2} & p_{3} & s_{1a} & s_{2a} \\
  \hline
  1 & 2 & 0 & 1 & 1 \\
  0 & 2 & 3 & 1 & 2 \\
  1 & 1 & 1 & 1 & 1 \\
\end{array}
\right)}~.~
$}
\eea
The corresponding toric diagram is shown in \fref{f_model_01_1_toric}.
The global symmetry of the brane tiling associated with $\mathbb{C}^3/\mathbb{Z}_{5}$ is given by,
\beal{es01a006}
SU(2) \times U(1) \times U(1)_R~.~
\eea
The global symmetry charges on the extremal GLSM fields $p_a$ are summarized in \tref{t_sym_geom_01}.

\begin{table}[H]
\centering
{\small
\begin{tabular}{|c|c|c|c|c|l|}
\hline
  & $SU(2)$ & $U(1)$ & $U(1)_R$ & $r_a$ & fugacities \\
\hline
  $p_{1}$ & $0$ & $-2$ & $r_{1}$ & $r_{1} = 2/3$ & $t_{1} = f^{-2} \bar{t}$ \\
  $p_{2}$ & $1$ & $1$ & $r_{2}$ & $r_{2} = 2/3$ & $t_{2} = x f \bar{t}$ \\
  $p_{3}$ & $-1$ & $1$ & $r_{3}$ & $r_{3} = 2/3$ & $t_{3} = x^{-1} f \bar{t}$ \\
\hline
\end{tabular}
}
\caption{Global symmetry charges on the extremal perfect matchings of Model~1 ($\mathbb{C}^3/\mathbb{Z}_{5}\,(1,1,3)$).}
\label{t_sym_geom_01}
\end{table}

The unrefined Hilbert series of the mesonic moduli space is given by,
\beal{es01a007}
\resizebox{\ifdim\width>\linewidth\linewidth\else\width\fi}{!}{$
g(t; \mathcal{M}^{mes}) \;=\; \frac{1 - t + 3 t^{3} - t^{4} + 3 t^{5} - t^{7} + t^{8}}{\left(1 - t\right) \left(1 - t^{5}\right)^{2}}~,~
$}
\eea
where $t$ is the fugacity counting degrees in extremal perfect matchings $p_a$, which corresponds to setting every fugacity $t_a$ in \tref{t_sym_geom_01} equal to $t$.
The plethystic logarithm of the refined Hilbert series takes the form,
\beal{es01a008}
PL\left[g(t_a;\mathcal{M}^{mes})\right] &=& t_{1} t_{3}^{2} + t_{1} t_{2} t_{3} + t_{1} t_{2}^{2} + t_{1}^{3} t_{3} + t_{1}^{3} t_{2} + t_{3}^{5} + t_{2} t_{3}^{4} + t_{2}^{2} t_{3}^{3} + t_{2}^{3} t_{3}^{2} \nonumber\\
&& {}+ t_{2}^{4} t_{3} + t_{2}^{5} + t_{1}^{5} \nonumber\\
&& {}- t_{1}^{2} t_{2}^{2} t_{3}^{2} - t_{1}^{4} t_{2}^{2} t_{3} - t_{1}^{4} t_{2} t_{3}^{2} - t_{1}^{6} t_{2}^{2} - t_{1}^{6} t_{2} t_{3} - t_{1}^{6} t_{3}^{2} - t_{1} t_{2}^{6} t_{3} \nonumber\\
&& {}- 2\, t_{1} t_{2}^{5} t_{3}^{2} - 2\, t_{1} t_{2}^{4} t_{3}^{3} - 2\, t_{1} t_{2}^{3} t_{3}^{4} - 2\, t_{1} t_{2}^{2} t_{3}^{5} - t_{1} t_{2} t_{3}^{6} - t_{1}^{3} t_{2}^{6} \nonumber\\
&& {}- 2\, t_{1}^{3} t_{2}^{5} t_{3} - 2\, t_{1}^{3} t_{2}^{4} t_{3}^{2} - 2\, t_{1}^{3} t_{2}^{3} t_{3}^{3} - 2\, t_{1}^{3} t_{2}^{2} t_{3}^{4} - 2\, t_{1}^{3} t_{2} t_{3}^{5} - t_{1}^{3} t_{3}^{6} \nonumber\\
&& {}- t_{1}^{5} t_{2}^{5} - t_{1}^{5} t_{2}^{4} t_{3} - t_{2}^{8} t_{3}^{2} - t_{1}^{5} t_{2}^{3} t_{3}^{2} - t_{2}^{7} t_{3}^{3} - 2\, t_{2}^{6} t_{3}^{4} - t_{1}^{5} t_{2}^{2} t_{3}^{3} \nonumber\\
&& {}- 2\, t_{2}^{5} t_{3}^{5} - 2\, t_{2}^{4} t_{3}^{6} - t_{1}^{5} t_{2} t_{3}^{4} - t_{2}^{3} t_{3}^{7} - t_{2}^{2} t_{3}^{8} - t_{1}^{5} t_{3}^{5} \nonumber\\
&& {}+ \dots~,~
\eea
where $t_a$ is the fugacity of the extremal perfect matching $p_a$, which can be expressed in terms of the global symmetry fugacities $x, f$ and $\bar{t}$ as shown in \tref{t_sym_geom_01}.

The first positive terms of the plethystic logarithm correspond to the generators of the mesonic moduli space, $A_{1}, \dots, A_{12}$. \tref{t_model_01_1} lists the generators in terms of the chiral fields of the brane tiling. The negative terms that follow correspond to the first order relations amongst the generators, which take the following~form,
\beal{es01a009}
&& A_{2}^{2} = A_{1} A_{3} ~,~ \quad A_{3} A_{4} = A_{2} A_{5} ~,~ \quad A_{2} A_{4} = A_{1} A_{5} ~,~ \quad A_{5}^{2} = A_{3} A_{12} ~,~ \nonumber\\
&& A_{4} A_{5} = A_{2} A_{12} ~,~ \quad A_{4}^{2} = A_{1} A_{12} ~,~ \quad A_{3} A_{10} = A_{2} A_{11} ~,~ \nonumber\\
&& A_{3} A_{9} = A_{2} A_{10} = A_{1} A_{11} ~,~ \quad A_{3} A_{8} = A_{2} A_{9} = A_{1} A_{10} ~,~ \nonumber\\
&& A_{3} A_{7} = A_{2} A_{8} = A_{1} A_{9} ~,~ \quad A_{3} A_{6} = A_{2} A_{7} = A_{1} A_{8} ~,~ \quad A_{2} A_{6} = A_{1} A_{7} ~,~ \nonumber\\
&& A_{5} A_{11} = A_{3}^{3} ~,~ \quad A_{5} A_{10} = A_{4} A_{11} = A_{2} A_{3}^{2} ~,~ \quad A_{5} A_{9} = A_{4} A_{10} = A_{1} A_{3}^{2} ~,~ \nonumber\\
&& A_{5} A_{8} = A_{4} A_{9} = A_{1} A_{2} A_{3} ~,~ \quad A_{5} A_{7} = A_{4} A_{8} = A_{1}^{2} A_{3} ~,~ \nonumber\\
&& A_{5} A_{6} = A_{4} A_{7} = A_{1}^{2} A_{2} ~,~ \quad A_{4} A_{6} = A_{1}^{3} ~,~ \quad A_{11} A_{12} = A_{3}^{2} A_{5} ~,~ \nonumber\\
&& A_{10} A_{12} = A_{2} A_{3} A_{5} ~,~ \quad A_{10}^{2} = A_{9} A_{11} ~,~ \quad A_{9} A_{12} = A_{1} A_{3} A_{5} ~,~ \nonumber\\
&& A_{9} A_{10} = A_{8} A_{11} ~,~ \quad A_{9}^{2} = A_{8} A_{10} = A_{7} A_{11} ~,~ \quad A_{8} A_{12} = A_{1} A_{2} A_{5} ~,~ \nonumber\\
&& A_{8} A_{9} = A_{7} A_{10} = A_{6} A_{11} ~,~ \quad A_{8}^{2} = A_{7} A_{9} = A_{6} A_{10} ~,~ \quad A_{7} A_{12} = A_{1}^{2} A_{5} ~,~ \nonumber\\
&& A_{7} A_{8} = A_{6} A_{9} ~,~ \quad A_{7}^{2} = A_{6} A_{8} ~,~ \quad A_{6} A_{12} = A_{1}^{2} A_{4}~.~
\eea

\begin{table}[H]
\centering
\resizebox{0.7\textwidth}{!}{
\begin{tabular}{|c|c|}
\hline
  Generators & Generators (in terms of chiral fields) \\
\hline
  $A_{1} = p_{1} p_{3}^{2}$ & $X^{1}_{12} X^{1}_{23} X_{31} = X^{1}_{12} X_{25} X^{1}_{51} = X^{1}_{23} X^{1}_{34} X_{42} = X^{1}_{34} X^{1}_{45} X_{53} = X^{1}_{45} X^{1}_{51} X_{14}$ \\
\hline
  $A_{2} = p_{1} p_{2} p_{3}$ & $\begin{array}{c}
    X^{1}_{12} X^{2}_{23} X_{31} = X^{1}_{12} X_{25} X^{2}_{51} = X^{1}_{23} X_{31} X^{2}_{12} = X^{1}_{23} X^{2}_{34} X_{42} = X^{1}_{34} X_{42} X^{2}_{23} \\
    = X^{1}_{34} X^{2}_{45} X_{53} = X^{1}_{45} X_{53} X^{2}_{34} = X^{1}_{45} X^{2}_{51} X_{14} = X^{1}_{51} X^{2}_{12} X_{25} \\
    = X^{1}_{51} X_{14} X^{2}_{45}
  \end{array}$ \\
\hline
  $A_{3} = p_{1} p_{2}^{2}$ & $X^{2}_{12} X^{2}_{23} X_{31} = X^{2}_{12} X_{25} X^{2}_{51} = X^{2}_{23} X^{2}_{34} X_{42} = X^{2}_{34} X^{2}_{45} X_{53} = X^{2}_{45} X^{2}_{51} X_{14}$ \\
\hline
  $A_{4} = p_{1}^{3} p_{3}$ & $\begin{array}{c}
    X^{1}_{12} X_{25} X_{53} X_{31} = X^{1}_{23} X_{31} X_{14} X_{42} = X^{1}_{34} X_{42} X_{25} X_{53} = X^{1}_{45} X_{53} X_{31} X_{14} \\
    = X^{1}_{51} X_{14} X_{42} X_{25}
  \end{array}$ \\
\hline
  $A_{5} = p_{1}^{3} p_{2}$ & $\begin{array}{c}
    X^{2}_{12} X_{25} X_{53} X_{31} = X^{2}_{23} X_{31} X_{14} X_{42} = X^{2}_{34} X_{42} X_{25} X_{53} = X^{2}_{45} X_{53} X_{31} X_{14} \\
    = X^{2}_{51} X_{14} X_{42} X_{25}
  \end{array}$ \\
\hline
  $A_{6} = p_{3}^{5}$ & $X^{1}_{12} X^{1}_{23} X^{1}_{34} X^{1}_{45} X^{1}_{51}$ \\
\hline
  $A_{7} = p_{2} p_{3}^{4}$ & $\begin{array}{c}
    X^{1}_{12} X^{1}_{23} X^{1}_{34} X^{1}_{45} X^{2}_{51} = X^{1}_{12} X^{1}_{23} X^{1}_{34} X^{2}_{45} X^{1}_{51} = X^{1}_{12} X^{1}_{23} X^{2}_{34} X^{1}_{45} X^{1}_{51} \\
    = X^{1}_{12} X^{2}_{23} X^{1}_{34} X^{1}_{45} X^{1}_{51} = X^{1}_{23} X^{1}_{34} X^{1}_{45} X^{1}_{51} X^{2}_{12}
  \end{array}$ \\
\hline
  $A_{8} = p_{2}^{2} p_{3}^{3}$ & $\begin{array}{c}
    X^{1}_{12} X^{1}_{23} X^{1}_{34} X^{2}_{45} X^{2}_{51} = X^{1}_{12} X^{1}_{23} X^{2}_{34} X^{1}_{45} X^{2}_{51} = X^{1}_{12} X^{1}_{23} X^{2}_{34} X^{2}_{45} X^{1}_{51} \\
    = X^{1}_{12} X^{2}_{23} X^{1}_{34} X^{1}_{45} X^{2}_{51} = X^{1}_{12} X^{2}_{23} X^{1}_{34} X^{2}_{45} X^{1}_{51} = X^{1}_{12} X^{2}_{23} X^{2}_{34} X^{1}_{45} X^{1}_{51} \\
    = X^{1}_{23} X^{1}_{34} X^{1}_{45} X^{2}_{51} X^{2}_{12} = X^{1}_{23} X^{1}_{34} X^{2}_{45} X^{1}_{51} X^{2}_{12} = X^{1}_{23} X^{2}_{34} X^{1}_{45} X^{1}_{51} X^{2}_{12} \\
    = X^{1}_{34} X^{1}_{45} X^{1}_{51} X^{2}_{12} X^{2}_{23}
  \end{array}$ \\
\hline
  $A_{9} = p_{2}^{3} p_{3}^{2}$ & $\begin{array}{c}
    X^{1}_{12} X^{1}_{23} X^{2}_{34} X^{2}_{45} X^{2}_{51} = X^{1}_{12} X^{2}_{23} X^{1}_{34} X^{2}_{45} X^{2}_{51} = X^{1}_{12} X^{2}_{23} X^{2}_{34} X^{1}_{45} X^{2}_{51} \\
    = X^{1}_{12} X^{2}_{23} X^{2}_{34} X^{2}_{45} X^{1}_{51} = X^{1}_{23} X^{1}_{34} X^{2}_{45} X^{2}_{51} X^{2}_{12} = X^{1}_{23} X^{2}_{34} X^{1}_{45} X^{2}_{51} X^{2}_{12} \\
    = X^{1}_{23} X^{2}_{34} X^{2}_{45} X^{1}_{51} X^{2}_{12} = X^{1}_{34} X^{1}_{45} X^{2}_{51} X^{2}_{12} X^{2}_{23} = X^{1}_{34} X^{2}_{45} X^{1}_{51} X^{2}_{12} X^{2}_{23} \\
    = X^{1}_{45} X^{1}_{51} X^{2}_{12} X^{2}_{23} X^{2}_{34}
  \end{array}$ \\
\hline
  $A_{10} = p_{2}^{4} p_{3}$ & $\begin{array}{c}
    X^{1}_{12} X^{2}_{23} X^{2}_{34} X^{2}_{45} X^{2}_{51} = X^{1}_{23} X^{2}_{34} X^{2}_{45} X^{2}_{51} X^{2}_{12} = X^{1}_{34} X^{2}_{45} X^{2}_{51} X^{2}_{12} X^{2}_{23} \\
    = X^{1}_{45} X^{2}_{51} X^{2}_{12} X^{2}_{23} X^{2}_{34} = X^{1}_{51} X^{2}_{12} X^{2}_{23} X^{2}_{34} X^{2}_{45}
  \end{array}$ \\
\hline
  $A_{11} = p_{2}^{5}$ & $X^{2}_{12} X^{2}_{23} X^{2}_{34} X^{2}_{45} X^{2}_{51}$ \\
\hline
  $A_{12} = p_{1}^{5}$ & $X_{14} X_{42} X_{25} X_{53} X_{31}$ \\
\hline
\end{tabular}}
\caption{Generators of the mesonic moduli space of Model~1 ($\mathbb{C}^3/\mathbb{Z}_{5}\,(1,1,3)$) in terms of the extremal perfect matchings and of the chiral fields.}
\label{t_model_01_1}
\end{table}

\subsection{Model 2: $\mathbb{C}^3/\mathbb{Z}_{6}~(1,1,4)$}\label{s_m2}
\input{model_02_1}

\subsection{Model 3: $\mathbb{C}^3/\mathbb{Z}_{8}~(1,3,4)$}\label{s_m3}
\input{model_03_1}

\subsection{Model 4: $\mathbb{C}^3/\mathbb{Z}_{10}~(1,4,5)$}\label{s_m4}
\input{model_04_1}

\subsection{Model 5: $\mathbb{C}^3/\mathbb{Z}_{2} \times \mathbb{Z}_{6}~(0,1,1)(1,0,5)$}\label{s_m5}
\input{model_05_1}

\section{Quadrilaterals}\label{s_c4}

\subsection{Model 6: $L^{1,5,2}$}\label{s_m6}

\begin{figure}[H]
\centering
\begin{subfigure}[b]{0.31\textwidth}
  \centering
  \includegraphics[width=\linewidth]{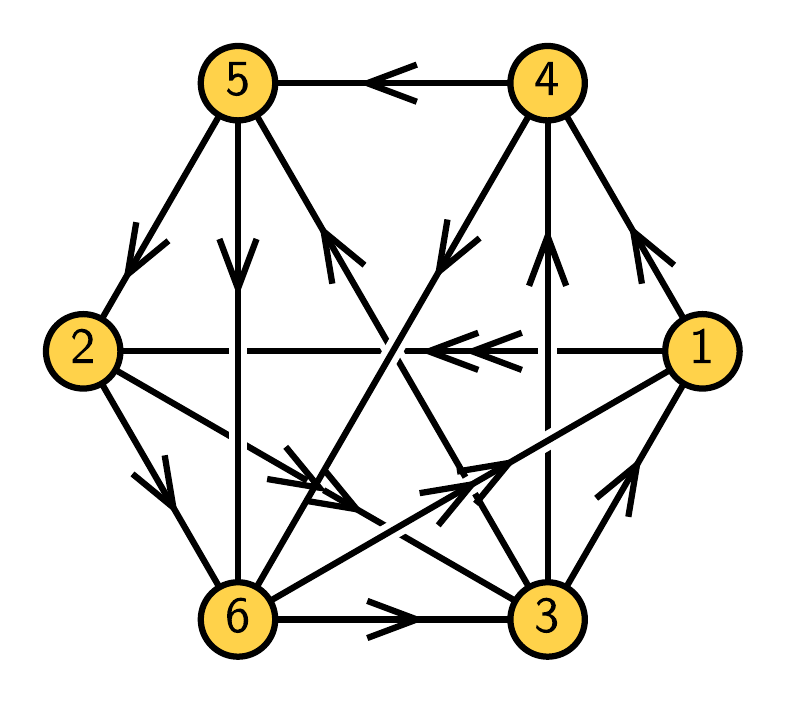}
  \caption{}
  \label{f_model_06_1_quiver}
\end{subfigure}
\hfill
\begin{subfigure}[b]{0.31\textwidth}
  \centering
  \resizebox{\linewidth}{!}{%
  \begin{tikzpicture}[x=1bp,y=1bp]
    \node[anchor=south west,inner sep=0pt,outer sep=0pt] at (0,0) {\includegraphics{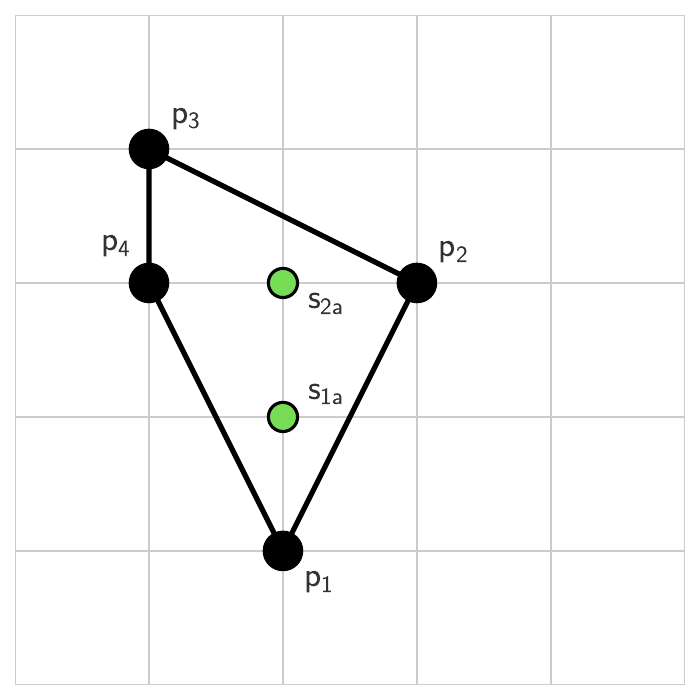}};
    \node[anchor=base west,inner sep=0pt,outer sep=0pt,text=black,overlay] at (147.94,121.33) {\font\pmcountfont=cmss10 at 16bp\pmcountfont 6};
    \node[anchor=base west,inner sep=0pt,outer sep=0pt,text=black,overlay] at (147.94,205.83) {\font\pmcountfont=cmss10 at 16bp\pmcountfont 6};
  \end{tikzpicture}}
  \caption{}
  \label{f_model_06_1_toric}
\end{subfigure}
\hfill
\begin{subfigure}[b]{0.31\textwidth}
  \centering
  \includegraphics[width=\linewidth]{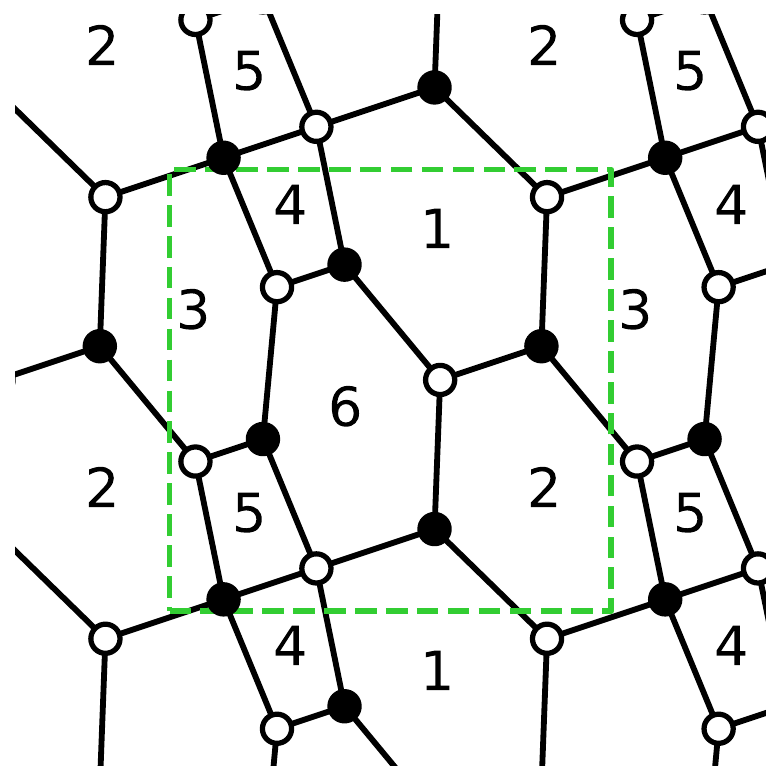}
  \caption{}
  \label{f_model_06_1_tiling}
\end{subfigure}
\caption{(a) Quiver, (b) toric diagram and (c) brane tiling for Model~6 ($L^{1,5,2}$). The numbers in (b) indicate the multiplicity of each internal point.}
\label{f_model_06_1}
\end{figure}

The superpotential of Model~6 ($L^{1,5,2}$) is given by,
\beal{es06a001}
\begin{aligned}
  W \;=\; & X^{2}_{12}\,X^{1}_{23}\,X_{31} + X^{2}_{23}\,X_{35}\,X_{52} + X_{34}\,X_{46}\,X_{63} \\
  & + X^{1}_{61}\,X^{1}_{12}\,X_{26} + X_{14}\,X_{45}\,X_{56}\,X^{2}_{61} - X^{2}_{12}\,X_{26}\,X^{2}_{61} \\
  & - X_{56}\,X_{63}\,X_{35} - X^{1}_{61}\,X_{14}\,X_{46} - X^{1}_{12}\,X^{2}_{23}\,X_{31} \\
  & - X^{1}_{23}\,X_{34}\,X_{45}\,X_{52} ~.~
\end{aligned}
\eea
Toric duality on the quadrilateral faces $4$ and $5$ of the brane tiling leads back to the same toric phase, which is therefore the only toric phase of the model.
The $P$-matrix takes the following form,
\beal{es06a002}
\resizebox{\ifdim\width>\linewidth\linewidth\else\width\fi}{!}{$
P \;=\; {\tiny
\left(
\begin{array}{c|cccc|cccccc|cccccc}
   & p_{1} & p_{2} & p_{3} & p_{4} & s_{11} & s_{12} & s_{13} & s_{14} & s_{15} & s_{16} & s_{21} & s_{22} & s_{23} & s_{24} & s_{25} & s_{26} \\
  \hline
  X^{1}_{12} & 0 & 1 & 0 & 0 & 1 & 0 & 0 & 0 & 0 & 0 & 1 & 1 & 1 & 0 & 0 & 0 \\
  X^{2}_{12} & 0 & 0 & 1 & 1 & 1 & 0 & 0 & 0 & 0 & 0 & 1 & 1 & 1 & 0 & 0 & 0 \\
  X_{14} & 1 & 0 & 0 & 0 & 1 & 1 & 1 & 0 & 0 & 0 & 1 & 0 & 0 & 0 & 0 & 0 \\
  X^{1}_{23} & 0 & 1 & 0 & 0 & 0 & 1 & 0 & 0 & 0 & 0 & 0 & 0 & 0 & 1 & 1 & 0 \\
  X^{2}_{23} & 0 & 0 & 1 & 1 & 0 & 1 & 0 & 0 & 0 & 0 & 0 & 0 & 0 & 1 & 1 & 0 \\
  X_{26} & 1 & 0 & 0 & 0 & 0 & 1 & 1 & 1 & 1 & 0 & 0 & 0 & 0 & 1 & 0 & 0 \\
  X_{31} & 1 & 0 & 0 & 0 & 0 & 0 & 1 & 1 & 1 & 1 & 0 & 0 & 0 & 0 & 0 & 1 \\
  X_{34} & 0 & 0 & 1 & 0 & 0 & 0 & 1 & 0 & 0 & 0 & 1 & 0 & 0 & 0 & 0 & 1 \\
  X_{35} & 0 & 1 & 0 & 0 & 0 & 0 & 1 & 1 & 0 & 0 & 1 & 1 & 0 & 0 & 0 & 1 \\
  X_{45} & 0 & 0 & 0 & 1 & 0 & 0 & 0 & 1 & 0 & 0 & 0 & 1 & 0 & 0 & 0 & 0 \\
  X_{46} & 0 & 1 & 0 & 0 & 0 & 0 & 0 & 1 & 1 & 0 & 0 & 1 & 1 & 1 & 0 & 0 \\
  X_{52} & 1 & 0 & 0 & 0 & 1 & 0 & 0 & 0 & 1 & 1 & 0 & 0 & 1 & 0 & 0 & 0 \\
  X_{56} & 0 & 0 & 1 & 0 & 0 & 0 & 0 & 0 & 1 & 0 & 0 & 0 & 1 & 1 & 0 & 0 \\
  X^{1}_{61} & 0 & 0 & 1 & 1 & 0 & 0 & 0 & 0 & 0 & 1 & 0 & 0 & 0 & 0 & 1 & 1 \\
  X^{2}_{61} & 0 & 1 & 0 & 0 & 0 & 0 & 0 & 0 & 0 & 1 & 0 & 0 & 0 & 0 & 1 & 1 \\
  X_{63} & 1 & 0 & 0 & 1 & 1 & 1 & 0 & 0 & 0 & 1 & 0 & 0 & 0 & 0 & 1 & 0 \\
\end{array}
\right)}~,~
$}
\eea
where $p_a$ are GLSM fields corresponding to the extremal vertices in the toric diagram.
The $F$-term charge matrix takes the following form,
\beal{es06a003}
\resizebox{\ifdim\width>\linewidth\linewidth\else\width\fi}{!}{$
Q_{F} \;=\; {\tiny
\left(
\begin{array}{cccc|cccccc|cccccc}
  p_{1} & p_{2} & p_{3} & p_{4} & s_{11} & s_{12} & s_{13} & s_{14} & s_{15} & s_{16} & s_{21} & s_{22} & s_{23} & s_{24} & s_{25} & s_{26} \\
  \hline
  0 & 0 & -1 & 1 & -1 & 0 & 0 & 0 & 0 & 0 & 1 & -1 & 1 & 0 & 0 & 0 \\
  1 & 0 & 0 & 0 & -1 & 0 & -1 & 0 & 0 & 0 & 1 & 0 & 0 & 0 & 0 & 0 \\
  0 & 0 & 0 & 0 & 1 & -1 & 0 & 0 & 0 & 0 & 0 & 0 & -1 & 1 & 0 & 0 \\
  0 & -1 & -1 & 0 & -1 & 0 & 0 & 0 & 0 & 0 & 1 & 0 & 1 & 0 & 1 & 0 \\
  0 & 0 & 0 & 0 & 0 & 0 & -1 & 1 & 0 & 0 & 1 & -1 & 0 & 0 & 0 & 0 \\
  0 & 0 & 0 & 0 & 0 & 0 & -1 & 0 & 1 & 0 & 1 & 0 & -1 & 0 & 0 & 0 \\
  0 & -1 & -1 & 0 & -1 & 1 & -1 & 0 & 0 & 0 & 1 & 0 & 1 & 0 & 0 & 1 \\
  0 & -1 & -1 & 0 & -2 & 1 & -1 & 0 & 0 & 1 & 2 & 0 & 1 & 0 & 0 & 0 \\
\end{array}
\right)}~,~
$}
\eea
and the $D$-term charge matrix takes the form,
\beal{es06a004}
\resizebox{\ifdim\width>\linewidth\linewidth\else\width\fi}{!}{$
Q_{D} \;=\; {\tiny
\left(
\begin{array}{cccc|cccccc|cccccc}
  p_{1} & p_{2} & p_{3} & p_{4} & s_{11} & s_{12} & s_{13} & s_{14} & s_{15} & s_{16} & s_{21} & s_{22} & s_{23} & s_{24} & s_{25} & s_{26} \\
  \hline
  0 & 1 & 1 & 0 & 1 & -1 & 1 & 0 & 0 & 0 & -2 & 0 & -1 & 0 & 0 & 0 \\
  0 & 0 & 0 & 0 & 1 & -1 & 0 & 0 & 0 & 0 & 0 & 0 & 0 & 0 & 0 & 0 \\
  0 & 0 & 0 & 0 & 0 & 1 & -1 & 0 & 0 & 0 & 0 & 0 & 0 & 0 & 0 & 0 \\
  0 & 0 & 0 & 0 & 0 & 0 & 0 & 0 & 0 & 0 & 1 & -1 & 0 & 0 & 0 & 0 \\
  0 & 0 & 0 & 0 & 0 & 0 & 0 & 0 & 0 & 0 & 0 & 1 & -1 & 0 & 0 & 0 \\
\end{array}
\right)}~.~
$}
\eea
The toric diagram is given by,
\beal{es06a005}
\resizebox{\ifdim\width>\linewidth\linewidth\else\width\fi}{!}{$
G_t \;=\; {\small
\left(
\begin{array}{cccc|c|c}
  p_{1} & p_{2} & p_{3} & p_{4} & s_{1a} & s_{2a} \\
  \hline
  1 & 2 & 0 & 0 & 1 & 1 \\
  0 & 2 & 3 & 2 & 1 & 2 \\
  1 & 1 & 1 & 1 & 1 & 1 \\
\end{array}
\right)}~.~
$}
\eea
The corresponding toric diagram is shown in \fref{f_model_06_1_toric}.
The global symmetry of the brane tiling associated with $L^{1,5,2}$ is given by,
\beal{es06a006}
U(1) \times U(1) \times U(1)_R~.~
\eea
The global symmetry charges on the extremal GLSM fields $p_a$ are summarized in \tref{t_sym_geom_06}.

\begin{table}[H]
\centering
{\small
\begin{tabular}{|c|c|c|c|c|l|}
\hline
  & $U(1)_{1}$ & $U(1)_{2}$ & $U(1)_R$ & $r_a$ & fugacities \\
\hline
  $p_{1}$ & $-1$ & $-1$ & $r_{1}$ & $r_{1} = 0.5932$ & $t_{1} = f_{1}^{-1} f_{2}^{-1} \bar{t}_{1}$ \\
  $p_{2}$ & $1$ & $0$ & $r_{2}$ & $r_{2} = 0.6322$ & $t_{2} = f_{1} \bar{t}_{2}$ \\
  $p_{3}$ & $0$ & $1$ & $r_{3}$ & $r_{3} = 0.4882$ & $t_{3} = f_{2} \bar{t}_{3}$ \\
  $p_{4}$ & $0$ & $0$ & $r_{4}$ & $r_{4} = 0.2863$ & $t_{4} = \bar{t}_{4}$ \\
\hline
\end{tabular}
}
\caption{Global symmetry charges on the extremal perfect matchings of Model~6 ($L^{1,5,2}$).}
\label{t_sym_geom_06}
\end{table}

The corresponding unrefined Hilbert series of the mesonic moduli space takes the form,
\beal{es06a007}
\resizebox{\ifdim\width>\linewidth\linewidth\else\width\fi}{!}{$
g(t; \mathcal{M}^{mes}) \;=\; \frac{1 + t^{3} + t^{4} + 2 t^{5} + 3 t^{6} + 2 t^{7} + 3 t^{8} + 3 t^{9} + 2 t^{10} + 3 t^{11} + 2 t^{12} + t^{13} + t^{14} + t^{17}}{\left(1 - t^{5}\right) \left(1 - t^{7}\right) \left(1 - t^{9}\right)}~,~
$}
\eea
where $t$ is the fugacity counting degrees in extremal perfect matchings $p_a$, which corresponds to setting every fugacity $t_a$ in \tref{t_sym_geom_06} equal to $t$.
The plethystic logarithm of the refined Hilbert series takes the form,
\beal{es06a008}
PL\left[g(t_a;\mathcal{M}^{mes})\right] &=& t_{1} t_{2}^{2} + t_{1} t_{2} t_{3} t_{4} + t_{2}^{4} t_{3} + t_{1} t_{3}^{2} t_{4}^{2} + t_{1}^{3} t_{2} t_{4} + t_{2}^{3} t_{3}^{2} t_{4} + t_{1}^{3} t_{3} t_{4}^{2} \nonumber\\
&& {}+ t_{2}^{2} t_{3}^{3} t_{4}^{2} + t_{1}^{5} t_{4}^{2} + t_{2} t_{3}^{4} t_{4}^{3} + t_{3}^{5} t_{4}^{4} \nonumber\\
&& {}- t_{1}^{2} t_{2}^{2} t_{3}^{2} t_{4}^{2} - t_{1}^{4} t_{2}^{2} t_{3} t_{4}^{2} - t_{1} t_{2}^{5} t_{3}^{2} t_{4} - t_{1}^{6} t_{2}^{2} t_{4}^{2} - t_{1}^{4} t_{2} t_{3}^{2} t_{4}^{3} - t_{1}^{3} t_{2}^{5} t_{3} t_{4} \nonumber\\
&& {}- 2\, t_{1} t_{2}^{4} t_{3}^{3} t_{4}^{2} - t_{1}^{6} t_{2} t_{3} t_{4}^{3} - 2\, t_{1}^{3} t_{2}^{4} t_{3}^{2} t_{4}^{2} - 2\, t_{1} t_{2}^{3} t_{3}^{4} t_{4}^{3} - t_{1}^{6} t_{3}^{2} t_{4}^{4} \nonumber\\
&& {}- 2\, t_{1}^{3} t_{2}^{3} t_{3}^{3} t_{4}^{3} - t_{2}^{6} t_{3}^{4} t_{4}^{2} - 2\, t_{1} t_{2}^{2} t_{3}^{5} t_{4}^{4} - t_{1}^{5} t_{2}^{4} t_{3} t_{4}^{2} - 2\, t_{1}^{3} t_{2}^{2} t_{3}^{4} t_{4}^{4} \nonumber\\
&& {}- t_{1}^{5} t_{2}^{3} t_{3}^{2} t_{4}^{3} - t_{2}^{5} t_{3}^{5} t_{4}^{3} - t_{1} t_{2} t_{3}^{6} t_{4}^{5} - t_{1}^{5} t_{2}^{2} t_{3}^{3} t_{4}^{4} - 2\, t_{2}^{4} t_{3}^{6} t_{4}^{4} \nonumber\\
&& {}- 2\, t_{1}^{3} t_{2} t_{3}^{5} t_{4}^{5} - t_{1}^{5} t_{2} t_{3}^{4} t_{4}^{5} - t_{2}^{3} t_{3}^{7} t_{4}^{5} - t_{1}^{3} t_{3}^{6} t_{4}^{6} - t_{2}^{2} t_{3}^{8} t_{4}^{6} - t_{1}^{5} t_{3}^{5} t_{4}^{6} \nonumber\\
&& {}+ \dots~,~
\eea
where each extremal perfect matching $p_a$ of the toric diagram in \fref{f_model_06_1_toric} contributes the fugacity $t_a$, which as shown in \tref{t_sym_geom_06} can be expressed in terms of the global symmetry fugacities $f_{1}, f_{2}$ and $\bar{t}_{1}, \dots, \bar{t}_{4}$.

The first positive terms of the plethystic logarithm correspond to the generators of the mesonic moduli space, $A_{1}, \dots, A_{11}$. The expressions of the generators in terms of the chiral fields are summarized in \tref{t_model_06_1}. The next negative terms correspond to the first order relations amongst the generators, which take the following~form,
\beal{es06a009}
&& A_{2}^{2} = A_{1} A_{4} ~,~ \quad A_{2} A_{5} = A_{1} A_{7} ~,~ \quad A_{2} A_{3} = A_{1} A_{6} ~,~ \quad A_{5}^{2} = A_{1} A_{9} ~,~ \nonumber\\
&& A_{4} A_{5} = A_{2} A_{7} ~,~ \quad A_{3} A_{5} = A_{1}^{2} A_{2} ~,~ \quad A_{3} A_{4} = A_{2} A_{6} = A_{1} A_{8} ~,~ \nonumber\\
&& A_{5} A_{7} = A_{2} A_{9} ~,~ \quad A_{5} A_{6} = A_{3} A_{7} = A_{1}^{2} A_{4} ~,~ \quad A_{4} A_{6} = A_{2} A_{8} = A_{1} A_{10} ~,~ \nonumber\\
&& A_{7}^{2} = A_{4} A_{9} ~,~ \quad A_{6} A_{7} = A_{5} A_{8} = A_{1} A_{2} A_{4} ~,~ \quad A_{6}^{2} = A_{3} A_{8} ~,~ \nonumber\\
&& A_{4} A_{8} = A_{2} A_{10} = A_{1} A_{11} ~,~ \quad A_{3} A_{9} = A_{1}^{2} A_{7} ~,~ \quad A_{7} A_{8} = A_{5} A_{10} = A_{1} A_{4}^{2} ~,~ \nonumber\\
&& A_{6} A_{9} = A_{1} A_{2} A_{7} ~,~ \quad A_{6} A_{8} = A_{3} A_{10} ~,~ \quad A_{4} A_{10} = A_{2} A_{11} ~,~ \nonumber\\
&& A_{8} A_{9} = A_{1} A_{4} A_{7} ~,~ \quad A_{8}^{2} = A_{6} A_{10} = A_{3} A_{11} ~,~ \quad A_{7} A_{10} = A_{5} A_{11} = A_{2} A_{4}^{2} ~,~ \nonumber\\
&& A_{9} A_{10} = A_{2} A_{4} A_{7} ~,~ \quad A_{8} A_{10} = A_{6} A_{11} ~,~ \quad A_{7} A_{11} = A_{4}^{3} ~,~ \quad A_{10}^{2} = A_{8} A_{11} ~,~ \nonumber\\
&& A_{9} A_{11} = A_{4}^{2} A_{7}~.~
\eea

\begin{table}[H]
\centering
\resizebox{0.7\textwidth}{!}{
\begin{tabular}{|c|c|}
\hline
  Generators & Generators (in terms of chiral fields) \\
\hline
  $A_{1} = p_{1} p_{2}^{2}$ & $X^{1}_{12} X^{1}_{23} X_{31} = X^{1}_{12} X_{26} X^{2}_{61} = X_{14} X_{46} X^{2}_{61} = X^{1}_{23} X_{35} X_{52}$ \\
\hline
  $A_{2} = p_{1} p_{2} p_{3} p_{4}$ & $\begin{array}{c}
    X^{1}_{12} X^{2}_{23} X_{31} = X^{1}_{12} X_{26} X^{1}_{61} = X^{2}_{12} X^{1}_{23} X_{31} = X^{2}_{12} X_{26} X^{2}_{61} \\
    = X_{14} X_{46} X^{1}_{61} = X^{2}_{23} X_{35} X_{52} = X_{34} X_{46} X_{63} = X_{35} X_{56} X_{63} \\
    = X_{14} X_{45} X_{56} X^{2}_{61} = X^{1}_{23} X_{34} X_{45} X_{52}
  \end{array}$ \\
\hline
  $A_{3} = p_{2}^{4} p_{3}$ & $X^{1}_{12} X^{1}_{23} X_{34} X_{46} X^{2}_{61} = X^{1}_{12} X^{1}_{23} X_{35} X_{56} X^{2}_{61}$ \\
\hline
  $A_{4} = p_{1} p_{3}^{2} p_{4}^{2}$ & $\begin{array}{c}
    X^{2}_{12} X^{2}_{23} X_{31} = X^{2}_{12} X_{26} X^{1}_{61} = X_{14} X_{45} X_{56} X^{1}_{61} = X^{2}_{23} X_{34} X_{45} X_{52} \\
    = X_{34} X_{45} X_{56} X_{63}
  \end{array}$ \\
\hline
  $A_{5} = p_{1}^{3} p_{2} p_{4}$ & $\begin{array}{c}
    X^{1}_{12} X_{26} X_{63} X_{31} = X_{14} X_{46} X_{63} X_{31} = X_{26} X_{63} X_{35} X_{52} \\
    = X_{14} X_{45} X_{52} X^{1}_{23} X_{31} = X_{14} X_{45} X_{52} X_{26} X^{2}_{61}
  \end{array}$ \\
\hline
  $A_{6} = p_{2}^{3} p_{3}^{2} p_{4}$ & $\begin{array}{c}
    X^{1}_{12} X^{1}_{23} X_{34} X_{46} X^{1}_{61} = X^{1}_{12} X^{1}_{23} X_{35} X_{56} X^{1}_{61} = X^{1}_{12} X^{2}_{23} X_{34} X_{46} X^{2}_{61} \\
    = X^{1}_{12} X^{2}_{23} X_{35} X_{56} X^{2}_{61} = X^{2}_{12} X^{1}_{23} X_{34} X_{46} X^{2}_{61} = X^{2}_{12} X^{1}_{23} X_{35} X_{56} X^{2}_{61} \\
    = X^{1}_{12} X^{1}_{23} X_{34} X_{45} X_{56} X^{2}_{61}
  \end{array}$ \\
\hline
  $A_{7} = p_{1}^{3} p_{3} p_{4}^{2}$ & $\begin{array}{c}
    X^{2}_{12} X_{26} X_{63} X_{31} = X_{14} X_{45} X_{52} X^{2}_{23} X_{31} = X_{14} X_{45} X_{52} X_{26} X^{1}_{61} \\
    = X_{14} X_{45} X_{56} X_{63} X_{31} = X_{26} X_{63} X_{34} X_{45} X_{52}
  \end{array}$ \\
\hline
  $A_{8} = p_{2}^{2} p_{3}^{3} p_{4}^{2}$ & $\begin{array}{c}
    X^{1}_{12} X^{2}_{23} X_{34} X_{46} X^{1}_{61} = X^{1}_{12} X^{2}_{23} X_{35} X_{56} X^{1}_{61} = X^{2}_{12} X^{1}_{23} X_{34} X_{46} X^{1}_{61} \\
    = X^{2}_{12} X^{1}_{23} X_{35} X_{56} X^{1}_{61} = X^{2}_{12} X^{2}_{23} X_{34} X_{46} X^{2}_{61} = X^{2}_{12} X^{2}_{23} X_{35} X_{56} X^{2}_{61} \\
    = X^{1}_{12} X^{1}_{23} X_{34} X_{45} X_{56} X^{1}_{61} = X^{1}_{12} X^{2}_{23} X_{34} X_{45} X_{56} X^{2}_{61} \\
    = X^{2}_{12} X^{1}_{23} X_{34} X_{45} X_{56} X^{2}_{61}
  \end{array}$ \\
\hline
  $A_{9} = p_{1}^{5} p_{4}^{2}$ & $X_{14} X_{45} X_{52} X_{26} X_{63} X_{31}$ \\
\hline
  $A_{10} = p_{2} p_{3}^{4} p_{4}^{3}$ & $\begin{array}{c}
    X^{2}_{12} X^{2}_{23} X_{34} X_{46} X^{1}_{61} = X^{2}_{12} X^{2}_{23} X_{35} X_{56} X^{1}_{61} = X^{1}_{12} X^{2}_{23} X_{34} X_{45} X_{56} X^{1}_{61} \\
    = X^{2}_{12} X^{1}_{23} X_{34} X_{45} X_{56} X^{1}_{61} = X^{2}_{12} X^{2}_{23} X_{34} X_{45} X_{56} X^{2}_{61}
  \end{array}$ \\
\hline
  $A_{11} = p_{3}^{5} p_{4}^{4}$ & $X^{2}_{12} X^{2}_{23} X_{34} X_{45} X_{56} X^{1}_{61}$ \\
\hline
\end{tabular}}
\caption{Generators of the mesonic moduli space of Model~6 ($L^{1,5,2}$) in terms of the extremal perfect matchings and of the chiral fields.}
\label{t_model_06_1}
\end{table}

\subsection{Model 7: $Y^{3,2}$}\label{s_m7}

\begin{figure}[H]
\centering
\begin{subfigure}[b]{0.31\textwidth}
  \centering
  \includegraphics[width=\linewidth]{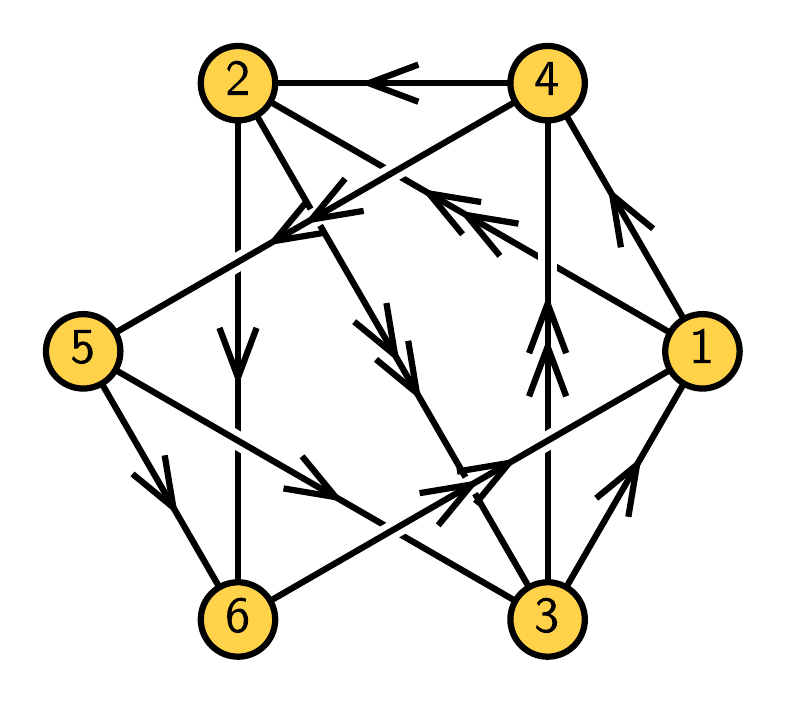}
  \caption{}
  \label{f_model_07_1_quiver}
\end{subfigure}
\hfill
\begin{subfigure}[b]{0.31\textwidth}
  \centering
  \resizebox{\linewidth}{!}{%
  \begin{tikzpicture}[x=1bp,y=1bp]
    \node[anchor=south west,inner sep=0pt,outer sep=0pt] at (0,0) {\includegraphics{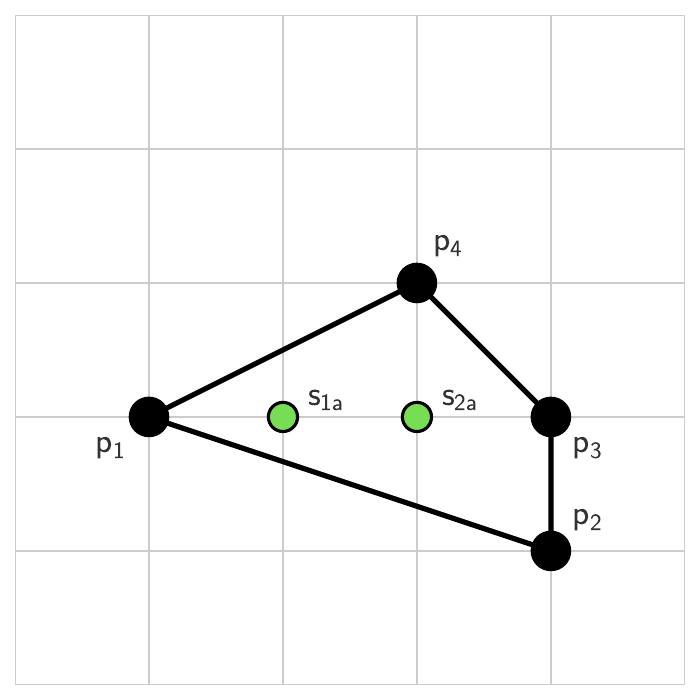}};
    \node[anchor=base west,inner sep=0pt,outer sep=0pt,text=black,overlay] at (147.94,121.33) {\font\pmcountfont=cmss10 at 16bp\pmcountfont 6};
    \node[anchor=base west,inner sep=0pt,outer sep=0pt,text=black,overlay] at (212.25,121.33) {\font\pmcountfont=cmss10 at 16bp\pmcountfont 8};
  \end{tikzpicture}}
  \caption{}
  \label{f_model_07_1_toric}
\end{subfigure}
\hfill
\begin{subfigure}[b]{0.31\textwidth}
  \centering
  \includegraphics[width=\linewidth]{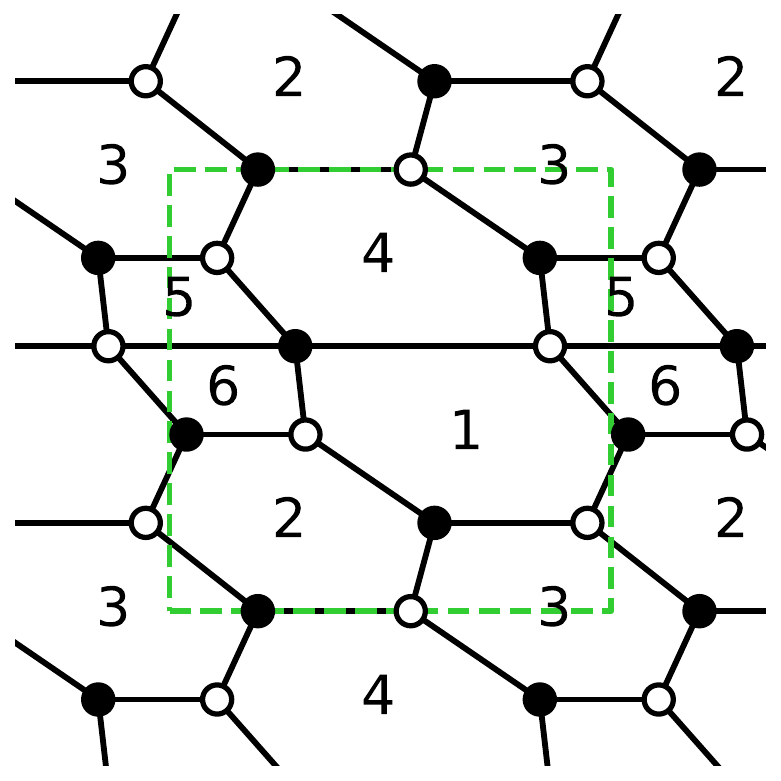}
  \caption{}
  \label{f_model_07_1_tiling}
\end{subfigure}
\caption{(a) Quiver, (b) toric diagram and (c) brane tiling for Model~7 ($Y^{3,2}$). The numbers in (b) indicate the multiplicity of each internal point.}
\label{f_model_07_1}
\end{figure}

The superpotential of Model~7 ($Y^{3,2}$) is given by,
\beal{es07a001}
\begin{aligned}
  W \;=\; & X^{2}_{12}\,X^{1}_{23}\,X_{31} + X^{2}_{23}\,X^{2}_{34}\,X_{42} + X^{1}_{34}\,X^{1}_{45}\,X_{53} \\
  & + X^{1}_{61}\,X^{1}_{12}\,X_{26} + X_{14}\,X^{2}_{45}\,X_{56}\,X^{2}_{61} - X^{2}_{12}\,X_{26}\,X^{2}_{61} \\
  & - X^{2}_{45}\,X_{53}\,X^{2}_{34} - X^{1}_{12}\,X^{2}_{23}\,X_{31} - X^{1}_{23}\,X^{1}_{34}\,X_{42} \\
  & - X^{1}_{45}\,X_{56}\,X^{1}_{61}\,X_{14} ~.~
\end{aligned}
\eea
Toric duality on the quadrilateral faces $5$ and $6$ of the brane tiling leads back to the same toric phase, which is therefore the only toric phase of the model.
Using the forward algorithm, we obtain the $P$-matrix as follows,
\beal{es07a002}
\resizebox{\ifdim\width>\linewidth\linewidth\else\width\fi}{!}{$
P \;=\; {\tiny
\left(
\begin{array}{c|cccc|cccccc|cccccccc}
   & p_{1} & p_{2} & p_{3} & p_{4} & s_{11} & s_{12} & s_{13} & s_{14} & s_{15} & s_{16} & s_{21} & s_{22} & s_{23} & s_{24} & s_{25} & s_{26} & s_{27} & s_{28} \\
  \hline
  X^{1}_{12} & 0 & 1 & 1 & 0 & 1 & 0 & 0 & 0 & 0 & 0 & 1 & 1 & 1 & 0 & 0 & 0 & 0 & 0 \\
  X^{2}_{12} & 0 & 0 & 1 & 1 & 1 & 0 & 0 & 0 & 0 & 0 & 1 & 1 & 1 & 0 & 0 & 0 & 0 & 0 \\
  X_{14} & 1 & 0 & 0 & 0 & 1 & 1 & 1 & 0 & 0 & 0 & 1 & 0 & 0 & 0 & 0 & 0 & 0 & 0 \\
  X^{1}_{23} & 0 & 1 & 0 & 0 & 0 & 1 & 0 & 0 & 0 & 0 & 0 & 0 & 0 & 1 & 1 & 1 & 0 & 0 \\
  X^{2}_{23} & 0 & 0 & 0 & 1 & 0 & 1 & 0 & 0 & 0 & 0 & 0 & 0 & 0 & 1 & 1 & 1 & 0 & 0 \\
  X_{26} & 1 & 0 & 0 & 0 & 0 & 1 & 1 & 1 & 1 & 0 & 0 & 0 & 0 & 1 & 1 & 0 & 1 & 0 \\
  X_{31} & 1 & 0 & 0 & 0 & 0 & 0 & 1 & 1 & 1 & 1 & 0 & 0 & 0 & 0 & 0 & 0 & 1 & 1 \\
  X^{1}_{34} & 0 & 0 & 1 & 1 & 0 & 0 & 1 & 0 & 0 & 0 & 1 & 0 & 0 & 0 & 0 & 0 & 1 & 1 \\
  X^{2}_{34} & 0 & 1 & 1 & 0 & 0 & 0 & 1 & 0 & 0 & 0 & 1 & 0 & 0 & 0 & 0 & 0 & 1 & 1 \\
  X_{42} & 1 & 0 & 0 & 0 & 1 & 0 & 0 & 1 & 1 & 1 & 0 & 1 & 1 & 0 & 0 & 0 & 0 & 0 \\
  X^{1}_{45} & 0 & 1 & 0 & 0 & 0 & 0 & 0 & 1 & 0 & 0 & 0 & 1 & 0 & 1 & 0 & 0 & 0 & 0 \\
  X^{2}_{45} & 0 & 0 & 0 & 1 & 0 & 0 & 0 & 1 & 0 & 0 & 0 & 1 & 0 & 1 & 0 & 0 & 0 & 0 \\
  X_{53} & 1 & 0 & 0 & 0 & 1 & 1 & 0 & 0 & 1 & 1 & 0 & 0 & 1 & 0 & 1 & 1 & 0 & 0 \\
  X_{56} & 0 & 0 & 1 & 0 & 0 & 0 & 0 & 0 & 1 & 0 & 0 & 0 & 1 & 0 & 1 & 0 & 1 & 0 \\
  X^{1}_{61} & 0 & 0 & 0 & 1 & 0 & 0 & 0 & 0 & 0 & 1 & 0 & 0 & 0 & 0 & 0 & 1 & 0 & 1 \\
  X^{2}_{61} & 0 & 1 & 0 & 0 & 0 & 0 & 0 & 0 & 0 & 1 & 0 & 0 & 0 & 0 & 0 & 1 & 0 & 1 \\
\end{array}
\right)}~,~
$}
\eea
where $p_a$ are GLSM fields corresponding to the extremal vertices in the toric diagram.
From the $P$-matrix we obtain the $F$-term charge matrix as follows,
\beal{es07a003}
\resizebox{\ifdim\width>\linewidth\linewidth\else\width\fi}{!}{$
Q_{F} \;=\; {\tiny
\left(
\begin{array}{cccc|cccccc|cccccccc}
  p_{1} & p_{2} & p_{3} & p_{4} & s_{11} & s_{12} & s_{13} & s_{14} & s_{15} & s_{16} & s_{21} & s_{22} & s_{23} & s_{24} & s_{25} & s_{26} & s_{27} & s_{28} \\
  \hline
  0 & 0 & -1 & 0 & -1 & 0 & 0 & 0 & 0 & 0 & 1 & 0 & 1 & 0 & 0 & 0 & 0 & 0 \\
  1 & 0 & 0 & 0 & -1 & 0 & -1 & 0 & 0 & 0 & 1 & 0 & 0 & 0 & 0 & 0 & 0 & 0 \\
  0 & 0 & 0 & 0 & 1 & -1 & 0 & 0 & 0 & 0 & 0 & -1 & 0 & 1 & 0 & 0 & 0 & 0 \\
  0 & 0 & -1 & 0 & 0 & -1 & 0 & 0 & 0 & 0 & 1 & 0 & 0 & 0 & 1 & 0 & 0 & 0 \\
  0 & -1 & 0 & -1 & -1 & 0 & 0 & 0 & 0 & 0 & 1 & 1 & 0 & 0 & 0 & 1 & 0 & 0 \\
  0 & 0 & -1 & 0 & 0 & 0 & -1 & 0 & 0 & 0 & 1 & 0 & 0 & 0 & 0 & 0 & 1 & 0 \\
  0 & 0 & 0 & 0 & 0 & 0 & -1 & 1 & 0 & 0 & 1 & -1 & 0 & 0 & 0 & 0 & 0 & 0 \\
  0 & 0 & -1 & 0 & -1 & 0 & -1 & 0 & 1 & 0 & 2 & 0 & 0 & 0 & 0 & 0 & 0 & 0 \\
  0 & -1 & 0 & -1 & -1 & 1 & -1 & 0 & 0 & 0 & 1 & 1 & 0 & 0 & 0 & 0 & 0 & 1 \\
  0 & -1 & 0 & -1 & -2 & 1 & -1 & 0 & 0 & 1 & 2 & 1 & 0 & 0 & 0 & 0 & 0 & 0 \\
\end{array}
\right)}~.~
$}
\eea
We also have the $D$-term charge matrix, which is given by,
\beal{es07a004}
\resizebox{\ifdim\width>\linewidth\linewidth\else\width\fi}{!}{$
Q_{D} \;=\; {\tiny
\left(
\begin{array}{cccc|cccccc|cccccccc}
  p_{1} & p_{2} & p_{3} & p_{4} & s_{11} & s_{12} & s_{13} & s_{14} & s_{15} & s_{16} & s_{21} & s_{22} & s_{23} & s_{24} & s_{25} & s_{26} & s_{27} & s_{28} \\
  \hline
  0 & 1 & 0 & 1 & 1 & -1 & 1 & 0 & 0 & 0 & -2 & -1 & 0 & 0 & 0 & 0 & 0 & 0 \\
  0 & 0 & 0 & 0 & 1 & -1 & 0 & 0 & 0 & 0 & 0 & 0 & 0 & 0 & 0 & 0 & 0 & 0 \\
  0 & 0 & 0 & 0 & 0 & 1 & -1 & 0 & 0 & 0 & 0 & 0 & 0 & 0 & 0 & 0 & 0 & 0 \\
  0 & 0 & 0 & 0 & 0 & 0 & 0 & 0 & 0 & 0 & 1 & -1 & 0 & 0 & 0 & 0 & 0 & 0 \\
  0 & 0 & -1 & 0 & -1 & 0 & 0 & 0 & 0 & 0 & 1 & 1 & 0 & 0 & 0 & 0 & 0 & 0 \\
\end{array}
\right)}~.~
$}
\eea
The toric diagram of the $Y^{3,2}$ model is given by,
\beal{es07a005}
\resizebox{\ifdim\width>\linewidth\linewidth\else\width\fi}{!}{$
G_t \;=\; {\small
\left(
\begin{array}{cccc|c|c}
  p_{1} & p_{2} & p_{3} & p_{4} & s_{1a} & s_{2a} \\
  \hline
  0 & 3 & 3 & 2 & 1 & 2 \\
  1 & 0 & 1 & 2 & 1 & 1 \\
  1 & 1 & 1 & 1 & 1 & 1 \\
\end{array}
\right)}~.~
$}
\eea
The corresponding toric diagram is shown in \fref{f_model_07_1_toric}.
The global symmetry of the brane tiling associated with $Y^{3,2}$ is given by,
\beal{es07a006}
SU(2) \times U(1) \times U(1)_R~.~
\eea
The global symmetry charges on the extremal GLSM fields $p_a$ are summarized in \tref{t_sym_geom_07}.

\begin{table}[H]
\centering
{\small
\begin{tabular}{|c|c|c|c|c|l|}
\hline
  & $SU(2)$ & $U(1)$ & $U(1)_R$ & $r_a$ & fugacities \\
\hline
  $p_{1}$ & $0$ & $-1$ & $r_{1}$ & $r_{1} = 0.6330$ & $t_{1} = f^{-1} \bar{t}_{1}$ \\
  $p_{2}$ & $1$ & $0$ & $r_{2}$ & $r_{2} = 0.5505$ & $t_{2} = x \bar{t}_{2}$ \\
  $p_{3}$ & $0$ & $1$ & $r_{3}$ & $r_{3} = 0.2660$ & $t_{3} = f \bar{t}_{3}$ \\
  $p_{4}$ & $-1$ & $0$ & $r_{4}$ & $r_{4} = 0.5505$ & $t_{4} = x^{-1} \bar{t}_{2}$ \\
\hline
\end{tabular}
}
\caption{Global symmetry charges on the extremal perfect matchings of Model~7 ($Y^{3,2}$).}
\label{t_sym_geom_07}
\end{table}

The unrefined Hilbert series of the mesonic moduli space is given by,
\beal{es07a007}
\resizebox{\ifdim\width>\linewidth\linewidth\else\width\fi}{!}{$
g(t; \mathcal{M}^{mes}) \;=\; \frac{1 + 4 t^{4} + 11 t^{8} + 4 t^{12} + t^{16}}{\left(1 - t^{4}\right) \left(1 - t^{8}\right)^{2}}~,~
$}
\eea
where $t$ is the fugacity counting degrees in extremal perfect matchings $p_a$, which corresponds to setting every fugacity $t_a$ in \tref{t_sym_geom_07} equal to $t$.
The plethystic logarithm of the refined Hilbert series takes the form,
\beal{es07a008}
PL\left[g(t_a;\mathcal{M}^{mes})\right] &=& t_{1} t_{3} t_{4}^{2} + t_{1} t_{2} t_{3} t_{4} + t_{1} t_{2}^{2} t_{3} + t_{1}^{3} t_{4} + t_{1}^{3} t_{2} + t_{3}^{3} t_{4}^{5} + t_{2} t_{3}^{3} t_{4}^{4} \nonumber\\
&& {}+ t_{2}^{2} t_{3}^{3} t_{4}^{3} + t_{2}^{3} t_{3}^{3} t_{4}^{2} + t_{2}^{4} t_{3}^{3} t_{4} + t_{2}^{5} t_{3}^{3} \nonumber\\
&& {}- t_{1}^{4} t_{2}^{2} t_{3} t_{4} - t_{1}^{4} t_{2} t_{3} t_{4}^{2} - t_{1}^{2} t_{2}^{2} t_{3}^{2} t_{4}^{2} - t_{1}^{3} t_{2}^{6} t_{3}^{3} - 2\, t_{1}^{3} t_{2}^{5} t_{3}^{3} t_{4} \nonumber\\
&& {}- 2\, t_{1}^{3} t_{2}^{4} t_{3}^{3} t_{4}^{2} - 2\, t_{1}^{3} t_{2}^{3} t_{3}^{3} t_{4}^{3} - 2\, t_{1}^{3} t_{2}^{2} t_{3}^{3} t_{4}^{4} - 2\, t_{1}^{3} t_{2} t_{3}^{3} t_{4}^{5} - t_{1}^{3} t_{3}^{3} t_{4}^{6} \nonumber\\
&& {}- t_{1} t_{2}^{6} t_{3}^{4} t_{4} - 2\, t_{1} t_{2}^{5} t_{3}^{4} t_{4}^{2} - 2\, t_{1} t_{2}^{4} t_{3}^{4} t_{4}^{3} - 2\, t_{1} t_{2}^{3} t_{3}^{4} t_{4}^{4} - 2\, t_{1} t_{2}^{2} t_{3}^{4} t_{4}^{5} \nonumber\\
&& {}- t_{1} t_{2} t_{3}^{4} t_{4}^{6} - t_{2}^{8} t_{3}^{6} t_{4}^{2} - t_{2}^{7} t_{3}^{6} t_{4}^{3} - 2\, t_{2}^{6} t_{3}^{6} t_{4}^{4} - 2\, t_{2}^{5} t_{3}^{6} t_{4}^{5} - 2\, t_{2}^{4} t_{3}^{6} t_{4}^{6} \nonumber\\
&& {}- t_{2}^{3} t_{3}^{6} t_{4}^{7} - t_{2}^{2} t_{3}^{6} t_{4}^{8} + \dots~,~
\eea
with $t_a$ the fugacity of $p_a$, one of the $4$ extremal perfect matchings at the corners of \fref{f_model_07_1_toric}. These fugacities can be expressed in terms of the global symmetry fugacities $x, f$ and $\bar{t}_{1}, \bar{t}_{2}, \bar{t}_{3}$ summarized in \tref{t_sym_geom_07}.

The positive terms at the start of the expansion correspond to the generators of the mesonic moduli space, which are $A_{1}, \dots, A_{11}$. \tref{t_model_07_1} lists the generators in terms of the chiral fields of the brane tiling. The negative terms that come after the positive ones correspond to the first order relations amongst the generators, which take the following~form,
\beal{es07a009}
&& A_{3} A_{4} = A_{2} A_{5} ~,~ \quad A_{2} A_{4} = A_{1} A_{5} ~,~ \quad A_{2}^{2} = A_{1} A_{3} ~,~ \quad A_{5} A_{11} = A_{3}^{3} ~,~ \nonumber\\
&& A_{5} A_{10} = A_{4} A_{11} = A_{2} A_{3}^{2} ~,~ \quad A_{5} A_{9} = A_{4} A_{10} = A_{1} A_{3}^{2} ~,~ \nonumber\\
&& A_{5} A_{8} = A_{4} A_{9} = A_{1} A_{2} A_{3} ~,~ \quad A_{5} A_{7} = A_{4} A_{8} = A_{1}^{2} A_{3} ~,~ \nonumber\\
&& A_{5} A_{6} = A_{4} A_{7} = A_{1}^{2} A_{2} ~,~ \quad A_{4} A_{6} = A_{1}^{3} ~,~ \quad A_{3} A_{10} = A_{2} A_{11} ~,~ \nonumber\\
&& A_{3} A_{9} = A_{2} A_{10} = A_{1} A_{11} ~,~ \quad A_{3} A_{8} = A_{2} A_{9} = A_{1} A_{10} ~,~ \nonumber\\
&& A_{3} A_{7} = A_{2} A_{8} = A_{1} A_{9} ~,~ \quad A_{3} A_{6} = A_{2} A_{7} = A_{1} A_{8} ~,~ \quad A_{2} A_{6} = A_{1} A_{7} ~,~ \nonumber\\
&& A_{10}^{2} = A_{9} A_{11} ~,~ \quad A_{9} A_{10} = A_{8} A_{11} ~,~ \quad A_{9}^{2} = A_{8} A_{10} = A_{7} A_{11} ~,~ \nonumber\\
&& A_{8} A_{9} = A_{7} A_{10} = A_{6} A_{11} ~,~ \quad A_{8}^{2} = A_{7} A_{9} = A_{6} A_{10} ~,~ \quad A_{7} A_{8} = A_{6} A_{9} ~,~ \nonumber\\
&& A_{7}^{2} = A_{6} A_{8}~.~
\eea

\begin{table}[H]
\centering
\resizebox{0.7\textwidth}{!}{
\begin{tabular}{|c|c|}
\hline
  Generators & Generators (in terms of chiral fields) \\
\hline
  $A_{1} = p_{1} p_{3} p_{4}^{2}$ & $\begin{array}{c}
    X^{2}_{12} X^{2}_{23} X_{31} = X^{2}_{12} X_{26} X^{1}_{61} = X^{2}_{23} X^{1}_{34} X_{42} = X^{1}_{34} X^{2}_{45} X_{53} \\
    = X_{14} X^{2}_{45} X_{56} X^{1}_{61}
  \end{array}$ \\
\hline
  $A_{2} = p_{1} p_{2} p_{3} p_{4}$ & $\begin{array}{c}
    X^{1}_{12} X^{2}_{23} X_{31} = X^{1}_{12} X_{26} X^{1}_{61} = X^{2}_{12} X^{1}_{23} X_{31} = X^{2}_{12} X_{26} X^{2}_{61} \\
    = X^{1}_{23} X^{1}_{34} X_{42} = X^{2}_{23} X^{2}_{34} X_{42} = X^{1}_{34} X^{1}_{45} X_{53} = X^{2}_{34} X^{2}_{45} X_{53} \\
    = X_{14} X^{1}_{45} X_{56} X^{1}_{61} = X_{14} X^{2}_{45} X_{56} X^{2}_{61}
  \end{array}$ \\
\hline
  $A_{3} = p_{1} p_{2}^{2} p_{3}$ & $\begin{array}{c}
    X^{1}_{12} X^{1}_{23} X_{31} = X^{1}_{12} X_{26} X^{2}_{61} = X^{1}_{23} X^{2}_{34} X_{42} = X^{2}_{34} X^{1}_{45} X_{53} \\
    = X_{14} X^{1}_{45} X_{56} X^{2}_{61}
  \end{array}$ \\
\hline
  $A_{4} = p_{1}^{3} p_{4}$ & $X_{14} X_{42} X^{2}_{23} X_{31} = X_{14} X_{42} X_{26} X^{1}_{61} = X_{14} X^{2}_{45} X_{53} X_{31}$ \\
\hline
  $A_{5} = p_{1}^{3} p_{2}$ & $X_{14} X_{42} X^{1}_{23} X_{31} = X_{14} X_{42} X_{26} X^{2}_{61} = X_{14} X^{1}_{45} X_{53} X_{31}$ \\
\hline
  $A_{6} = p_{3}^{3} p_{4}^{5}$ & $X^{2}_{12} X^{2}_{23} X^{1}_{34} X^{2}_{45} X_{56} X^{1}_{61}$ \\
\hline
  $A_{7} = p_{2} p_{3}^{3} p_{4}^{4}$ & $\begin{array}{c}
    X^{1}_{12} X^{2}_{23} X^{1}_{34} X^{2}_{45} X_{56} X^{1}_{61} = X^{2}_{12} X^{1}_{23} X^{1}_{34} X^{2}_{45} X_{56} X^{1}_{61} \\
    = X^{2}_{12} X^{2}_{23} X^{1}_{34} X^{1}_{45} X_{56} X^{1}_{61} = X^{2}_{12} X^{2}_{23} X^{1}_{34} X^{2}_{45} X_{56} X^{2}_{61} \\
    = X^{2}_{12} X^{2}_{23} X^{2}_{34} X^{2}_{45} X_{56} X^{1}_{61}
  \end{array}$ \\
\hline
  $A_{8} = p_{2}^{2} p_{3}^{3} p_{4}^{3}$ & $\begin{array}{c}
    X^{1}_{12} X^{1}_{23} X^{1}_{34} X^{2}_{45} X_{56} X^{1}_{61} = X^{1}_{12} X^{2}_{23} X^{1}_{34} X^{1}_{45} X_{56} X^{1}_{61} \\
    = X^{1}_{12} X^{2}_{23} X^{1}_{34} X^{2}_{45} X_{56} X^{2}_{61} = X^{1}_{12} X^{2}_{23} X^{2}_{34} X^{2}_{45} X_{56} X^{1}_{61} \\
    = X^{2}_{12} X^{1}_{23} X^{1}_{34} X^{1}_{45} X_{56} X^{1}_{61} = X^{2}_{12} X^{1}_{23} X^{1}_{34} X^{2}_{45} X_{56} X^{2}_{61} \\
    = X^{2}_{12} X^{1}_{23} X^{2}_{34} X^{2}_{45} X_{56} X^{1}_{61} = X^{2}_{12} X^{2}_{23} X^{1}_{34} X^{1}_{45} X_{56} X^{2}_{61} \\
    = X^{2}_{12} X^{2}_{23} X^{2}_{34} X^{1}_{45} X_{56} X^{1}_{61} = X^{2}_{12} X^{2}_{23} X^{2}_{34} X^{2}_{45} X_{56} X^{2}_{61}
  \end{array}$ \\
\hline
  $A_{9} = p_{2}^{3} p_{3}^{3} p_{4}^{2}$ & $\begin{array}{c}
    X^{1}_{12} X^{1}_{23} X^{1}_{34} X^{1}_{45} X_{56} X^{1}_{61} = X^{1}_{12} X^{1}_{23} X^{1}_{34} X^{2}_{45} X_{56} X^{2}_{61} \\
    = X^{1}_{12} X^{1}_{23} X^{2}_{34} X^{2}_{45} X_{56} X^{1}_{61} = X^{1}_{12} X^{2}_{23} X^{1}_{34} X^{1}_{45} X_{56} X^{2}_{61} \\
    = X^{1}_{12} X^{2}_{23} X^{2}_{34} X^{1}_{45} X_{56} X^{1}_{61} = X^{1}_{12} X^{2}_{23} X^{2}_{34} X^{2}_{45} X_{56} X^{2}_{61} \\
    = X^{2}_{12} X^{1}_{23} X^{1}_{34} X^{1}_{45} X_{56} X^{2}_{61} = X^{2}_{12} X^{1}_{23} X^{2}_{34} X^{1}_{45} X_{56} X^{1}_{61} \\
    = X^{2}_{12} X^{1}_{23} X^{2}_{34} X^{2}_{45} X_{56} X^{2}_{61} = X^{2}_{12} X^{2}_{23} X^{2}_{34} X^{1}_{45} X_{56} X^{2}_{61}
  \end{array}$ \\
\hline
  $A_{10} = p_{2}^{4} p_{3}^{3} p_{4}$ & $\begin{array}{c}
    X^{1}_{12} X^{1}_{23} X^{1}_{34} X^{1}_{45} X_{56} X^{2}_{61} = X^{1}_{12} X^{1}_{23} X^{2}_{34} X^{1}_{45} X_{56} X^{1}_{61} \\
    = X^{1}_{12} X^{1}_{23} X^{2}_{34} X^{2}_{45} X_{56} X^{2}_{61} = X^{1}_{12} X^{2}_{23} X^{2}_{34} X^{1}_{45} X_{56} X^{2}_{61} \\
    = X^{2}_{12} X^{1}_{23} X^{2}_{34} X^{1}_{45} X_{56} X^{2}_{61}
  \end{array}$ \\
\hline
  $A_{11} = p_{2}^{5} p_{3}^{3}$ & $X^{1}_{12} X^{1}_{23} X^{2}_{34} X^{1}_{45} X_{56} X^{2}_{61}$ \\
\hline
\end{tabular}}
\caption{Generators of the mesonic moduli space of Model~7 ($Y^{3,2}$) in terms of the extremal perfect matchings and of the chiral fields.}
\label{t_model_07_1}
\end{table}

\subsection{Model 8: $Y^{3,1}$}\label{s_m8}
\begin{figure}[H]
\centering
\begin{minipage}[c]{0.34\textwidth}
\resizebox{\linewidth}{!}{%
\begin{tikzpicture}[x=1bp,y=1bp]
  \node[anchor=south west,inner sep=0pt,outer sep=0pt] at (0,0) {\includegraphics{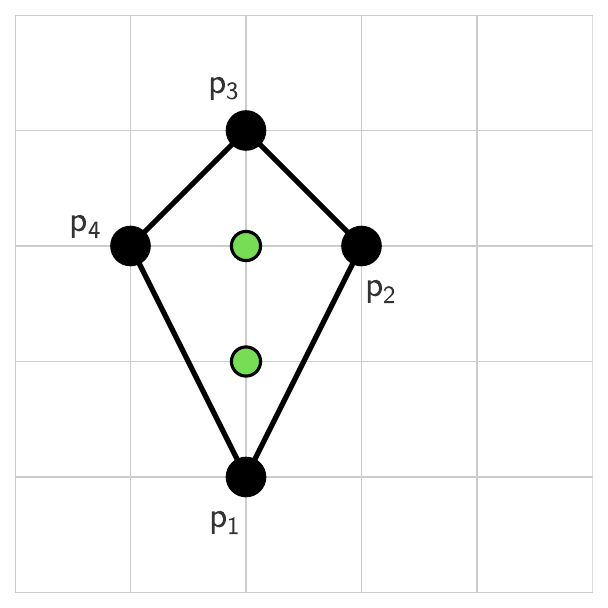}};
  \node[anchor=south west,inner sep=0pt,outer sep=0pt,overlay] at (124.52,121.63) {\includegraphics{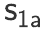}};
  \node[anchor=south west,inner sep=0pt,outer sep=0pt,overlay] at (128.22,175.47) {\includegraphics{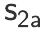}};
\end{tikzpicture}}
\end{minipage}%
\hspace{0.05\textwidth}%
{\small
\setlength{\tabcolsep}{4pt}
\begin{tabular}[c]{|c|c|c|}
\hline
\multirow{2}{*}{phase} & \multicolumn{2}{c|}{multiplicity} \\
\cline{2-3}
& \makebox[2.2em]{$s_{1a}$} & \makebox[2.2em]{$s_{2a}$} \\
\hline\hline
\hyperref[s_model_08_1]{1} & 6 & 7 \\
\hyperref[s_model_08_2]{2} & 7 & 9 \\
\hline
\end{tabular}
}
\caption{The toric diagram for Model~8 ($Y^{3,1}$).}
\label{f_geom_08}
\end{figure}

\begin{figure}[H]
\centering
\definecolor{dualityred}{rgb}{1,0.149,0}
\hypersetup{linkcolor=black}
\begin{tikzpicture}[line width=1.55bp]
\useasboundingbox (-0.338,-3.034) rectangle (0.338,0.320);
\draw[dualityred, preaction={draw=white, line width=4.55bp}] (0.000,-1.671) -- (0.000,-0.529);
\draw[dualityred, preaction={draw=white, line width=4.55bp}] (-0.301,-2.700) .. controls (-0.454,-3.118) and (0.454,-3.118) .. (0.301,-2.700);
\fill[dualityred] (0.000,-1.863) -- (0.072,-1.671) -- (-0.072,-1.671) -- cycle;
\fill[dualityred] (0.000,-0.337) -- (-0.072,-0.529) -- (0.072,-0.529) -- cycle;
\fill[dualityred] (-0.200,-2.537) -- (-0.363,-2.662) -- (-0.239,-2.738) -- cycle;
\fill[dualityred] (0.200,-2.537) -- (0.239,-2.738) -- (0.363,-2.662) -- cycle;
\node[draw, rectangle, fill=white, minimum size=0.6cm, inner sep=0pt, font=\sffamily\bfseries\footnotesize, line width=2.07bp] at (0.000,-2.200) {\hyperref[s_model_08_1]{1}};
\node[draw, rectangle, fill=white, minimum size=0.6cm, inner sep=0pt, font=\sffamily\bfseries\footnotesize, line width=2.07bp] at (0.000,0.000) {\hyperref[s_model_08_2]{2}};
\end{tikzpicture}
\caption{The Seiberg duality tree for Model~8 ($Y^{3,1}$). The node labels indicate the phase numbers.}
\label{f_geom_08_tree}
\end{figure}

\fref{f_geom_08} illustrates the toric diagram for the $Y^{3,1}$ model. The Seiberg duality tree of the $2$ toric phases associated with $Y^{3,1}$ is shown in \fref{f_geom_08_tree}.
The global symmetry of the brane tilings associated with $Y^{3,1}$ is given by,
\beal{es08a001}
SU(2) \times U(1) \times U(1)_R~.~
\eea
The global symmetry charges on the extremal GLSM fields $p_a$ are summarized in \tref{t_sym_geom_08}.

\begin{table}[H]
\centering
{\small
\begin{tabular}{|c|c|c|c|c|l|}
\hline
  & $SU(2)$ & $U(1)$ & $U(1)_R$ & $r_a$ & fugacities \\
\hline
  $p_{1}$ & $0$ & $-1$ & $r_{1}$ & $r_{1} = 0.5743$ & $t_{1} = f^{-1} \bar{t}_{1}$ \\
  $p_{2}$ & $1$ & $0$ & $r_{2}$ & $r_{2} = 0.5109$ & $t_{2} = x \bar{t}_{2}$ \\
  $p_{3}$ & $0$ & $1$ & $r_{3}$ & $r_{3} = 0.4040$ & $t_{3} = f \bar{t}_{3}$ \\
  $p_{4}$ & $-1$ & $0$ & $r_{4}$ & $r_{4} = 0.5109$ & $t_{4} = x^{-1} \bar{t}_{2}$ \\
\hline
\end{tabular}
}
\caption{Global symmetry charges on the extremal perfect matchings of Model~8 ($Y^{3,1}$).}
\label{t_sym_geom_08}
\end{table}

All toric phases associated with $Y^{3,1}$ share the same mesonic moduli space, whose unrefined Hilbert series takes the form,
\beal{es08a002}
\resizebox{\ifdim\width>\linewidth\linewidth\else\width\fi}{!}{$
g(t; \mathcal{M}^{mes}) \;=\; \frac{1 + 3 t^{4} + t^{5} + 3 t^{7} + 5 t^{8} - t^{9} + t^{11} - 5 t^{12} - 3 t^{13} - t^{15} - 3 t^{16} - t^{20}}{\left(1 - t^{5}\right)^{2} \left(1 - t^{7}\right)^{2}}~,~
$}
\eea
where $t$ is the fugacity counting degrees in extremal perfect matchings $p_a$, which corresponds to setting every fugacity $t_a$ in \tref{t_sym_geom_08} equal to $t$.
The plethystic logarithm of the refined Hilbert series takes the form,
\beal{es08a003}
PL\left[g(t_a;\mathcal{M}^{mes})\right] &=& t_{1} t_{3} t_{4}^{2} + t_{1} t_{2} t_{3} t_{4} + t_{1} t_{2}^{2} t_{3} + t_{1}^{3} t_{4}^{2} + t_{1}^{3} t_{2} t_{4} + t_{1}^{3} t_{2}^{2} + t_{3}^{3} t_{4}^{4} \nonumber\\
&& {}+ t_{2} t_{3}^{3} t_{4}^{3} + t_{2}^{2} t_{3}^{3} t_{4}^{2} + t_{2}^{3} t_{3}^{3} t_{4} + t_{2}^{4} t_{3}^{3} \nonumber\\
&& {}- t_{1}^{2} t_{2}^{2} t_{3}^{2} t_{4}^{2} - t_{1}^{4} t_{2}^{3} t_{3} t_{4} - 2\, t_{1}^{4} t_{2}^{2} t_{3} t_{4}^{2} - t_{1}^{4} t_{2} t_{3} t_{4}^{3} - t_{1}^{6} t_{2}^{2} t_{4}^{2} \nonumber\\
&& {}- t_{1} t_{2}^{5} t_{3}^{4} t_{4} - 2\, t_{1} t_{2}^{4} t_{3}^{4} t_{4}^{2} - 2\, t_{1} t_{2}^{3} t_{3}^{4} t_{4}^{3} - 2\, t_{1} t_{2}^{2} t_{3}^{4} t_{4}^{4} - t_{1} t_{2} t_{3}^{4} t_{4}^{5} \nonumber\\
&& {}- t_{1}^{3} t_{2}^{6} t_{3}^{3} - 2\, t_{1}^{3} t_{2}^{5} t_{3}^{3} t_{4} - 3\, t_{1}^{3} t_{2}^{4} t_{3}^{3} t_{4}^{2} - 3\, t_{1}^{3} t_{2}^{3} t_{3}^{3} t_{4}^{3} - 3\, t_{1}^{3} t_{2}^{2} t_{3}^{3} t_{4}^{4} \nonumber\\
&& {}- 2\, t_{1}^{3} t_{2} t_{3}^{3} t_{4}^{5} - t_{1}^{3} t_{3}^{3} t_{4}^{6} - t_{2}^{6} t_{3}^{6} t_{4}^{2} - t_{2}^{5} t_{3}^{6} t_{4}^{3} - 2\, t_{2}^{4} t_{3}^{6} t_{4}^{4} - t_{2}^{3} t_{3}^{6} t_{4}^{5} \nonumber\\
&& {}- t_{2}^{2} t_{3}^{6} t_{4}^{6} + \dots~,~
\eea
where $t_{1}, \dots, t_{4}$ correspond to the extremal perfect matchings $p_{1}, \dots, p_{4}$ of \fref{f_geom_08}, which can be expressed in terms of the global symmetry fugacities $x, f$ and $\bar{t}_{1}, \bar{t}_{2}, \bar{t}_{3}$ as summarized in \tref{t_sym_geom_08}.

The generators of the mesonic moduli space, $A_{1}, \dots, A_{11}$, correspond to the first positive terms of the plethystic logarithm. For each toric phase, the generators are given in \tref{t_model_08_1} and \tref{t_model_08_2} in terms of the chiral fields of the corresponding brane tiling. The negative terms that come next correspond to the first order relations amongst the generators, which take the following~form,
\beal{es08a004}
&& A_{2}^{2} = A_{1} A_{3} ~,~ \quad A_{3} A_{5} = A_{2} A_{6} ~,~ \quad A_{3} A_{4} = A_{2} A_{5} = A_{1} A_{6} ~,~ \nonumber\\
&& A_{2} A_{4} = A_{1} A_{5} ~,~ \quad A_{5}^{2} = A_{4} A_{6} ~,~ \quad A_{3} A_{10} = A_{2} A_{11} ~,~ \nonumber\\
&& A_{3} A_{9} = A_{2} A_{10} = A_{1} A_{11} ~,~ \quad A_{3} A_{8} = A_{2} A_{9} = A_{1} A_{10} ~,~ \nonumber\\
&& A_{3} A_{7} = A_{2} A_{8} = A_{1} A_{9} ~,~ \quad A_{2} A_{7} = A_{1} A_{8} ~,~ \quad A_{6} A_{11} = A_{3}^{3} ~,~ \nonumber\\
&& A_{6} A_{10} = A_{5} A_{11} = A_{2} A_{3}^{2} ~,~ \quad A_{6} A_{9} = A_{5} A_{10} = A_{4} A_{11} = A_{1} A_{3}^{2} ~,~ \nonumber\\
&& A_{6} A_{8} = A_{5} A_{9} = A_{4} A_{10} = A_{1} A_{2} A_{3} ~,~ \quad A_{6} A_{7} = A_{5} A_{8} = A_{4} A_{9} = A_{1}^{2} A_{3} ~,~ \nonumber\\
&& A_{5} A_{7} = A_{4} A_{8} = A_{1}^{2} A_{2} ~,~ \quad A_{4} A_{7} = A_{1}^{3} ~,~ \quad A_{10}^{2} = A_{9} A_{11} ~,~ \nonumber\\
&& A_{9} A_{10} = A_{8} A_{11} ~,~ \quad A_{9}^{2} = A_{8} A_{10} = A_{7} A_{11} ~,~ \quad A_{8} A_{9} = A_{7} A_{10} ~,~ \nonumber\\
&& A_{8}^{2} = A_{7} A_{9}~.~
\eea

\input{model_08_1}
\input{model_08_2}

\subsection{Model 9: $\mathcal{C}/\mathbb{Z}_{3}\,(1,1,2,2)$}\label{s_m9}
\begin{figure}[H]
\centering
\begin{minipage}[c]{0.34\textwidth}
\resizebox{\linewidth}{!}{%
\begin{tikzpicture}[x=1bp,y=1bp]
  \node[anchor=south west,inner sep=0pt,outer sep=0pt] at (0,0) {\includegraphics{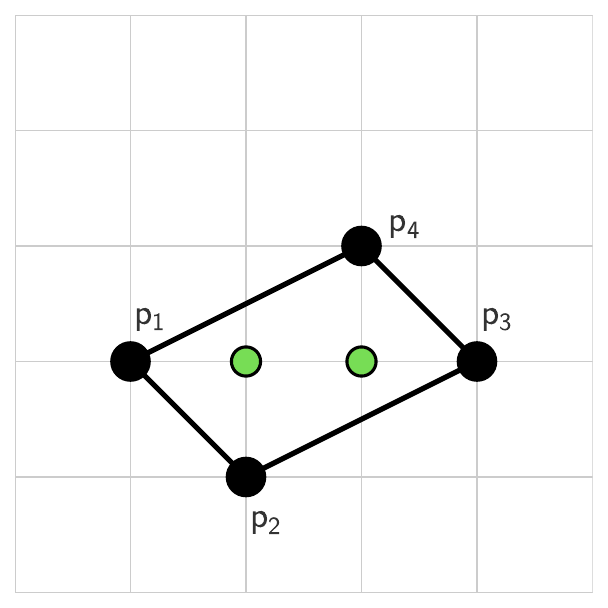}};
  \node[anchor=south west,inner sep=0pt,outer sep=0pt,overlay] at (128.22,120.03) {\includegraphics{figures/toric_label_s1a.pdf}};
  \node[anchor=south west,inner sep=0pt,outer sep=0pt,overlay] at (183.66,120.03) {\includegraphics{figures/toric_label_s2a.pdf}};
\end{tikzpicture}}
\end{minipage}%
\hspace{0.05\textwidth}%
{\small
\setlength{\tabcolsep}{4pt}
\begin{tabular}[c]{|c|c|c|}
\hline
\multirow{2}{*}{phase} & \multicolumn{2}{c|}{multiplicity} \\
\cline{2-3}
& \makebox[2.2em]{$s_{1a}$} & \makebox[2.2em]{$s_{2a}$} \\
\hline\hline
\hyperref[s_model_09_1]{1} & 6 & 6 \\
\hyperref[s_model_09_2]{2} & 7 & 7 \\
\hline
\end{tabular}
}
\caption{The toric diagram for Model~9 ($\mathcal{C}/\mathbb{Z}_{3}\,(1,1,2,2)$).}
\label{f_geom_09}
\end{figure}

\begin{figure}[H]
\centering
\definecolor{dualityred}{rgb}{1,0.149,0}
\hypersetup{linkcolor=black}
\begin{tikzpicture}[line width=1.55bp]
\useasboundingbox (-0.338,-2.520) rectangle (0.338,0.834);
\draw[dualityred, preaction={draw=white, line width=4.55bp}] (0.000,-1.671) -- (0.000,-0.529);
\draw[dualityred, preaction={draw=white, line width=4.55bp}] (0.301,0.500) .. controls (0.454,0.918) and (-0.454,0.918) .. (-0.301,0.500);
\fill[dualityred] (0.000,-1.863) -- (0.072,-1.671) -- (-0.072,-1.671) -- cycle;
\fill[dualityred] (0.000,-0.337) -- (-0.072,-0.529) -- (0.072,-0.529) -- cycle;
\fill[dualityred] (0.200,0.337) -- (0.363,0.462) -- (0.239,0.538) -- cycle;
\fill[dualityred] (-0.200,0.337) -- (-0.239,0.538) -- (-0.363,0.462) -- cycle;
\node[draw, rectangle, fill=white, minimum size=0.6cm, inner sep=0pt, font=\sffamily\bfseries\footnotesize, line width=2.07bp] at (0.000,-2.200) {\hyperref[s_model_09_1]{1}};
\node[draw, rectangle, fill=white, minimum size=0.6cm, inner sep=0pt, font=\sffamily\bfseries\footnotesize, line width=2.07bp] at (0.000,0.000) {\hyperref[s_model_09_2]{2}};
\end{tikzpicture}
\caption{The Seiberg duality tree for Model~9 ($\mathcal{C}/\mathbb{Z}_{3}\,(1,1,2,2)$). The node labels indicate the phase numbers.}
\label{f_geom_09_tree}
\end{figure}

The $\mathcal{C}/\mathbb{Z}_{3}$ model has a toric diagram shown in \fref{f_geom_09}. The Seiberg duality tree of the $2$ toric phases associated with $\mathcal{C}/\mathbb{Z}_{3}$ is shown in \fref{f_geom_09_tree}.
The global symmetry of the brane tilings associated with $\mathcal{C}/\mathbb{Z}_{3}$ is given by,
\beal{es09a001}
SU(2) \times U(1) \times U(1)_R~.~
\eea
The global symmetry charges on the extremal GLSM fields $p_a$ are summarized in \tref{t_sym_geom_09}.

\begin{table}[H]
\centering
{\small
\begin{tabular}{|c|c|c|c|c|l|}
\hline
  & $SU(2)$ & $U(1)$ & $U(1)_R$ & $r_a$ & fugacities \\
\hline
  $p_{1}$ & $0$ & $-1$ & $r_{1}$ & $r_{1} = 1/2$ & $t_{1} = f^{-1} \bar{t}$ \\
  $p_{2}$ & $1$ & $0$ & $r_{2}$ & $r_{2} = 1/2$ & $t_{2} = x \bar{t}$ \\
  $p_{3}$ & $0$ & $1$ & $r_{3}$ & $r_{3} = 1/2$ & $t_{3} = f \bar{t}$ \\
  $p_{4}$ & $-1$ & $0$ & $r_{4}$ & $r_{4} = 1/2$ & $t_{4} = x^{-1} \bar{t}$ \\
\hline
\end{tabular}
}
\caption{Global symmetry charges on the extremal perfect matchings of Model~9 ($\mathcal{C}/\mathbb{Z}_{3}\,(1,1,2,2)$).}
\label{t_sym_geom_09}
\end{table}

All toric phases associated with $\mathcal{C}/\mathbb{Z}_{3}$ share the same mesonic moduli space, whose unrefined Hilbert series takes the form,
\beal{es09a002}
\resizebox{\ifdim\width>\linewidth\linewidth\else\width\fi}{!}{$
g(t; \mathcal{M}^{mes}) \;=\; \frac{1 - t^{2} + 3 t^{4} + 3 t^{6} - t^{8} + t^{10}}{\left(1 - t^{2}\right) \left(1 - t^{6}\right)^{2}}~,~
$}
\eea
where $t$ is the fugacity counting degrees in extremal perfect matchings $p_a$, which corresponds to setting every fugacity $t_a$ in \tref{t_sym_geom_09} equal to $t$.
The plethystic logarithm of the refined Hilbert series takes the form,
\beal{es09a003}
PL\left[g(t_a;\mathcal{M}^{mes})\right] &=& t_{1} t_{3} t_{4}^{2} + t_{1} t_{2} t_{3} t_{4} + t_{1} t_{2}^{2} t_{3} + t_{3}^{3} t_{4}^{3} + t_{2} t_{3}^{3} t_{4}^{2} + t_{2}^{2} t_{3}^{3} t_{4} + t_{2}^{3} t_{3}^{3} \nonumber\\
&& {}+ t_{1}^{3} t_{4}^{3} + t_{1}^{3} t_{2} t_{4}^{2} + t_{1}^{3} t_{2}^{2} t_{4} + t_{1}^{3} t_{2}^{3} \nonumber\\
&& {}- t_{1}^{2} t_{2}^{2} t_{3}^{2} t_{4}^{2} - t_{1}^{4} t_{2}^{4} t_{3} t_{4} - 2\, t_{1}^{4} t_{2}^{3} t_{3} t_{4}^{2} - 2\, t_{1}^{4} t_{2}^{2} t_{3} t_{4}^{3} - t_{1} t_{2}^{4} t_{3}^{4} t_{4} \nonumber\\
&& {}- 2\, t_{1} t_{2}^{3} t_{3}^{4} t_{4}^{2} - 2\, t_{1} t_{2}^{2} t_{3}^{4} t_{4}^{3} - t_{1}^{4} t_{2} t_{3} t_{4}^{4} - t_{1} t_{2} t_{3}^{4} t_{4}^{4} - t_{1}^{6} t_{2}^{4} t_{4}^{2} \nonumber\\
&& {}- t_{1}^{6} t_{2}^{3} t_{4}^{3} - t_{1}^{6} t_{2}^{2} t_{4}^{4} - t_{1}^{3} t_{2}^{6} t_{3}^{3} - 2\, t_{1}^{3} t_{2}^{5} t_{3}^{3} t_{4} - 3\, t_{1}^{3} t_{2}^{4} t_{3}^{3} t_{4}^{2} \nonumber\\
&& {}- 4\, t_{1}^{3} t_{2}^{3} t_{3}^{3} t_{4}^{3} - 3\, t_{1}^{3} t_{2}^{2} t_{3}^{3} t_{4}^{4} - t_{2}^{4} t_{3}^{6} t_{4}^{2} - 2\, t_{1}^{3} t_{2} t_{3}^{3} t_{4}^{5} - t_{2}^{3} t_{3}^{6} t_{4}^{3} \nonumber\\
&& {}- t_{2}^{2} t_{3}^{6} t_{4}^{4} - t_{1}^{3} t_{3}^{3} t_{4}^{6} + \dots~,~
\eea
written in $t_{1}, \dots, t_{4}$, the fugacities of the extremal perfect matchings $p_{1}, \dots, p_{4}$ of \fref{f_geom_09}; each $t_a$ can be expressed in terms of the global symmetry fugacities $x, f$ and $\bar{t}$ as given in \tref{t_sym_geom_09}.

The positive terms with which the expansion begins correspond to the generators of the mesonic moduli space, namely $A_{1}, \dots, A_{11}$. The expressions of the generators in terms of the chiral fields of the individual toric phases are summarized in \tref{t_model_09_1} and \tref{t_model_09_2}. The negative terms that follow correspond to the first order relations amongst the generators, which take the following~form,
\beal{es09a004}
&& A_{2}^{2} = A_{1} A_{3} ~,~ \quad A_{3} A_{10} = A_{2} A_{11} ~,~ \quad A_{3} A_{9} = A_{2} A_{10} = A_{1} A_{11} ~,~ \nonumber\\
&& A_{3} A_{8} = A_{2} A_{9} = A_{1} A_{10} ~,~ \quad A_{3} A_{6} = A_{2} A_{7} ~,~ \quad A_{3} A_{5} = A_{2} A_{6} = A_{1} A_{7} ~,~ \nonumber\\
&& A_{3} A_{4} = A_{2} A_{5} = A_{1} A_{6} ~,~ \quad A_{2} A_{8} = A_{1} A_{9} ~,~ \quad A_{2} A_{4} = A_{1} A_{5} ~,~ \nonumber\\
&& A_{10}^{2} = A_{9} A_{11} ~,~ \quad A_{9} A_{10} = A_{8} A_{11} ~,~ \quad A_{9}^{2} = A_{8} A_{10} ~,~ \quad A_{7} A_{11} = A_{3}^{3} ~,~ \nonumber\\
&& A_{7} A_{10} = A_{6} A_{11} = A_{2} A_{3}^{2} ~,~ \quad A_{7} A_{9} = A_{6} A_{10} = A_{5} A_{11} = A_{1} A_{3}^{2} ~,~ \nonumber\\
&& A_{7} A_{8} = A_{6} A_{9} = A_{5} A_{10} = A_{4} A_{11} = A_{1} A_{2} A_{3} ~,~ \nonumber\\
&& A_{6} A_{8} = A_{5} A_{9} = A_{4} A_{10} = A_{1}^{2} A_{3} ~,~ \quad A_{6}^{2} = A_{5} A_{7} ~,~ \nonumber\\
&& A_{5} A_{8} = A_{4} A_{9} = A_{1}^{2} A_{2} ~,~ \quad A_{5} A_{6} = A_{4} A_{7} ~,~ \quad A_{5}^{2} = A_{4} A_{6} ~,~ \quad A_{4} A_{8} = A_{1}^{3}~.~
\eea

\input{model_09_1}
\input{model_09_2}

\subsection{Model 10: $X^{3,3}$}\label{s_m10}
\input{model_10_1}

\subsection{Model 11: $L^{2,5,3}$}\label{s_m11}
\begin{figure}[H]
\centering
\begin{minipage}[c]{0.34\textwidth}
\resizebox{\linewidth}{!}{%
\begin{tikzpicture}[x=1bp,y=1bp]
  \node[anchor=south west,inner sep=0pt,outer sep=0pt] at (0,0) {\includegraphics{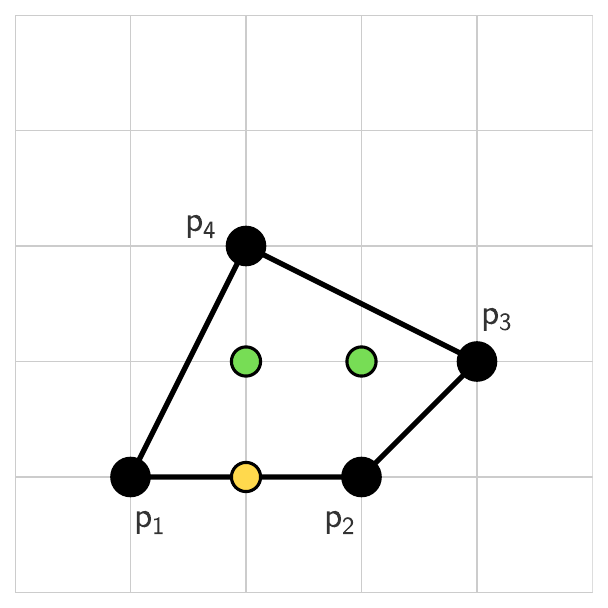}};
  \node[anchor=south west,inner sep=0pt,outer sep=0pt,overlay] at (128.22,120.03) {\includegraphics{figures/toric_label_s1a.pdf}};
  \node[anchor=south west,inner sep=0pt,outer sep=0pt,overlay] at (183.66,99.42) {\includegraphics{figures/toric_label_s2a.pdf}};
\end{tikzpicture}}
\end{minipage}%
\hspace{0.05\textwidth}%
{\small
\setlength{\tabcolsep}{4pt}
\begin{tabular}[c]{|c|c|c|}
\hline
\multirow{2}{*}{phase} & \multicolumn{2}{c|}{multiplicity} \\
\cline{2-3}
& \makebox[2.2em]{$s_{1a}$} & \makebox[2.2em]{$s_{2a}$} \\
\hline\hline
\hyperref[s_model_11_1]{1} & 7 & 7 \\
\hyperref[s_model_11_2]{2} & 7 & 8 \\
\hline
\end{tabular}
}
\caption{The toric diagram for Model~11 ($L^{2,5,3}$).}
\label{f_geom_11}
\end{figure}

\begin{figure}[H]
\centering
\definecolor{dualityred}{rgb}{1,0.149,0}
\hypersetup{linkcolor=black}
\begin{tikzpicture}[line width=1.55bp]
\useasboundingbox (-0.338,-3.034) rectangle (0.338,0.320);
\draw[dualityred, preaction={draw=white, line width=4.55bp}] (0.000,-1.671) -- (0.000,-0.529);
\draw[dualityred, preaction={draw=white, line width=4.55bp}] (-0.301,-2.700) .. controls (-0.454,-3.118) and (0.454,-3.118) .. (0.301,-2.700);
\fill[dualityred] (0.000,-1.863) -- (0.072,-1.671) -- (-0.072,-1.671) -- cycle;
\fill[dualityred] (0.000,-0.337) -- (-0.072,-0.529) -- (0.072,-0.529) -- cycle;
\fill[dualityred] (-0.200,-2.537) -- (-0.363,-2.662) -- (-0.239,-2.738) -- cycle;
\fill[dualityred] (0.200,-2.537) -- (0.239,-2.738) -- (0.363,-2.662) -- cycle;
\node[draw, rectangle, fill=white, minimum size=0.6cm, inner sep=0pt, font=\sffamily\bfseries\footnotesize, line width=2.07bp] at (0.000,-2.200) {\hyperref[s_model_11_1]{1}};
\node[draw, rectangle, fill=white, minimum size=0.6cm, inner sep=0pt, font=\sffamily\bfseries\footnotesize, line width=2.07bp] at (0.000,0.000) {\hyperref[s_model_11_2]{2}};
\end{tikzpicture}
\caption{The Seiberg duality tree for Model~11 ($L^{2,5,3}$). The node labels indicate the phase numbers.}
\label{f_geom_11_tree}
\end{figure}

\fref{f_geom_11} illustrates the toric diagram for the $L^{2,5,3}$ model. The Seiberg duality tree of the $2$ toric phases associated with $L^{2,5,3}$ is shown in \fref{f_geom_11_tree}.
The global symmetry of the brane tilings associated with $L^{2,5,3}$ is given by,
\beal{es11a001}
U(1) \times U(1) \times U(1)_R~.~
\eea
The global symmetry charges on the extremal GLSM fields $p_a$ are summarized in \tref{t_sym_geom_11}.

\begin{table}[H]
\centering
{\small
\begin{tabular}{|c|c|c|c|c|l|}
\hline
  & $U(1)_{1}$ & $U(1)_{2}$ & $U(1)_R$ & $r_a$ & fugacities \\
\hline
  $p_{1}$ & $-1$ & $-1$ & $r_{1}$ & $r_{1} = 0.5465$ & $t_{1} = f_{1}^{-1} f_{2}^{-1} \bar{t}_{1}$ \\
  $p_{2}$ & $1$ & $0$ & $r_{2}$ & $r_{2} = 0.3739$ & $t_{2} = f_{1} \bar{t}_{2}$ \\
  $p_{3}$ & $0$ & $1$ & $r_{3}$ & $r_{3} = 0.4868$ & $t_{3} = f_{2} \bar{t}_{3}$ \\
  $p_{4}$ & $0$ & $0$ & $r_{4}$ & $r_{4} = 0.5928$ & $t_{4} = \bar{t}_{4}$ \\
\hline
\end{tabular}
}
\caption{Global symmetry charges on the extremal perfect matchings of Model~11 ($L^{2,5,3}$).}
\label{t_sym_geom_11}
\end{table}

All toric phases associated with $L^{2,5,3}$ share the same mesonic moduli space, whose unrefined Hilbert series takes the form,
\beal{es11a002}
\resizebox{\ifdim\width>\linewidth\linewidth\else\width\fi}{!}{$
g(t; \mathcal{M}^{mes}) \;=\; \frac{1 + t^{4} + t^{5} + 2 t^{6} + 2 t^{7} - 2 t^{14} - 2 t^{15} - t^{16} - t^{17} - t^{21}}{\left(1 - t^{3}\right) \left(1 - t^{5}\right) \left(1 - t^{8}\right) \left(1 - t^{9}\right)}~,~
$}
\eea
where $t$ is the fugacity counting degrees in extremal perfect matchings $p_a$, which corresponds to setting every fugacity $t_a$ in \tref{t_sym_geom_11} equal to $t$.
The plethystic logarithm of the refined Hilbert series takes the form,
\beal{es11a003}
PL\left[g(t_a;\mathcal{M}^{mes})\right] &=& t_{3} t_{4}^{2} + t_{1} t_{2} t_{3} t_{4} + t_{1}^{2} t_{4}^{3} + t_{1}^{2} t_{2}^{2} t_{3} + t_{2}^{2} t_{3}^{3} t_{4} + t_{1}^{3} t_{2} t_{4}^{2} + t_{1} t_{2}^{3} t_{3}^{3} \nonumber\\
&& {}+ t_{1}^{4} t_{2}^{2} t_{4} + t_{1}^{5} t_{2}^{3} + t_{2}^{4} t_{3}^{5} \nonumber\\
&& {}- t_{1}^{2} t_{2}^{2} t_{3}^{2} t_{4}^{2} - t_{1}^{3} t_{2} t_{3} t_{4}^{4} - 2\, t_{1}^{4} t_{2}^{2} t_{3} t_{4}^{3} - t_{1} t_{2}^{3} t_{3}^{4} t_{4}^{2} - 2\, t_{1}^{5} t_{2}^{3} t_{3} t_{4}^{2} \nonumber\\
&& {}- t_{1}^{2} t_{2}^{4} t_{3}^{4} t_{4} - t_{1}^{2} t_{2}^{2} t_{3}^{3} t_{4}^{4} - t_{1}^{6} t_{2}^{2} t_{4}^{4} - 2\, t_{1}^{3} t_{2}^{3} t_{3}^{3} t_{4}^{3} - t_{2}^{4} t_{3}^{6} t_{4}^{2} - t_{1}^{6} t_{2}^{4} t_{3} t_{4} \nonumber\\
&& {}- t_{1}^{7} t_{2}^{3} t_{4}^{3} - 2\, t_{1}^{4} t_{2}^{4} t_{3}^{3} t_{4}^{2} - t_{1} t_{2}^{5} t_{3}^{6} t_{4} - t_{1}^{8} t_{2}^{4} t_{4}^{2} - 2\, t_{1}^{5} t_{2}^{5} t_{3}^{3} t_{4} - t_{1}^{2} t_{2}^{6} t_{3}^{6} \nonumber\\
&& {}- t_{1}^{2} t_{2}^{4} t_{3}^{5} t_{4}^{3} - t_{1}^{6} t_{2}^{6} t_{3}^{3} - t_{1}^{3} t_{2}^{5} t_{3}^{5} t_{4}^{2} - t_{1}^{4} t_{2}^{6} t_{3}^{5} t_{4} - t_{1}^{5} t_{2}^{7} t_{3}^{5} \nonumber\\
&& {}+ \dots~,~
\eea
where $t_a$ is attached to the extremal perfect matching $p_a$ of the $L^{2,5,3}$ toric diagram in \fref{f_geom_11}, and can be expressed in terms of the global symmetry fugacities $f_{1}, f_{2}$ and $\bar{t}_{1}, \dots, \bar{t}_{4}$ as shown in \tref{t_sym_geom_11}.

The first positive terms of the plethystic logarithm correspond to the generators of the mesonic moduli space, $A_{1}, \dots, A_{10}$. For each toric phase, the generators are given in \tref{t_model_11_1} and \tref{t_model_11_2} in terms of the chiral fields of the corresponding brane tiling. The negative terms that come after the positive ones correspond to the first order relations amongst the generators, which take the following~form,
\beal{es11a004}
&& A_{2}^{2} = A_{1} A_{4} ~,~ \quad A_{2} A_{3} = A_{1} A_{6} ~,~ \quad A_{3} A_{4} = A_{2} A_{6} = A_{1} A_{8} ~,~ \nonumber\\
&& A_{2} A_{5} = A_{1} A_{7} ~,~ \quad A_{4} A_{6} = A_{2} A_{8} = A_{1} A_{9} ~,~ \quad A_{4} A_{5} = A_{2} A_{7} ~,~ \nonumber\\
&& A_{3} A_{5} = A_{1}^{2} A_{4} ~,~ \quad A_{6}^{2} = A_{3} A_{8} ~,~ \quad A_{5} A_{6} = A_{3} A_{7} = A_{1} A_{2} A_{4} ~,~ \nonumber\\
&& A_{5}^{2} = A_{1} A_{10} ~,~ \quad A_{4} A_{8} = A_{2} A_{9} ~,~ \quad A_{6} A_{8} = A_{3} A_{9} ~,~ \nonumber\\
&& A_{6} A_{7} = A_{5} A_{8} = A_{1} A_{4}^{2} ~,~ \quad A_{5} A_{7} = A_{2} A_{10} ~,~ \quad A_{8}^{2} = A_{6} A_{9} ~,~ \nonumber\\
&& A_{7} A_{8} = A_{5} A_{9} = A_{2} A_{4}^{2} ~,~ \quad A_{7}^{2} = A_{4} A_{10} ~,~ \quad A_{3} A_{10} = A_{1} A_{2} A_{7} ~,~ \nonumber\\
&& A_{7} A_{9} = A_{4}^{3} ~,~ \quad A_{6} A_{10} = A_{1} A_{4} A_{7} ~,~ \quad A_{8} A_{10} = A_{2} A_{4} A_{7} ~,~ \nonumber\\
&& A_{9} A_{10} = A_{4}^{2} A_{7}~.~
\eea

\input{model_11_1}
\input{model_11_2}

\subsection{Model 12: $L^{1,3,2}/\mathbb{Z}_{2}\,(0,1,1,1,0,1,0,1,0)$}\label{s_m12}
\begin{figure}[H]
\centering
\begin{minipage}[c]{0.34\textwidth}
\resizebox{\linewidth}{!}{%
\begin{tikzpicture}[x=1bp,y=1bp]
  \node[anchor=south west,inner sep=0pt,outer sep=0pt] at (0,0) {\includegraphics{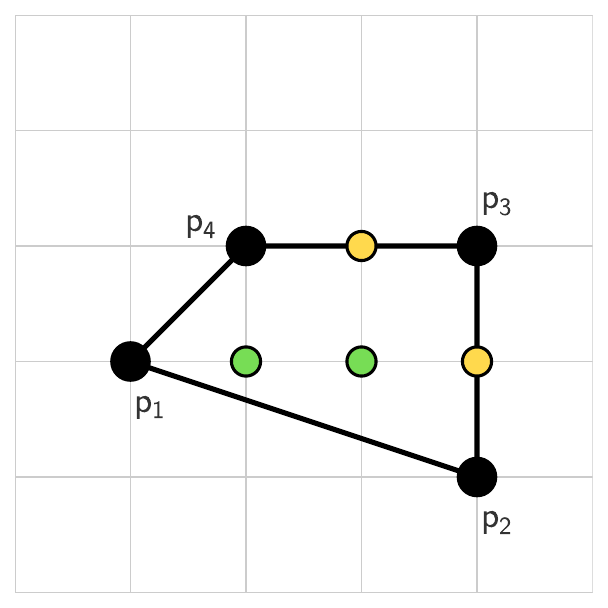}};
  \node[anchor=south west,inner sep=0pt,outer sep=0pt,overlay] at (128.22,120.03) {\includegraphics{figures/toric_label_s1a.pdf}};
  \node[anchor=south west,inner sep=0pt,outer sep=0pt,overlay] at (183.66,120.03) {\includegraphics{figures/toric_label_s2a.pdf}};
\end{tikzpicture}}
\end{minipage}%
\hspace{0.05\textwidth}%
{\small
\setlength{\tabcolsep}{4pt}
\begin{tabular}[c]{|c|c|c|}
\hline
\multirow{2}{*}{phase} & \multicolumn{2}{c|}{multiplicity} \\
\cline{2-3}
& \makebox[2.2em]{$s_{1a}$} & \makebox[2.2em]{$s_{2a}$} \\
\hline\hline
\hyperref[s_model_12_1]{1} & 8 & 13 \\
\hyperref[s_model_12_2]{2} & 8 & 14 \\
\hyperref[s_model_12_3]{3} & 9 & 14 \\
\hline
\end{tabular}
}
\caption{The toric diagram for Model~12 ($L^{1,3,2}/\mathbb{Z}_{2}\,(0,1,1,1,0,1,0,1,0)$).}
\label{f_geom_12}
\end{figure}

\begin{figure}[H]
\centering
\definecolor{dualityred}{rgb}{1,0.149,0}
\hypersetup{linkcolor=black}
\begin{tikzpicture}[line width=1.55bp]
\useasboundingbox (-1.270,-2.520) rectangle (1.270,0.320);
\draw[dualityred, preaction={draw=white, line width=4.55bp}] (-0.222,-0.514) -- (-0.728,-1.686);
\draw[dualityred, preaction={draw=white, line width=4.55bp}] (0.222,-0.514) -- (0.728,-1.686);
\fill[dualityred] (-0.145,-0.337) -- (-0.288,-0.485) -- (-0.155,-0.542) -- cycle;
\fill[dualityred] (-0.805,-1.863) -- (-0.662,-1.715) -- (-0.795,-1.658) -- cycle;
\fill[dualityred] (0.145,-0.337) -- (0.155,-0.542) -- (0.288,-0.485) -- cycle;
\fill[dualityred] (0.805,-1.863) -- (0.795,-1.658) -- (0.662,-1.715) -- cycle;
\node[draw, rectangle, fill=white, minimum size=0.6cm, inner sep=0pt, font=\sffamily\bfseries\footnotesize, line width=2.07bp] at (0.000,0.000) {\hyperref[s_model_12_1]{1}};
\node[draw, rectangle, fill=white, minimum size=0.6cm, inner sep=0pt, font=\sffamily\bfseries\footnotesize, line width=2.07bp] at (-0.950,-2.200) {\hyperref[s_model_12_2]{2}};
\node[draw, rectangle, fill=white, minimum size=0.6cm, inner sep=0pt, font=\sffamily\bfseries\footnotesize, line width=2.07bp] at (0.950,-2.200) {\hyperref[s_model_12_3]{3}};
\end{tikzpicture}
\caption{The Seiberg duality tree for Model~12 ($L^{1,3,2}/\mathbb{Z}_{2}\,(0,1,1,1,0,1,0,1,0)$). The node labels indicate the phase numbers.}
\label{f_geom_12_tree}
\end{figure}

The $L^{1,3,2}/\mathbb{Z}_{2}$ model has a toric diagram shown in \fref{f_geom_12}. The Seiberg duality tree of the $3$ toric phases associated with $L^{1,3,2}/\mathbb{Z}_{2}$ is shown in \fref{f_geom_12_tree}.
The global symmetry of the brane tilings associated with $L^{1,3,2}/\mathbb{Z}_{2}$ is given by,
\beal{es12a001}
U(1) \times U(1) \times U(1)_R~.~
\eea
The global symmetry charges on the extremal GLSM fields $p_a$ are summarized in \tref{t_sym_geom_12}.

\begin{table}[H]
\centering
{\small
\begin{tabular}{|c|c|c|c|c|l|}
\hline
  & $U(1)_{1}$ & $U(1)_{2}$ & $U(1)_R$ & $r_a$ & fugacities \\
\hline
  $p_{1}$ & $-1$ & $-1$ & $r_{1}$ & $r_{1} = 0.5259$ & $t_{1} = f_{1}^{-1} f_{2}^{-1} \bar{t}_{1}$ \\
  $p_{2}$ & $1$ & $0$ & $r_{2}$ & $r_{2} = 0.6056$ & $t_{2} = f_{1} \bar{t}_{2}$ \\
  $p_{3}$ & $0$ & $1$ & $r_{3}$ & $r_{3} = 0.5259$ & $t_{3} = f_{2} \bar{t}_{1}$ \\
  $p_{4}$ & $0$ & $0$ & $r_{4}$ & $r_{4} = 0.3426$ & $t_{4} = \bar{t}_{3}$ \\
\hline
\end{tabular}
}
\caption{Global symmetry charges on the extremal perfect matchings of Model~12 ($L^{1,3,2}/\mathbb{Z}_{2}\,(0,1,1,1,0,1,0,1,0)$).}
\label{t_sym_geom_12}
\end{table}

All toric phases associated with $L^{1,3,2}/\mathbb{Z}_{2}$ share the same mesonic moduli space, whose unrefined Hilbert series takes the form,
\beal{es12a002}
\resizebox{\ifdim\width>\linewidth\linewidth\else\width\fi}{!}{$
g(t; \mathcal{M}^{mes}) \;=\; \frac{1 + t^{4} + t^{5} + t^{7} + t^{8} + 2 t^{9} - 2 t^{11} - t^{12} - t^{13} - t^{15} - t^{16} - t^{20}}{\left(1 - t^{3}\right) \left(1 - t^{5}\right) \left(1 - t^{6}\right) \left(1 - t^{10}\right)}~,~
$}
\eea
where $t$ is the fugacity counting degrees in extremal perfect matchings $p_a$, which corresponds to setting every fugacity $t_a$ in \tref{t_sym_geom_12} equal to $t$.
The plethystic logarithm of the refined Hilbert series takes the form,
\beal{es12a003}
PL\left[g(t_a;\mathcal{M}^{mes})\right] &=& t_{1} t_{2}^{2} + t_{1} t_{2} t_{3} t_{4} + t_{1} t_{3}^{2} t_{4}^{2} + t_{1}^{3} t_{4}^{2} + t_{2}^{4} t_{3}^{2} + t_{2}^{3} t_{3}^{3} t_{4} + t_{2}^{2} t_{3}^{4} t_{4}^{2} \nonumber\\
&& {}+ t_{2} t_{3}^{5} t_{4}^{3} + t_{3}^{6} t_{4}^{4} \nonumber\\
&& {}- t_{1}^{2} t_{2}^{2} t_{3}^{2} t_{4}^{2} - t_{1} t_{2}^{5} t_{3}^{3} t_{4} - t_{1}^{3} t_{2}^{4} t_{3}^{2} t_{4}^{2} - 2\, t_{1} t_{2}^{4} t_{3}^{4} t_{4}^{2} - t_{1}^{3} t_{2}^{3} t_{3}^{3} t_{4}^{3} \nonumber\\
&& {}- 2\, t_{1} t_{2}^{3} t_{3}^{5} t_{4}^{3} - t_{1}^{3} t_{2}^{2} t_{3}^{4} t_{4}^{4} - 2\, t_{1} t_{2}^{2} t_{3}^{6} t_{4}^{4} - t_{2}^{6} t_{3}^{6} t_{4}^{2} - t_{1}^{3} t_{2} t_{3}^{5} t_{4}^{5} \nonumber\\
&& {}- t_{1} t_{2} t_{3}^{7} t_{4}^{5} - t_{2}^{5} t_{3}^{7} t_{4}^{3} - t_{1}^{3} t_{3}^{6} t_{4}^{6} - 2\, t_{2}^{4} t_{3}^{8} t_{4}^{4} - t_{2}^{3} t_{3}^{9} t_{4}^{5} - t_{2}^{2} t_{3}^{10} t_{4}^{6} \nonumber\\
&& {}+ \dots~,~
\eea
with $t_a$ the fugacity of the extremal perfect matching $p_a$ in \fref{f_geom_12}; $t_a$ can be expressed in terms of the global symmetry fugacities $f_{1}, f_{2}$ and $\bar{t}_{1}, \bar{t}_{2}, \bar{t}_{3}$ as given in \tref{t_sym_geom_12}.

The positive terms at the start of the expansion correspond to the generators of the mesonic moduli space, which are $A_{1}, \dots, A_{9}$. The expressions of the generators in terms of the chiral fields of the individual toric phases are summarized in \tref{t_model_12_1}, \tref{t_model_12_2} and \tref{t_model_12_3}. The negative terms that come next correspond to the first order relations amongst the generators, which take the following~form,
\beal{es12a004}
&& A_{2}^{2} = A_{1} A_{3} ~,~ \quad A_{2} A_{5} = A_{1} A_{6} ~,~ \quad A_{4} A_{5} = A_{1}^{2} A_{3} ~,~ \nonumber\\
&& A_{3} A_{5} = A_{2} A_{6} = A_{1} A_{7} ~,~ \quad A_{4} A_{6} = A_{1} A_{2} A_{3} ~,~ \quad A_{3} A_{6} = A_{2} A_{7} = A_{1} A_{8} ~,~ \nonumber\\
&& A_{4} A_{7} = A_{1} A_{3}^{2} ~,~ \quad A_{3} A_{7} = A_{2} A_{8} = A_{1} A_{9} ~,~ \quad A_{6}^{2} = A_{5} A_{7} ~,~ \nonumber\\
&& A_{4} A_{8} = A_{2} A_{3}^{2} ~,~ \quad A_{3} A_{8} = A_{2} A_{9} ~,~ \quad A_{6} A_{7} = A_{5} A_{8} ~,~ \quad A_{4} A_{9} = A_{3}^{3} ~,~ \nonumber\\
&& A_{7}^{2} = A_{6} A_{8} = A_{5} A_{9} ~,~ \quad A_{7} A_{8} = A_{6} A_{9} ~,~ \quad A_{8}^{2} = A_{7} A_{9}~.~
\eea

\input{model_12_1}
\input{model_12_2}
\input{model_12_3}

\subsection{Model 13: $L^{1,3,1}/\mathbb{Z}_{2}\,(1,1,1,1)$}\label{s_m13}
\begin{figure}[H]
\centering
\begin{minipage}[c]{0.34\textwidth}
\resizebox{\linewidth}{!}{%
\begin{tikzpicture}[x=1bp,y=1bp]
  \node[anchor=south west,inner sep=0pt,outer sep=0pt] at (0,0) {\includegraphics{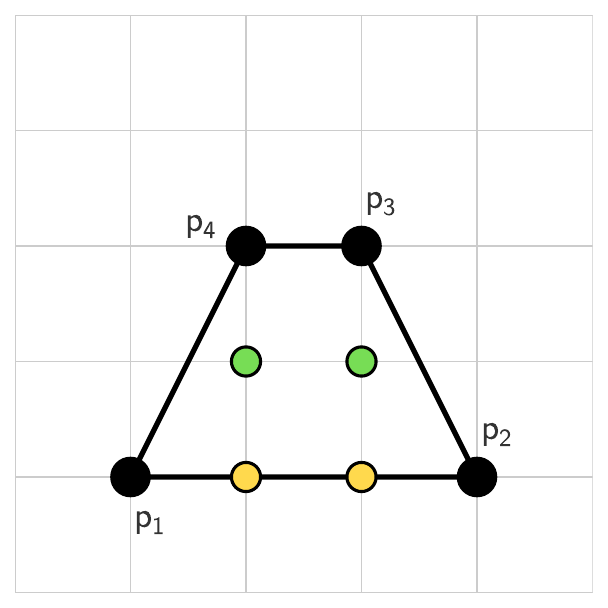}};
  \node[anchor=south west,inner sep=0pt,outer sep=0pt,overlay] at (128.22,120.03) {\includegraphics{figures/toric_label_s1a.pdf}};
  \node[anchor=south west,inner sep=0pt,outer sep=0pt,overlay] at (179.96,97.82) {\includegraphics{figures/toric_label_s2a.pdf}};
\end{tikzpicture}}
\end{minipage}%
\hspace{0.05\textwidth}%
{\small
\setlength{\tabcolsep}{4pt}
\begin{tabular}[c]{|c|c|c|}
\hline
\multirow{2}{*}{phase} & \multicolumn{2}{c|}{multiplicity} \\
\cline{2-3}
& \makebox[2.2em]{$s_{1a}$} & \makebox[2.2em]{$s_{2a}$} \\
\hline\hline
\hyperref[s_model_13_1]{1} & 8 & 8 \\
\hyperref[s_model_13_2]{2} & 9 & 9 \\
\hline
\end{tabular}
}
\caption{The toric diagram for Model~13 ($L^{1,3,1}/\mathbb{Z}_{2}\,(1,1,1,1)$).}
\label{f_geom_13}
\end{figure}

\begin{figure}[H]
\centering
\definecolor{dualityred}{rgb}{1,0.149,0}
\hypersetup{linkcolor=black}
\begin{tikzpicture}[line width=1.55bp]
\useasboundingbox (-0.320,-2.520) rectangle (0.320,0.320);
\draw[dualityred, preaction={draw=white, line width=4.55bp}] (0.000,-1.671) -- (0.000,-0.529);
\fill[dualityred] (0.000,-1.863) -- (0.072,-1.671) -- (-0.072,-1.671) -- cycle;
\fill[dualityred] (0.000,-0.337) -- (-0.072,-0.529) -- (0.072,-0.529) -- cycle;
\node[draw, rectangle, fill=white, minimum size=0.6cm, inner sep=0pt, font=\sffamily\bfseries\footnotesize, line width=2.07bp] at (0.000,-2.200) {\hyperref[s_model_13_1]{1}};
\node[draw, rectangle, fill=white, minimum size=0.6cm, inner sep=0pt, font=\sffamily\bfseries\footnotesize, line width=2.07bp] at (0.000,0.000) {\hyperref[s_model_13_2]{2}};
\end{tikzpicture}
\caption{The Seiberg duality tree for Model~13 ($L^{1,3,1}/\mathbb{Z}_{2}\,(1,1,1,1)$). The node labels indicate the phase numbers.}
\label{f_geom_13_tree}
\end{figure}

The toric diagram of the $L^{1,3,1}/\mathbb{Z}_{2}$ model is shown in \fref{f_geom_13}. The Seiberg duality tree of the $2$ toric phases associated with $L^{1,3,1}/\mathbb{Z}_{2}$ is shown in \fref{f_geom_13_tree}.
The global symmetry of the brane tilings associated with $L^{1,3,1}/\mathbb{Z}_{2}$ is given by,
\beal{es13a001}
U(1) \times U(1) \times U(1)_R~.~
\eea
The global symmetry charges on the extremal GLSM fields $p_a$ are summarized in \tref{t_sym_geom_13}.

\begin{table}[H]
\centering
{\small
\begin{tabular}{|c|c|c|c|c|l|}
\hline
  & $U(1)_{1}$ & $U(1)_{2}$ & $U(1)_R$ & $r_a$ & fugacities \\
\hline
  $p_{1}$ & $-1$ & $-1$ & $r_{1}$ & $r_{1} = 0.6076$ & $t_{1} = f_{1}^{-1} f_{2}^{-1} \bar{t}_{1}$ \\
  $p_{2}$ & $1$ & $0$ & $r_{2}$ & $r_{2} = 0.6076$ & $t_{2} = f_{1} \bar{t}_{1}$ \\
  $p_{3}$ & $0$ & $1$ & $r_{3}$ & $r_{3} = 0.3924$ & $t_{3} = f_{2} \bar{t}_{2}$ \\
  $p_{4}$ & $0$ & $0$ & $r_{4}$ & $r_{4} = 0.3924$ & $t_{4} = \bar{t}_{2}$ \\
\hline
\end{tabular}
}
\caption{Global symmetry charges on the extremal perfect matchings of Model~13 ($L^{1,3,1}/\mathbb{Z}_{2}\,(1,1,1,1)$).}
\label{t_sym_geom_13}
\end{table}

All toric phases associated with $L^{1,3,1}/\mathbb{Z}_{2}$ share the same mesonic moduli space, whose unrefined Hilbert series takes the form,
\beal{es13a002}
\resizebox{\ifdim\width>\linewidth\linewidth\else\width\fi}{!}{$
g(t; \mathcal{M}^{mes}) \;=\; \frac{1 - 2 t^{2} + 4 t^{4} - 2 t^{6} + t^{8}}{\left(1 - t^{2}\right)^{2} \left(1 - t^{8}\right)}~,~
$}
\eea
where $t$ is the fugacity counting degrees in extremal perfect matchings $p_a$, which corresponds to setting every fugacity $t_a$ in \tref{t_sym_geom_13} equal to $t$.
The plethystic logarithm of the refined Hilbert series takes the form,
\beal{es13a003}
PL\left[g(t_a;\mathcal{M}^{mes})\right] &=& t_{3}^{2} t_{4}^{2} + t_{1} t_{2} t_{3} t_{4} + t_{1}^{2} t_{2}^{2} + t_{2}^{3} t_{3}^{2} t_{4} + t_{1} t_{2}^{4} t_{3} + t_{1}^{3} t_{3} t_{4}^{2} + t_{1}^{4} t_{2} t_{4} \nonumber\\
&& {}+ t_{2}^{6} t_{3}^{2} + t_{1}^{6} t_{4}^{2} \nonumber\\
&& {}- t_{1}^{2} t_{2}^{2} t_{3}^{2} t_{4}^{2} - t_{1}^{5} t_{2}^{2} t_{3} t_{4}^{2} - t_{1}^{2} t_{2}^{5} t_{3}^{2} t_{4} - t_{1}^{4} t_{2} t_{3}^{2} t_{4}^{3} - t_{1} t_{2}^{4} t_{3}^{3} t_{4}^{2} - t_{1}^{8} t_{2}^{2} t_{4}^{2} \nonumber\\
&& {}- t_{1}^{7} t_{2} t_{3} t_{4}^{3} - t_{1}^{6} t_{3}^{2} t_{4}^{4} - t_{1}^{5} t_{2}^{5} t_{3} t_{4} - 2\, t_{1}^{4} t_{2}^{4} t_{3}^{2} t_{4}^{2} - t_{1}^{2} t_{2}^{8} t_{3}^{2} - t_{1}^{3} t_{2}^{3} t_{3}^{3} t_{4}^{3} \nonumber\\
&& {}- t_{1} t_{2}^{7} t_{3}^{3} t_{4} - t_{2}^{6} t_{3}^{4} t_{4}^{2} - t_{1}^{4} t_{2}^{7} t_{3}^{2} t_{4} - t_{1}^{3} t_{2}^{6} t_{3}^{3} t_{4}^{2} - t_{1}^{7} t_{2}^{4} t_{3} t_{4}^{2} - t_{1}^{6} t_{2}^{3} t_{3}^{2} t_{4}^{3} \nonumber\\
&& {}- t_{1}^{6} t_{2}^{6} t_{3}^{2} t_{4}^{2} + \dots~,~
\eea
where the $4$ fugacities $t_{1}, \dots, t_{4}$ grade each mesonic operator by its content in $p_{1}, \dots, p_{4}$ of \fref{f_geom_13}, and can be expressed in terms of the global symmetry fugacities $f_{1}, f_{2}$ and $\bar{t}_{1}, \bar{t}_{2}$ as shown in \tref{t_sym_geom_13}.

The generators of the mesonic moduli space, $A_{1}, \dots, A_{9}$, correspond to the first positive terms of the plethystic logarithm. In \tref{t_model_13_1} and \tref{t_model_13_2}, the generators are given in terms of the chiral fields of each toric phase. The negative terms that follow correspond to the first order relations amongst the generators, which take the following~form,
\beal{es13a004}
&& A_{2}^{2} = A_{1} A_{3} ~,~ \quad A_{3} A_{6} = A_{2} A_{7} ~,~ \quad A_{3} A_{4} = A_{2} A_{5} ~,~ \quad A_{2} A_{6} = A_{1} A_{7} ~,~ \nonumber\\
&& A_{2} A_{4} = A_{1} A_{5} ~,~ \quad A_{7}^{2} = A_{3} A_{9} ~,~ \quad A_{6} A_{7} = A_{2} A_{9} ~,~ \quad A_{6}^{2} = A_{1} A_{9} ~,~ \nonumber\\
&& A_{5} A_{7} = A_{2} A_{3}^{2} ~,~ \quad A_{5} A_{6} = A_{4} A_{7} = A_{1} A_{3}^{2} ~,~ \quad A_{5}^{2} = A_{3} A_{8} ~,~ \nonumber\\
&& A_{4} A_{6} = A_{1} A_{2} A_{3} ~,~ \quad A_{4} A_{5} = A_{2} A_{8} ~,~ \quad A_{4}^{2} = A_{1} A_{8} ~,~ \quad A_{7} A_{8} = A_{2} A_{3} A_{5} ~,~ \nonumber\\
&& A_{6} A_{8} = A_{1} A_{3} A_{5} ~,~ \quad A_{5} A_{9} = A_{2} A_{3} A_{7} ~,~ \quad A_{4} A_{9} = A_{1} A_{3} A_{7} ~,~ \nonumber\\
&& A_{8} A_{9} = A_{1} A_{3}^{3}~.~
\eea

\input{model_13_1}
\input{model_13_2}

\subsection{Model 14: $\mathcal{C}/\mathbb{Z}_{4}\,(1,2,3,2)$}\label{s_m14}
\begin{figure}[H]
\centering
\begin{minipage}[c]{0.34\textwidth}
\resizebox{\linewidth}{!}{%
\begin{tikzpicture}[x=1bp,y=1bp]
  \node[anchor=south west,inner sep=0pt,outer sep=0pt] at (0,0) {\includegraphics{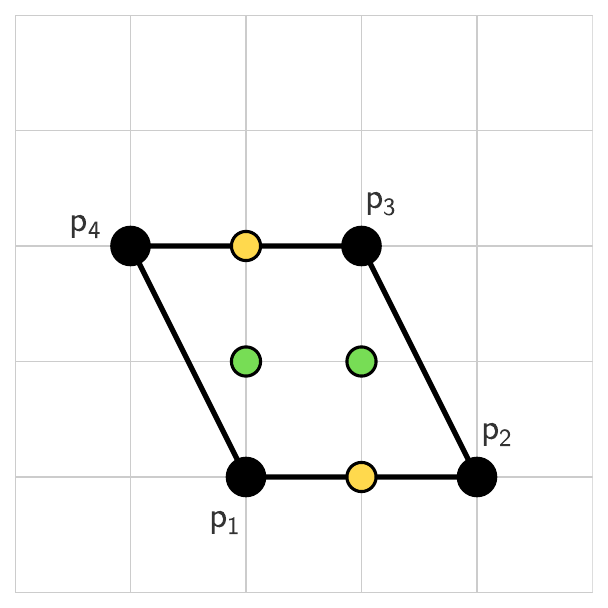}};
  \node[anchor=south west,inner sep=0pt,outer sep=0pt,overlay] at (128.22,120.03) {\includegraphics{figures/toric_label_s1a.pdf}};
  \node[anchor=south west,inner sep=0pt,outer sep=0pt,overlay] at (179.96,97.82) {\includegraphics{figures/toric_label_s2a.pdf}};
\end{tikzpicture}}
\end{minipage}%
\hspace{0.05\textwidth}%
{\small
\setlength{\tabcolsep}{4pt}
\begin{tabular}[c]{|c|c|c|}
\hline
\multirow{2}{*}{phase} & \multicolumn{2}{c|}{multiplicity} \\
\cline{2-3}
& \makebox[2.2em]{$s_{1a}$} & \makebox[2.2em]{$s_{2a}$} \\
\hline\hline
\hyperref[s_model_14_1]{1} & 8 & 8 \\
\hyperref[s_model_14_2]{2} & 8 & 8 \\
\hyperref[s_model_14_3]{3} & 9 & 9 \\
\hyperref[s_model_14_4]{4} & 10 & 10 \\
\hline
\end{tabular}
}
\caption{The toric diagram for Model~14 ($\mathcal{C}/\mathbb{Z}_{4}\,(1,2,3,2)$).}
\label{f_geom_14}
\end{figure}

\begin{figure}[H]
\centering
\definecolor{dualityred}{rgb}{1,0.149,0}
\hypersetup{linkcolor=black}
\begin{tikzpicture}[line width=1.55bp]
\useasboundingbox (-1.270,-4.720) rectangle (1.270,0.320);
\draw[dualityred, preaction={draw=white, line width=4.55bp}] (-0.728,-3.886) -- (-0.222,-2.714);
\draw[dualityred, preaction={draw=white, line width=4.55bp}] (0.000,-0.529) -- (0.000,-1.671);
\draw[dualityred, preaction={draw=white, line width=4.55bp}] (0.222,-2.714) -- (0.728,-3.886);
\fill[dualityred] (-0.805,-4.063) -- (-0.662,-3.915) -- (-0.795,-3.858) -- cycle;
\fill[dualityred] (-0.145,-2.537) -- (-0.288,-2.685) -- (-0.155,-2.742) -- cycle;
\fill[dualityred] (0.000,-0.337) -- (-0.072,-0.529) -- (0.072,-0.529) -- cycle;
\fill[dualityred] (0.000,-1.863) -- (0.072,-1.671) -- (-0.072,-1.671) -- cycle;
\fill[dualityred] (0.145,-2.537) -- (0.155,-2.742) -- (0.288,-2.685) -- cycle;
\fill[dualityred] (0.805,-4.063) -- (0.795,-3.858) -- (0.662,-3.915) -- cycle;
\node[draw, rectangle, fill=white, minimum size=0.6cm, inner sep=0pt, font=\sffamily\bfseries\footnotesize, line width=2.07bp] at (-0.950,-4.400) {\hyperref[s_model_14_1]{1}};
\node[draw, rectangle, fill=white, minimum size=0.6cm, inner sep=0pt, font=\sffamily\bfseries\footnotesize, line width=2.07bp] at (0.000,0.000) {\hyperref[s_model_14_2]{2}};
\node[draw, rectangle, fill=white, minimum size=0.6cm, inner sep=0pt, font=\sffamily\bfseries\footnotesize, line width=2.07bp] at (0.000,-2.200) {\hyperref[s_model_14_3]{3}};
\node[draw, rectangle, fill=white, minimum size=0.6cm, inner sep=0pt, font=\sffamily\bfseries\footnotesize, line width=2.07bp] at (0.950,-4.400) {\hyperref[s_model_14_4]{4}};
\end{tikzpicture}
\caption{The Seiberg duality tree for Model~14 ($\mathcal{C}/\mathbb{Z}_{4}\,(1,2,3,2)$). The node labels indicate the phase numbers.}
\label{f_geom_14_tree}
\end{figure}

\fref{f_geom_14} illustrates the toric diagram for the $\mathcal{C}/\mathbb{Z}_{4}$ model. The Seiberg duality tree of the $4$ toric phases associated with $\mathcal{C}/\mathbb{Z}_{4}$ is shown in \fref{f_geom_14_tree}.
The global symmetry of the brane tilings associated with $\mathcal{C}/\mathbb{Z}_{4}$ is given by,
\beal{es14a001}
U(1) \times U(1) \times U(1)_R~.~
\eea
The global symmetry charges on the extremal GLSM fields $p_a$ are summarized in \tref{t_sym_geom_14}.

\begin{table}[H]
\centering
{\small
\begin{tabular}{|c|c|c|c|c|l|}
\hline
  & $U(1)_{1}$ & $U(1)_{2}$ & $U(1)_R$ & $r_a$ & fugacities \\
\hline
  $p_{1}$ & $-1$ & $-1$ & $r_{1}$ & $r_{1} = 1/2$ & $t_{1} = f_{1}^{-1} f_{2}^{-1} \bar{t}$ \\
  $p_{2}$ & $1$ & $0$ & $r_{2}$ & $r_{2} = 1/2$ & $t_{2} = f_{1} \bar{t}$ \\
  $p_{3}$ & $0$ & $1$ & $r_{3}$ & $r_{3} = 1/2$ & $t_{3} = f_{2} \bar{t}$ \\
  $p_{4}$ & $0$ & $0$ & $r_{4}$ & $r_{4} = 1/2$ & $t_{4} = \bar{t}$ \\
\hline
\end{tabular}
}
\caption{Global symmetry charges on the extremal perfect matchings of Model~14 ($\mathcal{C}/\mathbb{Z}_{4}\,(1,2,3,2)$).}
\label{t_sym_geom_14}
\end{table}

All toric phases associated with $\mathcal{C}/\mathbb{Z}_{4}$ share the same mesonic moduli space, whose unrefined Hilbert series takes the form,
\beal{es14a002}
\resizebox{\ifdim\width>\linewidth\linewidth\else\width\fi}{!}{$
g(t; \mathcal{M}^{mes}) \;=\; \frac{1 - 2 t^{2} + 4 t^{4} - 2 t^{6} + t^{8}}{\left(1 - t^{2}\right)^{2} \left(1 - t^{8}\right)}~,~
$}
\eea
where $t$ is the fugacity counting degrees in extremal perfect matchings $p_a$, which corresponds to setting every fugacity $t_a$ in \tref{t_sym_geom_14} equal to $t$.
The plethystic logarithm of the refined Hilbert series takes the form,
\beal{es14a003}
PL\left[g(t_a;\mathcal{M}^{mes})\right] &=& t_{3}^{2} t_{4}^{2} + t_{1} t_{2} t_{3} t_{4} + t_{1}^{2} t_{2}^{2} + t_{2}^{2} t_{3}^{3} t_{4} + t_{1} t_{2}^{3} t_{3}^{2} + t_{1}^{2} t_{3} t_{4}^{3} + t_{1}^{3} t_{2} t_{4}^{2} \nonumber\\
&& {}+ t_{2}^{4} t_{3}^{4} + t_{1}^{4} t_{4}^{4} \nonumber\\
&& {}- t_{1}^{2} t_{2}^{2} t_{3}^{2} t_{4}^{2} - t_{1}^{4} t_{2}^{2} t_{3} t_{4}^{3} - t_{1}^{2} t_{2}^{4} t_{3}^{3} t_{4} - t_{1}^{3} t_{2} t_{3}^{2} t_{4}^{4} - t_{1} t_{2}^{3} t_{3}^{4} t_{4}^{2} - t_{1}^{6} t_{2}^{2} t_{4}^{4} \nonumber\\
&& {}- t_{1}^{5} t_{2} t_{3} t_{4}^{5} - t_{1}^{4} t_{3}^{2} t_{4}^{6} - t_{1}^{4} t_{2}^{4} t_{3}^{2} t_{4}^{2} - 2\, t_{1}^{3} t_{2}^{3} t_{3}^{3} t_{4}^{3} - t_{1}^{2} t_{2}^{6} t_{3}^{4} - t_{1}^{2} t_{2}^{2} t_{3}^{4} t_{4}^{4} \nonumber\\
&& {}- t_{1} t_{2}^{5} t_{3}^{5} t_{4} - t_{2}^{4} t_{3}^{6} t_{4}^{2} - t_{1}^{3} t_{2}^{5} t_{3}^{4} t_{4}^{2} - t_{1}^{2} t_{2}^{4} t_{3}^{5} t_{4}^{3} - t_{1}^{5} t_{2}^{3} t_{3}^{2} t_{4}^{4} - t_{1}^{4} t_{2}^{2} t_{3}^{3} t_{4}^{5} \nonumber\\
&& {}- t_{1}^{4} t_{2}^{4} t_{3}^{4} t_{4}^{4} + \dots~,~
\eea
in which $t_a$, the fugacity of the extremal perfect matching $p_a$ at a corner of \fref{f_geom_14}, can be expressed in terms of the global symmetry fugacities $f_{1}, f_{2}$ and $\bar{t}$ as shown in \tref{t_sym_geom_14}.

The positive terms with which the expansion begins correspond to the generators of the mesonic moduli space, namely $A_{1}, \dots, A_{9}$. For each toric phase, the generators are given in \tref{t_model_14_1}, \tref{t_model_14_2}, \tref{t_model_14_3} and \tref{t_model_14_4} in terms of the chiral fields of the corresponding brane tiling. The next negative terms correspond to the first order relations amongst the generators, which take the following~form,
\beal{es14a004}
&& A_{2}^{2} = A_{1} A_{3} ~,~ \quad A_{3} A_{6} = A_{2} A_{7} ~,~ \quad A_{3} A_{4} = A_{2} A_{5} ~,~ \quad A_{2} A_{6} = A_{1} A_{7} ~,~ \nonumber\\
&& A_{2} A_{4} = A_{1} A_{5} ~,~ \quad A_{7}^{2} = A_{3} A_{9} ~,~ \quad A_{6} A_{7} = A_{2} A_{9} ~,~ \quad A_{6}^{2} = A_{1} A_{9} ~,~ \nonumber\\
&& A_{5} A_{7} = A_{1} A_{3}^{2} ~,~ \quad A_{5} A_{6} = A_{4} A_{7} = A_{1} A_{2} A_{3} ~,~ \quad A_{5}^{2} = A_{3} A_{8} ~,~ \nonumber\\
&& A_{4} A_{6} = A_{1}^{2} A_{3} ~,~ \quad A_{4} A_{5} = A_{2} A_{8} ~,~ \quad A_{4}^{2} = A_{1} A_{8} ~,~ \quad A_{7} A_{8} = A_{1} A_{3} A_{5} ~,~ \nonumber\\
&& A_{6} A_{8} = A_{1} A_{2} A_{5} ~,~ \quad A_{5} A_{9} = A_{1} A_{3} A_{7} ~,~ \quad A_{4} A_{9} = A_{1} A_{2} A_{7} ~,~ \nonumber\\
&& A_{8} A_{9} = A_{1}^{2} A_{3}^{2}~.~
\eea

\input{model_14_1}
\input{model_14_2}
\input{model_14_3}
\input{model_14_4}

\subsection{Model 15: $L^{2,6,5}$}\label{s_m15}
\begin{figure}[H]
\centering
\begin{minipage}[c]{0.34\textwidth}
\resizebox{\linewidth}{!}{%
\begin{tikzpicture}[x=1bp,y=1bp]
  \node[anchor=south west,inner sep=0pt,outer sep=0pt] at (0,0) {\includegraphics{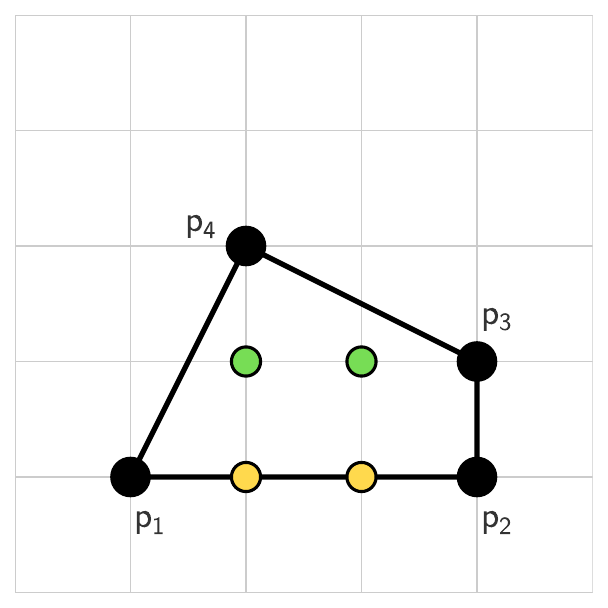}};
  \node[anchor=south west,inner sep=0pt,outer sep=0pt,overlay] at (128.22,120.03) {\includegraphics{figures/toric_label_s1a.pdf}};
  \node[anchor=south west,inner sep=0pt,outer sep=0pt,overlay] at (183.66,99.42) {\includegraphics{figures/toric_label_s2a.pdf}};
\end{tikzpicture}}
\end{minipage}%
\hspace{0.05\textwidth}%
{\small
\setlength{\tabcolsep}{4pt}
\begin{tabular}[c]{|c|c|c|}
\hline
\multirow{2}{*}{phase} & \multicolumn{2}{c|}{multiplicity} \\
\cline{2-3}
& \makebox[2.2em]{$s_{1a}$} & \makebox[2.2em]{$s_{2a}$} \\
\hline\hline
\hyperref[s_model_15_1]{1} & 8 & 9 \\
\hyperref[s_model_15_2]{2} & 8 & 11 \\
\hline
\end{tabular}
}
\caption{The toric diagram for Model~15 ($L^{2,6,5}$).}
\label{f_geom_15}
\end{figure}

\begin{figure}[H]
\centering
\definecolor{dualityred}{rgb}{1,0.149,0}
\hypersetup{linkcolor=black}
\begin{tikzpicture}[line width=1.55bp]
\useasboundingbox (-0.338,-3.034) rectangle (0.338,0.320);
\draw[dualityred, preaction={draw=white, line width=4.55bp}] (0.000,-1.671) -- (0.000,-0.529);
\draw[dualityred, preaction={draw=white, line width=4.55bp}] (-0.301,-2.700) .. controls (-0.454,-3.118) and (0.454,-3.118) .. (0.301,-2.700);
\fill[dualityred] (0.000,-1.863) -- (0.072,-1.671) -- (-0.072,-1.671) -- cycle;
\fill[dualityred] (0.000,-0.337) -- (-0.072,-0.529) -- (0.072,-0.529) -- cycle;
\fill[dualityred] (-0.200,-2.537) -- (-0.363,-2.662) -- (-0.239,-2.738) -- cycle;
\fill[dualityred] (0.200,-2.537) -- (0.239,-2.738) -- (0.363,-2.662) -- cycle;
\node[draw, rectangle, fill=white, minimum size=0.6cm, inner sep=0pt, font=\sffamily\bfseries\footnotesize, line width=2.07bp] at (0.000,-2.200) {\hyperref[s_model_15_1]{1}};
\node[draw, rectangle, fill=white, minimum size=0.6cm, inner sep=0pt, font=\sffamily\bfseries\footnotesize, line width=2.07bp] at (0.000,0.000) {\hyperref[s_model_15_2]{2}};
\end{tikzpicture}
\caption{The Seiberg duality tree for Model~15 ($L^{2,6,5}$). The node labels indicate the phase numbers.}
\label{f_geom_15_tree}
\end{figure}

The $L^{2,6,5}$ model has a toric diagram shown in \fref{f_geom_15}. The Seiberg duality tree of the $2$ toric phases associated with $L^{2,6,5}$ is shown in \fref{f_geom_15_tree}.
The global symmetry of the brane tilings associated with $L^{2,6,5}$ is given by,
\beal{es15a001}
U(1) \times U(1) \times U(1)_R~.~
\eea
The global symmetry charges on the extremal GLSM fields $p_a$ are summarized in \tref{t_sym_geom_15}.

\begin{table}[H]
\centering
{\small
\begin{tabular}{|c|c|c|c|c|l|}
\hline
  & $U(1)_{1}$ & $U(1)_{2}$ & $U(1)_R$ & $r_a$ & fugacities \\
\hline
  $p_{1}$ & $-1$ & $-1$ & $r_{1}$ & $r_{1} = 0.6058$ & $t_{1} = f_{1}^{-1} f_{2}^{-1} \bar{t}_{1}$ \\
  $p_{2}$ & $1$ & $0$ & $r_{2}$ & $r_{2} = 0.4707$ & $t_{2} = f_{1} \bar{t}_{2}$ \\
  $p_{3}$ & $0$ & $1$ & $r_{3}$ & $r_{3} = 0.3540$ & $t_{3} = f_{2} \bar{t}_{3}$ \\
  $p_{4}$ & $0$ & $0$ & $r_{4}$ & $r_{4} = 0.5695$ & $t_{4} = \bar{t}_{4}$ \\
\hline
\end{tabular}
}
\caption{Global symmetry charges on the extremal perfect matchings of Model~15 ($L^{2,6,5}$).}
\label{t_sym_geom_15}
\end{table}

All toric phases associated with $L^{2,6,5}$ share the same mesonic moduli space, whose unrefined Hilbert series takes the form,
\beal{es15a002}
\resizebox{\ifdim\width>\linewidth\linewidth\else\width\fi}{!}{$
g(t; \mathcal{M}^{mes}) \;=\; \frac{1 + t^{3} + t^{4} + t^{5} + 2 t^{6} + 2 t^{7} + 2 t^{8} + 2 t^{9} + 2 t^{10} + 2 t^{11} + 2 t^{12} + 2 t^{13} + t^{14} + t^{15} + t^{16} + t^{19}}{\left(1 - t^{5}\right) \left(1 - t^{7}\right) \left(1 - t^{11}\right)}~,~
$}
\eea
where $t$ is the fugacity counting degrees in extremal perfect matchings $p_a$, which corresponds to setting every fugacity $t_a$ in \tref{t_sym_geom_15} equal to $t$.
The plethystic logarithm of the refined Hilbert series takes the form,
\beal{es15a003}
PL\left[g(t_a;\mathcal{M}^{mes})\right] &=& t_{3} t_{4}^{2} + t_{1} t_{2} t_{3} t_{4} + t_{1}^{2} t_{2}^{2} t_{3} + t_{1}^{3} t_{4}^{2} + t_{1}^{4} t_{2} t_{4} + t_{2}^{3} t_{3}^{3} t_{4} + t_{1}^{5} t_{2}^{2} \nonumber\\
&& {}+ t_{1} t_{2}^{4} t_{3}^{3} + t_{2}^{6} t_{3}^{5} \nonumber\\
&& {}- t_{1}^{2} t_{2}^{2} t_{3}^{2} t_{4}^{2} - t_{1}^{4} t_{2} t_{3} t_{4}^{3} - 2\, t_{1}^{5} t_{2}^{2} t_{3} t_{4}^{2} - t_{1}^{6} t_{2}^{3} t_{3} t_{4} - t_{1} t_{2}^{4} t_{3}^{4} t_{4}^{2} \nonumber\\
&& {}- t_{1}^{8} t_{2}^{2} t_{4}^{2} - t_{1}^{3} t_{2}^{3} t_{3}^{3} t_{4}^{3} - t_{1}^{2} t_{2}^{5} t_{3}^{4} t_{4} - 2\, t_{1}^{4} t_{2}^{4} t_{3}^{3} t_{4}^{2} - 2\, t_{1}^{5} t_{2}^{5} t_{3}^{3} t_{4} \nonumber\\
&& {}- t_{2}^{6} t_{3}^{6} t_{4}^{2} - t_{1}^{6} t_{2}^{6} t_{3}^{3} - t_{1} t_{2}^{7} t_{3}^{6} t_{4} - t_{1}^{2} t_{2}^{8} t_{3}^{6} - t_{1}^{3} t_{2}^{6} t_{3}^{5} t_{4}^{2} - t_{1}^{4} t_{2}^{7} t_{3}^{5} t_{4} \nonumber\\
&& {}- t_{1}^{5} t_{2}^{8} t_{3}^{5} + \dots~,~
\eea
where every $t_a$ belongs to the extremal perfect matching $p_a$ of \fref{f_geom_15} and can be expressed in terms of the global symmetry fugacities $f_{1}, f_{2}$ and $\bar{t}_{1}, \dots, \bar{t}_{4}$, as listed in \tref{t_sym_geom_15}.

Each of the first $9$ positive terms corresponds to one generator of the mesonic moduli space, the generators being $A_{1}, \dots, A_{9}$. The expressions of the generators in terms of the chiral fields of the individual toric phases are summarized in \tref{t_model_15_1} and \tref{t_model_15_2}. The negative terms that come after the positive ones correspond to the first order relations amongst the generators, which take the following~form,
\beal{es15a004}
&& A_{2}^{2} = A_{1} A_{3} ~,~ \quad A_{2} A_{4} = A_{1} A_{5} ~,~ \quad A_{3} A_{4} = A_{2} A_{5} = A_{1} A_{7} ~,~ \nonumber\\
&& A_{3} A_{5} = A_{2} A_{7} ~,~ \quad A_{2} A_{6} = A_{1} A_{8} ~,~ \quad A_{5}^{2} = A_{4} A_{7} ~,~ \quad A_{4} A_{6} = A_{1} A_{2} A_{3} ~,~ \nonumber\\
&& A_{3} A_{6} = A_{2} A_{8} ~,~ \quad A_{5} A_{6} = A_{4} A_{8} = A_{1} A_{3}^{2} ~,~ \quad A_{6} A_{7} = A_{5} A_{8} = A_{2} A_{3}^{2} ~,~ \nonumber\\
&& A_{6}^{2} = A_{1} A_{9} ~,~ \quad A_{7} A_{8} = A_{3}^{3} ~,~ \quad A_{6} A_{8} = A_{2} A_{9} ~,~ \quad A_{8}^{2} = A_{3} A_{9} ~,~ \nonumber\\
&& A_{4} A_{9} = A_{1} A_{3} A_{8} ~,~ \quad A_{5} A_{9} = A_{2} A_{3} A_{8} ~,~ \quad A_{7} A_{9} = A_{3}^{2} A_{8}~.~
\eea

\input{model_15_1}
\input{model_15_2}

\subsection{Model 16: $L^{1,2,1}/\mathbb{Z}_{3}\,(0,1,1,2)$}\label{s_m16}
\begin{figure}[H]
\centering
\begin{minipage}[c]{0.34\textwidth}
\resizebox{\linewidth}{!}{%
\begin{tikzpicture}[x=1bp,y=1bp]
  \node[anchor=south west,inner sep=0pt,outer sep=0pt] at (0,0) {\includegraphics{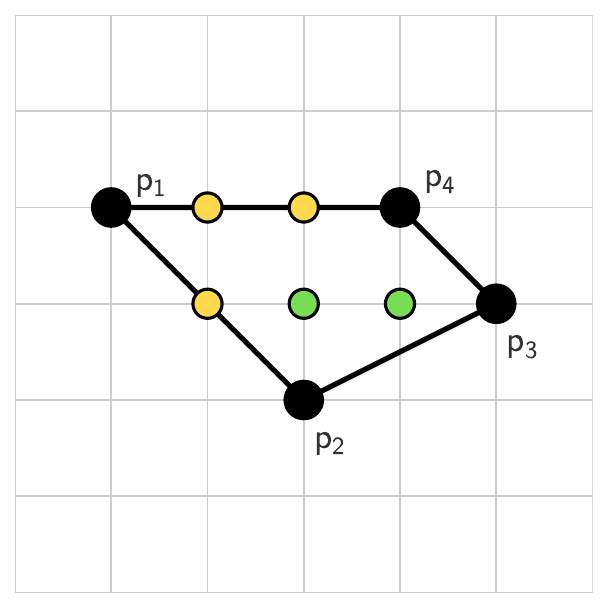}};
  \node[anchor=south west,inner sep=0pt,outer sep=0pt,overlay] at (113.94,147.75) {\includegraphics{figures/toric_label_s1a.pdf}};
  \node[anchor=south west,inner sep=0pt,outer sep=0pt,overlay] at (160.14,147.75) {\includegraphics{figures/toric_label_s2a.pdf}};
\end{tikzpicture}}
\end{minipage}%
\hspace{0.05\textwidth}%
{\small
\setlength{\tabcolsep}{4pt}
\begin{tabular}[c]{|c|c|c|}
\hline
\multirow{2}{*}{phase} & \multicolumn{2}{c|}{multiplicity} \\
\cline{2-3}
& \makebox[2.2em]{$s_{1a}$} & \makebox[2.2em]{$s_{2a}$} \\
\hline\hline
\hyperref[s_model_16_1]{1} & 15 & 9 \\
\hyperref[s_model_16_2]{2} & 17 & 10 \\
\hyperref[s_model_16_3]{3} & 17 & 12 \\
\hline
\end{tabular}
}
\caption{The toric diagram for Model~16 ($L^{1,2,1}/\mathbb{Z}_{3}\,(0,1,1,2)$).}
\label{f_geom_16}
\end{figure}

\begin{figure}[H]
\centering
\definecolor{dualityred}{rgb}{1,0.149,0}
\hypersetup{linkcolor=black}
\begin{tikzpicture}[line width=1.55bp]
\useasboundingbox (-0.717,-4.720) rectangle (0.717,0.320);
\draw[dualityred, preaction={draw=white, line width=4.55bp}] (0.000,-3.871) -- (0.000,-2.729);
\draw[dualityred, preaction={draw=white, line width=4.55bp}] (0.000,-1.671) -- (0.000,-0.529);
\draw[dualityred, preaction={draw=white, line width=4.55bp}] (0.526,-2.075) .. controls (1.025,-1.908) and (0.292,-1.175) .. (0.125,-1.674);
\fill[dualityred] (0.000,-4.063) -- (0.072,-3.871) -- (-0.072,-3.871) -- cycle;
\fill[dualityred] (0.000,-2.537) -- (-0.072,-2.729) -- (0.072,-2.729) -- cycle;
\fill[dualityred] (0.000,-1.863) -- (0.072,-1.671) -- (-0.072,-1.671) -- cycle;
\fill[dualityred] (0.000,-0.337) -- (-0.072,-0.529) -- (0.072,-0.529) -- cycle;
\fill[dualityred] (0.337,-2.107) -- (0.538,-2.146) -- (0.514,-2.004) -- cycle;
\fill[dualityred] (0.093,-1.863) -- (0.196,-1.686) -- (0.054,-1.662) -- cycle;
\node[draw, rectangle, fill=white, minimum size=0.6cm, inner sep=0pt, font=\sffamily\bfseries\footnotesize, line width=2.07bp] at (0.000,-4.400) {\hyperref[s_model_16_1]{1}};
\node[draw, rectangle, fill=white, minimum size=0.6cm, inner sep=0pt, font=\sffamily\bfseries\footnotesize, line width=2.07bp] at (0.000,-2.200) {\hyperref[s_model_16_2]{2}};
\node[draw, rectangle, fill=white, minimum size=0.6cm, inner sep=0pt, font=\sffamily\bfseries\footnotesize, line width=2.07bp] at (0.000,0.000) {\hyperref[s_model_16_3]{3}};
\end{tikzpicture}
\caption{The Seiberg duality tree for Model~16 ($L^{1,2,1}/\mathbb{Z}_{3}\,(0,1,1,2)$). The node labels indicate the phase numbers.}
\label{f_geom_16_tree}
\end{figure}

The toric diagram of the $L^{1,2,1}/\mathbb{Z}_{3}$ model is shown in \fref{f_geom_16}. The Seiberg duality tree of the $3$ toric phases associated with $L^{1,2,1}/\mathbb{Z}_{3}$ is shown in \fref{f_geom_16_tree}.
The global symmetry of the brane tilings associated with $L^{1,2,1}/\mathbb{Z}_{3}$ is given by,
\beal{es16a001}
U(1) \times U(1) \times U(1)_R~.~
\eea
The global symmetry charges on the extremal GLSM fields $p_a$ are summarized in \tref{t_sym_geom_16}.

\begin{table}[H]
\centering
{\small
\begin{tabular}{|c|c|c|c|c|l|}
\hline
  & $U(1)_{1}$ & $U(1)_{2}$ & $U(1)_R$ & $r_a$ & fugacities \\
\hline
  $p_{1}$ & $-1$ & $-1$ & $r_{1}$ & $r_{1} = 0.5774$ & $t_{1} = f_{1}^{-1} f_{2}^{-1} \bar{t}_{1}$ \\
  $p_{2}$ & $1$ & $0$ & $r_{2}$ & $r_{2} = 0.5774$ & $t_{2} = f_{1} \bar{t}_{1}$ \\
  $p_{3}$ & $0$ & $1$ & $r_{3}$ & $r_{3} = 0.4226$ & $t_{3} = f_{2} \bar{t}_{2}$ \\
  $p_{4}$ & $0$ & $0$ & $r_{4}$ & $r_{4} = 0.4226$ & $t_{4} = \bar{t}_{2}$ \\
\hline
\end{tabular}
}
\caption{Global symmetry charges on the extremal perfect matchings of Model~16 ($L^{1,2,1}/\mathbb{Z}_{3}\,(0,1,1,2)$).}
\label{t_sym_geom_16}
\end{table}

All toric phases associated with $L^{1,2,1}/\mathbb{Z}_{3}$ share the same mesonic moduli space, whose unrefined Hilbert series takes the form,
\beal{es16a002}
\resizebox{\ifdim\width>\linewidth\linewidth\else\width\fi}{!}{$
g(t; \mathcal{M}^{mes}) \;=\; \frac{1 - t^{2} + t^{4} + t^{5} + t^{6} - t^{8} + t^{10}}{\left(1 - t^{2}\right) \left(1 - t^{3}\right) \left(1 - t^{9}\right)}~,~
$}
\eea
where $t$ is the fugacity counting degrees in extremal perfect matchings $p_a$, which corresponds to setting every fugacity $t_a$ in \tref{t_sym_geom_16} equal to $t$.
The plethystic logarithm of the refined Hilbert series takes the form,
\beal{es16a003}
PL\left[g(t_a;\mathcal{M}^{mes})\right] &=& t_{2}^{2} t_{3} + t_{1} t_{2} t_{3} t_{4} + t_{1}^{2} t_{3} t_{4}^{2} + t_{3}^{3} t_{4}^{3} + t_{1}^{3} t_{2}^{3} + t_{1}^{4} t_{2}^{2} t_{4} + t_{1}^{5} t_{2} t_{4}^{2} \nonumber\\
&& {}+ t_{1}^{6} t_{4}^{3} \nonumber\\
&& {}- t_{1}^{2} t_{2}^{2} t_{3}^{2} t_{4}^{2} - t_{1}^{4} t_{2}^{4} t_{3} t_{4} - 2\, t_{1}^{5} t_{2}^{3} t_{3} t_{4}^{2} - t_{1}^{3} t_{2}^{3} t_{3}^{3} t_{4}^{3} - 2\, t_{1}^{6} t_{2}^{2} t_{3} t_{4}^{3} \nonumber\\
&& {}- t_{1}^{4} t_{2}^{2} t_{3}^{3} t_{4}^{4} - t_{1}^{7} t_{2} t_{3} t_{4}^{4} - t_{1}^{8} t_{2}^{4} t_{4}^{2} - t_{1}^{5} t_{2} t_{3}^{3} t_{4}^{5} - t_{1}^{9} t_{2}^{3} t_{4}^{3} - t_{1}^{6} t_{3}^{3} t_{4}^{6} \nonumber\\
&& {}- t_{1}^{10} t_{2}^{2} t_{4}^{4} + \dots~.~
\eea
Here the fugacity $t_a$ keeps track of the extremal perfect matching $p_a$ at a corner of \fref{f_geom_16}, and can be expressed in terms of the global symmetry fugacities $f_{1}, f_{2}$ and $\bar{t}_{1}, \bar{t}_{2}$ as shown in \tref{t_sym_geom_16}.

The first positive terms of the plethystic logarithm correspond to the generators of the mesonic moduli space, $A_{1}, \dots, A_{8}$. In \tref{t_model_16_1}, \tref{t_model_16_2} and \tref{t_model_16_3}, the generators are given in terms of the chiral fields of each toric phase. The negative terms that come next correspond to the first order relations amongst the generators, which take the following~form,
\beal{es16a004}
&& A_{2}^{2} = A_{1} A_{3} ~,~ \quad A_{2} A_{5} = A_{1} A_{6} ~,~ \quad A_{3} A_{5} = A_{2} A_{6} = A_{1} A_{7} ~,~ \nonumber\\
&& A_{4} A_{5} = A_{1} A_{2} A_{3} ~,~ \quad A_{3} A_{6} = A_{2} A_{7} = A_{1} A_{8} ~,~ \quad A_{4} A_{6} = A_{1} A_{3}^{2} ~,~ \nonumber\\
&& A_{3} A_{7} = A_{2} A_{8} ~,~ \quad A_{6}^{2} = A_{5} A_{7} ~,~ \quad A_{4} A_{7} = A_{2} A_{3}^{2} ~,~ \quad A_{6} A_{7} = A_{5} A_{8} ~,~ \nonumber\\
&& A_{4} A_{8} = A_{3}^{3} ~,~ \quad A_{7}^{2} = A_{6} A_{8}~.~
\eea

\input{model_16_1}
\input{model_16_2}
\input{model_16_3}

\subsection{Model 17: $L^{4,5,1}$}\label{s_m17}
\input{model_17_1}

\subsection{Model 18: $L^{2,3,2}/\mathbb{Z}_{2}\,(0,1,1,1)$}\label{s_m18}
\begin{figure}[H]
\centering
\begin{minipage}[c]{0.34\textwidth}
\resizebox{\linewidth}{!}{%
\begin{tikzpicture}[x=1bp,y=1bp]
  \node[anchor=south west,inner sep=0pt,outer sep=0pt] at (0,0) {\includegraphics{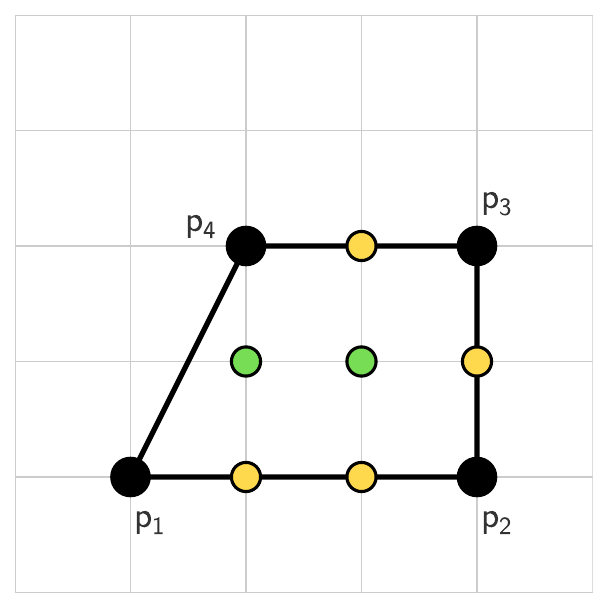}};
  \node[anchor=south west,inner sep=0pt,outer sep=0pt,overlay] at (128.22,120.03) {\includegraphics{figures/toric_label_s1a.pdf}};
  \node[anchor=south west,inner sep=0pt,outer sep=0pt,overlay] at (183.66,120.03) {\includegraphics{figures/toric_label_s2a.pdf}};
\end{tikzpicture}}
\end{minipage}%
\hspace{0.05\textwidth}%
{\small
\setlength{\tabcolsep}{4pt}
\begin{tabular}[c]{|c|c|c|}
\hline
\multirow{2}{*}{phase} & \multicolumn{2}{c|}{multiplicity} \\
\cline{2-3}
& \makebox[2.2em]{$s_{1a}$} & \makebox[2.2em]{$s_{2a}$} \\
\hline\hline
\hyperref[s_model_18_1]{1} & 10 & 20 \\
\hyperref[s_model_18_2]{2} & 11 & 23 \\
\hyperref[s_model_18_3]{3} & 10 & 20 \\
\hyperref[s_model_18_4]{4} & 11 & 23 \\
\hyperref[s_model_18_5]{5} & 12 & 22 \\
\hyperref[s_model_18_6]{6} & 12 & 31 \\
\hline
\end{tabular}
}
\caption{The toric diagram for Model~18 ($L^{2,3,2}/\mathbb{Z}_{2}\,(0,1,1,1)$).}
\label{f_geom_18}
\end{figure}

\begin{figure}[H]
\centering
\definecolor{dualityred}{rgb}{1,0.149,0}
\hypersetup{linkcolor=black}
\begin{tikzpicture}[line width=1.55bp]
\useasboundingbox (-3.170,-4.720) rectangle (3.170,0.320);
\draw[dualityred, preaction={draw=white, line width=4.55bp}] (-2.361,-2.578) -- (-0.489,-4.022);
\draw[dualityred, preaction={draw=white, line width=4.55bp}] (-2.361,-1.822) -- (-0.489,-0.378);
\draw[dualityred, preaction={draw=white, line width=4.55bp}] (-0.222,-3.886) -- (-0.728,-2.714);
\draw[dualityred, preaction={draw=white, line width=4.55bp}] (0.222,-3.886) -- (0.728,-2.714);
\draw[dualityred, preaction={draw=white, line width=4.55bp}] (2.361,-1.822) -- (0.489,-0.378);
\draw[dualityred, preaction={draw=white, line width=4.55bp}] (-0.222,-0.514) -- (-0.728,-1.686);
\draw[dualityred, preaction={draw=white, line width=4.55bp}] (0.222,-0.514) -- (0.728,-1.686);
\fill[dualityred] (-2.513,-2.460) -- (-2.405,-2.635) -- (-2.317,-2.520) -- cycle;
\fill[dualityred] (-0.337,-4.140) -- (-0.445,-3.965) -- (-0.533,-4.080) -- cycle;
\fill[dualityred] (-2.513,-1.940) -- (-2.317,-1.880) -- (-2.405,-1.765) -- cycle;
\fill[dualityred] (-0.337,-0.260) -- (-0.533,-0.320) -- (-0.445,-0.435) -- cycle;
\fill[dualityred] (-0.145,-4.063) -- (-0.155,-3.858) -- (-0.288,-3.915) -- cycle;
\fill[dualityred] (-0.805,-2.537) -- (-0.795,-2.742) -- (-0.662,-2.685) -- cycle;
\fill[dualityred] (0.145,-4.063) -- (0.288,-3.915) -- (0.155,-3.858) -- cycle;
\fill[dualityred] (0.805,-2.537) -- (0.662,-2.685) -- (0.795,-2.742) -- cycle;
\fill[dualityred] (2.513,-1.940) -- (2.405,-1.765) -- (2.317,-1.880) -- cycle;
\fill[dualityred] (0.337,-0.260) -- (0.445,-0.435) -- (0.533,-0.320) -- cycle;
\fill[dualityred] (-0.145,-0.337) -- (-0.288,-0.485) -- (-0.155,-0.542) -- cycle;
\fill[dualityred] (-0.805,-1.863) -- (-0.662,-1.715) -- (-0.795,-1.658) -- cycle;
\fill[dualityred] (0.145,-0.337) -- (0.155,-0.542) -- (0.288,-0.485) -- cycle;
\fill[dualityred] (0.805,-1.863) -- (0.795,-1.658) -- (0.662,-1.715) -- cycle;
\node[draw, rectangle, fill=white, minimum size=0.6cm, inner sep=0pt, font=\sffamily\bfseries\footnotesize, line width=2.07bp] at (-2.850,-2.200) {\hyperref[s_model_18_1]{1}};
\node[draw, rectangle, fill=white, minimum size=0.6cm, inner sep=0pt, font=\sffamily\bfseries\footnotesize, line width=2.07bp] at (0.000,-4.400) {\hyperref[s_model_18_2]{2}};
\node[draw, rectangle, fill=white, minimum size=0.6cm, inner sep=0pt, font=\sffamily\bfseries\footnotesize, line width=2.07bp] at (2.850,-2.200) {\hyperref[s_model_18_3]{3}};
\node[draw, rectangle, fill=white, minimum size=0.6cm, inner sep=0pt, font=\sffamily\bfseries\footnotesize, line width=2.07bp] at (0.000,0.000) {\hyperref[s_model_18_4]{4}};
\node[draw, rectangle, fill=white, minimum size=0.6cm, inner sep=0pt, font=\sffamily\bfseries\footnotesize, line width=2.07bp] at (-0.950,-2.200) {\hyperref[s_model_18_5]{5}};
\node[draw, rectangle, fill=white, minimum size=0.6cm, inner sep=0pt, font=\sffamily\bfseries\footnotesize, line width=2.07bp] at (0.950,-2.200) {\hyperref[s_model_18_6]{6}};
\end{tikzpicture}
\caption{The Seiberg duality tree for Model~18 ($L^{2,3,2}/\mathbb{Z}_{2}\,(0,1,1,1)$). The node labels indicate the phase numbers.}
\label{f_geom_18_tree}
\end{figure}

The $L^{2,3,2}/\mathbb{Z}_{2}$ model has a toric diagram shown in \fref{f_geom_18}. The Seiberg duality tree of the $6$ toric phases associated with $L^{2,3,2}/\mathbb{Z}_{2}$ is shown in \fref{f_geom_18_tree}.
The global symmetry of the brane tilings associated with $L^{2,3,2}/\mathbb{Z}_{2}$ is given by,
\beal{es18a001}
U(1) \times U(1) \times U(1)_R~.~
\eea
The global symmetry charges on the extremal GLSM fields $p_a$ are summarized in \tref{t_sym_geom_18}.

\begin{table}[H]
\centering
{\small
\begin{tabular}{|c|c|c|c|c|l|}
\hline
  & $U(1)_{1}$ & $U(1)_{2}$ & $U(1)_R$ & $r_a$ & fugacities \\
\hline
  $p_{1}$ & $-1$ & $-1$ & $r_{1}$ & $r_{1} = 0.5486$ & $t_{1} = f_{1}^{-1} f_{2}^{-1} \bar{t}_{1}$ \\
  $p_{2}$ & $1$ & $0$ & $r_{2}$ & $r_{2} = 0.5486$ & $t_{2} = f_{1} \bar{t}_{1}$ \\
  $p_{3}$ & $0$ & $1$ & $r_{3}$ & $r_{3} = 0.4514$ & $t_{3} = f_{2} \bar{t}_{2}$ \\
  $p_{4}$ & $0$ & $0$ & $r_{4}$ & $r_{4} = 0.4514$ & $t_{4} = \bar{t}_{2}$ \\
\hline
\end{tabular}
}
\caption{Global symmetry charges on the extremal perfect matchings of Model~18 ($L^{2,3,2}/\mathbb{Z}_{2}\,(0,1,1,1)$).}
\label{t_sym_geom_18}
\end{table}

All toric phases associated with $L^{2,3,2}/\mathbb{Z}_{2}$ share the same mesonic moduli space, whose unrefined Hilbert series takes the form,
\beal{es18a002}
\resizebox{\ifdim\width>\linewidth\linewidth\else\width\fi}{!}{$
g(t; \mathcal{M}^{mes}) \;=\; \frac{1 - t + 2 t^{4} - t^{5} - t^{6} + 2 t^{7} - t^{10} + t^{11}}{\left(1 - t\right) \left(1 - t^{4}\right) \left(1 - t^{10}\right)}~,~
$}
\eea
where $t$ is the fugacity counting degrees in extremal perfect matchings $p_a$, which corresponds to setting every fugacity $t_a$ in \tref{t_sym_geom_18} equal to $t$.
The plethystic logarithm of the refined Hilbert series takes the form,
\beal{es18a003}
PL\left[g(t_a;\mathcal{M}^{mes})\right] &=& t_{3}^{2} t_{4}^{2} + t_{1} t_{2} t_{3} t_{4} + t_{1}^{2} t_{2}^{2} + t_{1}^{3} t_{4}^{2} + t_{2}^{3} t_{3}^{3} t_{4} + t_{1} t_{2}^{4} t_{3}^{2} + t_{2}^{6} t_{3}^{4} \nonumber\\
&& {}- t_{1}^{2} t_{2}^{2} t_{3}^{2} t_{4}^{2} - t_{1}^{2} t_{2}^{5} t_{3}^{3} t_{4} - t_{1} t_{2}^{4} t_{3}^{4} t_{4}^{2} - t_{1}^{4} t_{2}^{4} t_{3}^{2} t_{4}^{2} - t_{1}^{3} t_{2}^{3} t_{3}^{3} t_{4}^{3} - t_{1}^{2} t_{2}^{8} t_{3}^{4} \nonumber\\
&& {}- t_{1} t_{2}^{7} t_{3}^{5} t_{4} - t_{2}^{6} t_{3}^{6} t_{4}^{2} - t_{1}^{3} t_{2}^{6} t_{3}^{4} t_{4}^{2} + \dots~,~
\eea
where $t_a$ is the fugacity of $p_a$ in \fref{f_geom_18} and can be expressed in terms of the global symmetry fugacities $f_{1}, f_{2}$ and $\bar{t}_{1}, \bar{t}_{2}$ as shown in \tref{t_sym_geom_18}.

The generators of the mesonic moduli space, $A_{1}, \dots, A_{7}$, correspond to the first positive terms of the plethystic logarithm. The expressions of the generators in terms of the chiral fields of the individual toric phases are summarized in \tref{t_model_18_1}, \tref{t_model_18_2}, \tref{t_model_18_3}, \tref{t_model_18_4}, \tref{t_model_18_5} and \tref{t_model_18_6}. The next negative terms correspond to the first order relations amongst the generators, which take the following~form,
\beal{es18a004}
&& A_{2}^{2} = A_{1} A_{3} ~,~ \quad A_{3} A_{5} = A_{2} A_{6} ~,~ \quad A_{2} A_{5} = A_{1} A_{6} ~,~ \quad A_{4} A_{6} = A_{1} A_{3}^{2} ~,~ \nonumber\\
&& A_{4} A_{5} = A_{1} A_{2} A_{3} ~,~ \quad A_{6}^{2} = A_{3} A_{7} ~,~ \quad A_{5} A_{6} = A_{2} A_{7} ~,~ \quad A_{5}^{2} = A_{1} A_{7} ~,~ \nonumber\\
&& A_{4} A_{7} = A_{1} A_{3} A_{6}~.~
\eea

\input{model_18_1}
\input{model_18_2}
\input{model_18_3}
\input{model_18_4}
\input{model_18_5}
\input{model_18_6}

\subsection{Model 19: $L^{1,4,1}/\mathbb{Z}_{2}\,(0,1,1,1)$}\label{s_m19}
\begin{figure}[H]
\centering
\begin{minipage}[c]{0.34\textwidth}
\resizebox{\linewidth}{!}{%
\begin{tikzpicture}[x=1bp,y=1bp]
  \node[anchor=south west,inner sep=0pt,outer sep=0pt] at (0,0) {\includegraphics{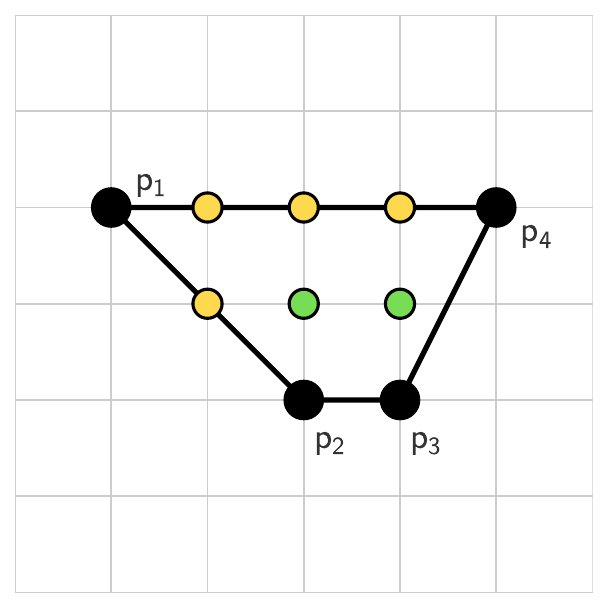}};
  \node[anchor=south west,inner sep=0pt,outer sep=0pt,overlay] at (113.94,147.75) {\includegraphics{figures/toric_label_s1a.pdf}};
  \node[anchor=south west,inner sep=0pt,outer sep=0pt,overlay] at (160.14,147.75) {\includegraphics{figures/toric_label_s2a.pdf}};
\end{tikzpicture}}
\end{minipage}%
\hspace{0.05\textwidth}%
{\small
\setlength{\tabcolsep}{4pt}
\begin{tabular}[c]{|c|c|c|}
\hline
\multirow{2}{*}{phase} & \multicolumn{2}{c|}{multiplicity} \\
\cline{2-3}
& \makebox[2.2em]{$s_{1a}$} & \makebox[2.2em]{$s_{2a}$} \\
\hline\hline
\hyperref[s_model_19_1]{1} & 20 & 10 \\
\hyperref[s_model_19_2]{2} & 23 & 11 \\
\hline
\end{tabular}
}
\caption{The toric diagram for Model~19 ($L^{1,4,1}/\mathbb{Z}_{2}\,(0,1,1,1)$).}
\label{f_geom_19}
\end{figure}

\begin{figure}[H]
\centering
\definecolor{dualityred}{rgb}{1,0.149,0}
\hypersetup{linkcolor=black}
\begin{tikzpicture}[line width=1.55bp]
\useasboundingbox (-0.320,-2.520) rectangle (0.320,0.320);
\draw[dualityred, preaction={draw=white, line width=4.55bp}] (0.000,-1.671) -- (0.000,-0.529);
\fill[dualityred] (0.000,-1.863) -- (0.072,-1.671) -- (-0.072,-1.671) -- cycle;
\fill[dualityred] (0.000,-0.337) -- (-0.072,-0.529) -- (0.072,-0.529) -- cycle;
\node[draw, rectangle, fill=white, minimum size=0.6cm, inner sep=0pt, font=\sffamily\bfseries\footnotesize, line width=2.07bp] at (0.000,-2.200) {\hyperref[s_model_19_1]{1}};
\node[draw, rectangle, fill=white, minimum size=0.6cm, inner sep=0pt, font=\sffamily\bfseries\footnotesize, line width=2.07bp] at (0.000,0.000) {\hyperref[s_model_19_2]{2}};
\end{tikzpicture}
\caption{The Seiberg duality tree for Model~19 ($L^{1,4,1}/\mathbb{Z}_{2}\,(0,1,1,1)$). The node labels indicate the phase numbers.}
\label{f_geom_19_tree}
\end{figure}

The toric diagram of the $L^{1,4,1}/\mathbb{Z}_{2}$ model is shown in \fref{f_geom_19}. The Seiberg duality tree of the $2$ toric phases associated with $L^{1,4,1}/\mathbb{Z}_{2}$ is shown in \fref{f_geom_19_tree}.
The global symmetry of the brane tilings associated with $L^{1,4,1}/\mathbb{Z}_{2}$ is given by,
\beal{es19a001}
U(1) \times U(1) \times U(1)_R~.~
\eea
The global symmetry charges on the extremal GLSM fields $p_a$ are summarized in \tref{t_sym_geom_19}.

\begin{table}[H]
\centering
{\small
\begin{tabular}{|c|c|c|c|c|l|}
\hline
  & $U(1)_{1}$ & $U(1)_{2}$ & $U(1)_R$ & $r_a$ & fugacities \\
\hline
  $p_{1}$ & $-1$ & $-1$ & $r_{1}$ & $r_{1} = 0.6228$ & $t_{1} = f_{1}^{-1} f_{2}^{-1} \bar{t}_{1}$ \\
  $p_{2}$ & $1$ & $0$ & $r_{2}$ & $r_{2} = 0.3772$ & $t_{2} = f_{1} \bar{t}_{2}$ \\
  $p_{3}$ & $0$ & $1$ & $r_{3}$ & $r_{3} = 0.3772$ & $t_{3} = f_{2} \bar{t}_{2}$ \\
  $p_{4}$ & $0$ & $0$ & $r_{4}$ & $r_{4} = 0.6228$ & $t_{4} = \bar{t}_{1}$ \\
\hline
\end{tabular}
}
\caption{Global symmetry charges on the extremal perfect matchings of Model~19 ($L^{1,4,1}/\mathbb{Z}_{2}\,(0,1,1,1)$).}
\label{t_sym_geom_19}
\end{table}

All toric phases associated with $L^{1,4,1}/\mathbb{Z}_{2}$ share the same mesonic moduli space, whose unrefined Hilbert series takes the form,
\beal{es19a002}
\resizebox{\ifdim\width>\linewidth\linewidth\else\width\fi}{!}{$
g(t; \mathcal{M}^{mes}) \;=\; \frac{1 - t + 2 t^{4} - t^{5} - t^{6} + 2 t^{7} - t^{10} + t^{11}}{\left(1 - t\right) \left(1 - t^{4}\right) \left(1 - t^{10}\right)}~,~
$}
\eea
where $t$ is the fugacity counting degrees in extremal perfect matchings $p_a$, which corresponds to setting every fugacity $t_a$ in \tref{t_sym_geom_19} equal to $t$.
The plethystic logarithm of the refined Hilbert series takes the form,
\beal{es19a003}
PL\left[g(t_a;\mathcal{M}^{mes})\right] &=& t_{2}^{2} t_{3}^{2} + t_{1} t_{2} t_{3} t_{4} + t_{1}^{2} t_{4}^{2} + t_{3} t_{4}^{4} + t_{1}^{4} t_{2}^{2} t_{3} + t_{1}^{5} t_{2} t_{4} + t_{1}^{8} t_{2}^{2} \nonumber\\
&& {}- t_{1}^{2} t_{2}^{2} t_{3}^{2} t_{4}^{2} - t_{1}^{6} t_{2}^{2} t_{3} t_{4}^{2} - t_{1}^{5} t_{2}^{3} t_{3}^{2} t_{4} - t_{1}^{5} t_{2} t_{3} t_{4}^{5} - t_{1}^{4} t_{2}^{2} t_{3}^{2} t_{4}^{4} \nonumber\\
&& {}- t_{1}^{10} t_{2}^{2} t_{4}^{2} - t_{1}^{9} t_{2}^{3} t_{3} t_{4} - t_{1}^{8} t_{2}^{4} t_{3}^{2} - t_{1}^{8} t_{2}^{2} t_{3} t_{4}^{4} + \dots~,~
\eea
where $t_a$ is the fugacity of the extremal perfect matching $p_a$ of \fref{f_geom_19} and can be expressed in terms of the global symmetry fugacities $f_{1}, f_{2}$ and $\bar{t}_{1}, \bar{t}_{2}$ as summarized in \tref{t_sym_geom_19}.

The positive terms with which the expansion begins correspond to the generators of the mesonic moduli space, namely $A_{1}, \dots, A_{7}$. In \tref{t_model_19_1} and \tref{t_model_19_2}, the generators are given in terms of the chiral fields of each toric phase. The negative terms that come after the positive ones correspond to the first order relations amongst the generators, which take the following~form,
\beal{es19a004}
&& A_{2}^{2} = A_{1} A_{3} ~,~ \quad A_{3} A_{5} = A_{2} A_{6} ~,~ \quad A_{2} A_{5} = A_{1} A_{6} ~,~ \quad A_{4} A_{6} = A_{2} A_{3}^{2} ~,~ \nonumber\\
&& A_{4} A_{5} = A_{1} A_{3}^{2} ~,~ \quad A_{6}^{2} = A_{3} A_{7} ~,~ \quad A_{5} A_{6} = A_{2} A_{7} ~,~ \quad A_{5}^{2} = A_{1} A_{7} ~,~ \nonumber\\
&& A_{4} A_{7} = A_{2} A_{3} A_{6}~.~
\eea

\input{model_19_1}
\input{model_19_2}

\subsection{Model 20: $L^{1,4,3}/\mathbb{Z}_{2}\,(1,1,0,1,0,1,0,0)$}\label{s_m20}
\begin{figure}[H]
\centering
\begin{minipage}[c]{0.34\textwidth}
\resizebox{\linewidth}{!}{%
\begin{tikzpicture}[x=1bp,y=1bp]
  \node[anchor=south west,inner sep=0pt,outer sep=0pt] at (0,0) {\includegraphics{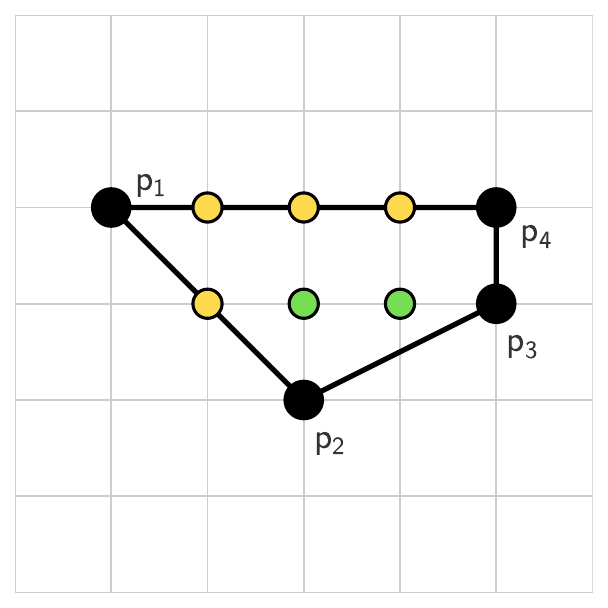}};
  \node[anchor=south west,inner sep=0pt,outer sep=0pt,overlay] at (155.94,127.14) {\includegraphics{figures/toric_label_s1a.pdf}};
  \node[anchor=south west,inner sep=0pt,outer sep=0pt,overlay] at (202.14,147.75) {\includegraphics{figures/toric_label_s2a.pdf}};
\end{tikzpicture}}
\end{minipage}%
\hspace{0.05\textwidth}%
{\small
\setlength{\tabcolsep}{4pt}
\begin{tabular}[c]{|c|c|c|}
\hline
\multirow{2}{*}{phase} & \multicolumn{2}{c|}{multiplicity} \\
\cline{2-3}
& \makebox[2.2em]{$s_{1a}$} & \makebox[2.2em]{$s_{2a}$} \\
\hline\hline
\hyperref[s_model_20_1]{1} & 20 & 13 \\
\hyperref[s_model_20_2]{2} & 21 & 12 \\
\hyperref[s_model_20_3]{3} & 21 & 16 \\
\hline
\end{tabular}
}
\caption{The toric diagram for Model~20 ($L^{1,4,3}/\mathbb{Z}_{2}\,(1,1,0,1,0,1,0,0)$).}
\label{f_geom_20}
\end{figure}

\begin{figure}[H]
\centering
\definecolor{dualityred}{rgb}{1,0.149,0}
\hypersetup{linkcolor=black}
\begin{tikzpicture}[line width=1.55bp]
\useasboundingbox (-0.320,-4.720) rectangle (0.320,0.320);
\draw[dualityred, preaction={draw=white, line width=4.55bp}] (0.000,-2.729) -- (0.000,-3.871);
\draw[dualityred, preaction={draw=white, line width=4.55bp}] (0.000,-1.671) -- (0.000,-0.529);
\fill[dualityred] (0.000,-2.537) -- (-0.072,-2.729) -- (0.072,-2.729) -- cycle;
\fill[dualityred] (0.000,-4.063) -- (0.072,-3.871) -- (-0.072,-3.871) -- cycle;
\fill[dualityred] (0.000,-1.863) -- (0.072,-1.671) -- (-0.072,-1.671) -- cycle;
\fill[dualityred] (0.000,-0.337) -- (-0.072,-0.529) -- (0.072,-0.529) -- cycle;
\node[draw, rectangle, fill=white, minimum size=0.6cm, inner sep=0pt, font=\sffamily\bfseries\footnotesize, line width=2.07bp] at (0.000,-2.200) {\hyperref[s_model_20_1]{1}};
\node[draw, rectangle, fill=white, minimum size=0.6cm, inner sep=0pt, font=\sffamily\bfseries\footnotesize, line width=2.07bp] at (0.000,-4.400) {\hyperref[s_model_20_2]{2}};
\node[draw, rectangle, fill=white, minimum size=0.6cm, inner sep=0pt, font=\sffamily\bfseries\footnotesize, line width=2.07bp] at (0.000,0.000) {\hyperref[s_model_20_3]{3}};
\end{tikzpicture}
\caption{The Seiberg duality tree for Model~20 ($L^{1,4,3}/\mathbb{Z}_{2}\,(1,1,0,1,0,1,0,0)$). The node labels indicate the phase numbers.}
\label{f_geom_20_tree}
\end{figure}

\fref{f_geom_20} illustrates the toric diagram for the $L^{1,4,3}/\mathbb{Z}_{2}$ model. The Seiberg duality tree of the $3$ toric phases associated with $L^{1,4,3}/\mathbb{Z}_{2}$ is shown in \fref{f_geom_20_tree}.
The global symmetry of the brane tilings associated with $L^{1,4,3}/\mathbb{Z}_{2}$ is given by,
\beal{es20a001}
U(1) \times U(1) \times U(1)_R~.~
\eea
The global symmetry charges on the extremal GLSM fields $p_a$ are summarized in \tref{t_sym_geom_20}.

\begin{table}[H]
\centering
{\small
\begin{tabular}{|c|c|c|c|c|l|}
\hline
  & $U(1)_{1}$ & $U(1)_{2}$ & $U(1)_R$ & $r_a$ & fugacities \\
\hline
  $p_{1}$ & $-1$ & $-1$ & $r_{1}$ & $r_{1} = 0.6223$ & $t_{1} = f_{1}^{-1} f_{2}^{-1} \bar{t}_{1}$ \\
  $p_{2}$ & $1$ & $0$ & $r_{2}$ & $r_{2} = 0.5698$ & $t_{2} = f_{1} \bar{t}_{2}$ \\
  $p_{3}$ & $0$ & $1$ & $r_{3}$ & $r_{3} = 0.3062$ & $t_{3} = f_{2} \bar{t}_{3}$ \\
  $p_{4}$ & $0$ & $0$ & $r_{4}$ & $r_{4} = 0.5016$ & $t_{4} = \bar{t}_{4}$ \\
\hline
\end{tabular}
}
\caption{Global symmetry charges on the extremal perfect matchings of Model~20 ($L^{1,4,3}/\mathbb{Z}_{2}\,(1,1,0,1,0,1,0,0)$).}
\label{t_sym_geom_20}
\end{table}

All toric phases associated with $L^{1,4,3}/\mathbb{Z}_{2}$ share the same mesonic moduli space, whose unrefined Hilbert series takes the form,
\beal{es20a002}
\resizebox{\ifdim\width>\linewidth\linewidth\else\width\fi}{!}{$
g(t; \mathcal{M}^{mes}) \;=\; \frac{1 + t^{4} + t^{5} + t^{7} + t^{9} - t^{11} - t^{13} - t^{15} - t^{16} - t^{20}}{\left(1 - t^{3}\right) \left(1 - t^{6}\right) \left(1 - t^{7}\right) \left(1 - t^{8}\right)}~,~
$}
\eea
where $t$ is the fugacity counting degrees in extremal perfect matchings $p_a$, which corresponds to setting every fugacity $t_a$ in \tref{t_sym_geom_20} equal to $t$.
The plethystic logarithm of the refined Hilbert series takes the form,
\beal{es20a003}
PL\left[g(t_a;\mathcal{M}^{mes})\right] &=& t_{2}^{2} t_{3} + t_{1} t_{2} t_{3} t_{4} + t_{1}^{2} t_{3} t_{4}^{2} + t_{1}^{4} t_{2}^{2} + t_{3}^{3} t_{4}^{4} + t_{1}^{5} t_{2} t_{4} + t_{1}^{6} t_{4}^{2} \nonumber\\
&& {}- t_{1}^{2} t_{2}^{2} t_{3}^{2} t_{4}^{2} - t_{1}^{5} t_{2}^{3} t_{3} t_{4} - 2\, t_{1}^{6} t_{2}^{2} t_{3} t_{4}^{2} - t_{1}^{7} t_{2} t_{3} t_{4}^{3} - t_{1}^{4} t_{2}^{2} t_{3}^{3} t_{4}^{4} \nonumber\\
&& {}- t_{1}^{10} t_{2}^{2} t_{4}^{2} - t_{1}^{5} t_{2} t_{3}^{3} t_{4}^{5} - t_{1}^{6} t_{3}^{3} t_{4}^{6} + \dots~,~
\eea
where $t_a$, the fugacity of $p_a$ in \fref{f_geom_20}, can be expressed in terms of the global symmetry fugacities $f_{1}, f_{2}$ and $\bar{t}_{1}, \dots, \bar{t}_{4}$ as shown in \tref{t_sym_geom_20}.

Each of the first $7$ positive terms corresponds to one generator of the mesonic moduli space, the generators being $A_{1}, \dots, A_{7}$. For each toric phase, the generators are given in \tref{t_model_20_1}, \tref{t_model_20_2} and \tref{t_model_20_3} in terms of the chiral fields of the corresponding brane tiling. The negative terms that come next correspond to the first order relations amongst the generators, which take the following~form,
\beal{es20a004}
&& A_{2}^{2} = A_{1} A_{3} ~,~ \quad A_{2} A_{4} = A_{1} A_{6} ~,~ \quad A_{3} A_{4} = A_{2} A_{6} = A_{1} A_{7} ~,~ \nonumber\\
&& A_{3} A_{6} = A_{2} A_{7} ~,~ \quad A_{4} A_{5} = A_{1} A_{3}^{2} ~,~ \quad A_{6}^{2} = A_{4} A_{7} ~,~ \quad A_{5} A_{6} = A_{2} A_{3}^{2} ~,~ \nonumber\\
&& A_{5} A_{7} = A_{3}^{3}~.~
\eea

\input{model_20_1}
\input{model_20_2}
\input{model_20_3}

\subsection{Model 21: $L^{5,6,1}$}\label{s_m21}
\input{model_21_1}

\subsection{Model 22: $L^{1,2,1}/\mathbb{Z}_{2} \times \mathbb{Z}_{2}\,(0,1,1,1)\,(1,0,1,0)$}\label{s_m22}
\input{geom_22}
\input{model_22_1}
\input{model_22_2}
\input{model_22_3}
\input{model_22_4}
\input{model_22_5}
\input{model_22_6}
\input{model_22_7}
\input{model_22_8}

\subsection{Model 23: $L^{1,5,1}/\mathbb{Z}_{2}\,(0,1,0,1)$}\label{s_m23}
\begin{figure}[H]
\centering
\begin{minipage}[c]{0.34\textwidth}
\resizebox{\linewidth}{!}{%
\begin{tikzpicture}[x=1bp,y=1bp]
  \node[anchor=south west,inner sep=0pt,outer sep=0pt] at (0,0) {\includegraphics{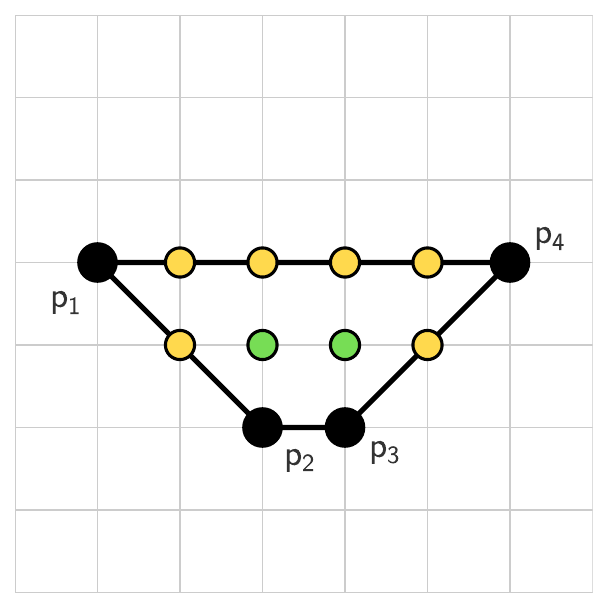}};
  \node[anchor=south west,inner sep=0pt,outer sep=0pt,overlay] at (132.94,128.55) {\includegraphics{figures/toric_label_s1a.pdf}};
  \node[anchor=south west,inner sep=0pt,outer sep=0pt,overlay] at (172.54,128.55) {\includegraphics{figures/toric_label_s2a.pdf}};
\end{tikzpicture}}
\end{minipage}%
\hspace{0.05\textwidth}%
{\small
\setlength{\tabcolsep}{4pt}
\begin{tabular}[c]{|c|c|c|}
\hline
\multirow{2}{*}{phase} & \multicolumn{2}{c|}{multiplicity} \\
\cline{2-3}
& \makebox[2.2em]{$s_{1a}$} & \makebox[2.2em]{$s_{2a}$} \\
\hline\hline
\hyperref[s_model_23_1]{1} & 30 & 30 \\
\hyperref[s_model_23_2]{2} & 34 & 34 \\
\hline
\end{tabular}
}
\caption{The toric diagram for Model~23 ($L^{1,5,1}/\mathbb{Z}_{2}\,(0,1,0,1)$).}
\label{f_geom_23}
\end{figure}

\begin{figure}[H]
\centering
\definecolor{dualityred}{rgb}{1,0.149,0}
\hypersetup{linkcolor=black}
\begin{tikzpicture}[line width=1.55bp]
\useasboundingbox (-0.320,-2.520) rectangle (0.320,0.320);
\draw[dualityred, preaction={draw=white, line width=4.55bp}] (0.000,-1.671) -- (0.000,-0.529);
\fill[dualityred] (0.000,-1.863) -- (0.072,-1.671) -- (-0.072,-1.671) -- cycle;
\fill[dualityred] (0.000,-0.337) -- (-0.072,-0.529) -- (0.072,-0.529) -- cycle;
\node[draw, rectangle, fill=white, minimum size=0.6cm, inner sep=0pt, font=\sffamily\bfseries\footnotesize, line width=2.07bp] at (0.000,-2.200) {\hyperref[s_model_23_1]{1}};
\node[draw, rectangle, fill=white, minimum size=0.6cm, inner sep=0pt, font=\sffamily\bfseries\footnotesize, line width=2.07bp] at (0.000,0.000) {\hyperref[s_model_23_2]{2}};
\end{tikzpicture}
\caption{The Seiberg duality tree for Model~23 ($L^{1,5,1}/\mathbb{Z}_{2}\,(0,1,0,1)$). The node labels indicate the phase numbers.}
\label{f_geom_23_tree}
\end{figure}

\fref{f_geom_23} illustrates the toric diagram for the $L^{1,5,1}/\mathbb{Z}_{2}$ model. The Seiberg duality tree of the $2$ toric phases associated with $L^{1,5,1}/\mathbb{Z}_{2}$ is shown in \fref{f_geom_23_tree}.
The global symmetry of the brane tilings associated with $L^{1,5,1}/\mathbb{Z}_{2}$ is given by,
\beal{es23a001}
U(1) \times U(1) \times U(1)_R~.~
\eea
The global symmetry charges on the extremal GLSM fields $p_a$ are summarized in \tref{t_sym_geom_23}.

\begin{table}[H]
\centering
{\small
\begin{tabular}{|c|c|c|c|c|l|}
\hline
  & $U(1)_{1}$ & $U(1)_{2}$ & $U(1)_R$ & $r_a$ & fugacities \\
\hline
  $p_{1}$ & $-1$ & $-1$ & $r_{1}$ & $r_{1} = 0.6319$ & $t_{1} = f_{1}^{-1} f_{2}^{-1} \bar{t}_{1}$ \\
  $p_{2}$ & $1$ & $0$ & $r_{2}$ & $r_{2} = 0.3681$ & $t_{2} = f_{1} \bar{t}_{2}$ \\
  $p_{3}$ & $0$ & $1$ & $r_{3}$ & $r_{3} = 0.3681$ & $t_{3} = f_{2} \bar{t}_{2}$ \\
  $p_{4}$ & $0$ & $0$ & $r_{4}$ & $r_{4} = 0.6319$ & $t_{4} = \bar{t}_{1}$ \\
\hline
\end{tabular}
}
\caption{Global symmetry charges on the extremal perfect matchings of Model~23 ($L^{1,5,1}/\mathbb{Z}_{2}\,(0,1,0,1)$).}
\label{t_sym_geom_23}
\end{table}

All toric phases associated with $L^{1,5,1}/\mathbb{Z}_{2}$ share the same mesonic moduli space, whose unrefined Hilbert series takes the form,
\beal{es23a002}
\resizebox{\ifdim\width>\linewidth\linewidth\else\width\fi}{!}{$
g(t; \mathcal{M}^{mes}) \;=\; \frac{1 - t^{2} + 2 t^{4} - t^{6} + t^{8}}{\left(1 - t^{2}\right) \left(1 - t^{4}\right) \left(1 - t^{6}\right)}~,~
$}
\eea
where $t$ is the fugacity counting degrees in extremal perfect matchings $p_a$, which corresponds to setting every fugacity $t_a$ in \tref{t_sym_geom_23} equal to $t$.
The plethystic logarithm of the refined Hilbert series takes the form,
\beal{es23a003}
PL\left[g(t_a;\mathcal{M}^{mes})\right] &=& t_{2}^{2} t_{3}^{2} + t_{1} t_{2} t_{3} t_{4} + t_{1}^{2} t_{4}^{2} + t_{3} t_{4}^{5} + t_{1}^{5} t_{2} \nonumber\\
&& {}- t_{1}^{2} t_{2}^{2} t_{3}^{2} t_{4}^{2} - t_{1}^{5} t_{2} t_{3} t_{4}^{5}~.~
\eea
As the mesonic moduli space is a complete intersection, the plethystic logarithm is finite. There is one fugacity $t_a$ for each of the $4$ extremal perfect matchings $p_a$ of the toric diagram in \fref{f_geom_23}, and each $t_a$ can be expressed in terms of the global symmetry fugacities $f_{1}, f_{2}$ and $\bar{t}_{1}, \bar{t}_{2}$ as given in \tref{t_sym_geom_23}.

The generators of the mesonic moduli space, $A_{1}, \dots, A_{5}$, correspond to the first positive terms of the plethystic logarithm. For each toric phase, the generators are given in \tref{t_model_23_1} and \tref{t_model_23_2} in terms of the chiral fields of the corresponding brane tiling. The negative terms that come after the positive ones correspond to the first order relations amongst the generators, which take the following~form,
\beal{es23a004}
&& A_{2}^{2} = A_{1} A_{3} ~,~ \quad A_{4} A_{5} = A_{2} A_{3}^{2}~.~
\eea

\input{model_23_1}
\input{model_23_2}

\subsection{Model 24: $\mathcal{C}/\mathbb{Z}_{6}\,(2,3,4,3)$}\label{s_m24}
\input{geom_24}
\input{model_24_1}
\input{model_24_2}
\input{model_24_3}
\input{model_24_4}
\input{model_24_5}
\input{model_24_6}
\input{model_24_7}
\input{model_24_8}
\input{model_24_9}
\input{model_24_10}
\input{model_24_11}

\section{Pentagons}\label{s_c5}

\subsection{Model 25: $X^{3,1}$}\label{s_m25}
\begin{figure}[H]
\centering
\begin{minipage}[c]{0.34\textwidth}
\resizebox{\linewidth}{!}{%
\begin{tikzpicture}[x=1bp,y=1bp]
  \node[anchor=south west,inner sep=0pt,outer sep=0pt] at (0,0) {\includegraphics{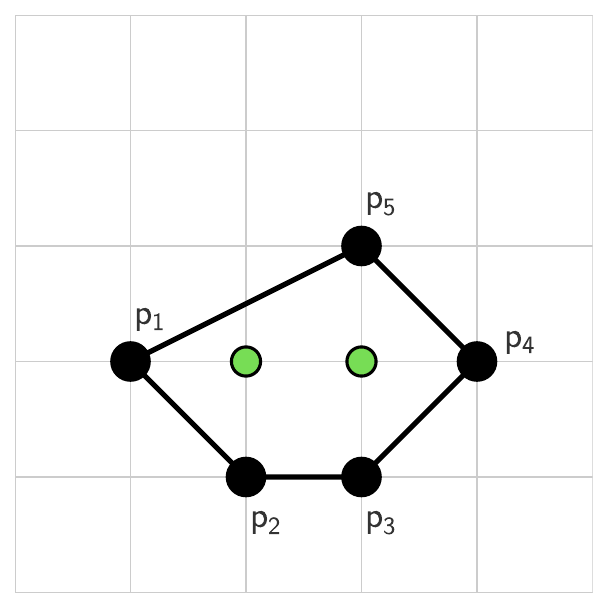}};
  \node[anchor=south west,inner sep=0pt,outer sep=0pt,overlay] at (128.22,120.03) {\includegraphics{figures/toric_label_s1a.pdf}};
  \node[anchor=south west,inner sep=0pt,outer sep=0pt,overlay] at (183.66,120.03) {\includegraphics{figures/toric_label_s2a.pdf}};
\end{tikzpicture}}
\end{minipage}%
\hspace{0.05\textwidth}%
{\small
\setlength{\tabcolsep}{4pt}
\begin{tabular}[c]{|c|c|c|}
\hline
\multirow{2}{*}{phase} & \multicolumn{2}{c|}{multiplicity} \\
\cline{2-3}
& \makebox[2.2em]{$s_{1a}$} & \makebox[2.2em]{$s_{2a}$} \\
\hline\hline
\hyperref[s_model_25_1]{1} & 7 & 8 \\
\hyperref[s_model_25_2]{2} & 8 & 8 \\
\hyperref[s_model_25_3]{3} & 8 & 9 \\
\hyperref[s_model_25_4]{4} & 9 & 10 \\
\hyperref[s_model_25_5]{5} & 8 & 11 \\
\hline
\end{tabular}
}
\caption{The toric diagram for Model~25 ($X^{3,1}$).}
\label{f_geom_25}
\end{figure}

\begin{figure}[H]
\centering
\definecolor{dualityred}{rgb}{1,0.149,0}
\hypersetup{linkcolor=black}
\begin{tikzpicture}[line width=1.55bp]
\useasboundingbox (-3.567,-3.620) rectangle (3.567,0.834);
\draw[dualityred, preaction={draw=white, line width=4.55bp}] (-0.489,-0.378) -- (-2.361,-1.822);
\draw[dualityred, preaction={draw=white, line width=4.55bp}] (-0.222,-0.514) -- (-0.728,-1.686);
\draw[dualityred, preaction={draw=white, line width=4.55bp}] (0.222,-0.514) -- (0.728,-1.686);
\draw[dualityred, preaction={draw=white, line width=4.55bp}] (0.489,-0.378) -- (2.361,-1.822);
\draw[dualityred, preaction={draw=white, line width=4.55bp}] (-2.321,-2.200) -- (-1.479,-2.200);
\draw[dualityred, preaction={draw=white, line width=4.55bp}] (-2.363,-2.605) .. controls (-1.198,-3.532) and (-0.702,-3.532) .. (0.463,-2.605);
\draw[dualityred, preaction={draw=white, line width=4.55bp}] (-0.502,-2.675) .. controls (0.704,-3.908) and (1.196,-3.908) .. (2.402,-2.675);
\draw[dualityred, preaction={draw=white, line width=4.55bp}] (1.479,-2.200) -- (2.321,-2.200);
\draw[dualityred, preaction={draw=white, line width=4.55bp}] (0.301,0.500) .. controls (0.454,0.918) and (-0.454,0.918) .. (-0.301,0.500);
\draw[dualityred, preaction={draw=white, line width=4.55bp}] (-2.975,-1.674) .. controls (-3.142,-1.175) and (-3.875,-1.908) .. (-3.376,-2.075);
\draw[dualityred, preaction={draw=white, line width=4.55bp}] (3.376,-2.075) .. controls (3.875,-1.908) and (3.142,-1.175) .. (2.975,-1.674);
\fill[dualityred] (-0.337,-0.260) -- (-0.533,-0.320) -- (-0.445,-0.435) -- cycle;
\fill[dualityred] (-2.513,-1.940) -- (-2.317,-1.880) -- (-2.405,-1.765) -- cycle;
\fill[dualityred] (-0.145,-0.337) -- (-0.288,-0.485) -- (-0.155,-0.542) -- cycle;
\fill[dualityred] (-0.805,-1.863) -- (-0.662,-1.715) -- (-0.795,-1.658) -- cycle;
\fill[dualityred] (0.145,-0.337) -- (0.155,-0.542) -- (0.288,-0.485) -- cycle;
\fill[dualityred] (0.805,-1.863) -- (0.795,-1.658) -- (0.662,-1.715) -- cycle;
\fill[dualityred] (0.337,-0.260) -- (0.445,-0.435) -- (0.533,-0.320) -- cycle;
\fill[dualityred] (2.513,-1.940) -- (2.405,-1.765) -- (2.317,-1.880) -- cycle;
\fill[dualityred] (-2.513,-2.200) -- (-2.321,-2.272) -- (-2.321,-2.128) -- cycle;
\fill[dualityred] (-1.287,-2.200) -- (-1.479,-2.128) -- (-1.479,-2.272) -- cycle;
\fill[dualityred] (-2.513,-2.483) -- (-2.409,-2.661) -- (-2.318,-2.548) -- cycle;
\fill[dualityred] (0.613,-2.483) -- (0.418,-2.548) -- (0.509,-2.661) -- cycle;
\fill[dualityred] (-0.636,-2.537) -- (-0.554,-2.726) -- (-0.450,-2.625) -- cycle;
\fill[dualityred] (2.536,-2.537) -- (2.350,-2.625) -- (2.454,-2.726) -- cycle;
\fill[dualityred] (1.287,-2.200) -- (1.479,-2.272) -- (1.479,-2.128) -- cycle;
\fill[dualityred] (2.513,-2.200) -- (2.321,-2.128) -- (2.321,-2.272) -- cycle;
\fill[dualityred] (0.200,0.337) -- (0.363,0.462) -- (0.239,0.538) -- cycle;
\fill[dualityred] (-0.200,0.337) -- (-0.239,0.538) -- (-0.363,0.462) -- cycle;
\fill[dualityred] (-2.943,-1.863) -- (-2.904,-1.662) -- (-3.046,-1.686) -- cycle;
\fill[dualityred] (-3.187,-2.107) -- (-3.364,-2.004) -- (-3.388,-2.146) -- cycle;
\fill[dualityred] (3.187,-2.107) -- (3.388,-2.146) -- (3.364,-2.004) -- cycle;
\fill[dualityred] (2.943,-1.863) -- (3.046,-1.686) -- (2.904,-1.662) -- cycle;
\node[draw, rectangle, fill=white, minimum size=0.6cm, inner sep=0pt, font=\sffamily\bfseries\footnotesize, line width=2.07bp] at (0.000,0.000) {\hyperref[s_model_25_1]{1}};
\node[draw, rectangle, fill=white, minimum size=0.6cm, inner sep=0pt, font=\sffamily\bfseries\footnotesize, line width=2.07bp] at (-2.850,-2.200) {\hyperref[s_model_25_2]{2}};
\node[draw, rectangle, fill=white, minimum size=0.6cm, inner sep=0pt, font=\sffamily\bfseries\footnotesize, line width=2.07bp] at (-0.950,-2.200) {\hyperref[s_model_25_3]{3}};
\node[draw, rectangle, fill=white, minimum size=0.6cm, inner sep=0pt, font=\sffamily\bfseries\footnotesize, line width=2.07bp] at (0.950,-2.200) {\hyperref[s_model_25_4]{4}};
\node[draw, rectangle, fill=white, minimum size=0.6cm, inner sep=0pt, font=\sffamily\bfseries\footnotesize, line width=2.07bp] at (2.850,-2.200) {\hyperref[s_model_25_5]{5}};
\end{tikzpicture}
\caption{The Seiberg duality tree for Model~25 ($X^{3,1}$). The node labels indicate the phase numbers.}
\label{f_geom_25_tree}
\end{figure}

The toric diagram of the $X^{3,1}$ model is shown in \fref{f_geom_25}. The Seiberg duality tree of the $5$ toric phases associated with $X^{3,1}$ is shown in \fref{f_geom_25_tree}.
The global symmetry of the brane tilings associated with $X^{3,1}$ is given by,
\beal{es25a001}
U(1) \times U(1) \times U(1)_R~.~
\eea
The global symmetry charges on the extremal GLSM fields $p_a$ are summarized in \tref{t_sym_geom_25}.

\begin{table}[H]
\centering
{\small
\begin{tabular}{|c|c|c|c|c|l|}
\hline
  & $U(1)_{1}$ & $U(1)_{2}$ & $U(1)_R$ & $r_a$ & fugacities \\
\hline
  $p_{1}$ & $-1$ & $-1$ & $r_{1}$ & $r_{1} = 0.4930$ & $t_{1} = f_{1}^{-1} f_{2}^{-1} \bar{t}_{1}$ \\
  $p_{2}$ & $1$ & $0$ & $r_{2}$ & $r_{2} = 0.3002$ & $t_{2} = f_{1} \bar{t}_{2}$ \\
  $p_{3}$ & $0$ & $1$ & $r_{3}$ & $r_{3} = 0.3247$ & $t_{3} = f_{2} \bar{t}_{3}$ \\
  $p_{4}$ & $0$ & $0$ & $r_{4}$ & $r_{4} = 0.3985$ & $t_{4} = \bar{t}_{4}$ \\
  $p_{5}$ & $0$ & $0$ & $r_{5}$ & $r_{5} = 0.4836$ & $t_{5} = \bar{t}_{5}$ \\
\hline
\end{tabular}
}
\caption{Global symmetry charges on the extremal perfect matchings of Model~25 ($X^{3,1}$).}
\label{t_sym_geom_25}
\end{table}

All toric phases associated with $X^{3,1}$ share the same mesonic moduli space, whose unrefined Hilbert series takes the form,
\beal{es25a002}
\resizebox{\ifdim\width>\linewidth\linewidth\else\width\fi}{!}{$
g(t; \mathcal{M}^{mes}) \;=\; \frac{\begin{array}{c} 1 + t^{4} + t^{5} + t^{6} + t^{7} + 2 t^{8} + 2 t^{9} + t^{10} + t^{11} + t^{12} \\ - t^{14} - t^{15} - t^{16} - 2 t^{17} - 2 t^{18} - t^{19} - t^{20} - t^{21} - t^{22} - t^{26} \end{array}}{\left(1 - t^{6}\right) \left(1 - t^{7}\right) \left(1 - t^{8}\right) \left(1 - t^{10}\right)}~,~
$}
\eea
where $t$ is the fugacity counting degrees in extremal perfect matchings $p_a$, which corresponds to setting every fugacity $t_a$ in \tref{t_sym_geom_25} equal to $t$.
The plethystic logarithm of the refined Hilbert series takes the form,
\beal{es25a003}
PL\left[g(t_a;\mathcal{M}^{mes})\right] &=& t_{1} t_{4} t_{5}^{2} + t_{1} t_{2} t_{3} t_{4} t_{5} + t_{1} t_{2}^{2} t_{3}^{2} t_{4} + t_{1}^{3} t_{2} t_{5}^{2} + t_{3} t_{4}^{3} t_{5}^{3} + t_{1}^{3} t_{2}^{2} t_{3} t_{5} \nonumber\\
&& {}+ t_{2} t_{3}^{2} t_{4}^{3} t_{5}^{2} + t_{1}^{3} t_{2}^{3} t_{3}^{2} + t_{2}^{2} t_{3}^{3} t_{4}^{3} t_{5} + t_{2}^{3} t_{3}^{4} t_{4}^{3} \nonumber\\
&& {}- t_{1}^{2} t_{2}^{2} t_{3}^{2} t_{4}^{2} t_{5}^{2} - t_{1}^{4} t_{2}^{2} t_{3} t_{4} t_{5}^{3} - 2\, t_{1}^{4} t_{2}^{3} t_{3}^{2} t_{4} t_{5}^{2} - t_{1} t_{2} t_{3}^{2} t_{4}^{4} t_{5}^{4} \nonumber\\
&& {}- t_{1}^{3} t_{2} t_{3} t_{4}^{3} t_{5}^{5} - t_{1}^{4} t_{2}^{4} t_{3}^{3} t_{4} t_{5} - 2\, t_{1} t_{2}^{2} t_{3}^{3} t_{4}^{4} t_{5}^{3} - t_{1}^{6} t_{2}^{4} t_{3}^{2} t_{5}^{2} \nonumber\\
&& {}- 2\, t_{1}^{3} t_{2}^{2} t_{3}^{2} t_{4}^{3} t_{5}^{4} - 2\, t_{1} t_{2}^{3} t_{3}^{4} t_{4}^{4} t_{5}^{2} - 3\, t_{1}^{3} t_{2}^{3} t_{3}^{3} t_{4}^{3} t_{5}^{3} - t_{1} t_{2}^{4} t_{3}^{5} t_{4}^{4} t_{5} \nonumber\\
&& {}- 3\, t_{1}^{3} t_{2}^{4} t_{3}^{4} t_{4}^{3} t_{5}^{2} - t_{2}^{2} t_{3}^{4} t_{4}^{6} t_{5}^{4} - 2\, t_{1}^{3} t_{2}^{5} t_{3}^{5} t_{4}^{3} t_{5} - t_{2}^{3} t_{3}^{5} t_{4}^{6} t_{5}^{3} - t_{2}^{4} t_{3}^{6} t_{4}^{6} t_{5}^{2} \nonumber\\
&& {}- t_{1}^{3} t_{2}^{6} t_{3}^{6} t_{4}^{3} + \dots~,~
\eea
where the fugacities $t_{1}, \dots, t_{5}$ correspond to the extremal perfect matchings $p_{1}, \dots, p_{5}$ of \fref{f_geom_25} and can be expressed in terms of the global symmetry fugacities $f_{1}, f_{2}$ and $\bar{t}_{1}, \dots, \bar{t}_{5}$ as given in \tref{t_sym_geom_25}.

Each of the first $10$ positive terms corresponds to one generator of the mesonic moduli space, the generators being $A_{1}, \dots, A_{10}$. In \tref{t_model_25_1}, \tref{t_model_25_2}, \tref{t_model_25_3}, \tref{t_model_25_4} and \tref{t_model_25_5}, the generators are given in terms of the chiral fields of each toric phase. The negative terms that follow correspond to the first order relations amongst the generators, which take the following~form,
\beal{es25a004}
&& A_{2}^{2} = A_{1} A_{3} ~,~ \quad A_{2} A_{4} = A_{1} A_{6} ~,~ \quad A_{3} A_{4} = A_{2} A_{6} = A_{1} A_{8} ~,~ \nonumber\\
&& A_{2} A_{5} = A_{1} A_{7} ~,~ \quad A_{4} A_{5} = A_{1}^{2} A_{2} ~,~ \quad A_{3} A_{6} = A_{2} A_{8} ~,~ \nonumber\\
&& A_{3} A_{5} = A_{2} A_{7} = A_{1} A_{9} ~,~ \quad A_{6}^{2} = A_{4} A_{8} ~,~ \quad A_{5} A_{6} = A_{4} A_{7} = A_{1}^{2} A_{3} ~,~ \nonumber\\
&& A_{3} A_{7} = A_{2} A_{9} = A_{1} A_{10} ~,~ \quad A_{6} A_{7} = A_{5} A_{8} = A_{4} A_{9} = A_{1} A_{2} A_{3} ~,~ \nonumber\\
&& A_{3} A_{9} = A_{2} A_{10} ~,~ \quad A_{7} A_{8} = A_{6} A_{9} = A_{4} A_{10} = A_{1} A_{3}^{2} ~,~ \quad A_{7}^{2} = A_{5} A_{9} ~,~ \nonumber\\
&& A_{8} A_{9} = A_{6} A_{10} = A_{2} A_{3}^{2} ~,~ \quad A_{7} A_{9} = A_{5} A_{10} ~,~ \quad A_{9}^{2} = A_{7} A_{10} ~,~ \nonumber\\
&& A_{8} A_{10} = A_{3}^{3}~.~
\eea

\input{model_25_1}
\input{model_25_2}
\input{model_25_3}
\input{model_25_4}
\input{model_25_5}

\subsection{Model 26: $X^{3,2}$}\label{s_m26}
\begin{figure}[H]
\centering
\begin{minipage}[c]{0.34\textwidth}
\resizebox{\linewidth}{!}{%
\begin{tikzpicture}[x=1bp,y=1bp]
  \node[anchor=south west,inner sep=0pt,outer sep=0pt] at (0,0) {\includegraphics{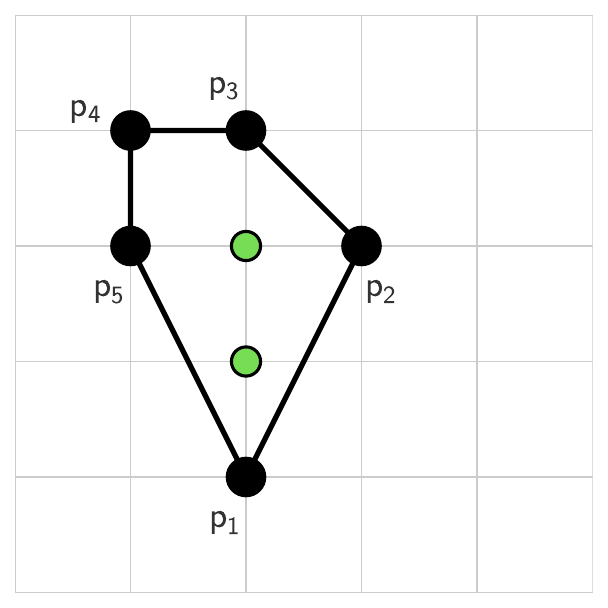}};
  \node[anchor=south west,inner sep=0pt,outer sep=0pt,overlay] at (124.52,121.63) {\includegraphics{figures/toric_label_s1a.pdf}};
  \node[anchor=south west,inner sep=0pt,outer sep=0pt,overlay] at (128.22,175.47) {\includegraphics{figures/toric_label_s2a.pdf}};
\end{tikzpicture}}
\end{minipage}%
\hspace{0.05\textwidth}%
{\small
\setlength{\tabcolsep}{4pt}
\begin{tabular}[c]{|c|c|c|}
\hline
\multirow{2}{*}{phase} & \multicolumn{2}{c|}{multiplicity} \\
\cline{2-3}
& \makebox[2.2em]{$s_{1a}$} & \makebox[2.2em]{$s_{2a}$} \\
\hline\hline
\hyperref[s_model_26_1]{1} & 7 & 9 \\
\hyperref[s_model_26_2]{2} & 7 & 10 \\
\hyperref[s_model_26_3]{3} & 8 & 11 \\
\hline
\end{tabular}
}
\caption{The toric diagram for Model~26 ($X^{3,2}$).}
\label{f_geom_26}
\end{figure}

\begin{figure}[H]
\centering
\definecolor{dualityred}{rgb}{1,0.149,0}
\hypersetup{linkcolor=black}
\begin{tikzpicture}[line width=1.55bp]
\useasboundingbox (-1.667,-2.917) rectangle (1.667,0.834);
\draw[dualityred, preaction={draw=white, line width=4.55bp}] (-0.421,-2.200) -- (0.421,-2.200);
\draw[dualityred, preaction={draw=white, line width=4.55bp}] (-0.728,-1.686) -- (-0.222,-0.514);
\draw[dualityred, preaction={draw=white, line width=4.55bp}] (0.728,-1.686) -- (0.222,-0.514);
\draw[dualityred, preaction={draw=white, line width=4.55bp}] (-1.476,-2.325) .. controls (-1.975,-2.492) and (-1.242,-3.225) .. (-1.075,-2.726);
\draw[dualityred, preaction={draw=white, line width=4.55bp}] (1.075,-2.726) .. controls (1.242,-3.225) and (1.975,-2.492) .. (1.476,-2.325);
\draw[dualityred, preaction={draw=white, line width=4.55bp}] (0.301,0.500) .. controls (0.454,0.918) and (-0.454,0.918) .. (-0.301,0.500);
\fill[dualityred] (-0.613,-2.200) -- (-0.421,-2.272) -- (-0.421,-2.128) -- cycle;
\fill[dualityred] (0.613,-2.200) -- (0.421,-2.128) -- (0.421,-2.272) -- cycle;
\fill[dualityred] (-0.805,-1.863) -- (-0.662,-1.715) -- (-0.795,-1.658) -- cycle;
\fill[dualityred] (-0.145,-0.337) -- (-0.288,-0.485) -- (-0.155,-0.542) -- cycle;
\fill[dualityred] (0.805,-1.863) -- (0.795,-1.658) -- (0.662,-1.715) -- cycle;
\fill[dualityred] (0.145,-0.337) -- (0.155,-0.542) -- (0.288,-0.485) -- cycle;
\fill[dualityred] (-1.287,-2.293) -- (-1.488,-2.254) -- (-1.464,-2.396) -- cycle;
\fill[dualityred] (-1.043,-2.537) -- (-1.146,-2.714) -- (-1.004,-2.738) -- cycle;
\fill[dualityred] (1.043,-2.537) -- (1.004,-2.738) -- (1.146,-2.714) -- cycle;
\fill[dualityred] (1.287,-2.293) -- (1.464,-2.396) -- (1.488,-2.254) -- cycle;
\fill[dualityred] (0.200,0.337) -- (0.363,0.462) -- (0.239,0.538) -- cycle;
\fill[dualityred] (-0.200,0.337) -- (-0.239,0.538) -- (-0.363,0.462) -- cycle;
\node[draw, rectangle, fill=white, minimum size=0.6cm, inner sep=0pt, font=\sffamily\bfseries\footnotesize, line width=2.07bp] at (-0.950,-2.200) {\hyperref[s_model_26_1]{1}};
\node[draw, rectangle, fill=white, minimum size=0.6cm, inner sep=0pt, font=\sffamily\bfseries\footnotesize, line width=2.07bp] at (0.950,-2.200) {\hyperref[s_model_26_2]{2}};
\node[draw, rectangle, fill=white, minimum size=0.6cm, inner sep=0pt, font=\sffamily\bfseries\footnotesize, line width=2.07bp] at (0.000,0.000) {\hyperref[s_model_26_3]{3}};
\end{tikzpicture}
\caption{The Seiberg duality tree for Model~26 ($X^{3,2}$). The node labels indicate the phase numbers.}
\label{f_geom_26_tree}
\end{figure}

\fref{f_geom_26} illustrates the toric diagram for the $X^{3,2}$ model. The Seiberg duality tree of the $3$ toric phases associated with $X^{3,2}$ is shown in \fref{f_geom_26_tree}.
The global symmetry of the brane tilings associated with $X^{3,2}$ is given by,
\beal{es26a001}
U(1) \times U(1) \times U(1)_R~.~
\eea
The global symmetry charges on the extremal GLSM fields $p_a$ are summarized in \tref{t_sym_geom_26}.

\begin{table}[H]
\centering
{\small
\begin{tabular}{|c|c|c|c|c|l|}
\hline
  & $U(1)_{1}$ & $U(1)_{2}$ & $U(1)_R$ & $r_a$ & fugacities \\
\hline
  $p_{1}$ & $-1$ & $-1$ & $r_{1}$ & $r_{1} = 0.5641$ & $t_{1} = f_{1}^{-1} f_{2}^{-1} \bar{t}_{1}$ \\
  $p_{2}$ & $1$ & $0$ & $r_{2}$ & $r_{2} = 0.5209$ & $t_{2} = f_{1} \bar{t}_{2}$ \\
  $p_{3}$ & $0$ & $1$ & $r_{3}$ & $r_{3} = 0.2619$ & $t_{3} = f_{2} \bar{t}_{3}$ \\
  $p_{4}$ & $0$ & $0$ & $r_{4}$ & $r_{4} = 0.3687$ & $t_{4} = \bar{t}_{4}$ \\
  $p_{5}$ & $0$ & $0$ & $r_{5}$ & $r_{5} = 0.2844$ & $t_{5} = \bar{t}_{5}$ \\
\hline
\end{tabular}
}
\caption{Global symmetry charges on the extremal perfect matchings of Model~26 ($X^{3,2}$).}
\label{t_sym_geom_26}
\end{table}

All toric phases associated with $X^{3,2}$ share the same mesonic moduli space, whose unrefined Hilbert series takes the form,
\beal{es26a002}
\resizebox{\ifdim\width>\linewidth\linewidth\else\width\fi}{!}{$
g(t; \mathcal{M}^{mes}) \;=\; \frac{1 + t^{4} + t^{5} + t^{6} + t^{8} + 2 t^{9} + t^{10} + t^{11} + 2 t^{12} - 2 t^{14} - t^{15} - t^{16} - 2 t^{17} - t^{18} - t^{20} - t^{21} - t^{22} - t^{26}}{\left(1 - t^{5}\right) \left(1 - t^{6}\right) \left(1 - t^{8}\right) \left(1 - t^{12}\right)}~,~
$}
\eea
where $t$ is the fugacity counting degrees in extremal perfect matchings $p_a$, which corresponds to setting every fugacity $t_a$ in \tref{t_sym_geom_26} equal to $t$.
The plethystic logarithm of the refined Hilbert series takes the form,
\beal{es26a003}
PL\left[g(t_a;\mathcal{M}^{mes})\right] &=& t_{1} t_{2}^{2} t_{3} + t_{1} t_{2} t_{3} t_{4} t_{5} + t_{1}^{3} t_{2} t_{5} + t_{1} t_{3} t_{4}^{2} t_{5}^{2} + t_{1}^{3} t_{4} t_{5}^{2} + t_{2}^{4} t_{3}^{3} t_{4} \nonumber\\
&& {}+ t_{2}^{3} t_{3}^{3} t_{4}^{2} t_{5} + t_{2}^{2} t_{3}^{3} t_{4}^{3} t_{5}^{2} + t_{2} t_{3}^{3} t_{4}^{4} t_{5}^{3} + t_{3}^{3} t_{4}^{5} t_{5}^{4} \nonumber\\
&& {}- t_{1}^{4} t_{2}^{2} t_{3} t_{4} t_{5}^{2} - t_{1}^{2} t_{2}^{2} t_{3}^{2} t_{4}^{2} t_{5}^{2} - t_{1}^{4} t_{2} t_{3} t_{4}^{2} t_{5}^{3} - t_{1}^{3} t_{2}^{5} t_{3}^{3} t_{4} t_{5} \nonumber\\
&& {}- t_{1} t_{2}^{5} t_{3}^{4} t_{4}^{2} t_{5} - 2\, t_{1}^{3} t_{2}^{4} t_{3}^{3} t_{4}^{2} t_{5}^{2} - 2\, t_{1} t_{2}^{4} t_{3}^{4} t_{4}^{3} t_{5}^{2} - 2\, t_{1}^{3} t_{2}^{3} t_{3}^{3} t_{4}^{3} t_{5}^{3} \nonumber\\
&& {}- 2\, t_{1} t_{2}^{3} t_{3}^{4} t_{4}^{4} t_{5}^{3} - 2\, t_{1}^{3} t_{2}^{2} t_{3}^{3} t_{4}^{4} t_{5}^{4} - 2\, t_{1} t_{2}^{2} t_{3}^{4} t_{4}^{5} t_{5}^{4} - 2\, t_{1}^{3} t_{2} t_{3}^{3} t_{4}^{5} t_{5}^{5} \nonumber\\
&& {}- t_{1} t_{2} t_{3}^{4} t_{4}^{6} t_{5}^{5} - t_{2}^{6} t_{3}^{6} t_{4}^{4} t_{5}^{2} - t_{1}^{3} t_{3}^{3} t_{4}^{6} t_{5}^{6} - t_{2}^{5} t_{3}^{6} t_{4}^{5} t_{5}^{3} - 2\, t_{2}^{4} t_{3}^{6} t_{4}^{6} t_{5}^{4} \nonumber\\
&& {}- t_{2}^{3} t_{3}^{6} t_{4}^{7} t_{5}^{5} - t_{2}^{2} t_{3}^{6} t_{4}^{8} t_{5}^{6} + \dots~.~
\eea
In this expansion $t_a$ is the fugacity of the extremal perfect matching $p_a$ in \fref{f_geom_26} and can be expressed in terms of the global symmetry fugacities $f_{1}, f_{2}$ and $\bar{t}_{1}, \dots, \bar{t}_{5}$ as shown in \tref{t_sym_geom_26}.

The first positive terms of the plethystic logarithm correspond to the generators of the mesonic moduli space, $A_{1}, \dots, A_{10}$. For each toric phase, the generators are given in \tref{t_model_26_1}, \tref{t_model_26_2} and \tref{t_model_26_3} in terms of the chiral fields of the corresponding brane tiling. The next negative terms correspond to the first order relations amongst the generators, which take the following~form,
\beal{es26a004}
&& A_{2} A_{3} = A_{1} A_{5} ~,~ \quad A_{2}^{2} = A_{1} A_{4} ~,~ \quad A_{3} A_{4} = A_{2} A_{5} ~,~ \quad A_{3} A_{6} = A_{1}^{2} A_{2} ~,~ \nonumber\\
&& A_{2} A_{6} = A_{1} A_{7} ~,~ \quad A_{5} A_{6} = A_{3} A_{7} = A_{1}^{2} A_{4} ~,~ \quad A_{4} A_{6} = A_{2} A_{7} = A_{1} A_{8} ~,~ \nonumber\\
&& A_{5} A_{7} = A_{3} A_{8} = A_{1} A_{2} A_{4} ~,~ \quad A_{4} A_{7} = A_{2} A_{8} = A_{1} A_{9} ~,~ \nonumber\\
&& A_{5} A_{8} = A_{3} A_{9} = A_{1} A_{4}^{2} ~,~ \quad A_{4} A_{8} = A_{2} A_{9} = A_{1} A_{10} ~,~ \nonumber\\
&& A_{5} A_{9} = A_{3} A_{10} = A_{2} A_{4}^{2} ~,~ \quad A_{4} A_{9} = A_{2} A_{10} ~,~ \quad A_{7}^{2} = A_{6} A_{8} ~,~ \nonumber\\
&& A_{5} A_{10} = A_{4}^{3} ~,~ \quad A_{7} A_{8} = A_{6} A_{9} ~,~ \quad A_{8}^{2} = A_{7} A_{9} = A_{6} A_{10} ~,~ \nonumber\\
&& A_{8} A_{9} = A_{7} A_{10} ~,~ \quad A_{9}^{2} = A_{8} A_{10}~.~
\eea

\input{model_26_1}
\input{model_26_2}
\input{model_26_3}

\subsection{Model 27: $H^{1,3,1,1}$}\label{s_m27}
\begin{figure}[H]
\centering
\begin{minipage}[c]{0.34\textwidth}
\resizebox{\linewidth}{!}{%
\begin{tikzpicture}[x=1bp,y=1bp]
  \node[anchor=south west,inner sep=0pt,outer sep=0pt] at (0,0) {\includegraphics{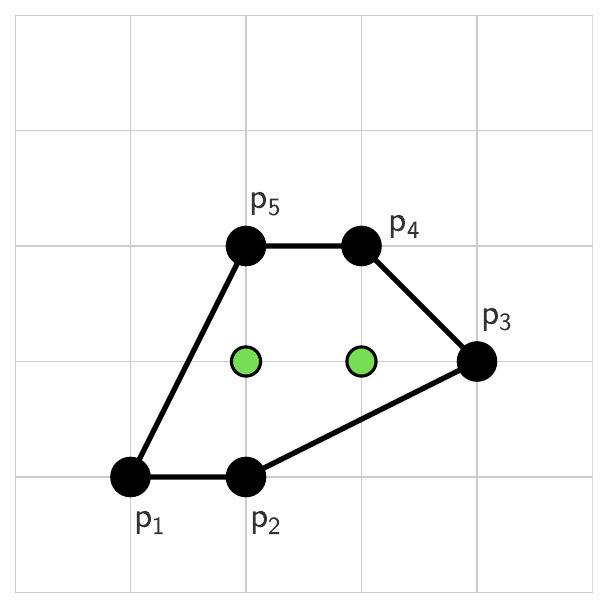}};
  \node[anchor=south west,inner sep=0pt,outer sep=0pt,overlay] at (128.22,120.03) {\includegraphics{figures/toric_label_s1a.pdf}};
  \node[anchor=south west,inner sep=0pt,outer sep=0pt,overlay] at (183.66,120.03) {\includegraphics{figures/toric_label_s2a.pdf}};
\end{tikzpicture}}
\end{minipage}%
\hspace{0.05\textwidth}%
{\small
\setlength{\tabcolsep}{4pt}
\begin{tabular}[c]{|c|c|c|}
\hline
\multirow{2}{*}{phase} & \multicolumn{2}{c|}{multiplicity} \\
\cline{2-3}
& \makebox[2.2em]{$s_{1a}$} & \makebox[2.2em]{$s_{2a}$} \\
\hline\hline
\hyperref[s_model_27_1]{1} & 7 & 7 \\
\hyperref[s_model_27_2]{2} & 7 & 8 \\
\hyperref[s_model_27_3]{3} & 8 & 9 \\
\hline
\end{tabular}
}
\caption{The toric diagram for Model~27 ($H^{1,3,1,1}$).}
\label{f_geom_27}
\end{figure}

\begin{figure}[H]
\centering
\definecolor{dualityred}{rgb}{1,0.149,0}
\hypersetup{linkcolor=black}
\begin{tikzpicture}[line width=1.55bp]
\useasboundingbox (-0.338,-5.234) rectangle (0.338,0.834);
\draw[dualityred, preaction={draw=white, line width=4.55bp}] (0.000,-3.871) -- (0.000,-2.729);
\draw[dualityred, preaction={draw=white, line width=4.55bp}] (0.000,-1.671) -- (0.000,-0.529);
\draw[dualityred, preaction={draw=white, line width=4.55bp}] (-0.301,-4.900) .. controls (-0.454,-5.318) and (0.454,-5.318) .. (0.301,-4.900);
\draw[dualityred, preaction={draw=white, line width=4.55bp}] (0.301,0.500) .. controls (0.454,0.918) and (-0.454,0.918) .. (-0.301,0.500);
\fill[dualityred] (0.000,-4.063) -- (0.072,-3.871) -- (-0.072,-3.871) -- cycle;
\fill[dualityred] (0.000,-2.537) -- (-0.072,-2.729) -- (0.072,-2.729) -- cycle;
\fill[dualityred] (0.000,-1.863) -- (0.072,-1.671) -- (-0.072,-1.671) -- cycle;
\fill[dualityred] (0.000,-0.337) -- (-0.072,-0.529) -- (0.072,-0.529) -- cycle;
\fill[dualityred] (-0.200,-4.737) -- (-0.363,-4.862) -- (-0.239,-4.938) -- cycle;
\fill[dualityred] (0.200,-4.737) -- (0.239,-4.938) -- (0.363,-4.862) -- cycle;
\fill[dualityred] (0.200,0.337) -- (0.363,0.462) -- (0.239,0.538) -- cycle;
\fill[dualityred] (-0.200,0.337) -- (-0.239,0.538) -- (-0.363,0.462) -- cycle;
\node[draw, rectangle, fill=white, minimum size=0.6cm, inner sep=0pt, font=\sffamily\bfseries\footnotesize, line width=2.07bp] at (0.000,-4.400) {\hyperref[s_model_27_1]{1}};
\node[draw, rectangle, fill=white, minimum size=0.6cm, inner sep=0pt, font=\sffamily\bfseries\footnotesize, line width=2.07bp] at (0.000,-2.200) {\hyperref[s_model_27_2]{2}};
\node[draw, rectangle, fill=white, minimum size=0.6cm, inner sep=0pt, font=\sffamily\bfseries\footnotesize, line width=2.07bp] at (0.000,0.000) {\hyperref[s_model_27_3]{3}};
\end{tikzpicture}
\caption{The Seiberg duality tree for Model~27 ($H^{1,3,1,1}$). The node labels indicate the phase numbers.}
\label{f_geom_27_tree}
\end{figure}

The $H^{1,3,1,1}$ model has a toric diagram shown in \fref{f_geom_27}. The Seiberg duality tree of the $3$ toric phases associated with $H^{1,3,1,1}$ is shown in \fref{f_geom_27_tree}.
The global symmetry of the brane tilings associated with $H^{1,3,1,1}$ is given by,
\beal{es27a001}
U(1) \times U(1) \times U(1)_R~.~
\eea
The global symmetry charges on the extremal GLSM fields $p_a$ are summarized in \tref{t_sym_geom_27}.

\begin{table}[H]
\centering
{\small
\begin{tabular}{|c|c|c|c|c|l|}
\hline
  & $U(1)_{1}$ & $U(1)_{2}$ & $U(1)_R$ & $r_a$ & fugacities \\
\hline
  $p_{1}$ & $-1$ & $-1$ & $r_{1}$ & $r_{1} = 0.4661$ & $t_{1} = f_{1}^{-1} f_{2}^{-1} \bar{t}_{1}$ \\
  $p_{2}$ & $1$ & $0$ & $r_{2}$ & $r_{2} = 0.2735$ & $t_{2} = f_{1} \bar{t}_{2}$ \\
  $p_{3}$ & $0$ & $1$ & $r_{3}$ & $r_{3} = 0.5208$ & $t_{3} = f_{2} \bar{t}_{3}$ \\
  $p_{4}$ & $0$ & $0$ & $r_{4}$ & $r_{4} = 0.2735$ & $t_{4} = \bar{t}_{2}$ \\
  $p_{5}$ & $0$ & $0$ & $r_{5}$ & $r_{5} = 0.4661$ & $t_{5} = \bar{t}_{1}$ \\
\hline
\end{tabular}
}
\caption{Global symmetry charges on the extremal perfect matchings of Model~27 ($H^{1,3,1,1}$).}
\label{t_sym_geom_27}
\end{table}

All toric phases associated with $H^{1,3,1,1}$ share the same mesonic moduli space, whose unrefined Hilbert series takes the form,
\beal{es27a002}
\resizebox{\ifdim\width>\linewidth\linewidth\else\width\fi}{!}{$
g(t; \mathcal{M}^{mes}) \;=\; \frac{1 + 2 t^{5} + 2 t^{7} + 2 t^{8} + 2 t^{10} + 2 t^{12} - 2 t^{13} - 2 t^{15} - 2 t^{17} - 2 t^{18} - 2 t^{20} - t^{25}}{\left(1 - t^{5}\right) \left(1 - t^{8}\right)^{2} \left(1 - t^{9}\right)}~,~
$}
\eea
where $t$ is the fugacity counting degrees in extremal perfect matchings $p_a$, which corresponds to setting every fugacity $t_a$ in \tref{t_sym_geom_27} equal to $t$.
The plethystic logarithm of the refined Hilbert series takes the form,
\beal{es27a003}
PL\left[g(t_a;\mathcal{M}^{mes})\right] &=& t_{3} t_{4}^{2} t_{5}^{2} + t_{1} t_{2} t_{3} t_{4} t_{5} + t_{1}^{2} t_{2}^{2} t_{3} + t_{2} t_{3}^{3} t_{4}^{2} t_{5} + t_{1} t_{2}^{2} t_{3}^{3} t_{4} + t_{1} t_{4}^{3} t_{5}^{4} \nonumber\\
&& {}+ t_{1}^{2} t_{2} t_{4}^{2} t_{5}^{3} + t_{1}^{3} t_{2}^{2} t_{4} t_{5}^{2} + t_{1}^{4} t_{2}^{3} t_{5} + t_{2}^{2} t_{3}^{5} t_{4}^{2} \nonumber\\
&& {}- t_{1}^{2} t_{2}^{2} t_{3}^{2} t_{4}^{2} t_{5}^{2} - t_{1}^{2} t_{2}^{3} t_{3}^{4} t_{4}^{2} t_{5} - t_{1} t_{2}^{2} t_{3}^{4} t_{4}^{3} t_{5}^{2} - t_{1}^{5} t_{2}^{4} t_{3} t_{4} t_{5}^{2} \nonumber\\
&& {}- 2\, t_{1}^{4} t_{2}^{3} t_{3} t_{4}^{2} t_{5}^{3} - 2\, t_{1}^{3} t_{2}^{2} t_{3} t_{4}^{3} t_{5}^{4} - t_{1}^{2} t_{2} t_{3} t_{4}^{4} t_{5}^{5} - t_{1}^{2} t_{2}^{4} t_{3}^{6} t_{4}^{2} \nonumber\\
&& {}- t_{1} t_{2}^{3} t_{3}^{6} t_{4}^{3} t_{5} - t_{2}^{2} t_{3}^{6} t_{4}^{4} t_{5}^{2} - t_{1}^{5} t_{2}^{5} t_{3}^{3} t_{4} t_{5} - 2\, t_{1}^{4} t_{2}^{4} t_{3}^{3} t_{4}^{2} t_{5}^{2} \nonumber\\
&& {}- 2\, t_{1}^{3} t_{2}^{3} t_{3}^{3} t_{4}^{3} t_{5}^{3} - 2\, t_{1}^{2} t_{2}^{2} t_{3}^{3} t_{4}^{4} t_{5}^{4} - t_{1} t_{2} t_{3}^{3} t_{4}^{5} t_{5}^{5} - t_{1}^{6} t_{2}^{4} t_{4}^{2} t_{5}^{4} - t_{1}^{5} t_{2}^{3} t_{4}^{3} t_{5}^{5} \nonumber\\
&& {}- t_{1}^{4} t_{2}^{2} t_{4}^{4} t_{5}^{6} - t_{1}^{4} t_{2}^{5} t_{3}^{5} t_{4}^{2} t_{5} - t_{1}^{3} t_{2}^{4} t_{3}^{5} t_{4}^{3} t_{5}^{2} - t_{1}^{2} t_{2}^{3} t_{3}^{5} t_{4}^{4} t_{5}^{3} - t_{1} t_{2}^{2} t_{3}^{5} t_{4}^{5} t_{5}^{4} \nonumber\\
&& {}+ \dots~,~
\eea
where $t_a$ is paired with the extremal perfect matching $p_a$ at a corner of \fref{f_geom_27}, and can be expressed in terms of the global symmetry fugacities $f_{1}, f_{2}$ and $\bar{t}_{1}, \bar{t}_{2}, \bar{t}_{3}$ as shown in \tref{t_sym_geom_27}.

The positive terms at the start of the expansion correspond to the generators of the mesonic moduli space, which are $A_{1}, \dots, A_{10}$. The expressions of the generators in terms of the chiral fields of the individual toric phases are summarized in \tref{t_model_27_1}, \tref{t_model_27_2} and \tref{t_model_27_3}. The negative terms that come after the positive ones correspond to the first order relations amongst the generators, which take the following~form,
\beal{es27a004}
&& A_{2}^{2} = A_{1} A_{3} ~,~ \quad A_{3} A_{4} = A_{2} A_{5} ~,~ \quad A_{2} A_{4} = A_{1} A_{5} ~,~ \quad A_{3} A_{8} = A_{2} A_{9} ~,~ \nonumber\\
&& A_{3} A_{7} = A_{2} A_{8} = A_{1} A_{9} ~,~ \quad A_{3} A_{6} = A_{2} A_{7} = A_{1} A_{8} ~,~ \quad A_{2} A_{6} = A_{1} A_{7} ~,~ \nonumber\\
&& A_{5}^{2} = A_{3} A_{10} ~,~ \quad A_{4} A_{5} = A_{2} A_{10} ~,~ \quad A_{4}^{2} = A_{1} A_{10} ~,~ \quad A_{5} A_{9} = A_{2} A_{3}^{2} ~,~ \nonumber\\
&& A_{5} A_{8} = A_{4} A_{9} = A_{1} A_{3}^{2} ~,~ \quad A_{5} A_{7} = A_{4} A_{8} = A_{1} A_{2} A_{3} ~,~ \nonumber\\
&& A_{5} A_{6} = A_{4} A_{7} = A_{1}^{2} A_{3} ~,~ \quad A_{4} A_{6} = A_{1}^{2} A_{2} ~,~ \quad A_{8}^{2} = A_{7} A_{9} ~,~ \nonumber\\
&& A_{7} A_{8} = A_{6} A_{9} ~,~ \quad A_{7}^{2} = A_{6} A_{8} ~,~ \quad A_{9} A_{10} = A_{2} A_{3} A_{5} ~,~ \quad A_{8} A_{10} = A_{1} A_{3} A_{5} ~,~ \nonumber\\
&& A_{7} A_{10} = A_{1} A_{2} A_{5} ~,~ \quad A_{6} A_{10} = A_{1}^{2} A_{5}~.~
\eea

\input{model_27_1}
\input{model_27_2}
\input{model_27_3}

\subsection{Model 28: $H^{1,4,1,1}$}\label{s_m28}
\begin{figure}[H]
\centering
\begin{minipage}[c]{0.34\textwidth}
\resizebox{\linewidth}{!}{%
\begin{tikzpicture}[x=1bp,y=1bp]
  \node[anchor=south west,inner sep=0pt,outer sep=0pt] at (0,0) {\includegraphics{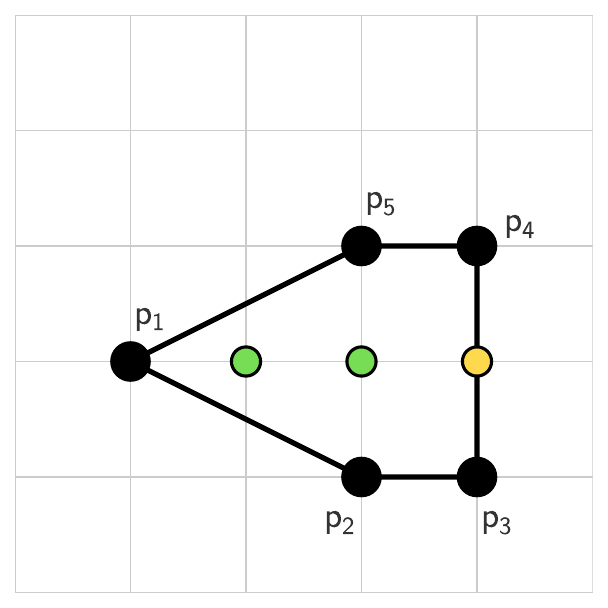}};
  \node[anchor=south west,inner sep=0pt,outer sep=0pt,overlay] at (128.22,120.03) {\includegraphics{figures/toric_label_s1a.pdf}};
  \node[anchor=south west,inner sep=0pt,outer sep=0pt,overlay] at (183.66,120.03) {\includegraphics{figures/toric_label_s2a.pdf}};
\end{tikzpicture}}
\end{minipage}%
\hspace{0.05\textwidth}%
{\small
\setlength{\tabcolsep}{4pt}
\begin{tabular}[c]{|c|c|c|}
\hline
\multirow{2}{*}{phase} & \multicolumn{2}{c|}{multiplicity} \\
\cline{2-3}
& \makebox[2.2em]{$s_{1a}$} & \makebox[2.2em]{$s_{2a}$} \\
\hline\hline
\hyperref[s_model_28_1]{1} & 8 & 13 \\
\hyperref[s_model_28_2]{2} & 8 & 14 \\
\hyperref[s_model_28_3]{3} & 9 & 14 \\
\hyperref[s_model_28_4]{4} & 10 & 16 \\
\hline
\end{tabular}
}
\caption{The toric diagram for Model~28 ($H^{1,4,1,1}$).}
\label{f_geom_28}
\end{figure}

\begin{figure}[H]
\centering
\definecolor{dualityred}{rgb}{1,0.149,0}
\hypersetup{linkcolor=black}
\begin{tikzpicture}[line width=1.55bp]
\useasboundingbox (-0.338,-6.920) rectangle (0.338,0.834);
\draw[dualityred, preaction={draw=white, line width=4.55bp}] (0.000,-4.929) -- (0.000,-6.071);
\draw[dualityred, preaction={draw=white, line width=4.55bp}] (0.000,-3.871) -- (0.000,-2.729);
\draw[dualityred, preaction={draw=white, line width=4.55bp}] (0.000,-1.671) -- (0.000,-0.529);
\draw[dualityred, preaction={draw=white, line width=4.55bp}] (0.301,0.500) .. controls (0.454,0.918) and (-0.454,0.918) .. (-0.301,0.500);
\fill[dualityred] (0.000,-4.737) -- (-0.072,-4.929) -- (0.072,-4.929) -- cycle;
\fill[dualityred] (0.000,-6.263) -- (0.072,-6.071) -- (-0.072,-6.071) -- cycle;
\fill[dualityred] (0.000,-4.063) -- (0.072,-3.871) -- (-0.072,-3.871) -- cycle;
\fill[dualityred] (0.000,-2.537) -- (-0.072,-2.729) -- (0.072,-2.729) -- cycle;
\fill[dualityred] (0.000,-1.863) -- (0.072,-1.671) -- (-0.072,-1.671) -- cycle;
\fill[dualityred] (0.000,-0.337) -- (-0.072,-0.529) -- (0.072,-0.529) -- cycle;
\fill[dualityred] (0.200,0.337) -- (0.363,0.462) -- (0.239,0.538) -- cycle;
\fill[dualityred] (-0.200,0.337) -- (-0.239,0.538) -- (-0.363,0.462) -- cycle;
\node[draw, rectangle, fill=white, minimum size=0.6cm, inner sep=0pt, font=\sffamily\bfseries\footnotesize, line width=2.07bp] at (0.000,-4.400) {\hyperref[s_model_28_1]{1}};
\node[draw, rectangle, fill=white, minimum size=0.6cm, inner sep=0pt, font=\sffamily\bfseries\footnotesize, line width=2.07bp] at (0.000,-6.600) {\hyperref[s_model_28_2]{2}};
\node[draw, rectangle, fill=white, minimum size=0.6cm, inner sep=0pt, font=\sffamily\bfseries\footnotesize, line width=2.07bp] at (0.000,-2.200) {\hyperref[s_model_28_3]{3}};
\node[draw, rectangle, fill=white, minimum size=0.6cm, inner sep=0pt, font=\sffamily\bfseries\footnotesize, line width=2.07bp] at (0.000,0.000) {\hyperref[s_model_28_4]{4}};
\end{tikzpicture}
\caption{The Seiberg duality tree for Model~28 ($H^{1,4,1,1}$). The node labels indicate the phase numbers.}
\label{f_geom_28_tree}
\end{figure}

The toric diagram of the $H^{1,4,1,1}$ model is shown in \fref{f_geom_28}. The Seiberg duality tree of the $4$ toric phases associated with $H^{1,4,1,1}$ is shown in \fref{f_geom_28_tree}.
The global symmetry of the brane tilings associated with $H^{1,4,1,1}$ is given by,
\beal{es28a001}
U(1) \times U(1) \times U(1)_R~.~
\eea
The global symmetry charges on the extremal GLSM fields $p_a$ are summarized in \tref{t_sym_geom_28}.

\begin{table}[H]
\centering
{\small
\begin{tabular}{|c|c|c|c|c|l|}
\hline
  & $U(1)_{1}$ & $U(1)_{2}$ & $U(1)_R$ & $r_a$ & fugacities \\
\hline
  $p_{1}$ & $-1$ & $-1$ & $r_{1}$ & $r_{1} = 0.5498$ & $t_{1} = f_{1}^{-1} f_{2}^{-1} \bar{t}_{1}$ \\
  $p_{2}$ & $1$ & $0$ & $r_{2}$ & $r_{2} = 0.2629$ & $t_{2} = f_{1} \bar{t}_{2}$ \\
  $p_{3}$ & $0$ & $1$ & $r_{3}$ & $r_{3} = 0.4622$ & $t_{3} = f_{2} \bar{t}_{3}$ \\
  $p_{4}$ & $0$ & $0$ & $r_{4}$ & $r_{4} = 0.4622$ & $t_{4} = \bar{t}_{3}$ \\
  $p_{5}$ & $0$ & $0$ & $r_{5}$ & $r_{5} = 0.2629$ & $t_{5} = \bar{t}_{2}$ \\
\hline
\end{tabular}
}
\caption{Global symmetry charges on the extremal perfect matchings of Model~28 ($H^{1,4,1,1}$).}
\label{t_sym_geom_28}
\end{table}

All toric phases associated with $H^{1,4,1,1}$ share the same mesonic moduli space, whose unrefined Hilbert series takes the form,
\beal{es28a002}
\resizebox{\ifdim\width>\linewidth\linewidth\else\width\fi}{!}{$
g(t; \mathcal{M}^{mes}) \;=\; \frac{1 + 3 t^{5} + 8 t^{10} + 3 t^{15} + t^{20}}{\left(1 - t^{5}\right) \left(1 - t^{10}\right)^{2}}~,~
$}
\eea
where $t$ is the fugacity counting degrees in extremal perfect matchings $p_a$, which corresponds to setting every fugacity $t_a$ in \tref{t_sym_geom_28} equal to $t$.
The plethystic logarithm of the refined Hilbert series takes the form,
\beal{es28a003}
PL\left[g(t_a;\mathcal{M}^{mes})\right] &=& t_{1} t_{4}^{2} t_{5}^{2} + t_{1} t_{2} t_{3} t_{4} t_{5} + t_{1} t_{2}^{2} t_{3}^{2} + t_{1}^{3} t_{2} t_{5} + t_{3} t_{4}^{5} t_{5}^{4} + t_{2} t_{3}^{2} t_{4}^{4} t_{5}^{3} \nonumber\\
&& {}+ t_{2}^{2} t_{3}^{3} t_{4}^{3} t_{5}^{2} + t_{2}^{3} t_{3}^{4} t_{4}^{2} t_{5} + t_{2}^{4} t_{3}^{5} t_{4} \nonumber\\
&& {}- t_{1}^{2} t_{2}^{2} t_{3}^{2} t_{4}^{2} t_{5}^{2} - t_{1}^{3} t_{2}^{5} t_{3}^{5} t_{4} t_{5} - t_{1}^{3} t_{2}^{4} t_{3}^{4} t_{4}^{2} t_{5}^{2} - t_{1}^{3} t_{2}^{3} t_{3}^{3} t_{4}^{3} t_{5}^{3} \nonumber\\
&& {}- t_{1}^{3} t_{2}^{2} t_{3}^{2} t_{4}^{4} t_{5}^{4} - t_{1}^{3} t_{2} t_{3} t_{4}^{5} t_{5}^{5} - t_{1} t_{2}^{5} t_{3}^{6} t_{4}^{2} t_{5} - 2\, t_{1} t_{2}^{4} t_{3}^{5} t_{4}^{3} t_{5}^{2} \nonumber\\
&& {}- 2\, t_{1} t_{2}^{3} t_{3}^{4} t_{4}^{4} t_{5}^{3} - 2\, t_{1} t_{2}^{2} t_{3}^{3} t_{4}^{5} t_{5}^{4} - t_{1} t_{2} t_{3}^{2} t_{4}^{6} t_{5}^{5} - t_{2}^{6} t_{3}^{8} t_{4}^{4} t_{5}^{2} - t_{2}^{5} t_{3}^{7} t_{4}^{5} t_{5}^{3} \nonumber\\
&& {}- 2\, t_{2}^{4} t_{3}^{6} t_{4}^{6} t_{5}^{4} - t_{2}^{3} t_{3}^{5} t_{4}^{7} t_{5}^{5} - t_{2}^{2} t_{3}^{4} t_{4}^{8} t_{5}^{6} + \dots~,~
\eea
where the fugacity $t_a$ of the extremal perfect matching $p_a$ in \fref{f_geom_28} can be expressed in terms of the global symmetry fugacities $f_{1}, f_{2}$ and $\bar{t}_{1}, \bar{t}_{2}, \bar{t}_{3}$, following the charges of $p_a$ listed in \tref{t_sym_geom_28}.

The generators of the mesonic moduli space, $A_{1}, \dots, A_{9}$, correspond to the first positive terms of the plethystic logarithm. In \tref{t_model_28_1}, \tref{t_model_28_2}, \tref{t_model_28_3} and \tref{t_model_28_4}, the generators are given in terms of the chiral fields of each toric phase. The negative terms that come next correspond to the first order relations amongst the generators, which take the following~form,
\beal{es28a004}
&& A_{2}^{2} = A_{1} A_{3} ~,~ \quad A_{4} A_{9} = A_{2} A_{3}^{2} ~,~ \quad A_{4} A_{8} = A_{1} A_{3}^{2} ~,~ \quad A_{4} A_{7} = A_{1} A_{2} A_{3} ~,~ \nonumber\\
&& A_{4} A_{6} = A_{1}^{2} A_{3} ~,~ \quad A_{4} A_{5} = A_{1}^{2} A_{2} ~,~ \quad A_{3} A_{8} = A_{2} A_{9} ~,~ \nonumber\\
&& A_{3} A_{7} = A_{2} A_{8} = A_{1} A_{9} ~,~ \quad A_{3} A_{6} = A_{2} A_{7} = A_{1} A_{8} ~,~ \nonumber\\
&& A_{3} A_{5} = A_{2} A_{6} = A_{1} A_{7} ~,~ \quad A_{2} A_{5} = A_{1} A_{6} ~,~ \quad A_{8}^{2} = A_{7} A_{9} ~,~ \nonumber\\
&& A_{7} A_{8} = A_{6} A_{9} ~,~ \quad A_{7}^{2} = A_{6} A_{8} = A_{5} A_{9} ~,~ \quad A_{6} A_{7} = A_{5} A_{8} ~,~ \quad A_{6}^{2} = A_{5} A_{7}~.~
\eea

\input{model_28_1}
\input{model_28_2}
\input{model_28_3}
\input{model_28_4}

\subsection{Model 29: $H^{2,2,1,1}$}\label{s_m29}
\input{geom_29}
\input{model_29_1}
\input{model_29_2}
\input{model_29_3}
\input{model_29_4}
\input{model_29_5}
\input{model_29_6}
\input{model_29_7}
\input{model_29_8}

\subsection{Model 30: $J^{1,4,3,2}$}\label{s_m30}
\begin{figure}[H]
\centering
\begin{minipage}[c]{0.34\textwidth}
\resizebox{\linewidth}{!}{%
\begin{tikzpicture}[x=1bp,y=1bp]
  \node[anchor=south west,inner sep=0pt,outer sep=0pt] at (0,0) {\includegraphics{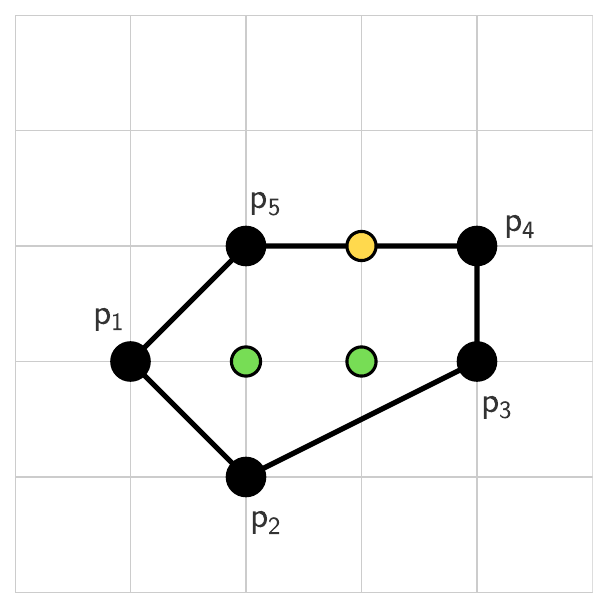}};
  \node[anchor=south west,inner sep=0pt,outer sep=0pt,overlay] at (128.22,120.03) {\includegraphics{figures/toric_label_s1a.pdf}};
  \node[anchor=south west,inner sep=0pt,outer sep=0pt,overlay] at (183.66,120.03) {\includegraphics{figures/toric_label_s2a.pdf}};
\end{tikzpicture}}
\end{minipage}%
\hspace{0.05\textwidth}%
{\small
\setlength{\tabcolsep}{4pt}
\begin{tabular}[c]{|c|c|c|}
\hline
\multirow{2}{*}{phase} & \multicolumn{2}{c|}{multiplicity} \\
\cline{2-3}
& \makebox[2.2em]{$s_{1a}$} & \makebox[2.2em]{$s_{2a}$} \\
\hline\hline
\hyperref[s_model_30_1]{1} & 9 & 9 \\
\hyperref[s_model_30_2]{2} & 9 & 9 \\
\hyperref[s_model_30_3]{3} & 10 & 9 \\
\hyperref[s_model_30_4]{4} & 12 & 10 \\
\hyperref[s_model_30_5]{5} & 11 & 11 \\
\hyperref[s_model_30_6]{6} & 11 & 12 \\
\hyperref[s_model_30_7]{7} & 12 & 12 \\
\hyperref[s_model_30_8]{8} & 14 & 10 \\
\hline
\end{tabular}
}
\caption{The toric diagram for Model~30 ($J^{1,4,3,2}$).}
\label{f_geom_30}
\end{figure}

\begin{figure}[H]
\centering
\definecolor{dualityred}{rgb}{1,0.149,0}
\hypersetup{linkcolor=black}
\begin{tikzpicture}[line width=1.55bp]
\useasboundingbox (-3.170,-7.270) rectangle (3.170,0.320);
\draw[dualityred, preaction={draw=white, line width=4.55bp}] (-1.367,-2.683) -- (-2.433,-3.917);
\draw[dualityred, preaction={draw=white, line width=4.55bp}] (-0.728,-1.686) -- (-0.222,-0.514);
\draw[dualityred, preaction={draw=white, line width=4.55bp}] (0.698,-4.902) .. controls (-0.099,-6.099) and (-1.801,-6.099) .. (-2.598,-4.902);
\draw[dualityred, preaction={draw=white, line width=4.55bp}] (0.950,-3.871) -- (0.950,-2.729);
\draw[dualityred, preaction={draw=white, line width=4.55bp}] (-2.346,-4.108) -- (0.446,-2.492);
\draw[dualityred, preaction={draw=white, line width=4.55bp}] (-2.361,-4.778) -- (-0.489,-6.222);
\draw[dualityred, preaction={draw=white, line width=4.55bp}] (-2.770,-4.926) .. controls (-2.078,-8.025) and (2.078,-8.025) .. (2.770,-4.926);
\draw[dualityred, preaction={draw=white, line width=4.55bp}] (1.367,-2.683) -- (2.433,-3.917);
\draw[dualityred, preaction={draw=white, line width=4.55bp}] (0.728,-1.686) -- (0.222,-0.514);
\draw[dualityred, preaction={draw=white, line width=4.55bp}] (0.533,-2.683) -- (-0.533,-3.917);
\draw[dualityred, preaction={draw=white, line width=4.55bp}] (-0.222,-6.086) -- (-0.728,-4.914);
\draw[dualityred, preaction={draw=white, line width=4.55bp}] (1.075,-4.926) .. controls (1.242,-5.425) and (1.975,-4.692) .. (1.476,-4.525);
\fill[dualityred] (-1.241,-2.537) -- (-1.421,-2.635) -- (-1.312,-2.730) -- cycle;
\fill[dualityred] (-2.559,-4.063) -- (-2.379,-3.965) -- (-2.488,-3.870) -- cycle;
\fill[dualityred] (-0.805,-1.863) -- (-0.662,-1.715) -- (-0.795,-1.658) -- cycle;
\fill[dualityred] (-0.145,-0.337) -- (-0.288,-0.485) -- (-0.155,-0.542) -- cycle;
\fill[dualityred] (0.798,-4.737) -- (0.637,-4.864) -- (0.760,-4.939) -- cycle;
\fill[dualityred] (-2.698,-4.737) -- (-2.660,-4.939) -- (-2.537,-4.864) -- cycle;
\fill[dualityred] (0.950,-4.063) -- (1.022,-3.871) -- (0.878,-3.871) -- cycle;
\fill[dualityred] (0.950,-2.537) -- (0.878,-2.729) -- (1.022,-2.729) -- cycle;
\fill[dualityred] (-2.513,-4.205) -- (-2.310,-4.171) -- (-2.383,-4.046) -- cycle;
\fill[dualityred] (0.613,-2.395) -- (0.410,-2.429) -- (0.483,-2.554) -- cycle;
\fill[dualityred] (-2.513,-4.660) -- (-2.405,-4.835) -- (-2.317,-4.720) -- cycle;
\fill[dualityred] (-0.337,-6.340) -- (-0.445,-6.165) -- (-0.533,-6.280) -- cycle;
\fill[dualityred] (-2.807,-4.737) -- (-2.841,-4.940) -- (-2.699,-4.912) -- cycle;
\fill[dualityred] (2.807,-4.737) -- (2.699,-4.912) -- (2.841,-4.940) -- cycle;
\fill[dualityred] (1.241,-2.537) -- (1.312,-2.730) -- (1.421,-2.635) -- cycle;
\fill[dualityred] (2.559,-4.063) -- (2.488,-3.870) -- (2.379,-3.965) -- cycle;
\fill[dualityred] (0.805,-1.863) -- (0.795,-1.658) -- (0.662,-1.715) -- cycle;
\fill[dualityred] (0.145,-0.337) -- (0.155,-0.542) -- (0.288,-0.485) -- cycle;
\fill[dualityred] (0.659,-2.537) -- (0.479,-2.635) -- (0.588,-2.730) -- cycle;
\fill[dualityred] (-0.659,-4.063) -- (-0.479,-3.965) -- (-0.588,-3.870) -- cycle;
\fill[dualityred] (-0.145,-6.263) -- (-0.155,-6.058) -- (-0.288,-6.115) -- cycle;
\fill[dualityred] (-0.805,-4.737) -- (-0.795,-4.942) -- (-0.662,-4.885) -- cycle;
\fill[dualityred] (1.043,-4.737) -- (1.004,-4.938) -- (1.146,-4.914) -- cycle;
\fill[dualityred] (1.287,-4.493) -- (1.464,-4.596) -- (1.488,-4.454) -- cycle;
\node[draw, rectangle, fill=white, minimum size=0.6cm, inner sep=0pt, font=\sffamily\bfseries\footnotesize, line width=2.07bp] at (-0.950,-2.200) {\hyperref[s_model_30_1]{1}};
\node[draw, rectangle, fill=white, minimum size=0.6cm, inner sep=0pt, font=\sffamily\bfseries\footnotesize, line width=2.07bp] at (0.950,-4.400) {\hyperref[s_model_30_2]{2}};
\node[draw, rectangle, fill=white, minimum size=0.6cm, inner sep=0pt, font=\sffamily\bfseries\footnotesize, line width=2.07bp] at (-2.850,-4.400) {\hyperref[s_model_30_3]{3}};
\node[draw, rectangle, fill=white, minimum size=0.6cm, inner sep=0pt, font=\sffamily\bfseries\footnotesize, line width=2.07bp] at (0.950,-2.200) {\hyperref[s_model_30_4]{4}};
\node[draw, rectangle, fill=white, minimum size=0.6cm, inner sep=0pt, font=\sffamily\bfseries\footnotesize, line width=2.07bp] at (0.000,-6.600) {\hyperref[s_model_30_5]{5}};
\node[draw, rectangle, fill=white, minimum size=0.6cm, inner sep=0pt, font=\sffamily\bfseries\footnotesize, line width=2.07bp] at (2.850,-4.400) {\hyperref[s_model_30_6]{6}};
\node[draw, rectangle, fill=white, minimum size=0.6cm, inner sep=0pt, font=\sffamily\bfseries\footnotesize, line width=2.07bp] at (0.000,0.000) {\hyperref[s_model_30_7]{7}};
\node[draw, rectangle, fill=white, minimum size=0.6cm, inner sep=0pt, font=\sffamily\bfseries\footnotesize, line width=2.07bp] at (-0.950,-4.400) {\hyperref[s_model_30_8]{8}};
\end{tikzpicture}
\caption{The Seiberg duality tree for Model~30 ($J^{1,4,3,2}$). The node labels indicate the phase numbers.}
\label{f_geom_30_tree}
\end{figure}

The $J^{1,4,3,2}$ model has a toric diagram shown in \fref{f_geom_30}. The Seiberg duality tree of the $8$ toric phases associated with $J^{1,4,3,2}$ is shown in \fref{f_geom_30_tree}.
The global symmetry of the brane tilings associated with $J^{1,4,3,2}$ is given by,
\beal{es30a001}
U(1) \times U(1) \times U(1)_R~.~
\eea
The global symmetry charges on the extremal GLSM fields $p_a$ are summarized in \tref{t_sym_geom_30}.

\begin{table}[H]
\centering
{\small
\begin{tabular}{|c|c|c|c|c|l|}
\hline
  & $U(1)_{1}$ & $U(1)_{2}$ & $U(1)_R$ & $r_a$ & fugacities \\
\hline
  $p_{1}$ & $-1$ & $-1$ & $r_{1}$ & $r_{1} = 0.3785$ & $t_{1} = f_{1}^{-1} f_{2}^{-1} \bar{t}_{1}$ \\
  $p_{2}$ & $1$ & $0$ & $r_{2}$ & $r_{2} = 0.4656$ & $t_{2} = f_{1} \bar{t}_{2}$ \\
  $p_{3}$ & $0$ & $1$ & $r_{3}$ & $r_{3} = 0.3785$ & $t_{3} = f_{2} \bar{t}_{1}$ \\
  $p_{4}$ & $0$ & $0$ & $r_{4}$ & $r_{4} = 0.3887$ & $t_{4} = \bar{t}_{3}$ \\
  $p_{5}$ & $0$ & $0$ & $r_{5}$ & $r_{5} = 0.3887$ & $t_{5} = \bar{t}_{3}$ \\
\hline
\end{tabular}
}
\caption{Global symmetry charges on the extremal perfect matchings of Model~30 ($J^{1,4,3,2}$).}
\label{t_sym_geom_30}
\end{table}

All toric phases associated with $J^{1,4,3,2}$ share the same mesonic moduli space, whose unrefined Hilbert series takes the form,
\beal{es30a002}
\resizebox{\ifdim\width>\linewidth\linewidth\else\width\fi}{!}{$
g(t; \mathcal{M}^{mes}) \;=\; \frac{1 - t^{3} + t^{5} + t^{6} + t^{7} + t^{8} - t^{10} - t^{11} - t^{12} - t^{13} + t^{15} - t^{18}}{\left(1 - t^{3}\right) \left(1 - t^{4}\right) \left(1 - t^{7}\right) \left(1 - t^{9}\right)}~,~
$}
\eea
where $t$ is the fugacity counting degrees in extremal perfect matchings $p_a$, which corresponds to setting every fugacity $t_a$ in \tref{t_sym_geom_30} equal to $t$.
The plethystic logarithm of the refined Hilbert series takes the form,
\beal{es30a003}
PL\left[g(t_a;\mathcal{M}^{mes})\right] &=& t_{1} t_{2}^{2} t_{3} + t_{1} t_{2} t_{3} t_{4} t_{5} + t_{1} t_{3} t_{4}^{2} t_{5}^{2} + t_{2}^{2} t_{3}^{3} t_{4}^{2} + t_{1}^{3} t_{2}^{2} t_{5}^{2} + t_{2} t_{3}^{3} t_{4}^{3} t_{5} \nonumber\\
&& {}+ t_{1}^{3} t_{2} t_{4} t_{5}^{3} + t_{3}^{3} t_{4}^{4} t_{5}^{2} + t_{1}^{3} t_{4}^{2} t_{5}^{4} \nonumber\\
&& {}- t_{1}^{2} t_{2}^{2} t_{3}^{2} t_{4}^{2} t_{5}^{2} - t_{1}^{4} t_{2}^{3} t_{3} t_{4} t_{5}^{3} - t_{1} t_{2}^{3} t_{3}^{4} t_{4}^{3} t_{5} - 2\, t_{1}^{4} t_{2}^{2} t_{3} t_{4}^{2} t_{5}^{4} \nonumber\\
&& {}- 2\, t_{1} t_{2}^{2} t_{3}^{4} t_{4}^{4} t_{5}^{2} - t_{1}^{3} t_{2}^{4} t_{3}^{3} t_{4}^{2} t_{5}^{2} - t_{1}^{4} t_{2} t_{3} t_{4}^{3} t_{5}^{5} - t_{1} t_{2} t_{3}^{4} t_{4}^{5} t_{5}^{3} \nonumber\\
&& {}- 2\, t_{1}^{3} t_{2}^{3} t_{3}^{3} t_{4}^{3} t_{5}^{3} - t_{1}^{6} t_{2}^{2} t_{4}^{2} t_{5}^{6} - 3\, t_{1}^{3} t_{2}^{2} t_{3}^{3} t_{4}^{4} t_{5}^{4} - t_{2}^{2} t_{3}^{6} t_{4}^{6} t_{5}^{2} \nonumber\\
&& {}- 2\, t_{1}^{3} t_{2} t_{3}^{3} t_{4}^{5} t_{5}^{5} - t_{1}^{3} t_{3}^{3} t_{4}^{6} t_{5}^{6} + \dots~,~
\eea
with $t_a$ the fugacity of $p_a$ in \fref{f_geom_30}, so that a generator $\prod_a p_a^{n_a}$ appears as $\prod_a t_a^{n_a}$; each $t_a$ can be expressed in terms of the global symmetry fugacities $f_{1}, f_{2}$ and $\bar{t}_{1}, \bar{t}_{2}, \bar{t}_{3}$ as shown in \tref{t_sym_geom_30}.

Each of the first $9$ positive terms corresponds to one generator of the mesonic moduli space, the generators being $A_{1}, \dots, A_{9}$. The expressions of the generators in terms of the chiral fields of the individual toric phases are summarized in \tref{t_model_30_1}, \tref{t_model_30_2}, \tref{t_model_30_3}, \tref{t_model_30_4}, \tref{t_model_30_5}, \tref{t_model_30_6}, \tref{t_model_30_7} and \tref{t_model_30_8}. The next negative terms correspond to the first order relations amongst the generators, which take the following~form,
\beal{es30a004}
&& A_{2}^{2} = A_{1} A_{3} ~,~ \quad A_{2} A_{5} = A_{1} A_{7} ~,~ \quad A_{2} A_{4} = A_{1} A_{6} ~,~ \nonumber\\
&& A_{3} A_{5} = A_{2} A_{7} = A_{1} A_{9} ~,~ \quad A_{3} A_{4} = A_{2} A_{6} = A_{1} A_{8} ~,~ \quad A_{4} A_{5} = A_{1}^{2} A_{3} ~,~ \nonumber\\
&& A_{3} A_{7} = A_{2} A_{9} ~,~ \quad A_{3} A_{6} = A_{2} A_{8} ~,~ \quad A_{5} A_{6} = A_{4} A_{7} = A_{1} A_{2} A_{3} ~,~ \nonumber\\
&& A_{7}^{2} = A_{5} A_{9} ~,~ \quad A_{6} A_{7} = A_{5} A_{8} = A_{4} A_{9} = A_{1} A_{3}^{2} ~,~ \quad A_{6}^{2} = A_{4} A_{8} ~,~ \nonumber\\
&& A_{7} A_{8} = A_{6} A_{9} = A_{2} A_{3}^{2} ~,~ \quad A_{8} A_{9} = A_{3}^{3}~.~
\eea

\input{model_30_1}
\input{model_30_2}
\input{model_30_3}
\input{model_30_4}
\input{model_30_5}
\input{model_30_6}
\input{model_30_7}
\input{model_30_8}

\subsection{Model 31: $G^{4,2}$}\label{s_m31}
\begin{figure}[H]
\centering
\begin{minipage}[c]{0.34\textwidth}
\resizebox{\linewidth}{!}{%
\begin{tikzpicture}[x=1bp,y=1bp]
  \node[anchor=south west,inner sep=0pt,outer sep=0pt] at (0,0) {\includegraphics{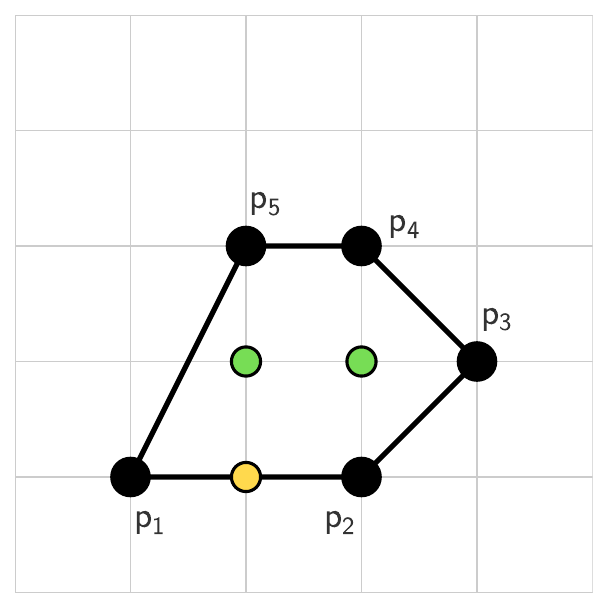}};
  \node[anchor=south west,inner sep=0pt,outer sep=0pt,overlay] at (128.22,120.03) {\includegraphics{figures/toric_label_s1a.pdf}};
  \node[anchor=south west,inner sep=0pt,outer sep=0pt,overlay] at (183.66,120.03) {\includegraphics{figures/toric_label_s2a.pdf}};
\end{tikzpicture}}
\end{minipage}%
\hspace{0.05\textwidth}%
{\small
\setlength{\tabcolsep}{4pt}
\begin{tabular}[c]{|c|c|c|}
\hline
\multirow{2}{*}{phase} & \multicolumn{2}{c|}{multiplicity} \\
\cline{2-3}
& \makebox[2.2em]{$s_{1a}$} & \makebox[2.2em]{$s_{2a}$} \\
\hline\hline
\hyperref[s_model_31_1]{1} & 8 & 9 \\
\hyperref[s_model_31_2]{2} & 8 & 11 \\
\hyperref[s_model_31_3]{3} & 8 & 9 \\
\hyperref[s_model_31_4]{4} & 8 & 11 \\
\hyperref[s_model_31_5]{5} & 9 & 10 \\
\hyperref[s_model_31_6]{6} & 9 & 11 \\
\hyperref[s_model_31_7]{7} & 9 & 14 \\
\hline
\end{tabular}
}
\caption{The toric diagram for Model~31 ($G^{4,2}$).}
\label{f_geom_31}
\end{figure}

\begin{figure}[H]
\centering
\definecolor{dualityred}{rgb}{1,0.149,0}
\hypersetup{linkcolor=black}
\begin{tikzpicture}[line width=1.55bp]
\useasboundingbox (-2.617,-5.117) rectangle (2.617,0.320);
\draw[dualityred, preaction={draw=white, line width=4.55bp}] (-1.900,-3.871) -- (-1.900,-2.729);
\draw[dualityred, preaction={draw=white, line width=4.55bp}] (-1.371,-4.400) -- (-0.529,-4.400);
\draw[dualityred, preaction={draw=white, line width=4.55bp}] (-1.483,-3.917) -- (-0.417,-2.683);
\draw[dualityred, preaction={draw=white, line width=4.55bp}] (-1.396,-4.108) -- (1.396,-2.492);
\draw[dualityred, preaction={draw=white, line width=4.55bp}] (-1.396,-2.492) -- (1.396,-4.108);
\draw[dualityred, preaction={draw=white, line width=4.55bp}] (-1.483,-1.717) -- (-0.417,-0.483);
\draw[dualityred, preaction={draw=white, line width=4.55bp}] (0.000,-3.871) -- (0.000,-2.729);
\draw[dualityred, preaction={draw=white, line width=4.55bp}] (0.417,-2.683) -- (1.483,-3.917);
\draw[dualityred, preaction={draw=white, line width=4.55bp}] (0.000,-1.671) -- (0.000,-0.529);
\draw[dualityred, preaction={draw=white, line width=4.55bp}] (1.900,-2.729) -- (1.900,-3.871);
\draw[dualityred, preaction={draw=white, line width=4.55bp}] (1.483,-1.717) -- (0.417,-0.483);
\draw[dualityred, preaction={draw=white, line width=4.55bp}] (-2.426,-4.525) .. controls (-2.925,-4.692) and (-2.192,-5.425) .. (-2.025,-4.926);
\fill[dualityred] (-1.900,-4.063) -- (-1.828,-3.871) -- (-1.972,-3.871) -- cycle;
\fill[dualityred] (-1.900,-2.537) -- (-1.972,-2.729) -- (-1.828,-2.729) -- cycle;
\fill[dualityred] (-1.563,-4.400) -- (-1.371,-4.472) -- (-1.371,-4.328) -- cycle;
\fill[dualityred] (-0.337,-4.400) -- (-0.529,-4.328) -- (-0.529,-4.472) -- cycle;
\fill[dualityred] (-1.609,-4.063) -- (-1.429,-3.965) -- (-1.538,-3.870) -- cycle;
\fill[dualityred] (-0.291,-2.537) -- (-0.471,-2.635) -- (-0.362,-2.730) -- cycle;
\fill[dualityred] (-1.563,-4.205) -- (-1.360,-4.171) -- (-1.433,-4.046) -- cycle;
\fill[dualityred] (1.563,-2.395) -- (1.360,-2.429) -- (1.433,-2.554) -- cycle;
\fill[dualityred] (-1.563,-2.395) -- (-1.433,-2.554) -- (-1.360,-2.429) -- cycle;
\fill[dualityred] (1.563,-4.205) -- (1.433,-4.046) -- (1.360,-4.171) -- cycle;
\fill[dualityred] (-1.609,-1.863) -- (-1.429,-1.765) -- (-1.538,-1.670) -- cycle;
\fill[dualityred] (-0.291,-0.337) -- (-0.471,-0.435) -- (-0.362,-0.530) -- cycle;
\fill[dualityred] (0.000,-4.063) -- (0.072,-3.871) -- (-0.072,-3.871) -- cycle;
\fill[dualityred] (0.000,-2.537) -- (-0.072,-2.729) -- (0.072,-2.729) -- cycle;
\fill[dualityred] (0.291,-2.537) -- (0.362,-2.730) -- (0.471,-2.635) -- cycle;
\fill[dualityred] (1.609,-4.063) -- (1.538,-3.870) -- (1.429,-3.965) -- cycle;
\fill[dualityred] (0.000,-1.863) -- (0.072,-1.671) -- (-0.072,-1.671) -- cycle;
\fill[dualityred] (0.000,-0.337) -- (-0.072,-0.529) -- (0.072,-0.529) -- cycle;
\fill[dualityred] (1.900,-2.537) -- (1.828,-2.729) -- (1.972,-2.729) -- cycle;
\fill[dualityred] (1.900,-4.063) -- (1.972,-3.871) -- (1.828,-3.871) -- cycle;
\fill[dualityred] (1.609,-1.863) -- (1.538,-1.670) -- (1.429,-1.765) -- cycle;
\fill[dualityred] (0.291,-0.337) -- (0.362,-0.530) -- (0.471,-0.435) -- cycle;
\fill[dualityred] (-2.237,-4.493) -- (-2.438,-4.454) -- (-2.414,-4.596) -- cycle;
\fill[dualityred] (-1.993,-4.737) -- (-2.096,-4.914) -- (-1.954,-4.938) -- cycle;
\node[draw, rectangle, fill=white, minimum size=0.6cm, inner sep=0pt, font=\sffamily\bfseries\footnotesize, line width=2.07bp] at (-1.900,-4.400) {\hyperref[s_model_31_1]{1}};
\node[draw, rectangle, fill=white, minimum size=0.6cm, inner sep=0pt, font=\sffamily\bfseries\footnotesize, line width=2.07bp] at (-1.900,-2.200) {\hyperref[s_model_31_2]{2}};
\node[draw, rectangle, fill=white, minimum size=0.6cm, inner sep=0pt, font=\sffamily\bfseries\footnotesize, line width=2.07bp] at (0.000,-4.400) {\hyperref[s_model_31_3]{3}};
\node[draw, rectangle, fill=white, minimum size=0.6cm, inner sep=0pt, font=\sffamily\bfseries\footnotesize, line width=2.07bp] at (0.000,-2.200) {\hyperref[s_model_31_4]{4}};
\node[draw, rectangle, fill=white, minimum size=0.6cm, inner sep=0pt, font=\sffamily\bfseries\footnotesize, line width=2.07bp] at (1.900,-2.200) {\hyperref[s_model_31_5]{5}};
\node[draw, rectangle, fill=white, minimum size=0.6cm, inner sep=0pt, font=\sffamily\bfseries\footnotesize, line width=2.07bp] at (1.900,-4.400) {\hyperref[s_model_31_6]{6}};
\node[draw, rectangle, fill=white, minimum size=0.6cm, inner sep=0pt, font=\sffamily\bfseries\footnotesize, line width=2.07bp] at (0.000,0.000) {\hyperref[s_model_31_7]{7}};
\end{tikzpicture}
\caption{The Seiberg duality tree for Model~31 ($G^{4,2}$). The node labels indicate the phase numbers.}
\label{f_geom_31_tree}
\end{figure}

The toric diagram of the $G^{4,2}$ model is shown in \fref{f_geom_31}. The Seiberg duality tree of the $7$ toric phases associated with $G^{4,2}$ is shown in \fref{f_geom_31_tree}.
The global symmetry of the brane tilings associated with $G^{4,2}$ is given by,
\beal{es31a001}
U(1) \times U(1) \times U(1)_R~.~
\eea
The global symmetry charges on the extremal GLSM fields $p_a$ are summarized in \tref{t_sym_geom_31}.

\begin{table}[H]
\centering
{\small
\begin{tabular}{|c|c|c|c|c|l|}
\hline
  & $U(1)_{1}$ & $U(1)_{2}$ & $U(1)_R$ & $r_a$ & fugacities \\
\hline
  $p_{1}$ & $-1$ & $-1$ & $r_{1}$ & $r_{1} = 0.5254$ & $t_{1} = f_{1}^{-1} f_{2}^{-1} \bar{t}_{1}$ \\
  $p_{2}$ & $1$ & $0$ & $r_{2}$ & $r_{2} = 0.3678$ & $t_{2} = f_{1} \bar{t}_{2}$ \\
  $p_{3}$ & $0$ & $1$ & $r_{3}$ & $r_{3} = 0.4065$ & $t_{3} = f_{2} \bar{t}_{3}$ \\
  $p_{4}$ & $0$ & $0$ & $r_{4}$ & $r_{4} = 0.2674$ & $t_{4} = \bar{t}_{4}$ \\
  $p_{5}$ & $0$ & $0$ & $r_{5}$ & $r_{5} = 0.4329$ & $t_{5} = \bar{t}_{5}$ \\
\hline
\end{tabular}
}
\caption{Global symmetry charges on the extremal perfect matchings of Model~31 ($G^{4,2}$).}
\label{t_sym_geom_31}
\end{table}

All toric phases associated with $G^{4,2}$ share the same mesonic moduli space, whose unrefined Hilbert series takes the form,
\beal{es31a002}
\resizebox{\ifdim\width>\linewidth\linewidth\else\width\fi}{!}{$
g(t; \mathcal{M}^{mes}) \;=\; \frac{1 + 2 t^{5} + t^{7} + 2 t^{8} + 2 t^{10} - 2 t^{12} + 2 t^{13} - 2 t^{15} - 2 t^{17} - t^{18} - 2 t^{20} - t^{25}}{\left(1 - t^{5}\right) \left(1 - t^{7}\right)^{2} \left(1 - t^{11}\right)}~,~
$}
\eea
where $t$ is the fugacity counting degrees in extremal perfect matchings $p_a$, which corresponds to setting every fugacity $t_a$ in \tref{t_sym_geom_31} equal to $t$.
The plethystic logarithm of the refined Hilbert series takes the form,
\beal{es31a003}
PL\left[g(t_a;\mathcal{M}^{mes})\right] &=& t_{3} t_{4}^{2} t_{5}^{2} + t_{1} t_{2} t_{3} t_{4} t_{5} + t_{1}^{2} t_{2}^{2} t_{3} + t_{1}^{2} t_{4}^{2} t_{5}^{3} + t_{1}^{3} t_{2} t_{4} t_{5}^{2} + t_{1}^{4} t_{2}^{2} t_{5} \nonumber\\
&& {}+ t_{2}^{2} t_{3}^{3} t_{4}^{2} t_{5} + t_{1} t_{2}^{3} t_{3}^{3} t_{4} + t_{2}^{4} t_{3}^{5} t_{4}^{2} \nonumber\\
&& {}- t_{1}^{2} t_{2}^{2} t_{3}^{2} t_{4}^{2} t_{5}^{2} - t_{1}^{5} t_{2}^{3} t_{3} t_{4} t_{5}^{2} - 2\, t_{1}^{4} t_{2}^{2} t_{3} t_{4}^{2} t_{5}^{3} - t_{1}^{3} t_{2} t_{3} t_{4}^{3} t_{5}^{4} \nonumber\\
&& {}- t_{1}^{2} t_{2}^{4} t_{3}^{4} t_{4}^{2} t_{5} - t_{1} t_{2}^{3} t_{3}^{4} t_{4}^{3} t_{5}^{2} - t_{1}^{6} t_{2}^{2} t_{4}^{2} t_{5}^{4} - t_{1}^{5} t_{2}^{5} t_{3}^{3} t_{4} t_{5} \nonumber\\
&& {}- 2\, t_{1}^{4} t_{2}^{4} t_{3}^{3} t_{4}^{2} t_{5}^{2} - 2\, t_{1}^{3} t_{2}^{3} t_{3}^{3} t_{4}^{3} t_{5}^{3} - t_{1}^{2} t_{2}^{2} t_{3}^{3} t_{4}^{4} t_{5}^{4} - t_{1}^{2} t_{2}^{6} t_{3}^{6} t_{4}^{2} \nonumber\\
&& {}- t_{1} t_{2}^{5} t_{3}^{6} t_{4}^{3} t_{5} - t_{2}^{4} t_{3}^{6} t_{4}^{4} t_{5}^{2} - t_{1}^{4} t_{2}^{6} t_{3}^{5} t_{4}^{2} t_{5} - t_{1}^{3} t_{2}^{5} t_{3}^{5} t_{4}^{3} t_{5}^{2} - t_{1}^{2} t_{2}^{4} t_{3}^{5} t_{4}^{4} t_{5}^{3} \nonumber\\
&& {}+ \dots~,~
\eea
where $t_a$ marks the extremal perfect matching $p_a$ of \fref{f_geom_31} and can be expressed in terms of the global symmetry fugacities $f_{1}, f_{2}$ and $\bar{t}_{1}, \dots, \bar{t}_{5}$ as shown in \tref{t_sym_geom_31}.

The first positive terms of the plethystic logarithm correspond to the generators of the mesonic moduli space, $A_{1}, \dots, A_{9}$. In \tref{t_model_31_1}, \tref{t_model_31_2}, \tref{t_model_31_3}, \tref{t_model_31_4}, \tref{t_model_31_5}, \tref{t_model_31_6} and \tref{t_model_31_7}, the generators are given in terms of the chiral fields of each toric phase. The negative terms that come after the positive ones correspond to the first order relations amongst the generators, which take the following~form,
\beal{es31a004}
&& A_{2}^{2} = A_{1} A_{3} ~,~ \quad A_{3} A_{5} = A_{2} A_{6} ~,~ \quad A_{3} A_{4} = A_{2} A_{5} = A_{1} A_{6} ~,~ \nonumber\\
&& A_{2} A_{4} = A_{1} A_{5} ~,~ \quad A_{3} A_{7} = A_{2} A_{8} ~,~ \quad A_{2} A_{7} = A_{1} A_{8} ~,~ \quad A_{5}^{2} = A_{4} A_{6} ~,~ \nonumber\\
&& A_{6} A_{8} = A_{2} A_{3}^{2} ~,~ \quad A_{6} A_{7} = A_{5} A_{8} = A_{1} A_{3}^{2} ~,~ \quad A_{5} A_{7} = A_{4} A_{8} = A_{1} A_{2} A_{3} ~,~ \nonumber\\
&& A_{4} A_{7} = A_{1}^{2} A_{3} ~,~ \quad A_{8}^{2} = A_{3} A_{9} ~,~ \quad A_{7} A_{8} = A_{2} A_{9} ~,~ \quad A_{7}^{2} = A_{1} A_{9} ~,~ \nonumber\\
&& A_{6} A_{9} = A_{2} A_{3} A_{8} ~,~ \quad A_{5} A_{9} = A_{1} A_{3} A_{8} ~,~ \quad A_{4} A_{9} = A_{1} A_{2} A_{8}~.~
\eea

\input{model_31_1}
\input{model_31_2}
\input{model_31_3}
\input{model_31_4}
\input{model_31_5}
\input{model_31_6}
\input{model_31_7}

\subsection{Model 32: $H^{1,3,1,2}$}\label{s_m32}
\input{geom_32}
\input{model_32_1}
\input{model_32_2}
\input{model_32_3}
\input{model_32_4}
\input{model_32_5}
\input{model_32_6}
\input{model_32_7}
\input{model_32_8}
\input{model_32_9}
\input{model_32_10}

\subsection{Model 33: $J^{4,3,1,1}$}\label{s_m33}
\input{geom_33}
\input{model_33_1}
\input{model_33_2}
\input{model_33_3}
\input{model_33_4}
\input{model_33_5}
\input{model_33_6}
\input{model_33_7}
\input{model_33_8}
\input{model_33_9}

\subsection{Model 34: $G^{4,3}$}\label{s_m34}
\begin{figure}[H]
\centering
\begin{minipage}[c]{0.34\textwidth}
\resizebox{\linewidth}{!}{%
\begin{tikzpicture}[x=1bp,y=1bp]
  \node[anchor=south west,inner sep=0pt,outer sep=0pt] at (0,0) {\includegraphics{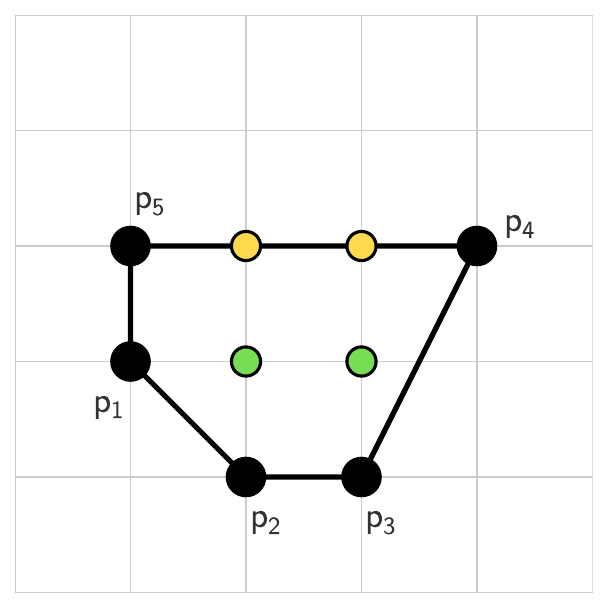}};
  \node[anchor=south west,inner sep=0pt,outer sep=0pt,overlay] at (128.22,120.03) {\includegraphics{figures/toric_label_s1a.pdf}};
  \node[anchor=south west,inner sep=0pt,outer sep=0pt,overlay] at (179.96,121.63) {\includegraphics{figures/toric_label_s2a.pdf}};
\end{tikzpicture}}
\end{minipage}%
\hspace{0.05\textwidth}%
{\small
\setlength{\tabcolsep}{4pt}
\begin{tabular}[c]{|c|c|c|}
\hline
\multirow{2}{*}{phase} & \multicolumn{2}{c|}{multiplicity} \\
\cline{2-3}
& \makebox[2.2em]{$s_{1a}$} & \makebox[2.2em]{$s_{2a}$} \\
\hline\hline
\hyperref[s_model_34_1]{1} & 14 & 9 \\
\hyperref[s_model_34_2]{2} & 14 & 9 \\
\hyperref[s_model_34_3]{3} & 13 & 10 \\
\hyperref[s_model_34_4]{4} & 16 & 10 \\
\hyperref[s_model_34_5]{5} & 19 & 10 \\
\hline
\end{tabular}
}
\caption{The toric diagram for Model~34 ($G^{4,3}$).}
\label{f_geom_34}
\end{figure}

\begin{figure}[H]
\centering
\definecolor{dualityred}{rgb}{1,0.149,0}
\hypersetup{linkcolor=black}
\begin{tikzpicture}[line width=1.55bp]
\useasboundingbox (-2.617,-5.234) rectangle (2.617,0.320);
\draw[dualityred, preaction={draw=white, line width=4.55bp}] (-1.371,-2.200) -- (-0.529,-2.200);
\draw[dualityred, preaction={draw=white, line width=4.55bp}] (-1.483,-2.683) -- (-0.417,-3.917);
\draw[dualityred, preaction={draw=white, line width=4.55bp}] (-1.767,-1.683) .. controls (-1.168,-0.372) and (1.168,-0.372) .. (1.767,-1.683);
\draw[dualityred, preaction={draw=white, line width=4.55bp}] (-1.483,-1.717) -- (-0.417,-0.483);
\draw[dualityred, preaction={draw=white, line width=4.55bp}] (0.000,-2.729) -- (0.000,-3.871);
\draw[dualityred, preaction={draw=white, line width=4.55bp}] (0.000,-1.671) -- (0.000,-0.529);
\draw[dualityred, preaction={draw=white, line width=4.55bp}] (0.417,-3.917) -- (1.483,-2.683);
\draw[dualityred, preaction={draw=white, line width=4.55bp}] (1.483,-1.717) -- (0.417,-0.483);
\draw[dualityred, preaction={draw=white, line width=4.55bp}] (-2.426,-2.325) .. controls (-2.925,-2.492) and (-2.192,-3.225) .. (-2.025,-2.726);
\draw[dualityred, preaction={draw=white, line width=4.55bp}] (0.500,-2.501) .. controls (0.918,-2.654) and (0.918,-1.746) .. (0.500,-1.899);
\draw[dualityred, preaction={draw=white, line width=4.55bp}] (-0.301,-4.900) .. controls (-0.454,-5.318) and (0.454,-5.318) .. (0.301,-4.900);
\fill[dualityred] (-1.563,-2.200) -- (-1.371,-2.272) -- (-1.371,-2.128) -- cycle;
\fill[dualityred] (-0.337,-2.200) -- (-0.529,-2.128) -- (-0.529,-2.272) -- cycle;
\fill[dualityred] (-1.609,-2.537) -- (-1.538,-2.730) -- (-1.429,-2.635) -- cycle;
\fill[dualityred] (-0.291,-4.063) -- (-0.362,-3.870) -- (-0.471,-3.965) -- cycle;
\fill[dualityred] (-1.835,-1.863) -- (-1.699,-1.709) -- (-1.834,-1.658) -- cycle;
\fill[dualityred] (1.835,-1.863) -- (1.834,-1.658) -- (1.699,-1.709) -- cycle;
\fill[dualityred] (-1.609,-1.863) -- (-1.429,-1.765) -- (-1.538,-1.670) -- cycle;
\fill[dualityred] (-0.291,-0.337) -- (-0.471,-0.435) -- (-0.362,-0.530) -- cycle;
\fill[dualityred] (0.000,-2.537) -- (-0.072,-2.729) -- (0.072,-2.729) -- cycle;
\fill[dualityred] (0.000,-4.063) -- (0.072,-3.871) -- (-0.072,-3.871) -- cycle;
\fill[dualityred] (0.000,-1.863) -- (0.072,-1.671) -- (-0.072,-1.671) -- cycle;
\fill[dualityred] (0.000,-0.337) -- (-0.072,-0.529) -- (0.072,-0.529) -- cycle;
\fill[dualityred] (0.291,-4.063) -- (0.471,-3.965) -- (0.362,-3.870) -- cycle;
\fill[dualityred] (1.609,-2.537) -- (1.429,-2.635) -- (1.538,-2.730) -- cycle;
\fill[dualityred] (1.609,-1.863) -- (1.538,-1.670) -- (1.429,-1.765) -- cycle;
\fill[dualityred] (0.291,-0.337) -- (0.362,-0.530) -- (0.471,-0.435) -- cycle;
\fill[dualityred] (-2.237,-2.293) -- (-2.438,-2.254) -- (-2.414,-2.396) -- cycle;
\fill[dualityred] (-1.993,-2.537) -- (-2.096,-2.714) -- (-1.954,-2.738) -- cycle;
\fill[dualityred] (0.337,-2.400) -- (0.462,-2.563) -- (0.538,-2.439) -- cycle;
\fill[dualityred] (0.337,-2.000) -- (0.538,-1.961) -- (0.462,-1.837) -- cycle;
\fill[dualityred] (-0.200,-4.737) -- (-0.363,-4.862) -- (-0.239,-4.938) -- cycle;
\fill[dualityred] (0.200,-4.737) -- (0.239,-4.938) -- (0.363,-4.862) -- cycle;
\node[draw, rectangle, fill=white, minimum size=0.6cm, inner sep=0pt, font=\sffamily\bfseries\footnotesize, line width=2.07bp] at (-1.900,-2.200) {\hyperref[s_model_34_1]{1}};
\node[draw, rectangle, fill=white, minimum size=0.6cm, inner sep=0pt, font=\sffamily\bfseries\footnotesize, line width=2.07bp] at (0.000,-2.200) {\hyperref[s_model_34_2]{2}};
\node[draw, rectangle, fill=white, minimum size=0.6cm, inner sep=0pt, font=\sffamily\bfseries\footnotesize, line width=2.07bp] at (0.000,-4.400) {\hyperref[s_model_34_3]{3}};
\node[draw, rectangle, fill=white, minimum size=0.6cm, inner sep=0pt, font=\sffamily\bfseries\footnotesize, line width=2.07bp] at (1.900,-2.200) {\hyperref[s_model_34_4]{4}};
\node[draw, rectangle, fill=white, minimum size=0.6cm, inner sep=0pt, font=\sffamily\bfseries\footnotesize, line width=2.07bp] at (0.000,0.000) {\hyperref[s_model_34_5]{5}};
\end{tikzpicture}
\caption{The Seiberg duality tree for Model~34 ($G^{4,3}$). The node labels indicate the phase numbers.}
\label{f_geom_34_tree}
\end{figure}

The toric diagram of the $G^{4,3}$ model is shown in \fref{f_geom_34}. The Seiberg duality tree of the $5$ toric phases associated with $G^{4,3}$ is shown in \fref{f_geom_34_tree}.
The global symmetry of the brane tilings associated with $G^{4,3}$ is given by,
\beal{es34a001}
U(1) \times U(1) \times U(1)_R~.~
\eea
The global symmetry charges on the extremal GLSM fields $p_a$ are summarized in \tref{t_sym_geom_34}.

\begin{table}[H]
\centering
{\small
\begin{tabular}{|c|c|c|c|c|l|}
\hline
  & $U(1)_{1}$ & $U(1)_{2}$ & $U(1)_R$ & $r_a$ & fugacities \\
\hline
  $p_{1}$ & $-1$ & $-1$ & $r_{1}$ & $r_{1} = 0.2422$ & $t_{1} = f_{1}^{-1} f_{2}^{-1} \bar{t}_{1}$ \\
  $p_{2}$ & $1$ & $0$ & $r_{2}$ & $r_{2} = 0.2891$ & $t_{2} = f_{1} \bar{t}_{2}$ \\
  $p_{3}$ & $0$ & $1$ & $r_{3}$ & $r_{3} = 0.3979$ & $t_{3} = f_{2} \bar{t}_{3}$ \\
  $p_{4}$ & $0$ & $0$ & $r_{4}$ & $r_{4} = 0.5869$ & $t_{4} = \bar{t}_{4}$ \\
  $p_{5}$ & $0$ & $0$ & $r_{5}$ & $r_{5} = 0.4839$ & $t_{5} = \bar{t}_{5}$ \\
\hline
\end{tabular}
}
\caption{Global symmetry charges on the extremal perfect matchings of Model~34 ($G^{4,3}$).}
\label{t_sym_geom_34}
\end{table}

All toric phases associated with $G^{4,3}$ share the same mesonic moduli space, whose unrefined Hilbert series takes the form,
\beal{es34a002}
\resizebox{\ifdim\width>\linewidth\linewidth\else\width\fi}{!}{$
g(t; \mathcal{M}^{mes}) \;=\; \frac{1 + 2 t^{5} + 2 t^{9} + 2 t^{10} - 2 t^{11} + 2 t^{14} - 2 t^{15} - 2 t^{16} - 2 t^{20} - t^{25}}{\left(1 - t^{5}\right) \left(1 - t^{6}\right)^{2} \left(1 - t^{13}\right)}~,~
$}
\eea
where $t$ is the fugacity counting degrees in extremal perfect matchings $p_a$, which corresponds to setting every fugacity $t_a$ in \tref{t_sym_geom_34} equal to $t$.
The plethystic logarithm of the refined Hilbert series takes the form,
\beal{es34a003}
PL\left[g(t_a;\mathcal{M}^{mes})\right] &=& t_{1} t_{4}^{2} t_{5}^{2} + t_{1} t_{2} t_{3} t_{4} t_{5} + t_{1} t_{2}^{2} t_{3}^{2} + t_{3} t_{4}^{4} t_{5} + t_{2} t_{3}^{2} t_{4}^{3} + t_{1}^{3} t_{2} t_{4} t_{5}^{4} \nonumber\\
&& {}+ t_{1}^{3} t_{2}^{2} t_{3} t_{5}^{3} + t_{1}^{5} t_{2}^{2} t_{5}^{6} \nonumber\\
&& {}- t_{1}^{2} t_{2}^{2} t_{3}^{2} t_{4}^{2} t_{5}^{2} - t_{1} t_{2}^{2} t_{3}^{3} t_{4}^{4} t_{5} - t_{1} t_{2} t_{3}^{2} t_{4}^{5} t_{5}^{2} - t_{1}^{4} t_{2}^{3} t_{3}^{2} t_{4} t_{5}^{4} \nonumber\\
&& {}- t_{1}^{4} t_{2}^{2} t_{3} t_{4}^{2} t_{5}^{5} - t_{1}^{3} t_{2}^{3} t_{3}^{3} t_{4}^{3} t_{5}^{3} - 2\, t_{1}^{3} t_{2}^{2} t_{3}^{2} t_{4}^{4} t_{5}^{4} - t_{1}^{3} t_{2} t_{3} t_{4}^{5} t_{5}^{5} \nonumber\\
&& {}- t_{1}^{6} t_{2}^{4} t_{3}^{2} t_{5}^{6} - t_{1}^{6} t_{2}^{3} t_{3} t_{4} t_{5}^{7} - t_{1}^{6} t_{2}^{2} t_{4}^{2} t_{5}^{8} - t_{1}^{5} t_{2}^{3} t_{3}^{2} t_{4}^{3} t_{5}^{6} - t_{1}^{5} t_{2}^{2} t_{3} t_{4}^{4} t_{5}^{7} \nonumber\\
&& {}+ \dots~,~
\eea
with the extremal perfect matchings $p_{1}, \dots, p_{5}$ of \fref{f_geom_34} carrying the fugacities $t_{1}, \dots, t_{5}$, which can be expressed in terms of the global symmetry fugacities $f_{1}, f_{2}$ and $\bar{t}_{1}, \dots, \bar{t}_{5}$ as shown in \tref{t_sym_geom_34}.

The positive terms with which the expansion begins correspond to the generators of the mesonic moduli space, namely $A_{1}, \dots, A_{8}$. In \tref{t_model_34_1}, \tref{t_model_34_2}, \tref{t_model_34_3}, \tref{t_model_34_4} and \tref{t_model_34_5}, the generators are given in terms of the chiral fields of each toric phase. The next negative terms correspond to the first order relations amongst the generators, which take the following~form,
\beal{es34a004}
&& A_{2}^{2} = A_{1} A_{3} ~,~ \quad A_{3} A_{4} = A_{2} A_{5} ~,~ \quad A_{2} A_{4} = A_{1} A_{5} ~,~ \quad A_{3} A_{6} = A_{2} A_{7} ~,~ \nonumber\\
&& A_{2} A_{6} = A_{1} A_{7} ~,~ \quad A_{5} A_{7} = A_{1} A_{2} A_{3} ~,~ \quad A_{5} A_{6} = A_{4} A_{7} = A_{1}^{2} A_{3} ~,~ \nonumber\\
&& A_{4} A_{6} = A_{1}^{2} A_{2} ~,~ \quad A_{7}^{2} = A_{3} A_{8} ~,~ \quad A_{6} A_{7} = A_{2} A_{8} ~,~ \quad A_{6}^{2} = A_{1} A_{8} ~,~ \nonumber\\
&& A_{5} A_{8} = A_{1} A_{2} A_{7} ~,~ \quad A_{4} A_{8} = A_{1}^{2} A_{7}~.~
\eea

\input{model_34_1}
\input{model_34_2}
\input{model_34_3}
\input{model_34_4}
\input{model_34_5}

\subsection{Model 35: $H^{2,3,1,1}$}\label{s_m35}
\input{geom_35}
\input{model_35_1}
\input{model_35_2}
\input{model_35_3}
\input{model_35_4}
\input{model_35_5}
\input{model_35_6}
\input{model_35_7}
\input{model_35_8}

\subsection{Model 36: $X^{2,1}/\mathbb{Z}_{2}\,(1,1,0,1,0,1,0,0)$}\label{s_m36}
\input{geom_36}
\input{model_36_1}
\input{model_36_2}
\input{model_36_3}
\input{model_36_4}
\input{model_36_5}
\input{model_36_6}
\input{model_36_7}
\input{model_36_8}
\input{model_36_9}
\input{model_36_10}
\input{model_36_11}
\input{model_36_12}
\input{model_36_13}
\input{model_36_14}
\input{model_36_15}
\input{model_36_16}
\input{model_36_17}
\input{model_36_18}
\input{model_36_19}

\subsection{Model 37: $G^{5,3}$}\label{s_m37}
\input{geom_37}
\input{model_37_1}
\input{model_37_2}
\input{model_37_3}
\input{model_37_4}
\input{model_37_5}
\input{model_37_6}
\input{model_37_7}
\input{model_37_8}
\input{model_37_9}
\input{model_37_10}
\input{model_37_11}
\input{model_37_12}

\subsection{Model 38: $J^{1,5,3,4}$}\label{s_m38}
\begin{figure}[H]
\centering
\begin{minipage}[c]{0.34\textwidth}
\resizebox{\linewidth}{!}{%
\begin{tikzpicture}[x=1bp,y=1bp]
  \node[anchor=south west,inner sep=0pt,outer sep=0pt] at (0,0) {\includegraphics{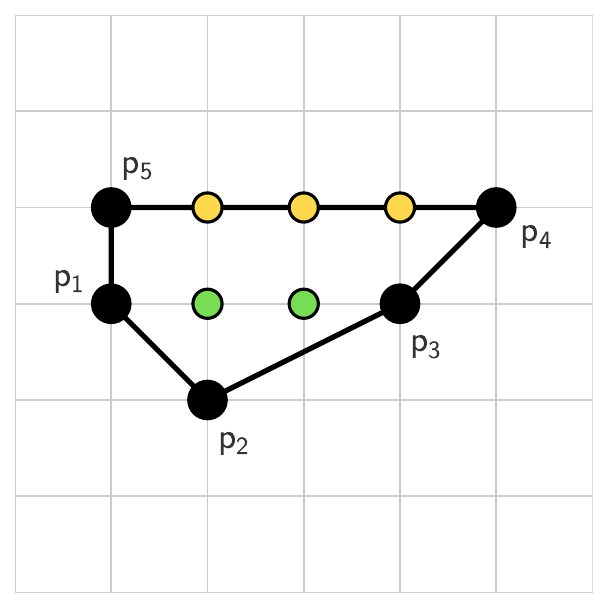}};
  \node[anchor=south west,inner sep=0pt,outer sep=0pt,overlay] at (109.74,127.14) {\includegraphics{figures/toric_label_s1a.pdf}};
  \node[anchor=south west,inner sep=0pt,outer sep=0pt,overlay] at (155.94,147.75) {\includegraphics{figures/toric_label_s2a.pdf}};
\end{tikzpicture}}
\end{minipage}%
\hspace{0.05\textwidth}%
{\small
\setlength{\tabcolsep}{4pt}
\begin{tabular}[c]{|c|c|c|}
\hline
\multirow{2}{*}{phase} & \multicolumn{2}{c|}{multiplicity} \\
\cline{2-3}
& \makebox[2.2em]{$s_{1a}$} & \makebox[2.2em]{$s_{2a}$} \\
\hline\hline
\hyperref[s_model_38_1]{1} & 15 & 15 \\
\hyperref[s_model_38_2]{2} & 18 & 16 \\
\hyperref[s_model_38_3]{3} & 17 & 17 \\
\hyperref[s_model_38_4]{4} & 19 & 19 \\
\hline
\end{tabular}
}
\caption{The toric diagram for Model~38 ($J^{1,5,3,4}$).}
\label{f_geom_38}
\end{figure}

\begin{figure}[H]
\centering
\definecolor{dualityred}{rgb}{1,0.149,0}
\hypersetup{linkcolor=black}
\begin{tikzpicture}[line width=1.55bp]
\useasboundingbox (-1.270,-4.720) rectangle (1.270,0.834);
\draw[dualityred, preaction={draw=white, line width=4.55bp}] (-0.728,-3.886) -- (-0.222,-2.714);
\draw[dualityred, preaction={draw=white, line width=4.55bp}] (0.222,-2.714) -- (0.728,-3.886);
\draw[dualityred, preaction={draw=white, line width=4.55bp}] (0.000,-1.671) -- (0.000,-0.529);
\draw[dualityred, preaction={draw=white, line width=4.55bp}] (0.301,0.500) .. controls (0.454,0.918) and (-0.454,0.918) .. (-0.301,0.500);
\fill[dualityred] (-0.805,-4.063) -- (-0.662,-3.915) -- (-0.795,-3.858) -- cycle;
\fill[dualityred] (-0.145,-2.537) -- (-0.288,-2.685) -- (-0.155,-2.742) -- cycle;
\fill[dualityred] (0.145,-2.537) -- (0.155,-2.742) -- (0.288,-2.685) -- cycle;
\fill[dualityred] (0.805,-4.063) -- (0.795,-3.858) -- (0.662,-3.915) -- cycle;
\fill[dualityred] (0.000,-1.863) -- (0.072,-1.671) -- (-0.072,-1.671) -- cycle;
\fill[dualityred] (0.000,-0.337) -- (-0.072,-0.529) -- (0.072,-0.529) -- cycle;
\fill[dualityred] (0.200,0.337) -- (0.363,0.462) -- (0.239,0.538) -- cycle;
\fill[dualityred] (-0.200,0.337) -- (-0.239,0.538) -- (-0.363,0.462) -- cycle;
\node[draw, rectangle, fill=white, minimum size=0.6cm, inner sep=0pt, font=\sffamily\bfseries\footnotesize, line width=2.07bp] at (-0.950,-4.400) {\hyperref[s_model_38_1]{1}};
\node[draw, rectangle, fill=white, minimum size=0.6cm, inner sep=0pt, font=\sffamily\bfseries\footnotesize, line width=2.07bp] at (0.000,-2.200) {\hyperref[s_model_38_2]{2}};
\node[draw, rectangle, fill=white, minimum size=0.6cm, inner sep=0pt, font=\sffamily\bfseries\footnotesize, line width=2.07bp] at (0.950,-4.400) {\hyperref[s_model_38_3]{3}};
\node[draw, rectangle, fill=white, minimum size=0.6cm, inner sep=0pt, font=\sffamily\bfseries\footnotesize, line width=2.07bp] at (0.000,0.000) {\hyperref[s_model_38_4]{4}};
\end{tikzpicture}
\caption{The Seiberg duality tree for Model~38 ($J^{1,5,3,4}$). The node labels indicate the phase numbers.}
\label{f_geom_38_tree}
\end{figure}

\fref{f_geom_38} illustrates the toric diagram for the $J^{1,5,3,4}$ model. The Seiberg duality tree of the $4$ toric phases associated with $J^{1,5,3,4}$ is shown in \fref{f_geom_38_tree}.
The global symmetry of the brane tilings associated with $J^{1,5,3,4}$ is given by,
\beal{es38a001}
U(1) \times U(1) \times U(1)_R~.~
\eea
The global symmetry charges on the extremal GLSM fields $p_a$ are summarized in \tref{t_sym_geom_38}.

\begin{table}[H]
\centering
{\small
\begin{tabular}{|c|c|c|c|c|l|}
\hline
  & $U(1)_{1}$ & $U(1)_{2}$ & $U(1)_R$ & $r_a$ & fugacities \\
\hline
  $p_{1}$ & $-1$ & $-1$ & $r_{1}$ & $r_{1} = 0.1993$ & $t_{1} = f_{1}^{-1} f_{2}^{-1} \bar{t}_{1}$ \\
  $p_{2}$ & $1$ & $0$ & $r_{2}$ & $r_{2} = 0.5077$ & $t_{2} = f_{1} \bar{t}_{2}$ \\
  $p_{3}$ & $0$ & $1$ & $r_{3}$ & $r_{3} = 0.1993$ & $t_{3} = f_{2} \bar{t}_{1}$ \\
  $p_{4}$ & $0$ & $0$ & $r_{4}$ & $r_{4} = 0.5468$ & $t_{4} = \bar{t}_{3}$ \\
  $p_{5}$ & $0$ & $0$ & $r_{5}$ & $r_{5} = 0.5468$ & $t_{5} = \bar{t}_{3}$ \\
\hline
\end{tabular}
}
\caption{Global symmetry charges on the extremal perfect matchings of Model~38 ($J^{1,5,3,4}$).}
\label{t_sym_geom_38}
\end{table}

All toric phases associated with $J^{1,5,3,4}$ share the same mesonic moduli space, whose unrefined Hilbert series takes the form,
\beal{es38a002}
\resizebox{\ifdim\width>\linewidth\linewidth\else\width\fi}{!}{$
g(t; \mathcal{M}^{mes}) \;=\; \frac{1 - t^{3} + t^{5} + t^{6} - t^{13} - t^{14} + t^{16} - t^{19}}{\left(1 - t^{3}\right) \left(1 - t^{4}\right) \left(1 - t^{8}\right) \left(1 - t^{9}\right)}~,~
$}
\eea
where $t$ is the fugacity counting degrees in extremal perfect matchings $p_a$, which corresponds to setting every fugacity $t_a$ in \tref{t_sym_geom_38} equal to $t$.
The plethystic logarithm of the refined Hilbert series takes the form,
\beal{es38a003}
PL\left[g(t_a;\mathcal{M}^{mes})\right] &=& t_{1} t_{2}^{2} t_{3} + t_{1} t_{2} t_{3} t_{4} t_{5} + t_{1} t_{3} t_{4}^{2} t_{5}^{2} + t_{2} t_{3}^{3} t_{4}^{4} + t_{1}^{3} t_{2} t_{5}^{4} + t_{3}^{3} t_{4}^{5} t_{5} \nonumber\\
&& {}+ t_{1}^{3} t_{4} t_{5}^{5} \nonumber\\
&& {}- t_{1}^{2} t_{2}^{2} t_{3}^{2} t_{4}^{2} t_{5}^{2} - t_{1}^{4} t_{2}^{2} t_{3} t_{4} t_{5}^{5} - t_{1} t_{2}^{2} t_{3}^{4} t_{4}^{5} t_{5} - t_{1}^{4} t_{2} t_{3} t_{4}^{2} t_{5}^{6} \nonumber\\
&& {}- t_{1} t_{2} t_{3}^{4} t_{4}^{6} t_{5}^{2} - t_{1}^{3} t_{2}^{2} t_{3}^{3} t_{4}^{4} t_{5}^{4} - 2\, t_{1}^{3} t_{2} t_{3}^{3} t_{4}^{5} t_{5}^{5} - t_{1}^{3} t_{3}^{3} t_{4}^{6} t_{5}^{6} \nonumber\\
&& {}+ \dots~.~
\eea
Here the fugacity $t_a$ of the extremal perfect matching $p_a$ at a corner of \fref{f_geom_38} can be expressed in terms of the global symmetry fugacities $f_{1}, f_{2}$ and $\bar{t}_{1}, \bar{t}_{2}, \bar{t}_{3}$, as fixed by the charges of $p_a$ in \tref{t_sym_geom_38}.

The generators of the mesonic moduli space, $A_{1}, \dots, A_{7}$, correspond to the first positive terms of the plethystic logarithm. For each toric phase, the generators are given in \tref{t_model_38_1}, \tref{t_model_38_2}, \tref{t_model_38_3} and \tref{t_model_38_4} in terms of the chiral fields of the corresponding brane tiling. The next negative terms correspond to the first order relations amongst the generators, which take the following~form,
\beal{es38a004}
&& A_{2}^{2} = A_{1} A_{3} ~,~ \quad A_{2} A_{5} = A_{1} A_{7} ~,~ \quad A_{2} A_{4} = A_{1} A_{6} ~,~ \quad A_{3} A_{5} = A_{2} A_{7} ~,~ \nonumber\\
&& A_{3} A_{4} = A_{2} A_{6} ~,~ \quad A_{4} A_{5} = A_{1} A_{3}^{2} ~,~ \quad A_{5} A_{6} = A_{4} A_{7} = A_{2} A_{3}^{2} ~,~ \nonumber\\
&& A_{6} A_{7} = A_{3}^{3}~.~
\eea

\input{model_38_1}
\input{model_38_2}
\input{model_38_3}
\input{model_38_4}

\subsection{Model 39: $H^{2,2,1,2}$}\label{s_m39}
\input{geom_39}
\input{model_39_1}
\input{model_39_2}
\input{model_39_3}
\input{model_39_4}
\input{model_39_5}
\input{model_39_6}
\input{model_39_7}
\input{model_39_8}
\input{model_39_9}
\input{model_39_10}
\input{model_39_11}
\input{model_39_12}
\input{model_39_13}
\input{model_39_14}
\input{model_39_15}
\input{model_39_16}
\input{model_39_17}
\input{model_39_18}
\input{model_39_19}
\input{model_39_20}
\input{model_39_21}
\input{model_39_22}
\input{model_39_23}
\input{model_39_24}
\input{model_39_25}

\subsection{Model 40: $G^{5,4}$}\label{s_m40}
\input{geom_40}
\input{model_40_1}
\input{model_40_2}
\input{model_40_3}
\input{model_40_4}
\input{model_40_5}
\input{model_40_6}
\input{model_40_7}

\section{Hexagons}\label{s_c6}

\subsection{Model 41: $Z^{3,1}$ ($H^{2,1,1,1}$)}\label{s_m41}
\input{geom_41}
\input{model_41_1}
\input{model_41_2}
\input{model_41_3}
\input{model_41_4}
\input{model_41_5}
\input{model_41_6}
\input{model_41_7}
\input{model_41_8}
\input{model_41_9}
\input{model_41_10}
\input{model_41_11}
\input{model_41_12}
\input{model_41_13}

\subsection{Model 42: $Z^{3,2}$}\label{s_m42}
\input{geom_42}
\input{model_42_1}
\input{model_42_2}
\input{model_42_3}
\input{model_42_4}
\input{model_42_5}
\input{model_42_6}
\input{model_42_7}
\input{model_42_8}
\input{model_42_9}
\input{model_42_10}
\input{model_42_11}
\input{model_42_12}
\input{model_42_13}

\subsection{Model 43: $H^{1,1,1,2}$}\label{s_m43}
\input{geom_43}
\input{model_43_1}
\input{model_43_2}
\input{model_43_3}
\input{model_43_4}
\input{model_43_5}
\input{model_43_6}
\input{model_43_7}
\input{model_43_8}
\input{model_43_9}
\input{model_43_10}
\input{model_43_11}
\input{model_43_12}
\input{model_43_13}
\input{model_43_14}
\input{model_43_15}
\input{model_43_16}
\input{model_43_17}
\input{model_43_18}
\input{model_43_19}
\input{model_43_20}
\input{model_43_21}
\input{model_43_22}
\input{model_43_23}
\input{model_43_24}
\input{model_43_25}
\input{model_43_26}
\input{model_43_27}
\input{model_43_28}
\input{model_43_29}
\input{model_43_30}
\input{model_43_31}
\input{model_43_32}
\input{model_43_33}
\input{model_43_34}
\input{model_43_35}
\input{model_43_36}
\input{model_43_37}
\input{model_43_38}
\input{model_43_39}
\input{model_43_40}
\input{model_43_41}
\input{model_43_42}

\subsection{Model 44: $N^{1,3,3,2}$}\label{s_m44}
\input{geom_44}
\input{model_44_1}
\input{model_44_2}
\input{model_44_3}
\input{model_44_4}
\input{model_44_5}
\input{model_44_6}
\input{model_44_7}
\input{model_44_8}
\input{model_44_9}
\input{model_44_10}
\input{model_44_11}
\input{model_44_12}
\input{model_44_13}
\input{model_44_14}
\input{model_44_15}
\input{model_44_16}
\input{model_44_17}
\input{model_44_18}
\input{model_44_19}
\input{model_44_20}
\input{model_44_21}
\input{model_44_22}
\input{model_44_23}
\input{model_44_24}
\input{model_44_25}
\input{model_44_26}
\input{model_44_27}
\input{model_44_28}
\input{model_44_29}
\input{model_44_30}
\input{model_44_31}
\input{model_44_32}
\input{model_44_33}

\subsection{Model 45: $N^{1,2,3,3}$}\label{s_m45}
\input{geom_45}
\input{model_45_1}
\input{model_45_2}
\input{model_45_3}
\input{model_45_4}
\input{model_45_5}
\input{model_45_6}
\input{model_45_7}
\input{model_45_8}
\input{model_45_9}
\input{model_45_10}
\input{model_45_11}
\input{model_45_12}
\input{model_45_13}
\input{model_45_14}
\input{model_45_15}
\input{model_45_16}
\input{model_45_17}
\input{model_45_18}
\input{model_45_19}
\input{model_45_20}
\input{model_45_21}
\input{model_45_22}
\input{model_45_23}

\section{Conclusions}\label{s_conclusions}
In this work, we have classified the brane tilings and the corresponding $4d$ $\mathcal{N}=1$ supersymmetric quiver gauge theories
associated with the toric Calabi-Yau $3$-folds given by convex lattice polygons with exactly two interior lattice points.
Up to $GL(2,\mathbb{Z})$ transformations and lattice translations, there are $45$ such polygons \cite{wei2012lattice},
consisting of $5$ triangles, $19$ quadrilaterals, $16$ pentagons and $5$ hexagons.
As for the $30$ brane tilings associated with the $16$ reflexive polygons \cite{Hanany:2012hi},
we obtained a brane tiling for each of the $45$ toric diagrams via Higgsing from brane tilings associated with abelian orbifolds of $\mathbb{C}^3$,
and then applied Seiberg duality on all quadrilateral faces until no new toric phase appeared.
Up to the equivalence of brane tilings defined in section \sref{s_conj}, we find $323$ inequivalent toric phases:
$5$ for the triangles, $57$ for the quadrilaterals, $137$ for the pentagons and $124$ for the hexagons.
Going from one to two interior points, the number of toric diagrams therefore increases from $16$ to $45$
and the number of toric phases from $30$ to $323$. Our classification completes the partial results in \cite{Bao:2020kji}.

The number of toric phases for each of the $45$ toric Calabi-Yau $3$-folds is summarized in \tref{t_conclusions}.
$10$ toric Calabi-Yau $3$-folds have a single toric phase, namely the $5$ triangles and the quadrilaterals of Models 6, 7, 10, 17 and 21.
The toric phases of the other $35$ toric Calabi-Yau 3-folds are related by Seiberg duality as shown by the Seiberg duality trees in sections \sref{s_c4} to \sref{s_c6}.
The largest number of toric phases is found for Model 43 ($H^{1,1,1,2}$) with $42$ toric phases,
followed by Model 44 ($N^{1,3,3,2}$) with $33$, Model 39 ($H^{2,2,1,2}$) with $25$ and Model 45 ($N^{1,2,3,3}$) with $23$ toric phases.
Our classification also contains toric phases with exactly the same quivers whose superpotentials are not related by \eref{es01a84}.
There are $8$ such pairs of toric phases, namely
Phases 1 and 2 of Model 36, Phases 11 and 12 of Model 39,
Phases 1 and 2, 6 and 7, 15 and 16, 17 and 18, and 22 and 23 of Model 43,
and Phases 6 and 7 of Model 44.

For every toric phase, we have presented the brane tiling with its quiver and superpotential,
the perfect matching matrix $P$ and the $F$- and $D$-term charge matrices, as well as the Hilbert series and the plethystic logarithm of the mesonic moduli space, whose generators we expressed as gauge invariant products of bifundamental chiral fields
with their mesonic flavor charges and superconformal $U(1)_R$ charges from $a$-maximization \cite{Butti:2005vn,Butti:2005ps}.
For $5$ of the $45$ toric Calabi-Yau $3$-folds, namely Models 1, 2, 7, 8 and 9,
the mesonic flavor symmetry is enhanced to $SU(2)\times U(1)$.
Moreover, in order to name all $45$ toric Calabi-Yau $3$-folds, we introduced the new families $G^{p,q}$, $J^{a,b,c,d}$ and $N^{a,b,c,d}$
in addition to the families and abelian orbifolds summarized in \tref{t_families}.

\begin{table}[ht!!]
\centering
\small
\setlength{\tabcolsep}{4pt}
\begin{tabular}{c@{\hspace{0.6cm}}c}
\begin{tabular}[t]{|c|c|c|c|}
\hline
\multirow{2}{*}{polygon}
& \multirow{2}{*}{model}
& \multicolumn{2}{c|}{toric phases} \\
\cline{3-4}
& & per model & total \\
\hline\hline
\multirow{5}{*}{triangles}
& \hyperref[s_m1]{1} & $1$
& \multirow{5}{*}{$5$} \\
\cline{2-3}
& \hyperref[s_m2]{2} & $1$ & \\
\cline{2-3}
& \hyperref[s_m3]{3} & $1$ & \\
\cline{2-3}
& \hyperref[s_m4]{4} & $1$ & \\
\cline{2-3}
& \hyperref[s_m5]{5} & $1$ & \\
\hline\hline
\multirow{19}{*}{
quadrilaterals}
& \hyperref[s_m6]{6} & $1$
& \multirow{19}{*}{$57$} \\
\cline{2-3}
& \hyperref[s_m7]{7} & $1$ & \\
\cline{2-3}
& \hyperref[s_m8]{8} & $2$ & \\
\cline{2-3}
& \hyperref[s_m9]{9} & $2$ & \\
\cline{2-3}
& \hyperref[s_m10]{10} & $1$ & \\
\cline{2-3}
& \hyperref[s_m11]{11} & $2$ & \\
\cline{2-3}
& \hyperref[s_m12]{12} & $3$ & \\
\cline{2-3}
& \hyperref[s_m13]{13} & $2$ & \\
\cline{2-3}
& \hyperref[s_m14]{14} & $4$ & \\
\cline{2-3}
& \hyperref[s_m15]{15} & $2$ & \\
\cline{2-3}
& \hyperref[s_m16]{16} & $3$ & \\
\cline{2-3}
& \hyperref[s_m17]{17} & $1$ & \\
\cline{2-3}
& \hyperref[s_m18]{18} & $6$ & \\
\cline{2-3}
& \hyperref[s_m19]{19} & $2$ & \\
\cline{2-3}
& \hyperref[s_m20]{20} & $3$ & \\
\cline{2-3}
& \hyperref[s_m21]{21} & $1$ & \\
\cline{2-3}
& \hyperref[s_m22]{22} & $8$ & \\
\cline{2-3}
& \hyperref[s_m23]{23} & $2$ & \\
\cline{2-3}
& \hyperref[s_m24]{24} & $11$ & \\
\hline
\end{tabular}
&
\begin{tabular}[t]{|c|c|c|c|}
\hline
\multirow{2}{*}{polygon}
& \multirow{2}{*}{model}
& \multicolumn{2}{c|}{toric phases} \\
\cline{3-4}
& & per model & total \\
\hline\hline
\multirow{16}{*}{
pentagons}
& \hyperref[s_m25]{25} & $5$
& \multirow{16}{*}{$137$} \\
\cline{2-3}
& \hyperref[s_m26]{26} & $3$ & \\
\cline{2-3}
& \hyperref[s_m27]{27} & $3$ & \\
\cline{2-3}
& \hyperref[s_m28]{28} & $4$ & \\
\cline{2-3}
& \hyperref[s_m29]{29} & $8$ & \\
\cline{2-3}
& \hyperref[s_m30]{30} & $8$ & \\
\cline{2-3}
& \hyperref[s_m31]{31} & $7$ & \\
\cline{2-3}
& \hyperref[s_m32]{32} & $10$ & \\
\cline{2-3}
& \hyperref[s_m33]{33} & $9$ & \\
\cline{2-3}
& \hyperref[s_m34]{34} & $5$ & \\
\cline{2-3}
& \hyperref[s_m35]{35} & $8$ & \\
\cline{2-3}
& \hyperref[s_m36]{36} & $19$ & \\
\cline{2-3}
& \hyperref[s_m37]{37} & $12$ & \\
\cline{2-3}
& \hyperref[s_m38]{38} & $4$ & \\
\cline{2-3}
& \hyperref[s_m39]{39} & $25$ & \\
\cline{2-3}
& \hyperref[s_m40]{40} & $7$ & \\
\hline\hline
\multirow{5}{*}{
hexagons}
& \hyperref[s_m41]{41} & $13$
& \multirow{5}{*}{$124$} \\
\cline{2-3}
& \hyperref[s_m42]{42} & $13$ & \\
\cline{2-3}
& \hyperref[s_m43]{43} & $42$ & \\
\cline{2-3}
& \hyperref[s_m44]{44} & $33$ & \\
\cline{2-3}
& \hyperref[s_m45]{45} & $23$ & \\
\hline
\end{tabular}
\end{tabular}
\caption{
The number of toric phases of each of the $45$ toric Calabi-Yau $3$-folds whose toric diagrams
are convex lattice polygons with two interior lattice points.
}
\label{t_conclusions}
\end{table}

The toric phases with isomorphic quivers in our classification provide new examples of quiver-invariant pairs of brane tilings,
whose $4d$ $\mathcal{N}=1$ theories have identical quivers and correspond to the same toric Calabi-Yau 3-fold 
but have different superpotentials \cite{Kho:2026geo,Lee:2026tju}.
Since quiver-invariant pairs of brane tilings can give rise to quiver-invariant triality in brane brick models under orbifold reduction \cite{Franco:2016fxm},
as observed recently in \cite{Ghim:2026fen},
it would also be interesting to search for further examples in the triality networks of the corresponding brane brick models
\cite{Bao:2024ygr,Ghim:2025zhs,Ghim:2024asj,Franco:2022gvl,Franco:2022isw,Franco:2023tyf,Carcamo:2025shw,Kwon:2026stu}.

Another direction is to use our classification and to investigate connections with other developments in the study of brane tilings,
including specular duality \cite{Hanany:2012vc} and brane tilings on higher genus Riemann surfaces \cite{Cremonesi:2013aba},
birational transformations \cite{Cremonesi:2023psg,Kho:2025fmp,Kho:2025jxk}
and brane tilings associated with generalized toric polygons (GTPs) \cite{Bourget:2023wlb,CarrenoBolla:2024fxy,Franco:2023flw,Arias-Tamargo:2024fjt,Kho:2026zwc}.
Further directions include exploring connections with dimer integrable systems \cite{Goncharov:2011hp,Franco:2011sz,Eager:2011dp,Kho:2025jxk,Kho:2025fmp,Kho:2026zwc}.
Since the number of interior points of a toric diagram equals the genus of the spectral curve and the number of independent Hamiltonians of the associated cluster integrable system \cite{Goncharov:2011hp},
the brane tilings classified in this work realize the dimer integrable systems with two independent Hamiltonians, with Seiberg duality acting as cluster mutations.
The toric Calabi-Yau 3-folds considered in this work also give rise to rank-two $5d$ $\mathcal{N}=1$ theories
that are realized by $(p,q)$ 5-brane webs of genus two \cite{Aharony:1997bh, Jefferson:2018irk, Hayashi:2018lyv, Saxena:2020ltf}.
The classification of brane tilings presented in this work may therefore also serve as a starting point for a systematic study of the BPS quivers and BPS spectra of rank-two
$5d$ $\mathcal{N}=1$ superconformal field theories \cite{Closset:2019juk}.
Other directions are bipartite field theories (BFTs) \cite{Franco:2012mm,Franco:2012wv,Kho:2024wcw}, and applications of machine learning to brane tilings and Seiberg duality \cite{Krefl:2017yox,Bao:2020nbi,Choi:2023rqg,Seong:2023njx,Seong:2024wkt,He:2024apw, Capuozzo:2024vdw, Armstrong-Williams:2024nzy, Heckman:2026xsi, Lee:2026avt},
for which the $323$ toric phases in our classification provide a natural dataset.
We leave these investigations to future work.

\section*{Acknowledgements}

R.-K. S. is supported by an Outstanding Young Scientist Grant (RS-2025-00516583) of the National Research
Foundation of Korea (NRF). He is also partly supported by the BK21 Program (“Next Generation Education Program for Mathematical Sciences”, 4299990414089) funded by the Ministry of Education in Korea and the National Research Foundation of Korea (NRF).
\textit{The author used AI tools (OpenAI ChatGPT and Anthropic Claude) to assist with checking computations and editing the manuscript. 
He takes full responsibility for the content of this paper.}
\\

\bibliographystyle{JHEP}
\bibliography{mybib}

\end{document}